\pdfoutput=1
\documentclass[preprint]{aastex} 

\newif\ifimage
\imagetrue

\newif\ifthesis
\thesisfalse

\usepackage{graphicx}
\usepackage{natbib}
\usepackage{afterpage}
\usepackage{lscape}
\usepackage{placeins} 
\usepackage{dpfloat}

\defcitealias{Ianj2012}{I2012}
\defcitealias{StilpGlobal}{Paper~I}
\defcitealias{StilpResolved}{Paper~II}
\defcitealias{StilpTimescales}{Paper~III}

\newcounter{subfig}

\newcommand{\um}{$\mu$m}
\newcommand{\hi}{\mbox{\rm H\,{\sc i}}}

\newcommand{\m}[1]{\ensuremath{M_\mathrm{#1}}}
\newcommand{\msun}{\ensuremath{\mathrm{M}_{\odot}}}

\newcommand{ \kms}{km~s$^{-1}$}

\newcommand{\halpha}{\mbox{\rm H$\alpha$}}
\newcommand{\degrees}{$^{\circ}$}

\newcommand{\mhi}{\ensuremath{M_{\mathrm{HI}}}}
\newcommand{\sfrsd}{\ensuremath{\Sigma_{\mathrm{SFR}}}}

\newcommand{\hisd}{\ensuremath{\Sigma_\mathrm{HI}}}

\newcommand{\sd}[1]{\ensuremath{\Sigma_{[1]}}}

\newcommand{\per}[2][1]{#2$^{-#1}$}
\newcommand{\sumlim}[2][]{\displaystyle\sum\limits_{#2}^{#1}}

\newcommand{\citeeg}[1]{\citep[e.g.,][]{#1}}

\newcommand{\vp}{\ensuremath{v_{\mathrm{peak}}}}

\newcommand{\scentral }{\ensuremath{\sigma_{\mathrm{central}}}}

\newcommand{\swing}{\ensuremath{ \sigma_{\mathrm{wings}}}}
\newcommand{\swingsq}{\ensuremath{ \sigma^2_{\mathrm{wings}}}}
\newcommand{\fw}{\ensuremath{f_{\mathrm{wings}}}}

\newcommand{\aw}{\ensuremath{a}}

\newcommand{\afull}{\ensuremath{a_\mathrm{full}}}

\newcommand{\baryonsd}{\ensuremath{\Sigma_\mathrm{baryon}}}

\newcommand{\hii}{\mbox{\rm H\,{\sc ii}}}

\newcommand{\figurepath}{./figures}
\newcommand{\tablepath}{./tables}

\shorttitle{Drivers of \hi{} Turbulence in Dwarf Galaxies}

\begin{document}

\title{Drivers of \hi{} Turbulence in Dwarf Galaxies}

\author{
Adrienne M. Stilp\altaffilmark{1},
Julianne J. Dalcanton\altaffilmark{1},
Evan Skillman\altaffilmark{2},
Steven R. Warren\altaffilmark{3},
J\"{u}rgen Ott\altaffilmark{4},
B\"{a}rbel Koribalski\altaffilmark{5}
}

\altaffiltext{1}{Department of Astronomy, University of Washington, Box 351580, Seattle, WA 98195,
USA}

\altaffiltext{2}{Minnesota Institute for Astrophysics, University of Minnesota, 116 Church St. SE,
Minneapolis, MN 55455, USA}

\altaffiltext{3}{Department of Astronomy, University of Maryland, CSS Bldg., Rm. 1024, Stadium Dr., College Park, MD 20742-2421}

\altaffiltext{4}{National Radio Astronomy Observatory, P.O. Box O, 1003
  Lopezville Road, Socorro, NM 87801, USA}

\altaffiltext{5}{Australia Telescope National Facility, CSIRO Astronomy and Space Science, PO Box 76,
Epping NSW 1710, Australia}

\ifthesis
\chapter{Drivers of \hi{} Turbulence in Dwarf Galaxies}
\label{resolved}
\fi

\ifthesis
\renewcommand{\figurepath}{./chapters/radial/figures}
\renewcommand{\tablepath}{./chapters/radial/tables}
\newcommand{\chapterpath}{./chapters/radial}
\else
\renewcommand{\figurepath}{.}
\renewcommand{\tablepath}{.}
\newcommand{\chapterpath}{.}
\fi

\ifthesis
\else
\begin{abstract}
\fi

Neutral hydrogen (\hi{}) velocity dispersions are believed to be set by turbulence in the interstellar medium (ISM).
Although turbulence is widely believed to be driven by star formation, recent studies have shown that this driving mechanism may not be dominant in regions of low star formation surface density (\sfrsd{}), such as found in dwarf galaxies or the outer regions of spirals.
We have generated average \hi{} line profiles in a number of nearby dwarfs and low-mass spirals by co-adding \hi{} spectra in subregions with either a common radius or \sfrsd{}.
We find that the individual spatially-resolved ``superprofiles'' are composed of a central narrow peak ($\sim 5 - 15$ \kms{}) with higher velocity wings to either side, similar to their global counterparts as calculated for the galaxy as a whole.
Under the assumption that the central peak reflects the \hi{} turbulent velocity dispersion, we compare measures of \hi{} kinematics determined from the superprofiles to local ISM properties, including surface mass densities and measures of star formation.
The shape of the wings of the superprofiles do not show any correlation with local ISM properties, which indicates that they may be an intrinsic feature of \hi{} line-of-sight spectra.
On the other hand, the \hi{} velocity dispersion is correlated most strongly with baryonic and \hi{} surface mass density, which points at a gravitational origin for turbulence, but it is unclear which, if any, gravitational instabilities are able to operate efficiently in these systems.
Star formation energy is typically produced at a level sufficient to drive \hi{} turbulent motions at realistic coupling efficiencies in regimes where $\sfrsd{} \gtrsim 10^{-4}$ \msun{} \per{yr} \per[2]{kpc}, as is typically found in inner spiral disks.
At low star formation intensities, on the other hand, star formation cannot supply enough energy to drive the observed turbulence, nor does it uniquely determine the turbulent velocity dispersion.
Nevertheless, even at low intensity, star formation does appear to provide a lower threshold for \hi{} velocity dispersions.
We find a pronounced decrease in coupling efficiency with increasing \sfrsd{}, which would be consistent with a picture where star formation couples to the ISM with constant efficiency, but that less of that energy is found in the neutral phase at higher \sfrsd{}.
We have examined a number of potential drivers of \hi{} turbulence, including star formation, gravitational instabilities, the magneto-rotational instability, and accretion-driven turbulence, and found that, individually, \emph{none} of these drivers is capable of driving the observed levels of turbulence in the low \sfrsd{} regime.
We discuss possible solutions to this conundrum.

\ifthesis
\else
\end{abstract}

\fi

\section{Introduction}
\label{resolved::sec:intro}
The local velocity dispersion of neutral hydrogen (\hi{}) provides a good tracer of the small-scale kinematics of the interstellar medium (ISM) in disk galaxies.
The velocity dispersions of \hi{} clouds tend to range between $5 - 15$~\kms{} across a wide range of galaxy and ISM properties \citeeg{Dickey1990, vanZee1999, Tamburro2009}.
These velocity dispersions are thought to be turbulent in nature \citeeg{MacLow2004} because they are much greater than the velocity dispersions expected for the stable thermal temperatures in the ISM \citep[1 \kms{} and 7 \kms{} for the cold and warm neutral phases; e.g., ][]{Wolfire1995}.
\hi{} velocity dispersions also tend to either remain constant or decrease with increasing galaxy radius in spirals and large dwarfs \citep{Dickey1990, Boulanger1992, Petric2007, Tamburro2009}.

The origin of turbulent \hi{} velocity dispersions remains uncertain, but many studies attribute \hi{}turbulence to star formation and resulting supernova explosions \cite[SNe; e.g.,][]{MacLow2004, Tamburro2009, Joung2009}.
This relationship seems to hold in the central regions of spiral galaxies, but breaks down at large radii where star formation intensity drops dramatically while \hi{} velocity dispersions remain relatively constant \citeeg{Boulanger1992, vanZee1999, Tamburro2009}.
For some massive spiral disks, \hi{} velocity dispersions at large radii have been tentatively attributed to turbulence induced by non-stellar sources, such as the magneto-rotational instability \citep[MRI; e.g., ][]{Sellwood1999, Zhang2012}.
This outer disk regime exhibits star formation rate intensities and \hi{} velocity dispersions similar to those found in dwarf galaxies, which tend to have solid-body rotation curves \citeeg{Oh2011} and therefore lack the shear required for the MRI to function efficiently.
Additionally, \hi{} velocity dispersions in the outskirts of some dwarfs are not necessarily correlated with optical features or star forming regions \citeeg{Hunter1999, Hunter2001}.
Studies of the relationship between star formation and \hi{} velocity dispersions in dwarf galaxies may therefore help address the question of what provides the energy to drive \hi{} turbulence, particularly in regions where other proposed turbulence drivers are inefficient.

\ifthesis
\citetalias{StilpGlobal}
\else
\citet[][hereafter Paper~I]{StilpGlobal}
\fi
presented a method to characterize the average \hi{} kinematics in dwarf galaxies by co-adding individual line-of-sight profiles after removal of the rotational velocity for a single galaxy, as also used by \citet{Ianj2012}.
These ``superprofiles'' were composed of a central peak with higher-velocity wings to either side.
We interpreted the central peak of the superprofile as representative of the average \hi{} turbulent kinematics, with the higher velocity wings representing anomalous motions such as expanding \hi{} holes or other bulk flows.
Our conclusions in \citetalias{StilpGlobal} were limited by the fact that the superprofiles were generated on global scales, whereas \hi{} velocity dispersions are known to vary across the disk.

In this paper, we extend the technique presented in \citetalias{StilpGlobal} to analyze subregions of these same galaxies.
By extending our analysis to carefully-chosen subregions, we can essentially increase the dynamic range of various quantities, such as \sfrsd{}, \hisd{}, and \sfrsd{} / \hisd{}, which were forced to galaxy-wide averages in our earlier analysis.
In contrast, many of the proposed drivers of turbulence are local phenomena.
This approach therefore provides a more direct assessment of which parameters influence \hi{} kinematics.

We first compute superprofiles in radial annuli to facilitate comparison with \citet{Tamburro2009}, who found that star formation does not provide enough energy to drive \hi{} velocity dispersions outside the optical radius $r_{25}$ of spiral galaxies.
However, radius is not necessarily a good proxy for local ISM properties like \sfrsd{} and \hisd{} in the low-mass dwarfs in our sample.
We therefore also generate superprofiles in regions of constant \sfrsd{} to maximize the sensitivity of the effects that star formation may have on \hi{} kinematics.
Our study is complementary to that of \citet{Tamburro2009}, as we focus on lower-mass galaxies and isolate regions with similar star formation surface density.

The layout of this chapter is as follows.
In \S~\ref{resolved::sec:data}, we discuss the data used to generate spatially-resolved superprofiles and other galaxy properties.
In \S~\ref{resolved::sec:analysis}, we give a brief overview of the method used to generate the superprofiles, present the spatially-resolved superprofiles, and address their robustness.
In \S~\ref{resolved::sec:correlations}, we compare the superprofile parameters to galaxy physical properties.
In \S~\ref{resolved::sec:discussion}, we then discuss the relevant correlations and compare star formation energy to \hi{} energy.
Finally, we summarize the conclusions in \S~\ref{resolved::sec:conclusions}.
Figures of the spatially-resolved superprofiles for the entire sample are presented 
\ifthesis
at the end of the chapter in \S
\else
in Appendix
\fi
\ref{resolved::sec:sp-figures} as a general reference. 

\section{Data}
\label{resolved::sec:data}

We use \hi{} data from the Very Large Array ACS Nearby Galaxy Survey Treasury Program \citep[``VLA-ANGST'';][]{Ott2012} and The \hi{} Nearby Galaxy Survey \citep[``THINGS'';][]{Walter2008}.
Following \citetalias{StilpGlobal}, we convolve these data to a common physical resolution of 200~pc to ensure that we are sampling ISM properties on the same physical scale for our entire sample.
All spatially-resolved ancillary data have also been convolved to this resolution to ensure a robust comparison between \hi{} kinematics and other ISM properties.
We assume distances as listed in Table 
\ifthesis
~\ref{tab:global--sample}
\else
1
\fi
 of \citetalias{StilpGlobal}, as compiled from \citet{Ott2012}, \citet{Walter2008}, and \citet{Dalcanton2009}.

\subsection{Initial Sample Selection}

We select our analysis sample for this paper from a subset of galaxies in \citetalias{StilpGlobal}.
Briefly, the selection criteria used in \citetalias{StilpGlobal} are:
\begin{enumerate}
\item Instrumental angular resolution smaller than 200~pc, to avoid artificially broadening \hi{} line-of-sight spectra at coarser resolution.
\item Velocity resolution $\Delta v \leq 2.6$ \kms{}, to resolve the width of the \hi{} line-of-sight spectra.
\item Inclination $i < 70$\degrees{}, to avoid broadening \hi{} line-of-sight spectra with rotation.
\item No noticeable contamination from the Milky Way or a companion, to ensure that detected \hi{} belongs to each galaxy.
\item More than 10 independent beams across the galaxy above a signal to noise threshold of $S/N > 5$, to allow for accurate determination of the peak of \hi{} line-of-sight spectra.
\item Available ancillary far ultraviolet (FUV, GALEX) data, to uniformly measure SFRs.
\end{enumerate}

In this paper, we apply additional selection criteria to the spatially-resolved superprofiles to ensure that they are robust.
We have empirically found that superprofiles with fewer than 5 contributing independent beams are too noisy to accurately determine superprofile parameters, which eliminates four low-mass galaxies (DDO~125, M81~DwB, NGC~4163, and GR~8) from the \citetalias{StilpGlobal} sample.
Second, the superprofiles in the other disks of some of the larger galaxies exhibit ``clean bowls'' that hinder accurate parameterization.
These clean bowls are due to missing short-spacings at the VLA, and can present as negative flux on either side of the central peak in the superprofiles generated at large radii for some galaxies.
We eliminate these superprofiles from our analysis, and note that they usually occur past $2 r_{25}$.

General properties of the final sample are given in Table~\ref{tab:resolved--sample-full}.
Galaxies are listed in order of decreasing total baryonic mass (\m{baryon,tot}).
We list:
(1) the galaxy name;
(2) the \hi{} survey from which data were taken;
(3-4) the position in J2000 coordinates;
(5) distance in Mpc;
(6) inclination in degrees;
(7) total baryonic mass, \m{baryon,tot};
(8) total \hi{} mass, \m{HI,tot};
(9) SFR as determined from FUV+24\um{} emission;
(10) the peak \sfrsd{} included in this analysis for each galaxy;
(11) the optical radius at a $B$-band surface brightness of 25 mag \per{arcsec} ($r_{25}$);
and (12) de Vaucouleurs T-type.
All references are given in \citetalias{StilpGlobal}, with the exception of the inclination for Sextans~B.
The $i = 52$\degrees{} value given in \citetalias{StilpGlobal} is a poor match to the \hi{} morphology; we adopt $i = 30$\degrees{}, which is a much better match to the properties of the \hi{} disk.

\ifthesis
\afterpage{\input{\tablepath/table-sample-full}}
\else
\afterpage{

\begin{deluxetable}{llcccccccccc}
\tabletypesize{\scriptsize}
\rotate
\tablewidth{0pt}
\tablecaption{Resolved \hi{} kinematics sample. \label{tab:resolved--sample-full}}
\tablehead{
  \colhead{1} &
  \colhead{2} &
  \colhead{3} &
  \colhead{4} &
  \colhead{5} &
  \colhead{6} &
  \colhead{7} &
  \colhead{8} &
  \colhead{9} &
  \colhead{10} &
  \colhead{11} &
  \colhead{12} \\

  \colhead{Galaxy} &
  \colhead{Survey} & 
  \colhead{RA} & 
  \colhead{Dec} & 
  \colhead{Distance} & 
  \colhead{$i$} & 
  \colhead{\m{baryon,tot}}  & 
  \colhead{\m{HI,tot}} & 
  \colhead{SFR} & 
  \colhead{Peak $\Sigma_\mathrm{SFR}$} & 
  \colhead{$r_{25}$} & 
  \colhead{Type} \\
  
  \colhead{} & 
  \colhead{} & 
  \colhead{(hh:mm:ss)} & 
  \colhead{(dd:mm:ss)} & 
  \colhead{(Mpc)} & 
  \colhead{(\degrees{})} & 
  \colhead{(log \msun{})} &
  \colhead{(log \msun{})} & 
  \colhead{($10^{-3}$ M$_\mathrm{\odot}$ \per{yr})} &
  \colhead{($10^{-3}$ M$_\mathrm{\odot}$ \per{yr} \per[2]{kpc})} &
  \colhead{(kpc)} &  
  \colhead{} \\

}

\startdata
NGC 7793  &  THINGS  &  23:57:49.7  &  -32:35:28  &  3.90  &  50  &  9.7  &  8.9  &  258.88  &  52.76  &  5.9  &   7  \\ 
IC 2574  &  THINGS  &  10:28:27.7  &  +68:24:59  &  3.79  &  55  &  9.4  &  9.1  &  71.72  &  53.32  &  7.1  &   9  \\ 
NGC 4214  &  THINGS  &  12:15:39.2  &  +36:19:37  &  3.04  &  44  &  9.2  &  8.6  &  128.09  &  323.92  &  3.0  &  10  \\ 
Ho II  &  THINGS  &  08:19:05.0  &  +70:43:12  &  3.38  &  49  &  9.0  &  8.8  &  45.40  &  39.13  &  3.2  &  10  \\ 
NGC 2366  &  THINGS  &  07:28:53.4  &  +69:12:51  &  3.21  &  63  &  9.0  &  8.8  &  53.98  &  274.74  &  2.0  &  10  \\ 
DDO 154  &  THINGS  &  12:54:05.9  &  +27:09:10  &  4.30  &  66  &  8.7  &  8.6  &  5.79  &  4.90  &  1.2  &  10  \\ 
Ho I  &  THINGS  &  09:40:32.3  &  +71:10:56  &  3.90  &  13  &  8.4  &  8.2  &  6.04  &  5.11  &  1.9  &  10  \\ 
NGC 4190  &  VLA-ANGST  &  12:13:44.6  &  +36:38:00  &  3.50  &  41  &  8.1  &  7.7  &  5.19  &  15.96  &  0.9  &  10  \\ 
NGC 3741  &  VLA-ANGST  &  11:36:06.4  &  +45:17:07  &  3.24  &  64  &  8.1  &  7.9  &  4.53  &  19.56  &  0.9  &  10  \\ 
Sextans A  &  VLA-ANGST  &  10:11:00.8  &  -04:41:34  &  1.38  &  36  &  8.0  &  7.8  &  5.47  &  11.67  &  1.1  &  10  \\ 
DDO 53  &  THINGS  &  08:34:07.2  &  +66:10:54  &  3.61  &  27  &  8.0  &  7.8  &  2.97  &  16.87  &  0.4  &  10  \\ 
DDO 190  &  VLA-ANGST  &  14:24:43.5  &  +44:31:33  &  2.79  &  30  &  8.0  &  7.6  &  3.30  &  6.45  &  0.7  &  10  \\ 
Sextans B  &  VLA-ANGST  &  10:00:00.1  &  +05:19:56  &  1.39  &  30  &  7.9  &  7.6  &  1.99  &  5.74  &  1.0  &  10  \\ 
DDO 99  &  VLA-ANGST  &  11:50:53.0  &  +38:52:50  &  2.59  &  60  &  7.9  &  7.7  &  2.29  &  6.30  &  1.5  &  10  \\ 
UGCA 292  &  VLA-ANGST  &  12:38:40.0  &  +32:46:00  &  3.62  &  16  &  7.8  &  7.6  &  1.10  &  4.84  &  0.5  &  10  \\ 
UGC 4483  &  VLA-ANGST  &  08:37:03.0  &  +69:46:31  &  3.41  &  42  &  7.7  &  7.5  &  1.99  &  14.13  &  0.6  &  10  \\ 
DDO 181  &  VLA-ANGST  &  13:39:53.8  &  +40:44:21  &  3.14  &  50  &  7.7  &  7.4  &  2.71  &  8.77  &  1.1  &  10  \\ 
UGC 8833  &  VLA-ANGST  &  13:54:48.7  &  +35:50:15  &  3.08  &  33  &  7.4  &  7.1  &  0.62  &  3.32  &  0.4  &  10  \\ 
DDO 187  &  VLA-ANGST  &  14:15:56.5  &  +23:03:19  &  2.21  &  55  &  7.3  &  7.1  &  0.44  &  1.71  &  0.5  &  10  \\ 
\enddata

\tablecomments{The sample. Galaxies are listed in order of decreasing \m{baryon,tot}. All references are as given in \citet{StilpGlobal}. (1) Galaxy name. (2) \hi{} survey. (3-4) Position in J2000 coordinates. (5) Distance in Mpc. (6) Inclination. (7) \m{baryon, tot} in log \msun{}. (8) \m{HI, tot} in log \msun{}. (9) SFR. (10) Peak $\Sigma_\mathrm{SFR}$ included in this analysis. (10) $B$-band $r_{25}. $(11) de Vaucouleurs T-type.}

\end{deluxetable}
}
\fi

\subsection{Converting Ancillary Data to Physical Properties}

Detailed information about the data used and methodology for deriving galaxy physical properties are given in \citetalias{StilpGlobal}.
For this study in particular, we focus on the SFR surface density (\sfrsd{}); the star formation rate per available \hi{} mass (\sfrsd{} / \hisd{}); and the \hi{} and baryonic surface densities (\hisd{}, \baryonsd{}).
Briefly, we calculate \sfrsd{} using FUV and 24\um{} data from the Local Volume Legacy Survey \citep[``LVL'';][]{Dale2009} and the methodology outlined in \citet{Leroy2008} and \citetalias{StilpGlobal}.
We derive stellar surface mass density from LVL 3.6\um{} data and apply the conversion factor given in \citet{Leroy2008} and \citetalias{StilpGlobal}.
Finally, we derive baryonic surface mass density by combining the \hi{} gas mass, including a factor of 1.36 correction for helium, with the stellar mass.
All surface densities are inclination-corrected.

For each superprofile subregion, we calculate \hisd{}, \baryonsd{}, and $\sfrsd{}$ by taking the total \mhi{}, \m{baryon}, and SFR in that subregion divided by its inclination-corrected area.
We also calculate $\sfrsd{} / \hisd{}$ by measuring the total SFR in that subregion divided by the total \hi{} mass in that subregion.
In the outer regions of Sextans~B, the low levels of star formation plus noise conspire to produce negative total SFRs in three resolved superprofiles.
In these cases, we artificially set $\sfrsd{} = 10^{-6}$ \msun{} \per{yr} \per[2]{kpc} and recalculate other SF-related quantities using that value.

\begin{figure}
\centering
\includegraphics[width=3.5in]{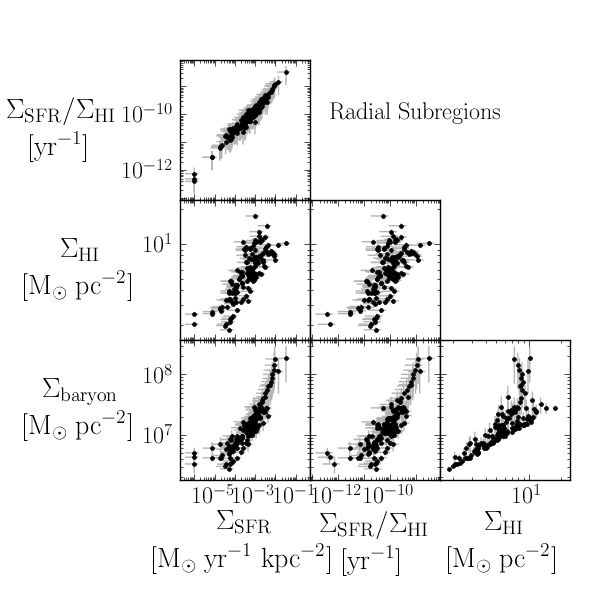}\\
\includegraphics[width=3.5in]{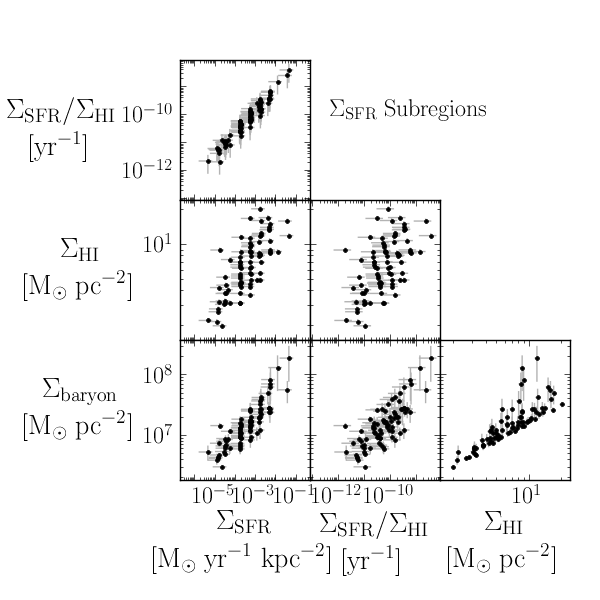}
\caption[Correlations between Physical ISM Properties]{ Correlations between measured ISM properties (\sfrsd{}, $\sfrsd{} / \hisd{}$, \hisd{}, \baryonsd{}) in subregions with constant radius (left) or constant star formation intensity (right). All axes are shown with log scaling, and the scaling of each individual panel is the same in the top and bottom panels.
The resolved superprofiles are described further in \S~\ref{resolved::sec:analysis--spatially-resolved}.
The quantized lines in the left panel are an artifact of the \sfrsd{} binning.
\label{resolved::fig:phys-prop-corr}
}
\end{figure}

We note that local ISM properties tend to be correlated; for example, regions with higher \sfrsd{} tend to have higher \hisd{}.
In Figure~\ref{resolved::fig:phys-prop-corr}, we show the correlations between \sfrsd{}, $\sfrsd{} / \hisd{}$, \hisd{}, and \baryonsd{} as measured in subregions of constant radius (left) or constant star formation intensity (right).
All of the physical properties under consideration are correlated with each other, so a correlation with one may be causally due to another.

\section{Analysis}
\label{resolved::sec:analysis}

To estimate \hi{} kinematics on spatially-resolved scales, we generate superprofiles in multiple subregions for each sample galaxy.
We first give a brief overview of the methodology for generating and parameterizing the superprofiles in \S~\ref{resolved::sec:analysis--generation} and \S~\ref{resolved::sec:analysis--parameterization}.
We then discuss in detail how we generate superprofiles in spatially-resolved subregions for the sample in \S~\ref{resolved::sec:analysis--spatially-resolved}.

\subsection{Overview of the Superprofile Generation Procedure}
\label{resolved::sec:analysis--generation}

A full discussion of the method used to derive superprofiles is given in \citetalias{StilpGlobal}, but we review the process here for clarity.
We first measure the line-of-sight velocity of the peak (\vp{}) of each line-of-sight spectrum, by fitting a Gauss-Hermite polynomials using the standard data cube.
We include only the line-of-sight spectra that have $S/N > 5$, where $S/N$ is defined as the ratio between the maximum of the Gauss-Hermite polynomial and the noise in a single channel; simulations show that this $S/N$ threshold results in an uncertainty of $< 2$ \kms{} in the measured \vp{}.
Finally, we shift each selected line-of-sight spectrum in the flux-rescaled data cube by its \vp{}.
During this process, we interpolate the spectra by a factor of 10, because \vp{} can often be measured to a better accuracy than the width of a channel.
These shifted spectra are then co-added to produce a flux-weighted average \hi{} spectrum, or ``superprofile.''

As discussed in \citetalias{StilpGlobal}, we estimate the noise on each pixel of the superprofiles using:
\begin{equation}
\sigma_\mathrm{SP} = \sigma_\mathrm{chan} 
\times \sqrt{ \frac{N_\mathrm{pix}}{ N_\mathrm{pix/beam}} }
\times \frac{F_\mathrm{rescaled}}{F_\mathrm{standard}},
\label{resolved::eqn:noise-jvm}
\end{equation}
where $\sigma_\mathrm{chan}$ is the \emph{rms}  noise in a single channel, $N_\mathrm{pix} / N_\mathrm{pix/beam}$ is the number of independent resolution elements contributing to each superprofile point, and $F_\mathrm{rescaled} / F_\mathrm{standard}$ is a factor that approximates the flux-rescaling process in the flux-rescaled data cube.

\subsection{Overview of Superprofile Parameterization}
\label{resolved::sec:analysis--parameterization}

\begin{figure}[hbp]
\centering
\includegraphics{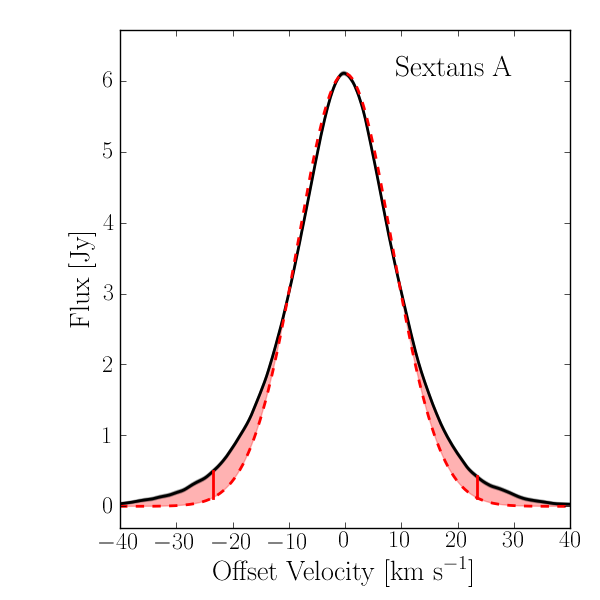}
\caption[Global superprofile for Sextans A]{The global superprofile for Sextans~A, from \citetalias{StilpGlobal}.
The superprofile itself is shown as the thick black line.
The uncertainties are smaller than the width of the line.
The dashed red line shows the HWHM-scaled Gaussian model.
The shaded red regions between the dashed red line and the superprofile represent \hi{} in the wings of the profile, \fw{}.
The solid red lines at $\swing \sim \pm 23$ \kms{} represent the excess-flux-weighted root mean square velocity of the wings.
Because the superprofiles are the analog of integrated line profiles but with each contributing spectra shifted by its \vp{}, we have shown the y-axis in Jy. 
 \label{resolved::fig:superprofile-global}}
\end{figure}

We show an example of a global superprofile in Figure~\ref{resolved::fig:superprofile-global} from \citetalias{StilpGlobal}.
The superprofile itself is shown as a thick black line; the uncertainties are smaller than the width of the superprofile line.
The central peak is largely Gaussian, with only $\sim 5$\% deviations, with excess flux in the wings of the superprofile.
In \citetalias{StilpGlobal}, we proposed that the superprofiles could be parameterized as a turbulent peak, with wings to either side representing higher-velocity \hi{}.
We adopt the same physical interpretation in this study.

We first parameterize the central peak of the superprofile with a Gaussian profile whose amplitude and half-width half-maximum (HWHM) are matched to those of the superprofile.
We adopt the standard deviation of this Gaussian profile as the width of the central peak (\scentral{}).
The HWHM-scaled Gaussian profile for the global superprofile of Sextans~A is shown as a dashed red line in Figure~\ref{resolved::fig:superprofile-global}.

In addition to a Gaussian core, the superprofiles tend to have high velocity gas in excess of a single Gaussian extrapolation of the central peak.
This excess high-velocity gas is shown in Figure~\ref{resolved::fig:superprofile-global} as the transparent red region between the HWHM-scaled Gaussian fit and the superprofile itself.
We estimate the fraction of higher-velocity gas in the wings of the superprofile using:
\begin{equation}
\fw{} = \frac{ \sumlim{|v| > \mathrm{HWHM}} \left[ S(v) - G(v) \right]} { \sumlim{|v| > 0} S (v) },
\label{resolved::eqn:fw}
\end{equation}
where $v$ is the velocity, $S(v)$ is the superprofile intensity at $v$, and $G(v)$ is the value of the HWHM-scaled Gaussian at $v$.

We characterize the typical velocity of the wings using the \emph{rms} velocity of excess flux in the wings:
\begin{equation}
\swingsq{} = \frac{ \sumlim{ | v | > \mathrm{HWHM}}  \left[ S (v) - G(v) \right] v^2 }  { \sumlim{ | v | > \mathrm{HWHM}} \left[ S (v) - G(v) \right] }.
\label{resolved::eqn:swing}
\end{equation}
The \swingsq{} parameter provides an estimate of the energy per unit mass in the wings of the superprofile.
In Figure~\ref{resolved::fig:superprofile-global}, \swing{} is shown to either side of the profile as a solid vertical red line between the HWHM-scaled Gaussian profile and the superprofile.

Finally, we quantify the asymmetry of the entire superprofile around the peak using:
\begin{equation}
a_\mathrm{full} = \frac{ \sumlim{v}{ \sqrt{ \left( S(v) - S ( - v) \right)^2 } } }  { \sumlim{v}{ S(v)} }.
\label{resolved::eqn:afull}
\end{equation}
This quantity differs from the asymmetry parameter \aw{} used in \citetalias{StilpGlobal}, which was calculated only for the wings ($|v| > \mathrm{HWHM}$).
In this study, we instead use the full asymmetry (\afull{}) because the asymmetry of the wings alone is much more sensitive to noise, which has a larger effect on superprofiles calculated for galaxy subregions as they include less flux.
Although the asymmetry parameter in this study may be tracing different effects than that used in \citetalias{StilpGlobal}, the change to \afull{} is not extreme because the majority of the global superprofile asymmetry was due to the wings.
We have verified that the global superprofiles presented in \citetalias{StilpGlobal} follow the same trends with \afull{} as they do with \aw{}.

The uncertainties on the superprofile parameters are discussed in detail in 
\ifthesis
\S
\else
Appendix
\fi
\ref{resolved::sec:robustness}.

\subsection{Spatially-Resolved Superprofiles}
\label{resolved::sec:analysis--spatially-resolved}

In \citetalias{StilpGlobal}, we included all selected line-of-sight spectra in the superprofile to derive an average \hi{} spectrum for the entire galaxy.
In this study, we generate superprofiles in subregions for each galaxy, determined either by radius or by local \sfrsd{}.

First, we derive superprofiles in subregions of constant radius for the sample galaxies.
We note that it is unlikely that radius itself is the actual driver of any trends, because \hi{} kinematic properties are local and have no special knowledge about their location in the galaxy.
Instead, any radial correlations are more likely to reflect the fact that other galaxy properties that do affect the local ISM also correlate with radius, such as \sfrsd{} and \hisd{}, in large disks.
We include this analysis because it facilitates comparison with \citet{Tamburro2009}, who worked in radial annuli for more massive galaxies from the THINGS survey, although with a different methodology than we present here.
We describe the superprofile generation on radial scales in more detail in \S~\ref{resolved::sec:analysis--radial}.

Second, we generate superprofiles in regions determined by the local \sfrsd{}.
Because star formation is a commonly-cited candidate in the literature for the driver of \hi{} turbulence, this choice allows us to directly explore how the superprofiles change as a function of local star formation properties.
We discuss the \sfrsd{} superprofiles in more detail in \S~\ref{resolved::sec:analysis--sfr}.

\subsubsection{Superprofiles in Radial Subregions}
\label{resolved::sec:analysis--radial}

First, we explore how the superprofiles behave as a function of radius.
For the nine larger galaxies in our sample, we generate superprofiles in radially-resolved bins whose widths are chosen to be approximately two times the beam area, rounded to the nearest 0.05 $r_\mathrm{25}$ after correcting for inclination.
These bin widths correspond to physical sizes of $200 - 500$~pc.
For the 10 smaller dwarf galaxies with either small sizes or poorly-defined inclinations and position angles, it is impossible to generate reliable annuli with high signal-to-noise.
In these cases, we instead generate superprofiles inside and outside $r_{25}$ (typically $0.4 - 1.5$ kpc; Table~\ref{tab:resolved--sample-full}).

In Table~\ref{tab:resolved--sample-res}, we list the properties of resolved superprofiles for the sample.
We list
(1) the galaxy name;
(2) the number of beams across the entire galaxy at $S/N > 5$;
(3) the number of superprofiles in radial subregions that meet the selection criteria for the galaxy;
(4) the maximum radius at which we were able to generate a superprofile;
(5) the step size for radial annuli;
and (6) the number of superprofiles in \sfrsd{} subregions that meet the selection criteria for the galaxy.
For those galaxies where we have generated radial superprofiles only inside and outside $r_{25}$, we do not have values for $r_\mathrm{max}$ or $\Delta r$.

\ifthesis
\afterpage{\input{\tablepath/table-sample-res}}
\else
\afterpage{\begin{deluxetable}{lcccccc}
\tablewidth{0pt}
\tablecaption{Spatially-resolved Superprofiles. \label{tab:resolved--sample-res}}

\tablehead{

  \colhead{1} &
  \colhead{2} &
  \colhead{3} &
  \colhead{4} &
  \colhead{5} &
  \colhead{6} &
  \colhead{7} \\

  \colhead{Galaxy} & 
  \colhead{$\Delta v$} & 
  \colhead{$N_\mathrm{beams}$} &
  \colhead{$N_\mathrm{radial}$} & 
  \colhead{$r_\mathrm{max}$} & 
  \colhead{$\Delta r$} & 
  \colhead{$N_\mathrm{SF}$} \\
  
  \colhead{} &
  \colhead{(\kms{})} &
  \colhead{}  &
  \colhead{}  &
  \colhead{($r_{25}$)} & 
  \colhead{($r_{25}$)} & 
  \colhead{}
}

\startdata
NGC 7793  &  2.6  &  909  &  21  &  1.25  &  0.05  &  6\\ 
IC 2574  &  2.6  &  2442  &  25  &  1.25  &  0.05  &  5\\ 
NGC 4214  &  1.3  &  1557  &  20  &  2.10  &  0.10  &  6\\ 
Ho II  &  2.6  &  924  &  17  &  1.80  &  0.10  &  5\\ 
NGC 2366  &  2.6  &  892  &  12  &  2.40  &  0.20  &  6\\ 
DDO 154  &  2.6  &  681  &  8  &  3.20  &  0.40  &  4\\ 
Ho I  &  2.6  &  161  &  9  &  1.30  &  0.10  &  4\\ 
NGC 4190  &  1.3  &  44  &  2  &  ...   & ...   &  4\\ 
NGC 3741  &  1.3  &  111  &  2  &  ...   & ...   &  3\\ 
Sextans A  &  1.3  &  214  &  7  &  1.60  &  0.20  &  5\\ 
DDO 53  &  2.6  &  92  &  2  &  ...   & ...   &  5\\ 
DDO 190  &  2.6  &  76  &  2  &  ...   & ...   &  5\\ 
Sextans B  &  1.3  &  259  &  9  &  2.00  &  0.20  &  4\\ 
DDO 99  &  1.3  &  91  &  2  &  ...   & ...   &  4\\ 
UGCA 292  &  0.6  &  45  &  2  &  ...   & ...   &  4\\ 
UGC 4483  &  2.6  &  54  &  2  &  ...   & ...   &  3\\ 
DDO 181  &  1.3  &  53  &  2  &  ...   & ...   &  3\\ 
UGC 8833  &  2.6  &  30  &  2  &  ...   & ...   &  2\\ 
DDO 187  &  1.3  &  22  &  2  &  ...   & ...   &  3\\ 
\enddata

\tablecomments{Superprofiles in spatially-resolved subregions.  (1) Galaxy name. (2) Velocity resolution. (3) Total number of beams above S/N$ > 5$ for the entire galaxy. (4) Number of radially-resolved superprofiles for this galaxy. (5) Maximum radius for superprofile generation (radial annuli only). (6) Radial step size (radial annuli only). (7) Number of constant-\sfrsd{} superprofiles for this galaxy. }

\end{deluxetable}
}
\fi

As an example, we show the behavior of the \hi{} superprofiles as a function of radius for NGC 7793 in Figures~\ref{resolved::fig:superprofiles-radial-n7793-a} - \ref{resolved::fig:superprofiles-radial-n7793-c}.
In Figure~\ref{resolved::fig:superprofiles-radial-n7793-a}, we show the annuli in which we have generated superprofiles superimposed on the \hi{} column density ($N_{HI}$) map.
The solid black outline indicates the region within which all 200~pc-smoothed pixels in the \hi{} data cube have $S/N > 5$.
The solid colored lines mark the midpoint of each annulus, and the corresponding shaded color regions show the pixels that contribute to that annulus.
The beam is shown in the lower left corner, and the physical resolution is indicated by a scale bar in the lower right corner.

In Figure~\ref{resolved::fig:superprofiles-radial-n7793-b}, we show the superprofiles that correspond to the annuli in Figure~\ref{resolved::fig:superprofiles-radial-n7793-a}.
In the upper left panel, we plot the observed superprofiles.
To highlight the cumulative contribution from each radial annulus to the total superprofile, each resolved superprofile plotted also includes the superprofiles at all smaller radii.
In the lower left panel, we have normalized each radial superprofile to the same peak amplitude.
The superprofiles are clearly wider in the central regions of NGC~7793 than in the outskirts.
This behavior has previously been observed in other galaxies but using the \hi{} second moment \citep{Boulanger1992, Petric2007, Tamburro2009}.
In the upper right panel, we have normalized the superprofiles by their amplitude and HWHM, as determined from the HWHM-scaled Gaussian model (thick black line).
As seen in \citetalias{StilpGlobal}, the shape of these normalized superprofiles is typically very similar.
In the lower right panel, we show the residuals of the superprofiles minus the HWHM-scaled Gaussian model.
When compared to a Gaussian profile, the superprofiles typically have more flux in the wings and are peakier in the center.
This behavior was clearly seen in the global superprofiles from \citetalias{StilpGlobal}, but clearly persists when analyzed on smaller spatial scales.

In Figure~\ref{resolved::fig:superprofiles-radial-n7793-c}, we show the behavior of the superprofile parameters for NGC~7793 as a function of radius, normalized to $r_{25}$.
Again, the color of each point corresponds to the same colored annulus in Figure~\ref{resolved::fig:superprofiles-radial-n7793-a}.
The left-hand panels show \scentral{} (upper) and \swing{} (lower), and the right-hand panels show \fw{} (upper) and \afull{} (lower).

\ifthesis
\renewcommand{\thefigure}{\arabic{chapter}.\arabic{figure}\alph{subfig}}
\else
\renewcommand{\thefigure}{\arabic{figure}\alph{subfig}}
\fi

\setcounter{subfig}{1}
\begin{figure}[p]
\ifthesis
\begin{leftfullpage}
\fi
\centering
\includegraphics[width=4in]{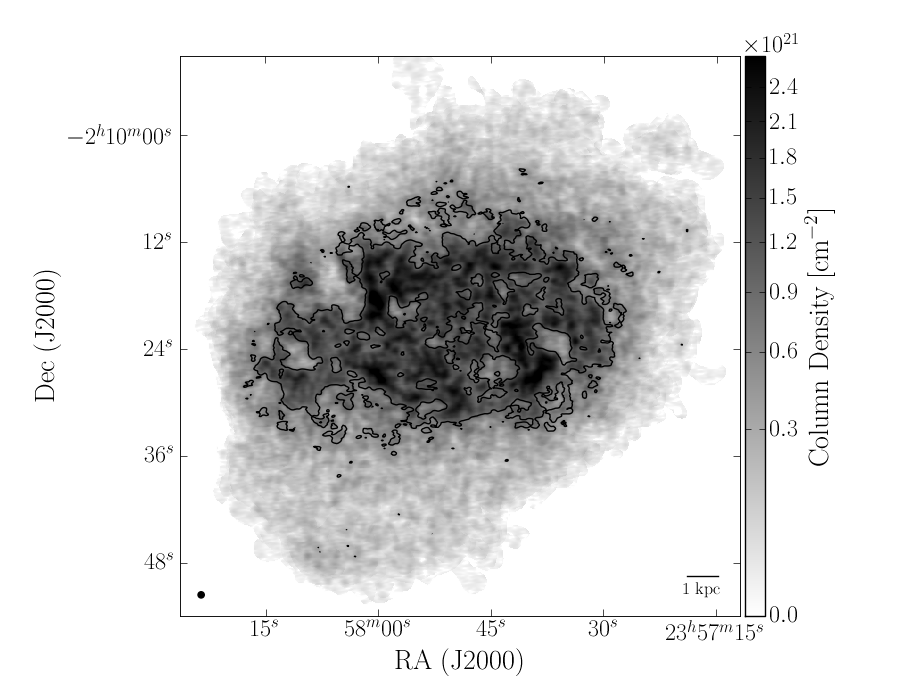}
\vskip -0.1in
\includegraphics[width=4in]{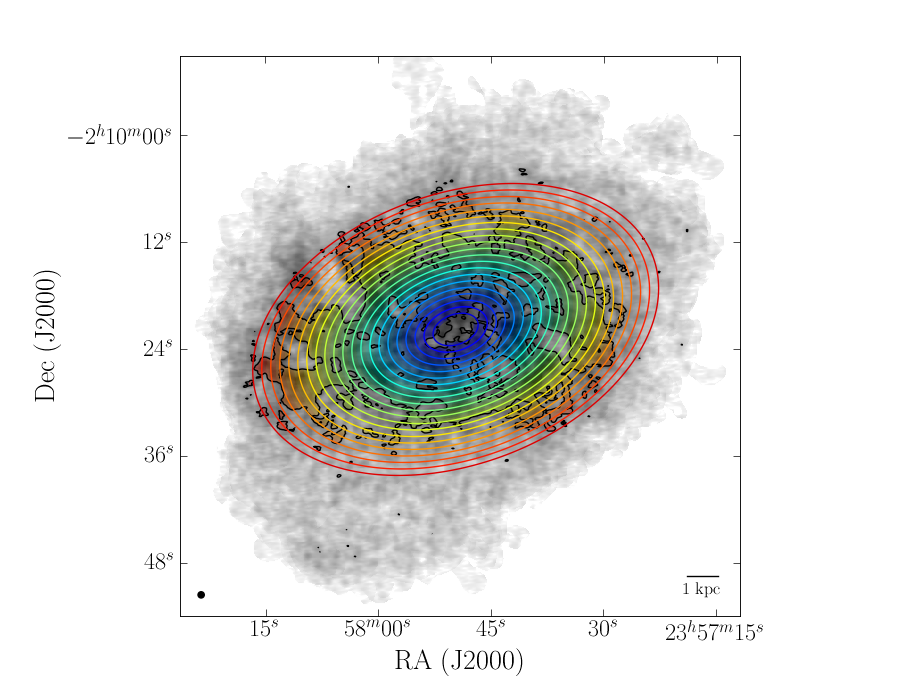}
\caption[Superprofile radial annuli for NGC 7793]{Radial annuli in which radial superprofiles are generated for NGC~7793. 
In both panels, the background greyscale shows \hisd{}, and the solid black line represents the $S/N > 5$ threshold where we can accurately measure \vp{}. 
In the lower panel, the colored solid lines represent the average radius of each annulus, and the corresponding shaded regions of the same color indicate which pixels have contributed to each radial superprofile.
\label{resolved::fig:superprofiles-radial-n7793-a} }
\ifthesis
\end{leftfullpage}
\fi
\end{figure}

\addtocounter{figure}{-1}
\addtocounter{subfig}{1}
\begin{figure}
\centering
\includegraphics[height=2.7in]{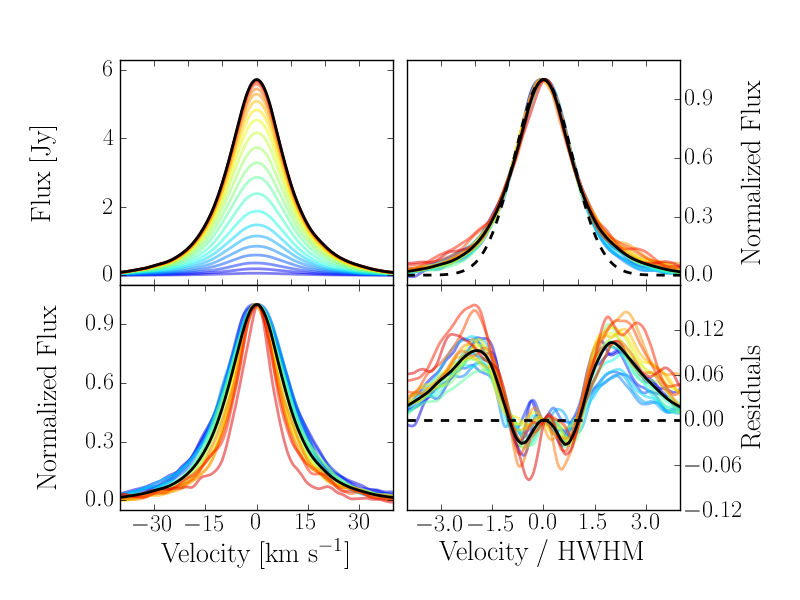}
\caption[Radial superprofiles in NGC~7793]{The radial superprofiles in NGC~7793, where colors indicate the corresponding radial annuli in the previous figure.
The left hand panels show the raw superprofiles (upper left) and the superprofiles normalized to the same peak flux (lower left).
The right hand panels show the flux-normalized superprofiles scaled by the HWHM (upper right) and the flux-normalized superprofiles minus the model of the Gaussian core (lower right).
In all panels, the solid black line represents the global superprofile. In the left panels, we have shown the HWHM-scaled Gaussian model as the dashed black line.
\label{resolved::fig:superprofiles-radial-n7793-b}
}
\end{figure}
\addtocounter{figure}{-1}
\addtocounter{subfig}{1}
\begin{figure}
\centering
\includegraphics[height=2.7in]{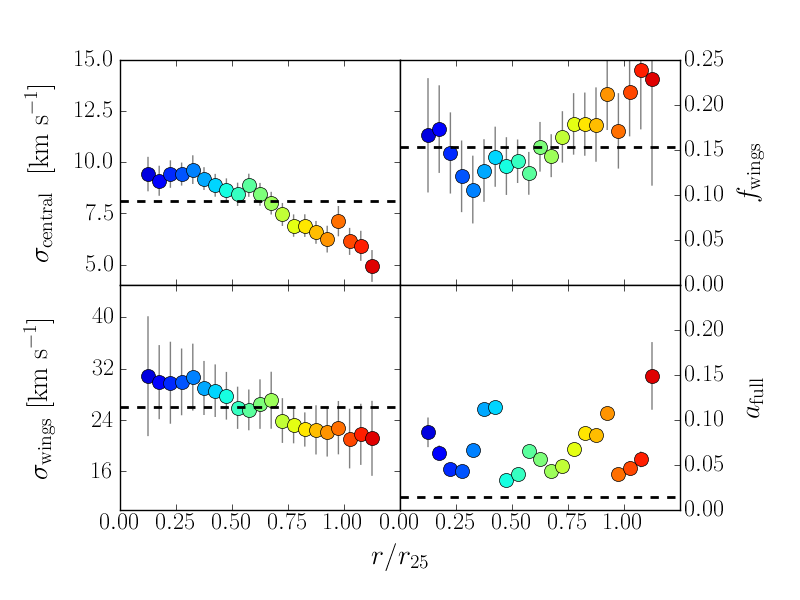}
\caption[Variations of superprofile parameters as a function of radius in NGC~7793]{Variation of the superprofile parameters as a function of normalized radius for NGC~7793.
The dashed black line shows the parameter value for the global superprofile \citepalias{StilpGlobal}.
We have plotted the equivalent \afull{} value of the global superprofile instead of \aw{}, as measured in \citetalias{StilpGlobal}.
The asymmetry of spatially-resolved profiles is higher than that of the global superprofile because smaller-scale asymmetries do not average out.
\label{resolved::fig:superprofiles-radial-n7793-c}
}
\end{figure}

\ifthesis
\renewcommand{\thefigure}{\arabic{chapter}.\arabic{figure}}
\else
\renewcommand{\thefigure}{\arabic{figure}}
\fi

\begin{figure}
\centering
\includegraphics[width=5.5in]{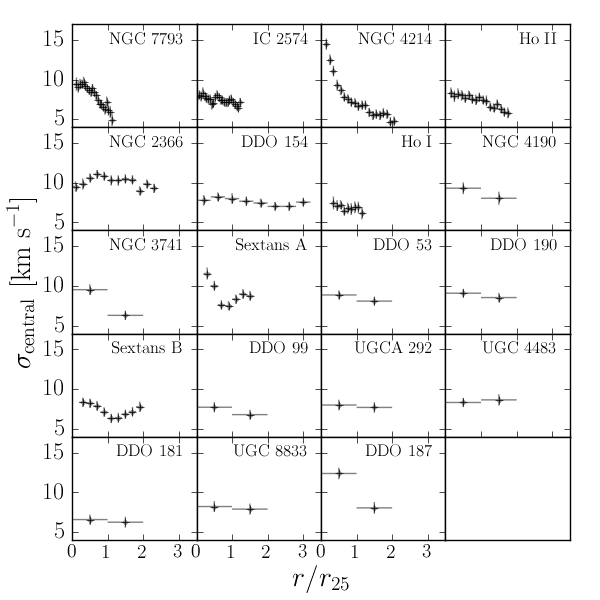}
\caption[\scentral{} as a function of radius for the sample]{The velocity dispersion, \scentral{}, of the superprofiles generated in regions of constant radius.
Each panel represents a single galaxy, and galaxies are ordered by decreasing \m{baryon,tot}.
Within one panel, each point represents one superprofile.
For those galaxies where we have generated superprofiles inside and outside $r_{25}$, we plot the two superprofiles at $r / r_{25} = 0.5$ and $1.5$, respectively.
The velocity dispersion decreases smoothly with radius for the more massive galaxies, but shows erratic behavior for lower mass galaxies.
\label{resolved::fig:radial-panels}
}
\end{figure}

In Figure~\ref{resolved::fig:radial-panels}, we show the behavior of \scentral{} with radius for the sample.
Inspection of these figures shows that the more massive galaxies, such as NGC~4214 and NGC~7793, tend to show declining \scentral{} and \swing{} with increasing radius.
These trends are similar to those discussed by \citet{Tamburro2009}, who found that \hi{} second moment declined with radius in the more massive galaxies of their sample.
In contrast, the parameters for the lower mass galaxies (e.g., Sextans A) do not vary as smoothly with radius as those in larger galaxies.
Other galaxy properties, such as SFR or \hisd{}, do not exhibit smooth radial trends in these dwarfs, so we would not necessarily expect smooth radial trends in the superprofile parameters of lower mass galaxies.
This behavior reaffirms our expectation that \hi{} kinematics are not determined by radius itself, but by other parameters that tend to correlate with radius in higher-mass galaxies.

We show the radially-resolved superprofiles for all sample galaxies at the end of the
\ifthesis
chapter in \S
\else
paper in Appendix
\fi
\ref{resolved::sec:sp-figures}, in order of decreasing \m{baryon,tot}.
For those 10 dwarf galaxies where radial analysis is impossible, we plot the superprofiles inside and outside $r_{25}$, colored respectively by blue and red.

\subsubsection{\sfrsd{}-determined Superprofiles}
\label{resolved::sec:analysis--sfr}

Because star formation is often connected to \hi{} kinematics in the literature, we also generate superprofiles based on local values of the inclination-corrected star formation rate intensity \sfrsd{}.
This choice allows us to characterize how \hi{} kinematics change as the star formation rate varies across the galaxy in a more direct way than the radial method.

\begin{samepage}
We have adopted six \sfrsd{} bins for our analysis:
\begin{equation}
\begin{array}{lcccl}
          & & \sfrsd{} &<& 10^{-4} \; \mathrm{M}_\odot \; \mathrm{yr}^{-1} \; \mathrm{kpc}^{-2} \\
10^{-4}   &<& \sfrsd{} &<& 10^{-3.5} \\
10^{-3.5} &<& \sfrsd{} &<& 10^{-3} \\
10^{-3}   &<& \sfrsd{} &<& 10^{-2.5} \\
10^{-2.5} &<& \sfrsd{} &<& 10^{-2} \\
10^{-2}   &<& \sfrsd{} & & \\
\end{array}
\label{resolved::eqn:sfr-ranges}
\end{equation}
\end{samepage}
These values span the observed \sfrsd{} range of our sample, but are small compared to typical \sfrsd{} values for larger spirals, which can exceed $10^{-1}$ \msun{} \per[2]{kpc} \per{yr} in the central regions \citeeg{Leroy2008}.

For each galaxy, we generate superprofiles using pixels whose local \sfrsd{} falls in each bin.
We note that star formation rates in some of the lower-mass galaxies do not span the full \sfrsd{} range sampled by our bins, and thus will have fewer than six superprofiles.

\begin{figure}[tp]
\centering
\includegraphics[width=6in]{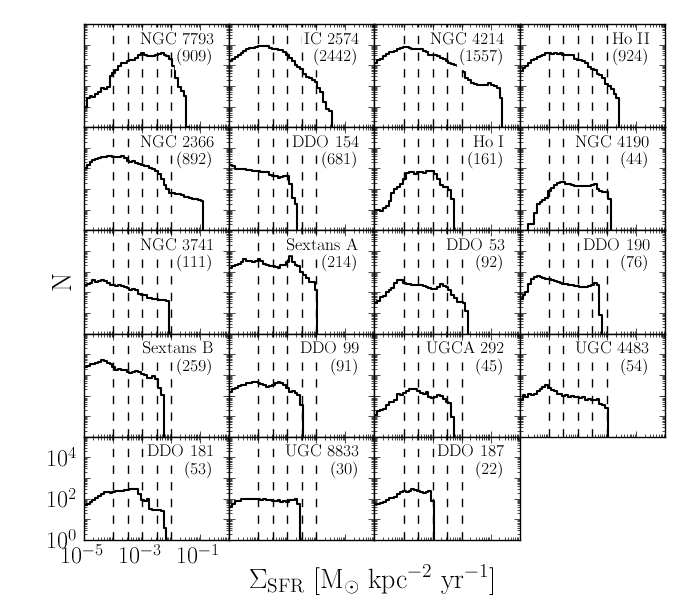}
\caption[\sfrsd{} values for the sample]{ Histograms of inclination-corrected \sfrsd{} values in each galaxy, in order of decreasing \m{baryon,tot}.
Each panel represents the distribution of \sfrsd{} values for pixels above the \hi{} $S/N > 5$ threshold for a single galaxy.
The number in parentheses below the galaxy name indicates the number of independent resolution elements above the $S/N > 5$ threshold for the entire galaxy.
Note that not all galaxies have valid pixels in each bin.
\label{resolved::fig:sfrsd-histogram}}
\end{figure}

In Figure~\ref{resolved::fig:sfrsd-histogram}, we show histograms of the \sfrsd{} values for the galaxies in our sample, ordered by decreasing \m{baryon,tot}.
Each panel shows the distribution of inclination-corrected \sfrsd{} values for pixels that fall above the \hi{} $S/N > 5$ threshold for a single galaxy.
The \sfrsd{} bin edges given in Equation~\ref{resolved::eqn:sfr-ranges} are displayed as dashed vertical black lines.

In Figures~\ref{resolved::fig:superprofiles-sfr-n7793-a} - \ref{resolved::fig:superprofiles-sfr-n7793-c}, we show an example of the \sfrsd{} superprofiles for NGC~7793.
The layout is similar to that used in the figures for the radially-resolved superprofiles (Figures~\ref{resolved::fig:superprofiles-radial-n7793-a} - \ref{resolved::fig:superprofiles-radial-n7793-c}), with a few exceptions.
in Figure~\ref{resolved::fig:superprofiles-sfr-n7793-a}, the greyscale image is the \sfrsd{} map measured with FUV+24\um{} emission instead of \hisd{} as was used in Figure~\ref{resolved::fig:superprofiles-radial-n7793-a}.
We have also plotted the parameters in Figure~\ref{resolved::fig:superprofiles-sfr-n7793-c} as a function of \sfrsd{} instead of radius.

\ifthesis
\renewcommand{\thefigure}{\arabic{chapter}.\arabic{figure}\alph{subfig}}
\else
\renewcommand{\thefigure}{\arabic{figure}\alph{subfig}}
\fi

\setcounter{subfig}{1}
\begin{figure}[p]
\ifthesis
\begin{leftfullpage}
\fi
\centering
\includegraphics[width=4in]{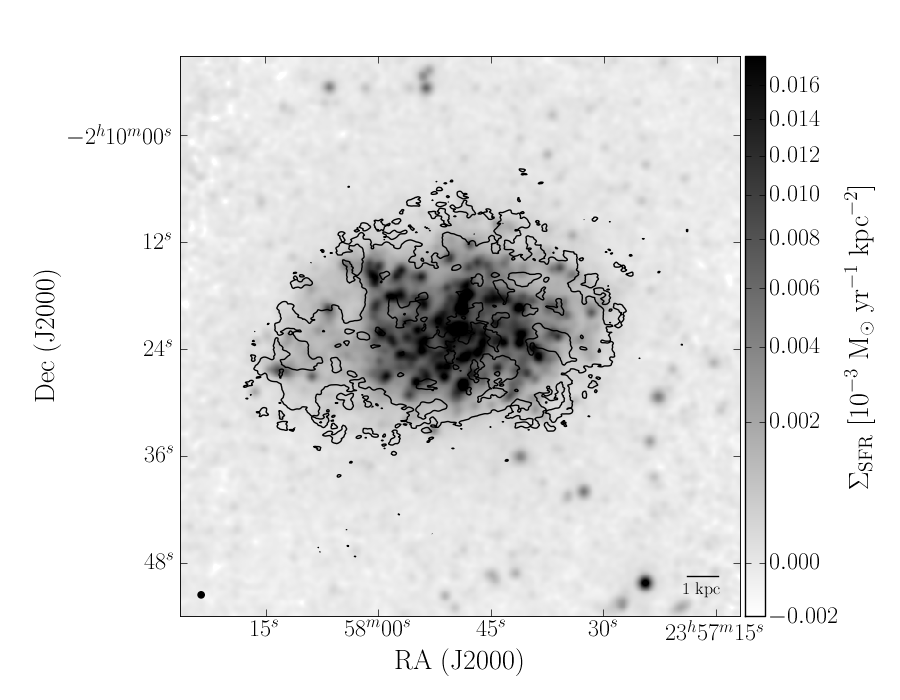}
\includegraphics[width=4in]{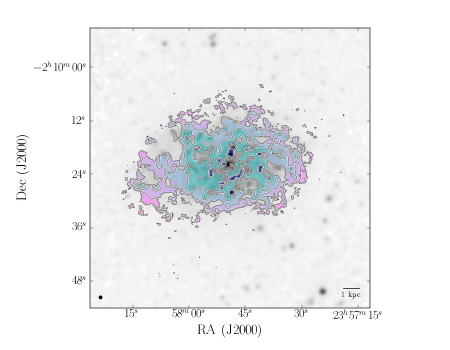}
\caption[Superprofile \sfrsd{} annuli for NGC 7793]{\sfrsd{} regions in which superprofiles are generated for NGC~7793. 
In both panels, the background greyscale shows \sfrsd{}, and the solid black line represents the $S/N > 5$ threshold where we can accurately measure \vp{}. 
In the lower panel, the colored regions show which pixels have contributed to each \sfrsd{} superprofile.
\label{resolved::fig:superprofiles-sfr-n7793-a} }
\ifthesis
\end{leftfullpage}
\fi
\end{figure}

\addtocounter{figure}{-1}
\addtocounter{subfig}{1}
\begin{figure}
\centering
\includegraphics[height=2.7in]{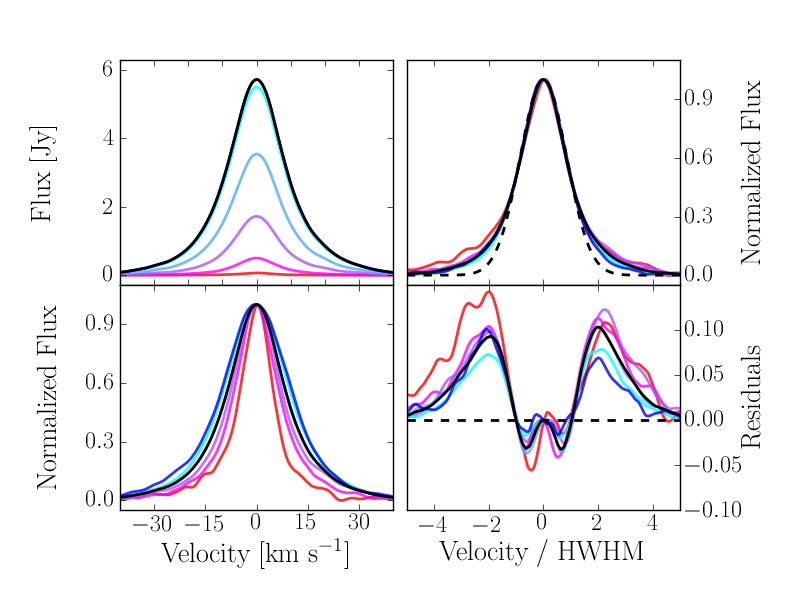}
\caption[\sfrsd{} superprofiles in NGC~7793]{The \sfrsd{} superprofiles in NGC~7793, where colors indicate the corresponding regions in the previous figure.
The left hand panels show the raw superprofiles (upper left) and the superprofiles normalized to the same peak flux (lower left).
The right hand panels show the flux-normalized superprofiles scaled by the HWHM (upper right) and the flux-normalized superprofiles minus the model of the Gaussian core (lower right). In all panels, the solid black line represents the global superprofile. In the left panels, we have shown the HWHM-scaled Gaussian model as the dashed black line.
\label{resolved::fig:superprofiles-sfr-n7793-b}
}
\end{figure}
\addtocounter{figure}{-1}
\addtocounter{subfig}{1}
\begin{figure}
\centering
\includegraphics[height=2.7in]{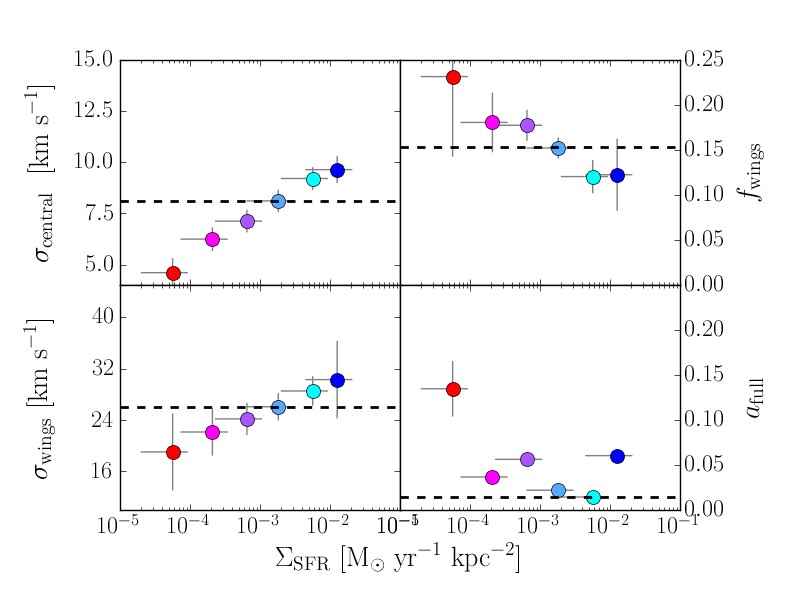}
\caption[Variation of the \sfrsd{} superprofile parameters as a function of \sfrsd{} for NGC~7793]{Variation of the \sfrsd{} superprofile parameters as a function of \sfrsd{} for NGC~7793.
The solid dashed line shows the parameter value for the global superprofile \citepalias{StilpGlobal}.
As before, we have plotted the equivalent \afull{} value of the global superprofile instead of \aw{}, as measured in \citetalias{StilpGlobal}.
Both \scentral{} and \swing{} increase with increasing \sfrsd{}, while \fw{} decreases with increasing \sfrsd{}.
The trends between \afull{} and \sfrsd{} is less smooth.
\label{resolved::fig:superprofiles-sfr-n7793-c}
}
\end{figure}

\ifthesis
\renewcommand{\thefigure}{\arabic{chapter}.\arabic{figure}}
\else
\renewcommand{\thefigure}{\arabic{figure}}
\fi

\begin{figure}
\centering
\includegraphics[width=5.5in]{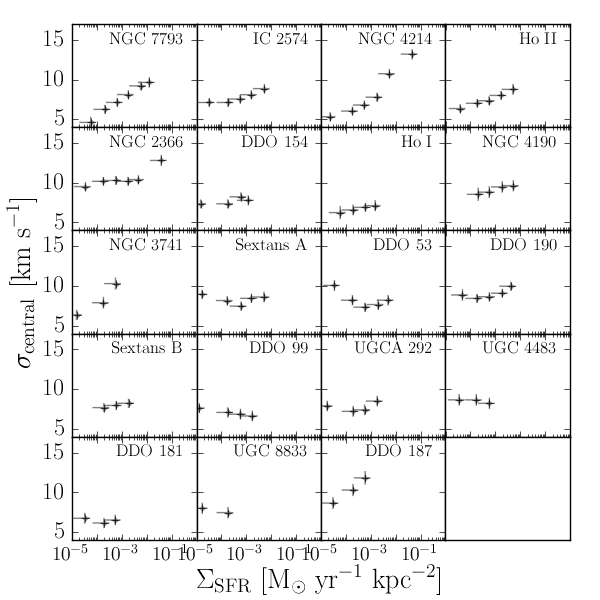}
\caption[\scentral{} as a function of \sfrsd{} for the sample]{The velocity dispersion, \scentral{}, of the superprofiles generated in regions of constant \sfrsd{}.
Each panel represents a single galaxy, and galaxies are ordered by decreasing \m{baryon,tot}.
Within one panel, each point represents one superprofile.
The velocity dispersion increases smoothly with \sfrsd{} for the more massive galaxies, but shows erratic behavior for lower mass galaxies.
\label{resolved::fig:panels-sfr}
}
\end{figure}

In Figure~\ref{resolved::fig:panels-sfr}, we highlight the behavior of \scentral{} as a function of radius in the sample.
As with the radial superprofiles, many of the more massive galaxies in the sample have parameters that vary smoothly as a function \sfrsd{}.
In particular, both \scentral{} and \swing{} tend to increase with increasing \sfrsd{} in the more massive galaxies (IC~2574, NGC~4214, Ho~II, NGC~3741), but this trend is not ubiquitous; a number of the galaxies show no strong trend or the opposite trend.

The \sfrsd{} superprofiles for the sample are shown in 
\ifthesis
\S~\ref{resolved::sec:sp-figures--sfrsd}.
\else
Appendix~\ref{resolved::sec:sp-figures--sfrsd}.
\fi

\ifthesis
\input{chapters/radial/robustness}
\fi

\section{\hi{} Kinematics as a Function of Local Galaxy Properties}
\label{resolved::sec:correlations}

In \citetalias{StilpGlobal}, we compared global superprofile parameters to galaxy physical properties, averaged over the entire galaxy.
In this paper, we assess the behavior of superprofile parameters in regions of constant projected radius or \sfrsd{}, averaged over 200 pc scales.
By deriving superprofiles in these subregions, we increase our sensitivity to any dependence of \hi{} kinematics on local galaxy properties (i.e., \hisd{}, \sfrsd{}, \baryonsd{}).

To determine whether a correlation exists between a superprofile parameter and a physical property, we use the Spearman rank correlation coefficient.
As described in \citetalias{StilpGlobal}, this statistic tests for a monotonically increasing ($r_S > 0$) or decreasing ($r_S < 0$) relationship between the two variables.
The corresponding $p_S$ value gives the probability of finding an $r_S$ value equal to or more extreme than the observed $r_S$ from a random sample.
We choose $p_S = 0.01$ as a threshold for finding a significant correlation.
The $r_S$ and $p_S$ values are given in Tables~\ref{tab:resolved--correlations-radial} for the radially-resolved superprofiles and in Table~\ref{tab:resolved--correlations-sfr} for the \sfrsd{} superprofiles.

For the radial superprofiles, the larger galaxies in our sample have far more independent points than the smaller galaxies.
In fact, as seen in Table~\ref{tab:resolved--sample-res} the four most massive dwarfs (NGC~7793, IC~2574, NGC~4214, and Ho~II) contribute over 50\% of the points to the radial correlation plots, whereas the smallest dwarfs typically have only two radial bins each (inside and outside $r_{25}$).
This overweighting of the larger galaxies is decreased when examining correlations in the \sfrsd{} superprofiles, for which higher mass and lower mass galaxies contribute similar numbers of points.
The superprofiles in subregions of constant \sfrsd{} may therefore trace a more representative result for the entire sample compared to the superprofiles from radial subregions.
In both the figures and text of this section, we discuss separately the superprofiles from smaller galaxies, with points only inside and outside $r_{25}$, and those from larger galaxies, with points from radial annuli.

We explore how various local properties correlate 
first with measures of \hi{} velocity (\scentral{} and \swing{}; \S~\ref{resolved::sec:correlations--v})
and second with measures of superprofile shape (\fw{} and \afull{}; \S~\ref{resolved::sec:correlations--shape}).
We now present the different correlations, but largely defer discussion of their physical interpretation until \S~\ref{resolved::sec:discussion}.
The correlation coefficient $r_s$ and corresponding $p_s$ values for all correlations are given in Tables~\ref{tab:resolved--correlations-radial} - \ref{tab:resolved--correlations-sfr}.

\subsection{\hi{} Velocities}
\label{resolved::sec:correlations--v}

We start by exploring the correlations of \scentral{} and \swing{} with a variety of local properties, keeping in mind that we expect some degree of correlation between \scentral{} and \swing{}, as \swing{} is restricted to values greater than \scentral{} by definition (Equation~\ref{resolved::eqn:swing}).

\subsubsection{\hi{} Velocities versus Surface Mass Density}
\label{resolved::sec:correlations--v--surface-mass-density}

In \citetalias{StilpGlobal}, we found that \scentral{} and \swing{} were correlated most strongly with \hisd{}, when averaged over the entire galaxy.
In this section, we assess whether a similar correlation with \hisd{} holds locally when the properties of the \hi{} kinematics are derived in subregions of constant normalized radius or \sfrsd{}.

\begin{figure}
\centering
\includegraphics[width=6.5in]{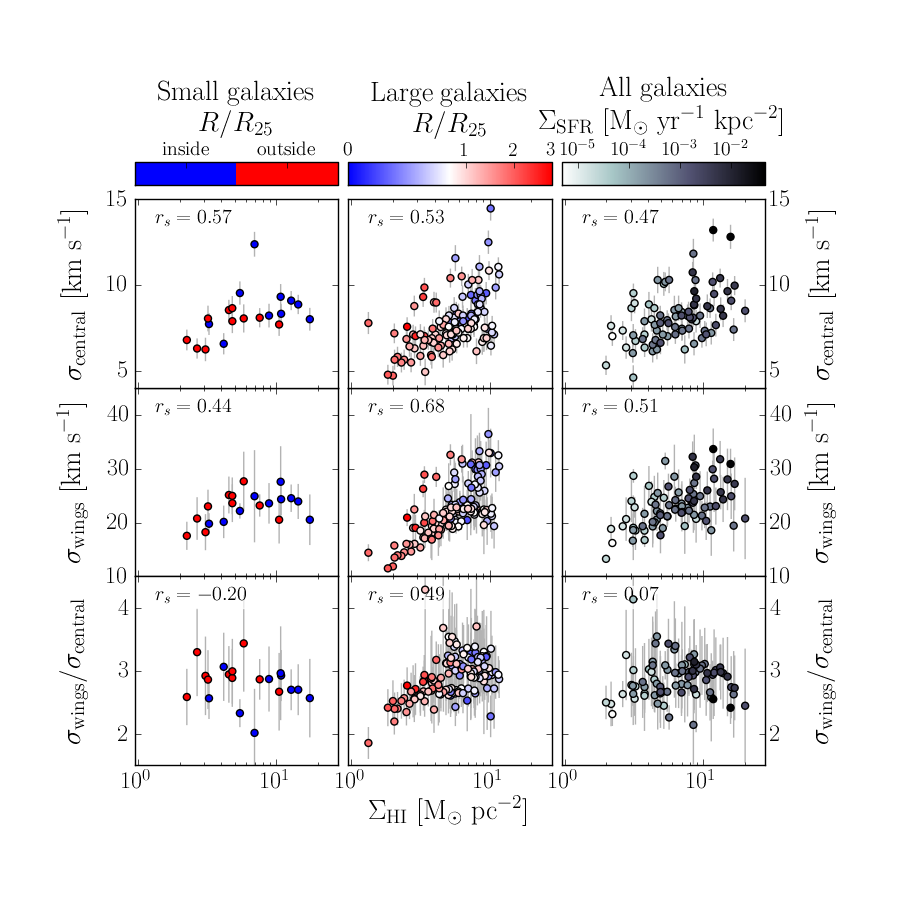}
\caption[\hi{} Velocities versus \hisd{}]{\hi{} velocity measurements versus \hisd{} for spatially-resolved superprofiles.
Each point indicates the \hisd{} and \hi{} velocity measurement derived within one independent superprofile.
The left-hand panels show \scentral{}, \swing{}, and $\swing{} / \scentral{}$ for superprofiles in generated inside and outside $r_{25}$, for the smaller galaxies. The middle panels show the same parameters, but for superprofiles generated in radial annuli for the larger galaxies.
Blue indicates central regions, and red indicates outer regions.
The right-hand panels show the same measurements, but now for superprofiles in \sfrsd{} subregions.
In these panels, points are color-coded by the \sfrsd{} bin into which they fall, with black indicating high \sfrsd{} and white indicating low \sfrsd{}.
Both \scentral{} and \swing{} show trends with \hisd{}, though they are more pronounced in the radial superprofiles.
\label{resolved::fig:velocities-sigma-hi}
}
\end{figure}

In Figure~\ref{resolved::fig:velocities-sigma-hi}, we show \scentral{}, \swing{}, and $\swing{} / \scentral{}$ as a function of \hisd{}, calculated in spatially-resolved subregions.
The left-hand panels show superprofiles calculated inside and outside $r_{25}$ for the smaller galaxies, and the middle panels show the superprofiles calculated in radial subregions of the larger galaxies.
The points are color-coded by the normalized radius of the subregion ($r / r_{25}$, with blue indicating the central regions and red indicating the outskirts.

The right-hand panels of Figure~\ref{resolved::fig:velocities-sigma-hi} show the same parameters as those on the right, but for superprofiles calculated in \sfrsd{} subregions.
The points are color-coded by the average \sfrsd{} of the region (Equation~\ref{resolved::eqn:sfr-ranges}), with white indicating low \sfrsd{} and black indicating high \sfrsd{}.
Both large and small galaxies contribute similar numbers of points to these panels compared to the superprofiles in radial subregions, so we do not separate the large and small galaxies.

We find significant correlations between \hisd{} and both \scentral{} and \swing{} for superprofiles generated in the $r_{25}$, radial, and \sfrsd{} subregions.
In all cases, neither \scentral{} nor \swing{} is strongly determined by \hisd{}; the observed spread in \scentral{} at a given \hisd{} is comparable to the mean.
We also find that the ratio between \swing{} and \scentral{} shows little structure with \hisd{}.
Typically values are $\swing{} / \scentral{} \sim 2 - 3$.

\begin{figure}
\centering
\includegraphics[width=6.5in]{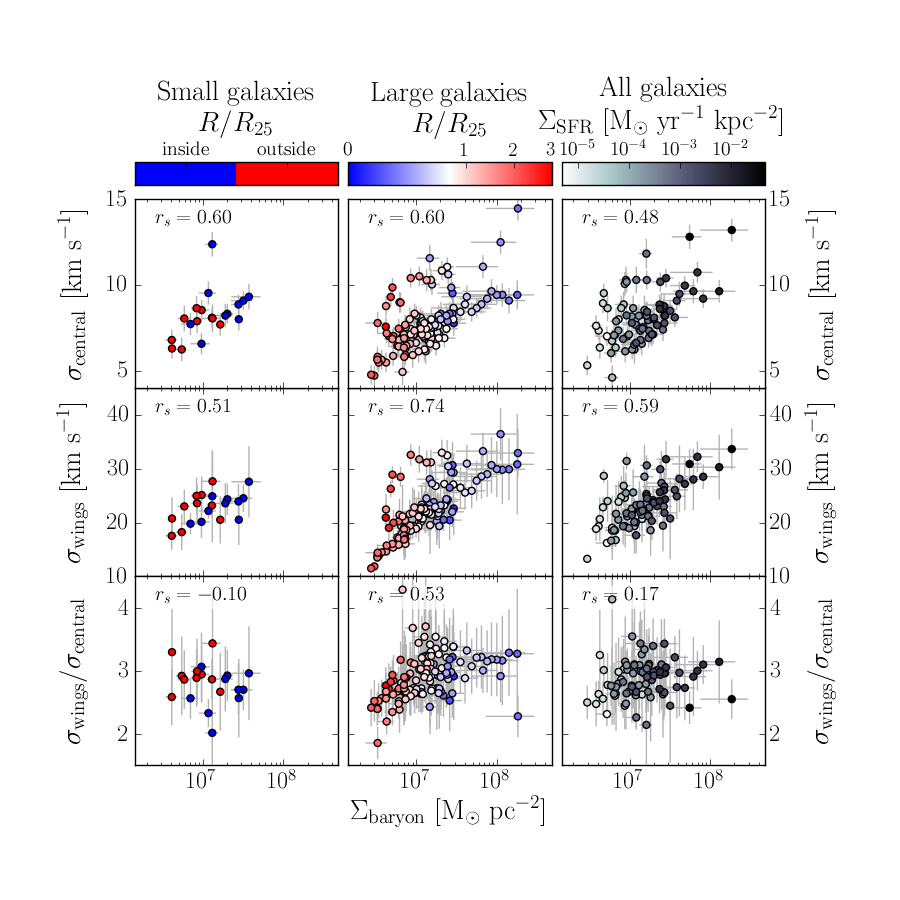}
\caption[\hi{} Velocities versus \baryonsd{}]{\hi{} velocity measurements versus \baryonsd{} for spatially-resolved superprofiles.
Each point indicates the \baryonsd{} and \hi{} velocity measurement derived within one independent superprofile.
The left-hand panels show \scentral{}, \swing{}, and $\swing{} / \scentral{}$ for superprofiles in generated inside and outside $r_{25}$, for the smaller galaxies. The middle panels show the same parameters, but for superprofiles generated in radial annuli for the larger galaxies.
Blue indicates central regions, and red indicates outer regions.
The right-hand panels show the same measurements, but now for superprofiles in \sfrsd{} subregions.
In these panels, points are color-coded by the \sfrsd{} bin into which they fall, with black indicating high \sfrsd{} and white indicating low \sfrsd{}.
Both \scentral{} and \swing{} show trends with \baryonsd{}, though they are more pronounced in the radial superprofiles.
\label{resolved::fig:velocities-sigma-baryon}
}
\end{figure}

Next, we examine correlations between \hi{} velocities and \baryonsd{} in Figure~\ref{resolved::fig:velocities-sigma-baryon}.
The figure layout is the same as Figure~\ref{resolved::fig:velocities-sigma-hi}, but for correlations with \baryonsd{} instead of \hisd{}.
The correlations between \scentral{} and \swing{} are modestly stronger with \baryonsd{} than with \hisd{}, but again, the spread is comparable to the range of the correlation.
We also find that \scentral{} shows a lower bound that increases with increasing \baryonsd{}.
Regions with high \baryonsd{} do not have low \scentral{} values, while regions with low \baryonsd{} can have either low or high \scentral{}.
Similar behavior is seen for \swing{}, which may be a reflection of the fact that regions with higher \scentral{} values must by definition have higher \swing{} values.

\subsubsection{\hi{} Velocities versus Star Formation}
\label{resolved::sec:correlations--v--sf}

We now address the behavior of \hi{} velocities as a function of local star formation measures.
We consider two measures of star formation: the local star formation intensity (\sfrsd{}) and the ratio of available star formation energy to the \hi{} mass that the energy couples to ($\hisd{} / \sfrsd{}$).
In \citetalias{StilpGlobal}, we did not find a correlation between \scentral{} and $\langle \sfrsd{} \rangle$, averaged over the entire galaxy, unless higher-mass galaxies were included.

\begin{figure}
\centering
\includegraphics[width=6.5in]{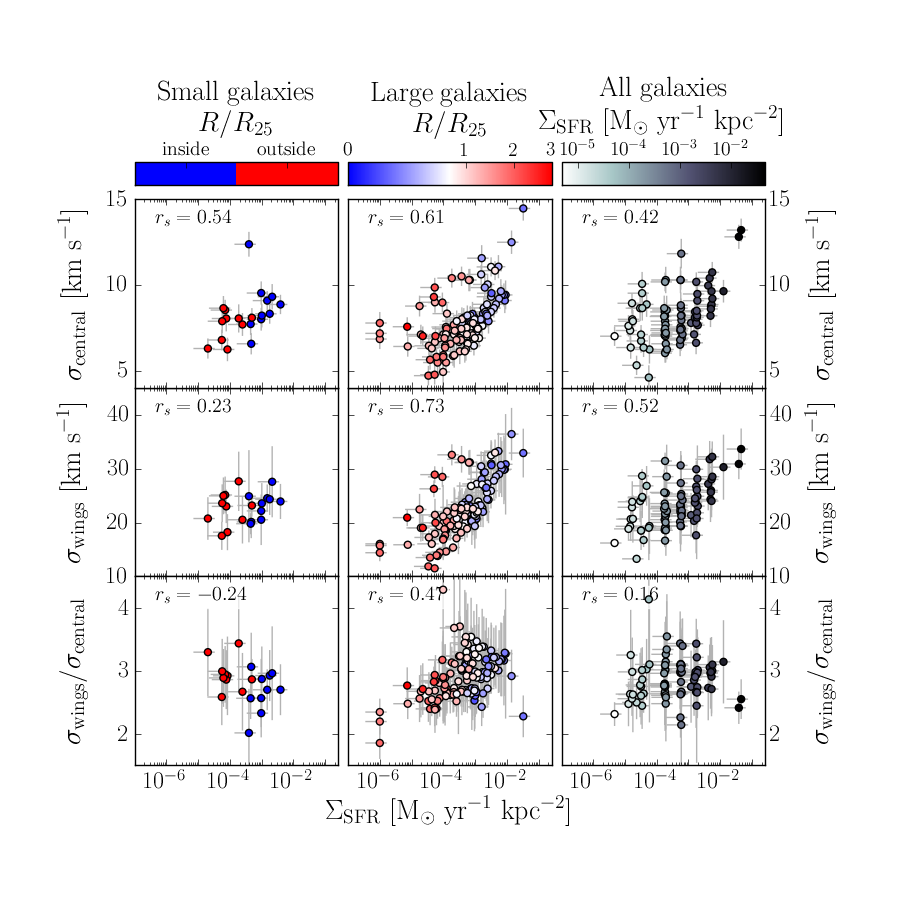}
\caption[\hi{} Velocities versus \sfrsd{}]{\hi{} velocity measurements versus \sfrsd{} for spatially-resolved superprofiles.
Each point indicates the \sfrsd{} and \hi{} velocity measurement derived within one independent superprofile.
The left-hand panels show \scentral{}, \swing{}, and $\swing{} / \scentral{}$ for superprofiles in generated inside and outside $r_{25}$, for the smaller galaxies. The middle panels show the same parameters, but for superprofiles generated in radial annuli for the larger galaxies.
Blue indicates central regions, and red indicates outer regions.
The right-hand panels show the same measurements, but now for superprofiles in \sfrsd{} subregions.
In these panels, points are color-coded by the \sfrsd{} bin into which they fall, with black indicating high \sfrsd{} and white indicating low \sfrsd{}.
The vertical lines of points in the left-hand panels are an artifact of the \sfrsd{} binning; the horizontal jitter is due to a different distribution of \sfrsd{} values in each bin.
Both \scentral{} and \swing{} show trends with \sfrsd{}, though they are more pronounced in the radial superprofiles.
We also observe a lower limit to \scentral{} that increases with \sfrsd{}.
\label{resolved::fig:velocities-sigma-sfr}
}
\end{figure}

In Figure~\ref{resolved::fig:velocities-sigma-sfr} we show the behavior of \hi{} superprofile velocities as a function of \sfrsd{}.
The layout of the figure is the same as for Figure~\ref{resolved::fig:velocities-sigma-hi}.
We find that \sfrsd{} is correlated with both \scentral{} and \swing{} for superprofiles in the $r_{25}$, radial, and \sfrsd{} subregions.
The observed correlation between \sfrsd{} and \scentral{} in the \sfrsd{} subregions is strongly driven by the three highest \sfrsd{} bins.
When we calculate $r_S$ for the lower three bins ($\sfrsd{} < 10^{-3}$ M$_\odot{}$ \per{yr} \per[2]{kpc}), the correlations disappear entirely ($r_S = 0.02$ and $p_S$ = 0.86, compared to $r_s = 0.42$ and $p_s < 0.001$ for the full range).
At low \sfrsd{}, either \hi{} kinematics do not appear to be influenced by star formation, or the measurements of \sfrsd{} have insufficient accuracy to reveal a correlation.

Even though it is strongly driven by the high \sfrsd{} regions, the correlation between \sfrsd{} and \scentral{} is new compared to \citetalias{StilpGlobal}.
The local correlation was lost when averaged over the entire disk, as regions with both high and low star formation intensities were combined.
In this case, we have sampled the \hi{} kinematics and \sfrsd{} on local scales and therefore would expect a stronger correlation between these properties, if they are connected.
Interestingly, regions of with very low star formation intensity ($\sfrsd{} < 10^{-3}$ M$_\odot{}$ \per{yr} \per[2]{kpc}) can still exhibit \scentral{} values of $\sim 10$ \kms{}, comparable to those seen in higher \sfrsd{} regions.
The \scentral{} and \swing{} again show a lower bound at a given \sfrsd{}, similar to that seen with \baryonsd{}.

\begin{figure}
\centering
\includegraphics[width=6.5in]{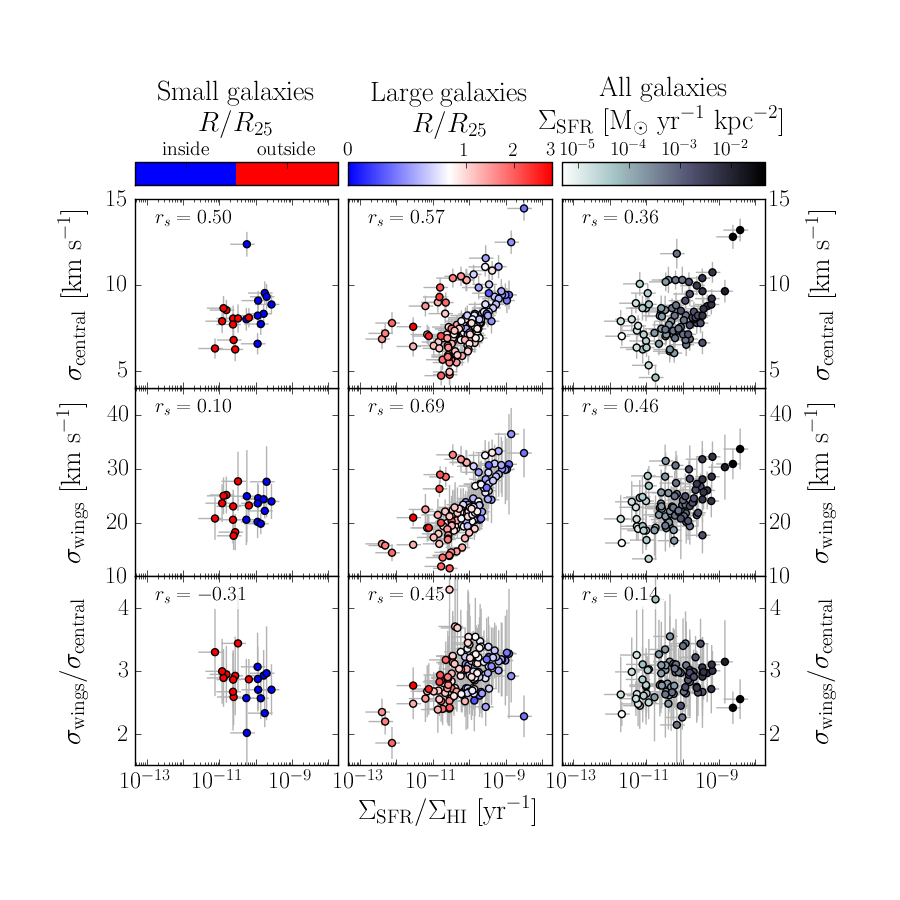}
\caption[\hi{} Velocities versus $\sfrsd{} / \hisd{}$]{\hi{} velocity measurements versus $\sfrsd{} / \hisd{}$ for spatially-resolved superprofiles.
Each point indicates the $\sfrsd{} / \hisd{}$ and \hi{} velocity measurement derived within one independent superprofile.
The left-hand panels show \scentral{}, \swing{}, and $\swing{} / \scentral{}$ for superprofiles in generated inside and outside $r_{25}$, for the smaller galaxies. The middle panels show the same parameters, but for superprofiles generated in radial annuli for the larger galaxies.
Blue indicates central regions, and red indicates outer regions.
The right-hand panels show the same measurements, but now for superprofiles in \sfrsd{} subregions.
In these panels, points are color-coded by the \sfrsd{} bin into which they fall, with black indicating high \sfrsd{} and white indicating low \sfrsd{}.
Both \scentral{} and \swing{} show trends with $\sfrsd{} / \hisd{}$.
\label{resolved::fig:velocities-sfe}
}
\end{figure}

We next compare the \hi{} velocities to $\sfrsd{} / \hisd{}$ in Figure~\ref{resolved::fig:velocities-sfe}.
As stated before, this measurement provides an estimate of the energy available from star formation per unit gas mass, rather than a strict star formation efficiency measurement.
Again, we find that \scentral{} and \swing{} are correlated with $\sfrsd{} / \hisd{}$, but with weaker correlation coefficients compared to \sfrsd{}.
The connection between star formation and \scentral{} is not one-to-one; \scentral{} values at low $\sfrsd{} / \hisd{}$ can be as high as 10 \kms{}.
Additionally, \scentral{} exhibits an increasing lower bound with increasing $\sfrsd{} / \hisd{}$, which could imply a minimum efficiency at which star formation affects the surrounding \hi{}.

\subsection{Superprofile Shapes}
\label{resolved::sec:correlations--shape}

We now address whether the shapes of the spatially-resolved superprofiles are connected to local ISM properties.
In particular, we examine the fraction of gas in the wings of the superprofiles (\fw{}) and the asymmetry of the superprofiles (\afull{}).
In \citetalias{StilpGlobal}, we found that \fw{} increased with $\mathrm{SFR} / \mhi{}$, averaged on global scales, and that the asymmetry of the profiles, measured with \aw{}, decreased with increasing total SFR and galaxy mass.
We now assess whether similar correlations hold for superprofiles generated in radial bins or regions of similar \sfrsd{}.
As with the \hi{} velocities discussed in \S~\ref{resolved::sec:correlations--v}, we examine correlations both with surface mass density (\S~\ref{resolved::sec:correlations--shape--surface-mass}) and with measures of star formation (\S~\ref{resolved::sec:correlations--shape--sfr}).

We note that the \fw{} and \afull{} parameters may not be tracing the same effects in the resolved superprofiles as in the global superprofiles from \citetalias{StilpGlobal}.
In the global superprofiles, line-of-sight spectra with a range of velocity dispersions were averaged together.
The final superprofile was therefore a composite of \hi{} profiles with different velocity dispersions.
Some \hi{} in the wings of those superprofiles could have been due to \hi{} line-of-sight spectra with larger velocity dispersions compared to the average.
Second, we measure the full asymmetry of the superprofiles (\afull{}) in this paper instead of the asymmetry of the wings (\aw{}).
This difference is unlikely to be major, since the measurement of asymmetry in the global superprofiles was strongly weighted by the wings.

\subsubsection{Surface Mass Density}
\label{resolved::sec:correlations--shape--surface-mass}

We start by examining correlations between superprofile shape and surface mass density.
We note that in \citetalias{StilpGlobal}, we found no correlations between surface density and either \fw{} or the wing asymmetry (as measured with \aw{}), but any local correlations may have been obscured when averaging over the entire disk.

\begin{figure}
\centering
\includegraphics[width=6.5in]{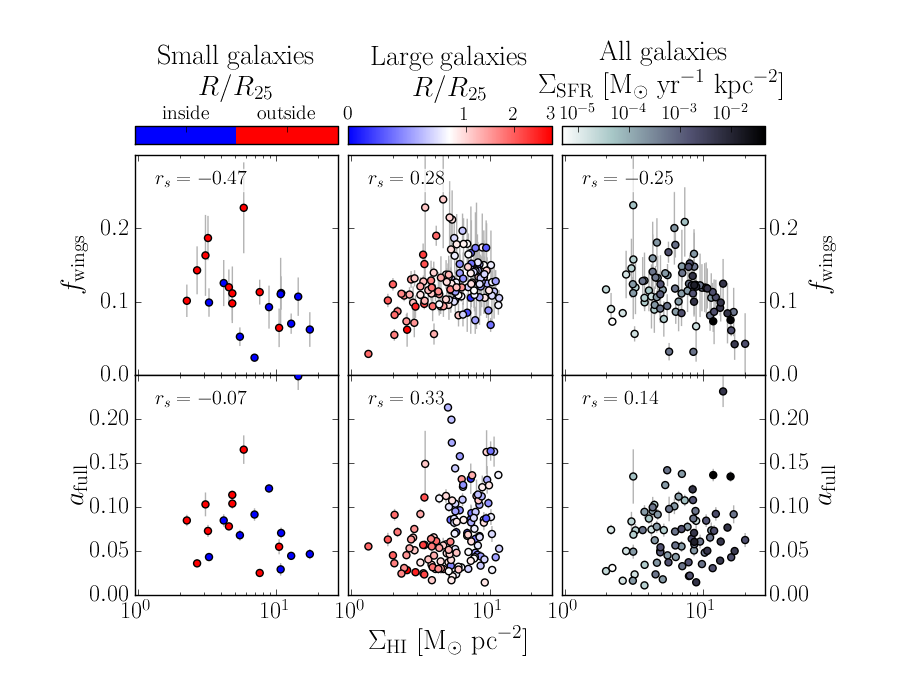}
\caption[Superprofile shape parameters versus \sfrsd{}]{\hi{} superprofile shape parameters versus \hisd{} for spatially-resolved superprofiles.
Each point indicates the \hisd{} measurement and superprofile shape parameter derived for one independent superprofile.
The left-hand panels show \fw{} and \afull{} for superprofiles in generated inside and outside $r_{25}$, for the smaller galaxies. The middle panels show the same parameters, but for superprofiles generated in radial annuli for the larger galaxies.
Blue indicates central regions, and red indicates outer regions.
The right-hand panels show the same measurements, but now for superprofiles in \sfrsd{} subregions.
In these panels, points are color-coded by the \sfrsd{} bin into which they fall, with black indicating high \sfrsd{} and white indicating low \sfrsd{}.
\label{resolved::fig:shapes-sigma-hi}
}
\end{figure}

In Figure~\ref{resolved::fig:shapes-sigma-hi} we show the superprofile shape parameters as a function of \hisd{}.
The layout is similar to Figure~\ref{resolved::fig:velocities-sigma-hi}, but we now plot \fw{} and \afull{} for superprofiles derived in subregions of constant radius or \sfrsd{}.
Very little structure is apparent in either \fw{} or \afull{} with increasing \hisd{}.
We note that \afull{} has a $p$-value $p_S < 0.01$ for the radial subregions, but the corresponding correlation coefficient is very low ($r_S = 0.26$), indicating that the two quantities are not strongly correlated.
Moreover, the correlation coefficients for the $r_{25}$ subregions, in smaller galaxies, have the opposite sign than that of the radial subregions for larger galaxies.
The two quantities are not significantly correlated for the \sfrsd{} subregions.

\begin{figure}
\centering
\includegraphics[width=6.5in]{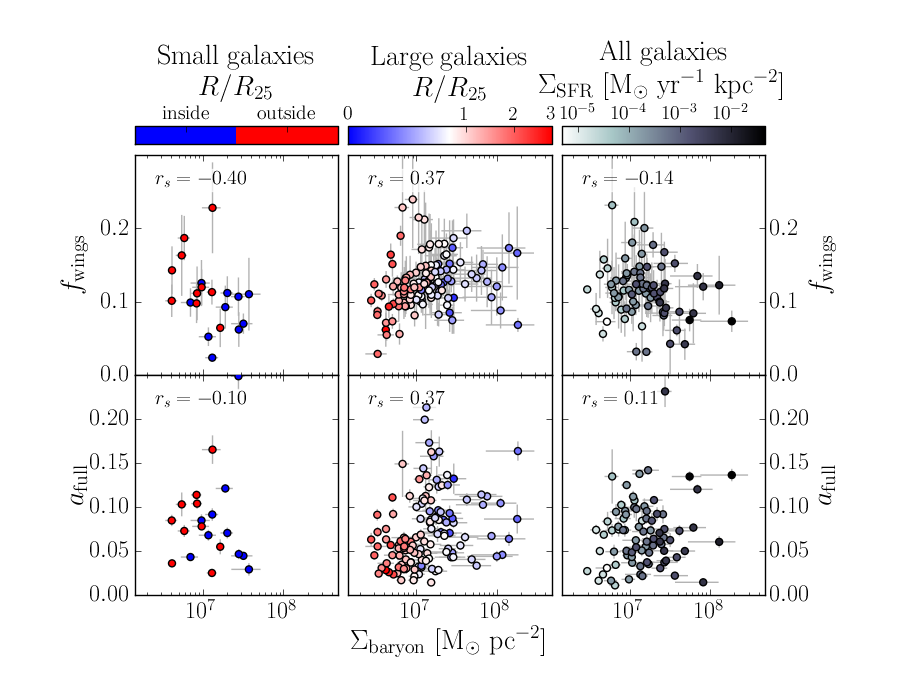}
\caption[Superprofile shape parameters versus \baryonsd{}]{\hi{} superprofile shape parameters versus \baryonsd{} for spatially-resolved superprofiles.
Each point indicates the \baryonsd{} measurement and \hi{} superprofile parameter derived for one independent superprofile.
The left-hand panels show \fw{} and \afull{} for superprofiles in generated inside and outside $r_{25}$, for the smaller galaxies. The middle panels show the same parameters, but for superprofiles generated in radial annuli for the larger galaxies.
Blue indicates central regions, and red indicates outer regions.
The right-hand panels show the same measurements, but now for superprofiles in \sfrsd{} subregions.
In these panels, points are color-coded by the \sfrsd{} bin into which they fall, with black indicating high \sfrsd{} and white indicating low \sfrsd{}.
\label{resolved::fig:shapes-sigma-baryon}
}
\end{figure}

In Figure~\ref{resolved::fig:shapes-sigma-baryon}, we show the same figure but for \baryonsd{} instead of \hisd{}.
As with \hisd{}, the superprofile shapes are not strongly influenced by \baryonsd{}.
The radial subregions of larger galaxies show significant correlations as measured by their $p_S$ value, but as before, the corresponding correlation coefficients themselves are low and have the opposite sign of the $r_{25}$ subregions of smaller galaxies.
The correlations disappear entirely for the \sfrsd{} subregions.
Because neither \hisd{} nor \baryonsd{} show strong or consistent correlations with superprofile shape, surface mass density does not drive the properties of the velocity of the wings.

\subsubsection{Star Formation}
\label{resolved::sec:correlations--shape--sfr}

We now compare the superprofile shapes with measures of star formation. As before, we examine correlations with both \sfrsd{} and $\sfrsd{} / \hisd{}$.

\begin{figure}
\centering
\includegraphics[width=6.5in]{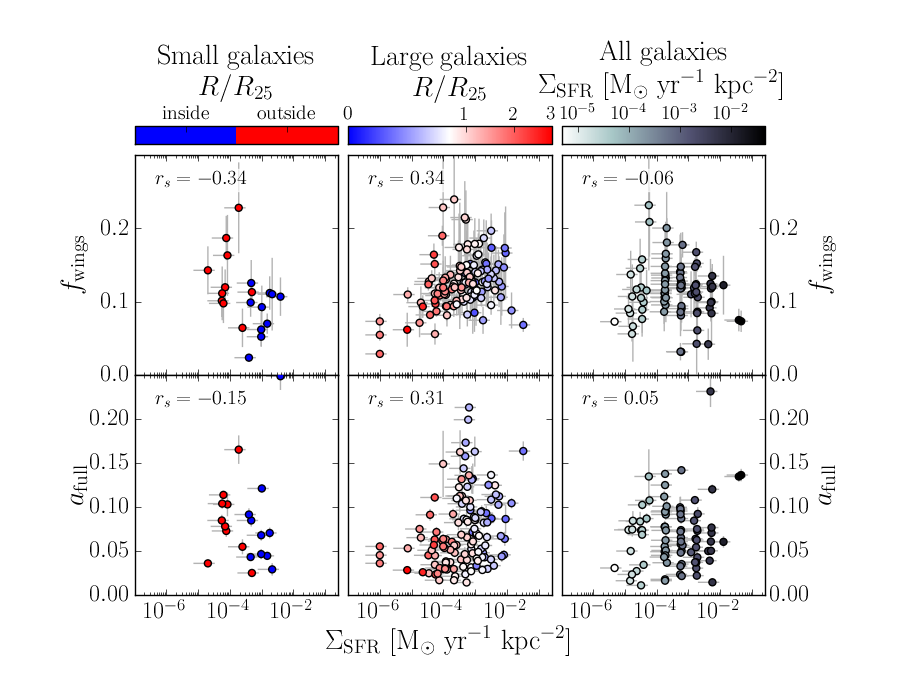}
\caption[Superprofile shape parameters versus \sfrsd{}]{\hi{} superprofile shape parameters versus \sfrsd{} for spatially-resolved superprofiles.
Each point indicates the \sfrsd{} measurement and \hi{} superprofile parameter derived for one independent superprofile.
The left-hand panels show \fw{} and \afull{} for superprofiles in generated inside and outside $r_{25}$, for the smaller galaxies. The middle panels show the same parameters, but for superprofiles generated in radial annuli for the larger galaxies.
Blue indicates central regions, and red indicates outer regions.
The right-hand panels show the same measurements, but now for superprofiles in \sfrsd{} subregions.
In these panels, points are color-coded by the \sfrsd{} bin into which they fall, with black indicating high \sfrsd{} and white indicating low \sfrsd{}.
The vertical lines of points are an artifact of the \sfrsd{} binning; the horizontal jitter is due to a different distribution of \sfrsd{} values in each bin.
\label{resolved::fig:shapes-sigma-sfr}
}
\end{figure}

In Figure~\ref{resolved::fig:shapes-sigma-sfr}, we show \fw{} and \afull{} for superprofiles generated in radial and \sfrsd{} subregions.
The layout is the same as Figure~\ref{resolved::fig:shapes-sigma-hi}.
The \hi{} superprofiles tend to have $\fw{} \sim 0.1$, which suggests that average \hi{} spectra are typically non-Gaussian in most ISM conditions.

Again, there are no strong correlations between these parameters and \sfrsd{}, indicating that there is no monotonic relationship between superprofile shape and star formation intensity.
For the superprofiles from constant radial annuli or \sfrsd{}, the smallest values of \fw{} and \afull{} do appear to occur in regions of low star formation intensity ($\sfrsd{} < 10^4$ M$_\odot{}$ \per{yr} \per[2]{kpc}), with higher values at moderate \sfrsd{}.
This trend is reversed in the smaller galaxies, which have superprofiles inside and outside $r_{25}$.
We also observe a tentative upper bound to \fw{} that decreases with increasing \sfrsd{}, but this behavior may reflect the lower bound between \scentral{} and \sfrsd{} (\S~\ref{resolved::sec:correlations--v--sf}).

\begin{figure}
\centering
\includegraphics[width=6.5in]{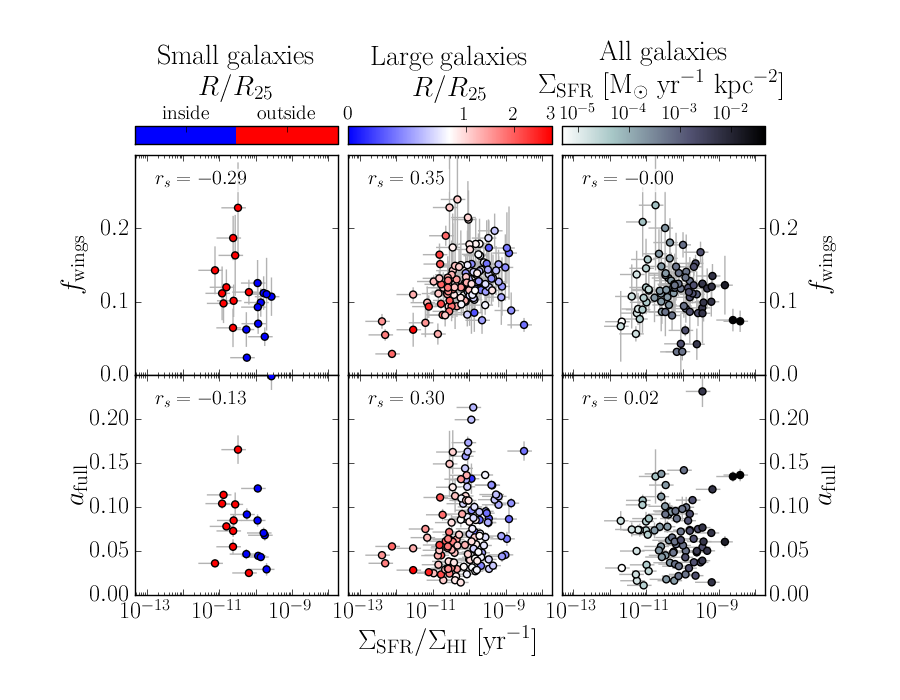}
\caption[Superprofile shape parameters versus $\sfrsd{} / \hisd{}$]{\hi{} superprofile shape parameters versus $\sfrsd{} / \hisd{}$ for spatially-resolved superprofiles.
Each point indicates the $\sfrsd{} / \hisd{}$ measurement and \hi{} superprofile parameter derived for one independent superprofile.
The left-hand panels show \fw{} and \afull{} for superprofiles in generated inside and outside $r_{25}$, for the smaller galaxies. The middle panels show the same parameters, but for superprofiles generated in radial annuli for the larger galaxies.
Blue indicates central regions, and red indicates outer regions.
The right-hand panels show the same measurements, but now for superprofiles in \sfrsd{} subregions.
In these panels, points are color-coded by the \sfrsd{} bin into which they fall, with black indicating high \sfrsd{} and white indicating low \sfrsd{}.
\label{resolved::fig:shapes-sfe}
}
\end{figure}

In Figure~\ref{resolved::fig:shapes-sfe}, we show the superprofile shapes versus $\sfrsd{} / \hisd{}$.
As with the previous figures, there are no obvious trends between these parameters and $\sfrsd{} / \hisd{}$.
A similar upper bound for \fw{} as a function of $\sfrsd{} / \hisd{}$ is present, but it could again be an artifact of the lower bound in the \scentral{} versus $\sfrsd{} / \hisd{}$ figure.

\ifthesis
\input{\tablepath/table-correlations-radial}
\input{\tablepath/table-correlations-sfr}
\else
\begin{deluxetable}{lcc|cc|cc|cc}
\tablewidth{0pt}
\tablecaption{Correlation coefficients for superprofiles derived in subregions of constant radius, for the large galaxies only. \label{tab:resolved--correlations-radial}} 
\tablehead{
  \colhead{} &
  \multicolumn{2}{c}{ \scentral{}} &
  \multicolumn{2}{c}{ \swing{} } &
  \multicolumn{2}{c}{ \fw{} } &
  \multicolumn{2}{c}{ \afull{} } \\

  \colhead{Property} &
  \colhead{$r_s$} &
  \colhead{$p_s$} &
  \colhead{$r_s$} &
  \colhead{$p_s$} &
  \colhead{$r_s$} &
  \colhead{$p_s$} &
  \colhead{$r_s$} &
  \colhead{$p_s$}

}

\startdata
$\Sigma_\mathrm{HI}$  &  \textbf{0.530}  &  \textbf{$<$0.001}  &  \textbf{0.675}  &  \textbf{$<$0.001}  &  \textbf{0.282}  &  \textbf{0.001}  &  \textbf{0.333}  &  \textbf{$<$0.001}  \\  
$\Sigma_\mathrm{baryon}$  &  \textbf{0.599}  &  \textbf{$<$0.001}  &  \textbf{0.745}  &  \textbf{$<$0.001}  &  \textbf{0.367}  &  \textbf{$<$0.001}  &  \textbf{0.373}  &  \textbf{$<$0.001}  \\  
$\sfrsd{}$  &  \textbf{0.612}  &  \textbf{$<$0.001}  &  \textbf{0.732}  &  \textbf{$<$0.001}  &  \textbf{0.336}  &  \textbf{$<$0.001}  &  \textbf{0.315}  &  \textbf{$<$0.001}  \\  
$\Sigma_\mathrm{SFR} / \Sigma_\mathrm{HI}$  &  \textbf{0.571}  &  \textbf{$<$0.001}  &  \textbf{0.695}  &  \textbf{$<$0.001}  &  \textbf{0.346}  &  \textbf{$<$0.001}  &  \textbf{0.299}  &  \textbf{$<$0.001}  \\  
\enddata

\tablecomments{Spearman correlation coefficient $r_S$ and probability $p_S$ between superprofile parameters and physical properties for the superprofiles in regions of constant radius, for the large galaxies only. Significant correlations are shown in bold.}
\end{deluxetable}

\begin{deluxetable}{lcc|cc|cc|cc}
\tablewidth{0pt}
\tablecaption{Correlation coefficients for superprofiles derived in subregions of constant radius, for the small galaxies only. \label{tab:resolved--correlations-r25}} 
\tablehead{
  \colhead{} &
  \multicolumn{2}{c}{ \scentral{}} &
  \multicolumn{2}{c}{ \swing{} } &
  \multicolumn{2}{c}{ \fw{} } &
  \multicolumn{2}{c}{ \afull{} } \\

  \colhead{Property} &
  \colhead{$r_s$} &
  \colhead{$p_s$} &
  \colhead{$r_s$} &
  \colhead{$p_s$} &
  \colhead{$r_s$} &
  \colhead{$p_s$} &
  \colhead{$r_s$} &
  \colhead{$p_s$}

}

\startdata
$\Sigma_\mathrm{HI}$  &  \textbf{0.565}  &  \textbf{0.009}  &  0.438  &  0.054  &  -0.469  &  0.037  &  -0.068  &  0.777  \\  
$\Sigma_\mathrm{baryon}$  &  \textbf{0.600}  &  \textbf{0.005}  &  0.514  &  0.020  &  -0.397  &  0.083  &  -0.096  &  0.686  \\  
$\sfrsd{}$  &  0.543  &  0.013  &  0.226  &  0.339  &  -0.341  &  0.141  &  -0.146  &  0.539  \\  
$\Sigma_\mathrm{SFR} / \Sigma_\mathrm{HI}$  &  0.499  &  0.025  &  0.104  &  0.663  &  -0.289  &  0.217  &  -0.126  &  0.596  \\  
\enddata

\tablecomments{Spearman correlation coefficient $r_S$ and probability $p_S$ between superprofile parameters and physical properties for the superprofiles generated inside and outside $r_{25}$, for the small galaxies only. Significant correlations are shown in bold.}
\end{deluxetable}

\begin{deluxetable}{lcc|cc|cc|cc}
\tablewidth{0pt}
\tablecaption{Correlation coefficients for superprofiles derived in subregions of constant radius, for all sample galaxies. \label{tab:resolved--correlations-radial-all}} 
\tablehead{
  \colhead{} &
  \multicolumn{2}{c}{ \scentral{}} &
  \multicolumn{2}{c}{ \swing{} } &
  \multicolumn{2}{c}{ \fw{} } &
  \multicolumn{2}{c}{ \afull{} } \\

  \colhead{Property} &
  \colhead{$r_s$} &
  \colhead{$p_s$} &
  \colhead{$r_s$} &
  \colhead{$p_s$} &
  \colhead{$r_s$} &
  \colhead{$p_s$} &
  \colhead{$r_s$} &
  \colhead{$p_s$}

}

\startdata
$\Sigma_\mathrm{HI}$  &  \textbf{0.537}  &  \textbf{$<$0.001}  &  \textbf{0.638}  &  \textbf{$<$0.001}  &  0.140  &  0.090  &  \textbf{0.263}  &  \textbf{0.001}  \\  
$\Sigma_\mathrm{baryon}$  &  \textbf{0.590}  &  \textbf{$<$0.001}  &  \textbf{0.714}  &  \textbf{$<$0.001}  &  \textbf{0.251}  &  \textbf{0.002}  &  \textbf{0.309}  &  \textbf{$<$0.001}  \\  
$\sfrsd{}$  &  \textbf{0.577}  &  \textbf{$<$0.001}  &  \textbf{0.678}  &  \textbf{$<$0.001}  &  \textbf{0.242}  &  \textbf{0.003}  &  \textbf{0.250}  &  \textbf{0.002}  \\  
$\Sigma_\mathrm{SFR} / \Sigma_\mathrm{HI}$  &  \textbf{0.518}  &  \textbf{$<$0.001}  &  \textbf{0.626}  &  \textbf{$<$0.001}  &  \textbf{0.271}  &  \textbf{$<$0.001}  &  \textbf{0.234}  &  \textbf{0.004}  \\  
\enddata

\tablecomments{Spearman correlation coefficient $r_S$ and probability $p_S$ between superprofile parameters and physical properties for the superprofiles in regions of constant radius, for all sample galaxies. Significant correlations are shown in bold.}
\end{deluxetable}

\begin{deluxetable}{lcc|cc|cc|cc}
\tablewidth{0pt}
\tablecaption{Correlation coefficients for superprofiles derived in subregions of constant \sfrsd{}. \label{tab:resolved--correlations-sfr}} 
\tablehead{
  \colhead{} &
  \multicolumn{2}{c}{ \scentral{}} &
  \multicolumn{2}{c}{ \swing{} } &
  \multicolumn{2}{c}{ \fw{} } &
  \multicolumn{2}{c}{ \afull{} } \\

  \colhead{Property} &
  \colhead{$r_s$} &
  \colhead{$p_s$} &
  \colhead{$r_s$} &
  \colhead{$p_s$} &
  \colhead{$r_s$} &
  \colhead{$p_s$} &
  \colhead{$r_s$} &
  \colhead{$p_s$}

}

\startdata

$\Sigma_\mathrm{HI}$  &  \textbf{0.466}  &  \textbf{$<$0.001}  &  \textbf{0.507}  &  \textbf{$<$0.001}  &  -0.249  &  0.025  &  0.137  &  0.224  \\  
$\Sigma_\mathrm{baryon}$  &  \textbf{0.484}  &  \textbf{$<$0.001}  &  \textbf{0.587}  &  \textbf{$<$0.001}  &  -0.142  &  0.205  &  0.107  &  0.341  \\  
$\sfrsd{}$  &  \textbf{0.422}  &  \textbf{$<$0.001}  &  \textbf{0.525}  &  \textbf{$<$0.001}  &  -0.058  &  0.605  &  0.052  &  0.645  \\  
$\Sigma_\mathrm{SFR} / \Sigma_\mathrm{HI}$  &  \textbf{0.358}  &  \textbf{0.001}  &  \textbf{0.463}  &  \textbf{$<$0.001}  &  -0.004  &  0.975  &  0.023  &  0.836  \\  

\enddata

\tablecomments{Spearman correlation coefficient $r_S$ and probability $p_S$ between superprofile parameters and physical properties for the \sfrsd{} superprofiles. Significant correlations are shown in bold.}
\end{deluxetable}

\fi

\section{Discussion}
\label{resolved::sec:discussion}

In \S~\ref{resolved::sec:correlations}, we calculated the correlations between superprofile parameters and local ISM properties.
The $r_S$ correlation coefficients and corresponding $p_S$ significance values for the correlations discussed in \S~\ref{resolved::sec:correlations} are listed in Table~\ref{tab:resolved--correlations-radial} for the radial subregions and in Table~\ref{tab:resolved--correlations-sfr} for the \sfrsd{} subregions.
In general, the correlations are weaker for the \sfrsd{} subregions compared to the radial or $r_{25}$ subregions.

We find that measures of \hi{} turbulent velocity, traced by \scentral{}, and wing velocity, traced by \swing{}, are correlated with all physical properties we have inspected (\hisd{}, \baryonsd{}, \sfrsd{}, and $\sfrsd{} / \hisd{}$).
The \hi{} turbulent velocities also tend to show a lower bound for a given \sfrsd{} or \baryonsd{}.
Individual galaxies can show smooth well-defined trends with increasing radius (Figure~\ref{resolved::fig:radial-panels}) or \sfrsd{} (Figure~\ref{resolved::fig:panels-sfr}), as seen previously in \citet{Tamburro2009}, but the galaxy-to-galaxy variation in the average value of these trends is large.

Because \swing{} and \scentral{} are correlated, we focus primarily on the correlations between the width of the central peak (\scentral{}) and local ISM properties.
The parameters describing the shapes of the superprofiles, \fw{} and \afull{}, do not show strong correlations with any of these properties, though there are weak correlations with measures of local SFR.

In this section, we discuss the physical meaning behind these correlations.
We focus primarily on correlations between the central peak of the superprofile and ISM properties, in \S~\ref{resolved::sec:discussion--v}.
We then briefly discuss correlations with \swing{}, \fw{}, and \afull{}, in \S~\ref{resolved::sec:discussion--shape}.

\subsection{The Central \hi{} peak}
\label{resolved::sec:discussion--v}

\hi{} turbulent velocities are often thought to be caused by feedback from star formation.
\citet{Tamburro2009} found that star formation provides enough energy to drive \hi{} turbulence inside $r_{25}$ for relatively massive disk galaxies, but that the relationship between \sfrsd{} and turbulence breaks down both in the outskirts of spiral galaxies or in dwarf galaxies where star formation rates are low.
For our sample, which is dominated by systems in the latter regime, we find that the width of the central peak of the \hi{} velocity dispersion correlates most strongly with \baryonsd{}, and less strongly with measures of star formation.
We first discuss the origins and limitations of the correlation of \scentral{} with \sfrsd{}, and then discuss the stronger correlation with \baryonsd{}.

\subsubsection{Turbulent \hi{} Kinematics and Star Formation}
\label{resolved::sec:discussion--v--sfr}

We find significant correlations between \scentral{} and \sfrsd{}, for both types of subregions (Figure~\ref{resolved::fig:velocities-sigma-sfr}, Tables~\ref{tab:resolved--correlations-radial} and \ref{tab:resolved--correlations-sfr}).
The observed \scentral{}-\sfrsd{} correlation is strongest in the radial subregions ($r_s = 0.61$), where ISM properties tend to track smoothly with radius.
The correlation is weaker for smaller galaxies (inside and outside $r_{25}$; $r_s = 0.54$), but the by-eye scatter of this correlation is relatively large.
For the \sfrsd{}-based subregions, the correlation between \scentral{} and \sfrsd{} is weaker and is only seen when the highest SFR intensity bins ($\sfrsd{} > 10^{-3}$ M$_\odot{}$ \per{yr} \per[2]{kpc}) are included.
The $\sfrsd{} < 10^{-3}$ M$_\odot{}$ \per{yr} \per[2]{kpc} threshold occurs at approximately $r_{25}$ in the largest dwarfs in our sample, as seen both in Figure~\ref{resolved::fig:velocities-sigma-sfr} and in \citet{Tamburro2009} for \hi{} kinematics inside $r_{25}$ in larger galaxies, so the fact that the correlations break down at around this \sfrsd{} mirrors the \citet{Tamburro2009} results.

These results indicate that at low \sfrsd{}, \hi{} kinematics do not appear to be strictly driven by star formation properties.
Superprofiles generated in regions of low star formation can show velocity dispersions as high as 10 \kms{}, values similar to their counterparts in regions of higher star formation.

Even at higher \sfrsd{}, the \hi{} velocity dispersions are not 
uniquely defined by \sfrsd{}.
The relationship between \scentral{} and \sfrsd{} has large scatter at low star formation intensities and inconsistent behavior within individual galaxies.
Although the more massive galaxies in the sample tend to show increasing \scentral{} values with increasing \sfrsd{}, the behavior of \scentral{} compared to \sfrsd{} is more erratic in the low mass dwarfs.
Some have no strong trends between \scentral{} and \sfrsd{} (e.g., DDO~154, Sextans~A, UGCA~292), while others show the complete opposite trend as seen in the more massive spirals (e.g., DDO~53), with decreasing \scentral{} with increasing \sfrsd{}.

There are a number of reasons why \scentral{} may not couple strongly to the local \sfrsd{}.
First, the discrepancy may be due to a mismatch between the timescale of the FUV+24\um{} star formation tracer and the timescale on which star formation energy is injected.
We assess this possibility in 
\ifthesis
Chapter~\ref{angst} of this thesis
\else
Paper III of this series \citep{StilpTimescales}
\fi
using time-resolved star formation histories.
It may also be due to a mismatch in physical scale.
If SF drives turbulence over an area larger than a subregion, then the turbulence and SFR may appear decoupled.
In such a case, we may see stronger trends for galaxy-wide superprofiles (as in \citetalias{StilpGlobal} or \citetalias{StilpTimescales}) where all the SF and turbulent energy is considered.
The discrepancy may also originate from the fact that star formation energy may couple more strongly to \hi{} energy instead of turbulent velocity alone.

While the local star formation intensity does not appear to set the \hi{} velocity dispersion, it does appear to provide a floor below which \scentral{} does not fall.
This behavior can be seen as the clear lower bound in Figure~\ref{resolved::fig:velocities-sigma-sfr}, which approximately follows
\begin{equation}
\scentral \sim 6.8 \; \mathrm{km \; s}^{-1} \left( \frac{ \sfrsd{} }{ 10^{-3} \, \msun{} \; \mathrm{yr}^{-1} \; \mathrm{kpc}^{-2} } \right) ^{0.14}.
\end{equation}
The local star formation intensity may influence the smallest allowed velocity dispersion in a region, even if it does not solely drive the \hi{} velocity dispersions.

\subsubsection{Turbulent \hi{} Energy}
\label{resolved::sec:discussion--v--hi-energy}

In addition to examining the correlations between \scentral{} and \sfrsd{}, we also compare the turbulent \hi{} energy to the energy provided by SF.
This comparison provides a more robust connection between ISM properties and SF because the star formation energy may couple at similar efficiencies in regions that have different \hi{} masses.

As in \citetalias{StilpGlobal}, we estimate the energy contained in the central \hi{} peak, $E_\mathrm{HI,central}$ as:
\begin{equation}
E_\mathrm{HI,central} = \frac{3}{2} M_\mathrm{SP} (1 - \fw{}) (1 - f_\mathrm{cold}) \sigma_\mathrm{central}^2.
\label{resolved::eqn:ehi-central}
\end{equation}
Here, $M_\mathrm{SP}$ is the total mass of the superprofile in \msun{}, given by $M_\mathrm{SP} = 2.36 \times 10^5 D^2 F_\mathrm{SP} \Delta v$, for a distance $D$ in Mpc and an integrated superprofile flux $F_\mathrm{SP} \Delta v$ in Jy \kms{}.
The factor of $3/2$ accounts for motion in three dimensions, assuming isotropy.
The total mass in the central peak is given by $M_\mathrm{SP} (1 - \fw{})$.
We exclude \hi{} in the wings of each superprofile because it does not follow the Gaussian velocity structure of the central peak.
We estimate that a fraction $f_\mathrm{cold} = 0.15$ of this mass is cold \hi{} and has kinematics that are poorly represented by \scentral{}.
The true amount and kinematics of cold \hi{} are unknown, but limits on $f_\mathrm{cold}$ in dwarf galaxies range between $1 - 20$\% \citeeg{Young2003, Warren2012}.
This value also matches the approximate fraction of cold \hi{} in the SMC \citep[$< 15$\%; ][]{Dickey2000}.
Because the kinematics of cold \hi{} are likely not turbulent and may not be connected to star formation in the same way as warm \hi{}, we exclude this estimated fraction of cold gas from our energy calculation.

If the \hi{} is turbulent, its energy can be dissipated approximately over one turbulent timescale, $\tau_D$, defined by \citet{MacLow1999} as:
\begin{equation}
\tau_D \simeq 9.8 \; \mathrm{Myr} 
\left( \frac{\lambda}{ 100 \; \mathrm{pc} } \right)
\left( \frac{ \sigma }{ 10 \; \mathrm{ km \; s^{-1}} } \right)^{-1}
\end{equation}
where $\lambda$ is the turbulent driving scale and $\sigma$ is the \hi{} velocity dispersion.
Following \citetalias{StilpGlobal}, we have approximated $\lambda = 100$ pc and $\sigma = \scentral{}$.
To maintain the observed turbulent kinetic energy, sufficient energy must be replenished over this timescale.

\subsubsection{Available Star Formation Energy}
\label{resolved::sec:discussion--v--sf-energy}

We now compare the turbulent kinetic \hi{} energy to that provided by star formation, $E_\mathrm{SF}$, over one turbulent timescale, $\tau_D$:
\begin{equation}
E_\mathrm{SF} = \eta_\mathrm{SN} \left( \mathrm{SFR} \times \tau_D \right) 10^{51} \; \mathrm{ergs},
\label{resolved::eqn:e-sf}
\end{equation}
Here, $\eta_\mathrm{SN}$ is the number of SN per unit stellar mass formed; we adopt $\eta_\mathrm{SN} = 1.3 \times 10^{-2}$ SN \per{\msun{}} based on a \citet{Kroupa2001} IMF with a 120 \msun{} upper mass limit.
This equation assumes that the star formation rate is constant across a $10 - 20$ Myr interval, so it may be problematic for dwarf galaxies, which often have non-uniform recent SFHs \citeeg{McQuinn2010, Weisz2011}.
It also assumes the SFR is averaged over the past $\sim10 - 100$ Myr, without allowing for variations in SFH.
Because star formation can provide energy for up to $\sim 50$ Myr after a burst, these assumption may not be strongly in error, but we refer the reader to the third paper in this series \citepalias{StilpTimescales}.
We also note that the stochastic sampling of the IMF and cluster mass function in regions of low star formation can introduce scatter into SFR estimates from FUV or \halpha{} tracers \citep{Weisz2012}.
Compared with \halpha{}, FUV estimates of star formation are more likely to provide correct estimates of the SFR as they rely on both O and B stars instead of only those with $M \gtrsim 15$ \msun{} \citep{Lee2011, Kennicutt2012}.

\subsubsection{Efficiency of Coupling Star Formation Energy to Kinetic \hi{} Energy}
\label{resolved::sec:discussion--v--sf-efficiency}

With estimates of both $E_\mathrm{HI,central}$ and $E_\mathrm{SF}$, we next derive the range of efficiencies that are consistent with our measurements.
If star formation is the only driver of \hi{} velocity dispersions, then the conversion efficiency from star formation energy to \hi{} kinetic energy is given by $\epsilon \equiv E_\mathrm{HI,central} / E_\mathrm{SF}$.
The actual efficiency could be lower if additional drivers also contribute to the kinetic \hi{} energy.
The conversion efficiency has been limited by a number of various studies, including those focusing on \hi{} holes \citeeg{Weisz2009, Warren2012} and simulations \citeeg{TenorioTagle1991, Thornton1998}, but estimates of $\epsilon$ vary widely in these studies, from as low as $< 1$\% to up to $50$\%.

\begin{figure}
\centering
\includegraphics[width=6.5in]{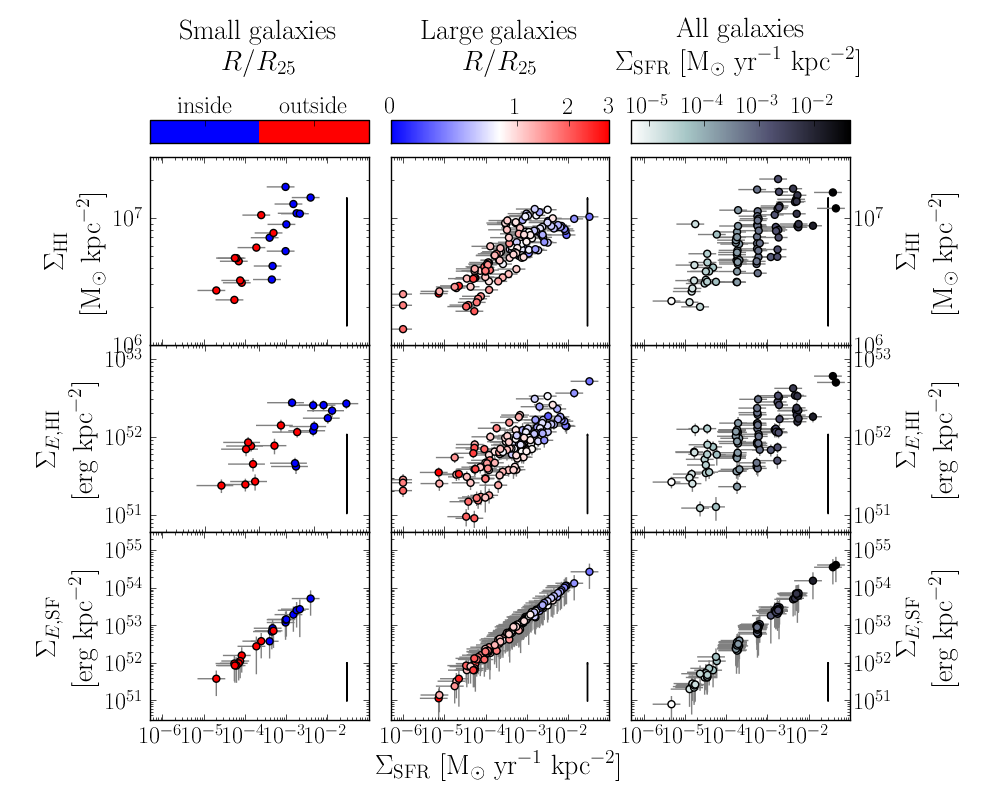}
\caption[\hisd{}, $\Sigma_\mathrm{E,HI}$, and $\Sigma_\mathrm{E,SF}$ as a function of \sfrsd{}]{ \hisd{} (upper panels), $\Sigma_\mathrm{E,HI}$ (middle panels), and $\Sigma_\mathrm{E,SF}$ (lower panels) as a function of \sfrsd{}.
Each point represents one spatially-resolved superprofile.
The left-hand panels show measurements for superprofiles in radial subregions (colored by normalized radius), and the right-hand panels show measurements for superprofiles in subregions of constant \sfrsd{} (colored by \sfrsd{}).
The vertical black line in the lower right corner of each panel represents one order of magnitude, indicating that $\Sigma_\mathrm{E,SF}$ covers a much larger range than either \hisd{} or $\Sigma_\mathrm{E,HI}$.
\label{resolved::fig:energy-sfrsd}
}
\end{figure}

We start by examining the behavior of $E_\mathrm{HI, central}$ and $E_\mathrm{SF}$ as a function of \sfrsd{} in Figure~\ref{resolved::fig:energy-sfrsd}.
For this figure, we normalize the energies by area, as larger radial annuli often have a larger area of contributing star formation but with lower intensities.
We plot \hisd{} (top panels), the energy surface density of the \hi{} central peak ($\Sigma_{E,\mathrm{HI, central}}$; middle panels), and the star formation energy surface density ($\Sigma_{E,\mathrm{SF}}$; bottom panels) as a function of \sfrsd{}.
The left-hand panels represent the superprofiles generated inside and outside $r_{25}$, for the smaller galaxies, while the middle panels show those in radial annuli for larger galaxies; points are colored by $r / r_{25}$.
The right-hand panels show the superprofiles generated in subregions of constant \sfrsd{}; points are colored by \sfrsd{}.

Figure~\ref{resolved::fig:energy-sfrsd} shows that $\Sigma_\mathrm{E,SF}$ and $\Sigma_\mathrm{E,HI}$ both increase with increasing \sfrsd{}, but their ranges span very different orders of magnitude.
To indicate this difference in range, we have also plotted a black vertical line indicating one order of magnitude in the lower right corner of each panel. 
The \hi{} energy surface density, which depends on both \hisd{} and \scentral{}, covers $\lesssim 2$ orders of magnitude, while the star formation energy density spans $\sim 4$ orders of magnitude. 
If star formation couples to \hi{} with a universal efficiency, then we expect both $\Sigma_\mathrm{E,SF}$ and $\Sigma_\mathrm{E,HI}$ to span approximately the same number of orders of magnitude.

\begin{figure}
\centering
\includegraphics[width=6.5in]{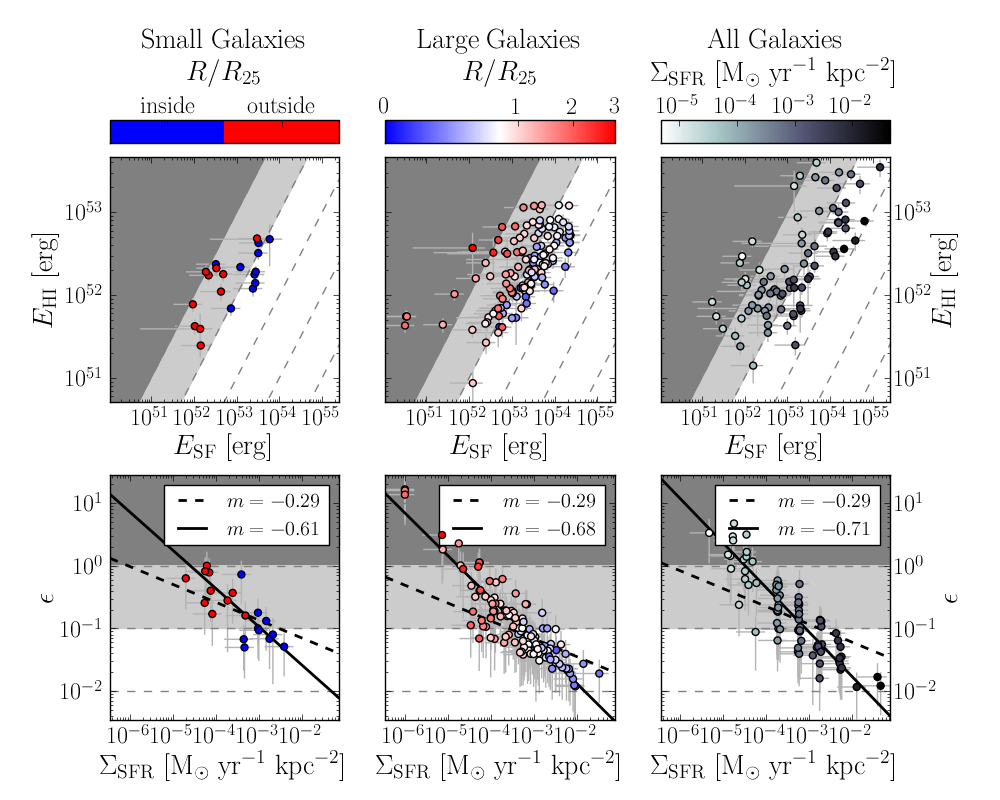}
\caption[Comparison of Energies and Coupling Efficiencies for Resolved Superprofiles]{ Comparison of \hi{} and SF energies and coupling efficiencies for the spatially-resolved sample.
The upper panels show $E_\mathrm{HI}$ versus $E_\mathrm{SF}$, and the lower panels show $\epsilon \equiv E_\mathrm{HI} /E_\mathrm{SF}$.
The left panels represent measurements from the superprofiles in radial subregions (colored by normalized radius), and the right panels show measurements from superprofiles in subregions of constant \sfrsd{} (colored by \sfrsd{}).
The shaded regions represent regions where $\epsilon > 1$ (dark grey) and $\epsilon > 0.1$ (light grey).
In the lower panels, we have overlaid the scaling relation between $\epsilon$ and \sfrsd{} from simulations by \citet{Joung2009} as the thick dashed line.
We also overlay the best-fit scaling relation to our data as the black solid line.
In our data, $\epsilon$ has a steeper dependence on \sfrsd{} than in the \citet{Joung2009} simulations, which were for higher star formation intensities and more massive galaxies.
\label{resolved::fig:energy-efficiency}
}
\end{figure}

We can recast the data from Figure~\ref{resolved::fig:energy-sfrsd} by examining the correlation between $E_\mathrm{SF}$ and $E_\mathrm{HI}$ directly.
We plot this comparison in Figure~\ref{resolved::fig:energy-efficiency} for both the correlation between $E_\mathrm{SF}$ and $E_\mathrm{HI}$ and for the inferred energy coupling efficiency $\epsilon \equiv E_\mathrm{HI} / E_\mathrm{SF}$.
Each point represents a single superprofile, and points are colored by $r / r_\mathrm{25}$ for the $r_{25}$ and radial subregions (left and middle panels) and by \sfrsd{} for the \sfrsd{} subregions (right panels).
The top panels show $E_\mathrm{HI}$ versus $E_\mathrm{SF}$, while the bottom panels show $\epsilon$ versus \sfrsd{}.
Unphysical or theoretically-forbidden efficiencies ($\epsilon > 1$ or $\epsilon > 0.1$) are shown in dark or light grey, and lines of constant efficiency are shown as dashed grey lines.

The data in Figure~\ref{resolved::fig:energy-efficiency} clearly show that the coupling efficiency $\epsilon$ is lowest at the high SFR intensities characteristic of the inner disks of galaxies.
At low SFR intensities ($\sfrsd{} \lesssim 10^{-4}$), $\epsilon$ often has theoretically-forbidden \citep[$\epsilon > 0.1$][]{Thornton1998} or unphysical values ($\epsilon > 1$).
If star formation were the sole driver of \hi{} velocity dispersions, we might expect to see efficiencies that were constant.
Instead, the strong apparent increase of efficiency with decreasing \sfrsd{} may simply be due to the decrease of star formation energy with \sfrsd{} while \hi{} energy remains relatively constant due to another driving mechanism unrelated to SF.
It also seems unlikely that the efficiencies conspire to produce the observed well-defined trend.

The observed trend of decreasing efficiencies with increasing \sfrsd{} has also been seen in numerical simulations by \citet{Joung2009}\footnote{The exponent on this scaling is different than that published in \citet{Joung2009}, but the author agreed with the change (Joung, private communication).}, who found that $\epsilon \propto \Sigma_\mathrm{SFR}^{-0.29}$.
We show this scaling as the thick dashed black line in the lower panels of Figure~\ref{resolved::fig:energy-efficiency}, with the caveat that we have normalized the relation to match our data range.
The \citet{Joung2009} scaling relation from the simulations is shallower than the observed relationship between \sfrsd{} and $\epsilon$ in our sample.
We also overlay the linear regression to the data as the solid black line, as given by $\log{\epsilon} = m \log{ ( \sfrsd{} / \Sigma_0 ) } + b$;
 the best-fit regression corresponds to 
$\epsilon = ( \Sigma_\mathrm{SFR} / \Sigma_0 )^{-0.61 \pm 0.09}$,
$\epsilon = ( \Sigma_\mathrm{SFR} / \Sigma_0 )^{-0.68 \pm 0.03}$,
and
$\epsilon = ( \Sigma_\mathrm{SFR} / \Sigma_0 )^{-0.71 \pm 0.04}$
for the superprofiles generated in subregions of constant radius or star formation intensity, respectively.
The normalization is 
$\Sigma_0 = (5.8 \pm 1.3) \times 10^{-4}$ \msun{} \per{yr} \per[2]{kpc} for the $r_{25}$ subregions; 
$\Sigma_0 = (6.0 \pm 1.3) \times 10^{-4}$ \msun{} \per{yr} \per[2]{kpc} for the radial annuli; and 
$\Sigma_0 = (6.4 \pm 1.8) \times 10^{-4}$ \msun{} \per{yr} \per[2]{kpc} for the constant \sfrsd{} subregions.
For all types of subregion, the relation between \sfrsd{} and $\epsilon$ is steeper than the relationship found by \citet{Joung2009}.
This discrepancy may exist because the \citet{Joung2009} simulations focused on more intense star formation regimes than our sample, spanning $1 - 512$ times the SFR of the Milky Way.
In addition, the scaling relations adopted in the \citet{Joung2009} simulations may not be appropriate for lower-mass galaxies like those in our sample.
For example, \citet{Joung2009} assumed that $\sfrsd{} \propto \Sigma_\mathrm{gas}^{1.4}$, a value appropriate for the larger spirals in their simulations, but in the lower-mass SMC, \citet{Bolatto2011} found a much steeper relation of $\sfrsd{} \propto \Sigma_\mathrm{gas}^{2.2}$, a scaling more appropriate for our low-mass sample.
In spite of the different SFR regimes and input assumptions, the observed behavior of declining efficiency with increasing \sfrsd{} is still qualitatively similar to the \citet{Joung2009} simulations.

A declining coupling efficiency with increasing \sfrsd{} is not theoretically unreasonable.
The ISM is multiphase, and the neutral medium we trace here is probably less directly coupled to star formation than the dense molecular phase, which precedes star formation, or the hotter ionized phase, which is an immediate result of star formation.
For example, if the energy of star formation, in the form of stellar winds and SN feedback, goes primarily into shock heating and producing ionized outflows, then any coupling to the neutral medium must occur primarily on the interfaces between the expanding SN-driven bubbles and the neutral gas reservoir, suggesting a relatively weak coupling consistent with what is seen here.
At high enough \sfrsd{}, SN-driven outflows may be strong enough to break out of the galactic disk, allowing more of the star formation energy to escape \citeeg{Dekel1986}.
Similarly, \citet{Monaco2004} find different efficiencies are associated with SNe that are contained within the disk compared with those that undergo blowout.
In addition, \citet{Lopez2011} have shown that hot gas in the intense star forming region 30 Doradus can leak out of the surrounding shell, consistent with this picture.

The star formation energy may also couple more strongly to the warm ionized phase of the ISM.
Indeed, studies of \halpha{} in dwarf galaxies indicate that the ionized gas shows both larger velocity dispersions and a larger dynamic range compared with \hi{} \citeeg{vanEymeren2010, Moiseev2012}.
For example, Sextans~A contains a shell of ionized gas with expansion velocities of $\sim 60$ \kms{}, much higher than surrounding \hi{} velocity dispersions \citep{Martin1998}.
One could imagine that as the SFR intensity decreases, a smaller fraction of the SN-driven bubbles actively undergo blowout, thus guaranteeing that most of the SN energy is deposited in the disk and increasing the coupling efficiency between star formation and \hi{} energies.
Indeed, extraplanar diffuse ionized gas has been seen primarily in spirals whose disks exceed a given \sfrsd{} threshold \citep{Rossa2003}, consistent with the picture that a stronger coupling to the ionized phase (and therefore a weaker coupling to \hi{}) exists at higher \sfrsd{}.

The turbulent dissipation timescale in dwarf galaxies could potentially be longer than the fiducial $\sim 10$~Myr found by \citet{MacLow1999}.
At low SFRs, SNe are sparsely distributed across the disk, implying that the driving scale of turbulence in these systems could be larger as remnants could propagate farther through the disk.
The increased scale heights of dwarf galaxies compared to larger spiral disks mean that SNe are still contained within the disk to larger radii than in more massive galaxies.
The scale height therefore provides an upper limit to the driving scale, as energy can escape from the disk once the radius of the SNe is equal to the scale height.
The scale heights of dwarf galaxies are a few times larger than more massive spiral counterparts \citeeg{Banerjee2011}.
This difference only provides a factor of a few increase in the energy available from star formation, which is not enough to compensate for implied efficiencies of $\epsilon \gtrsim 1$.

If star formation has nothing to do with setting turbulent \hi{} velocity dispersions, we would also expect to see declining efficiencies with increasing \sfrsd{}.
For a fixed \hi{} turbulent energy, we would expect $\epsilon \propto \Sigma_\mathrm{E,SF}^{-1}$, implying a similar scaling between $\epsilon$ and \sfrsd{}.
This scaling is steeper than the observed $\epsilon \propto \Sigma_\mathrm{SFR}^{-0.64}$ from Figure~\ref{resolved::fig:energy-efficiency}, so it is likely that star formation does have some influence on \hi{} turbulent velocity dispersions.
However, the efficiencies reach unphysically high values at low star formation intensities, indicating that \sfrsd{} cannot be the sole driver of \hi{} velocity dispersions.
This scaling is also hampered by the fact that local ISM conditions also scale with star formation; for example, \hisd{} and \sfrsd{} are correlated.
We explore other drivers of turbulence in \S~\ref{resolved::sec:discussion--correlations--surface-mass-density} and \ref{resolved::sec:discussion--correlations--other-drivers}.

\subsubsection{Turbulent \hi{} Kinematics and Surface Mass Density}
\label{resolved::sec:discussion--correlations--surface-mass-density}

Although the \hi{} velocity dispersion correlates with the local \sfrsd{} at high star formation intensities, it is even more strongly correlated with \baryonsd{} (Tables~\ref{tab:resolved--correlations-radial} and \ref{tab:resolved--correlations-sfr}).
The mismatch between star formation energy and \hi{} energy, especially at low star formation intensities, supports the idea that \hi{} velocity dispersions are driven at least partially by other processes.

The fact that \scentral{} correlates most strongly with \baryonsd{} may indicate that \hi{} velocity dispersions are driven by some type of gravitational instability in the low \sfrsd{}, low \baryonsd{} regime that dominates our sample.
However, the solid-body rotation curves of low-mass galaxies do not provide the large shearing motions that are thought to be necessary to drive most gravitational instabilities \citeeg{Kim2007, Agertz2009}.
A number of these gravitational instabilities also require the presence of strong spiral arms \citeeg{Roberts1969, Lin1964}, which are likewise not present in the majority of our sample.
In addition, our earlier analysis of the global kinematics \citepalias{StilpGlobal} showed that one of these gravitational instabilities, as outlined by \citet{Wada2002}, does not provide enough energy to drive \hi{} velocity dispersions.

Simple energy scaling arguments for gravitational instabilities suggests that \scentral{} should increase as $\Sigma_\mathrm{baryon}^{0.5}$ for an isothermal disk with fixed scale height \citeeg{vanDerKruit1981}.
A MCMC fit \citep[see][]{ForemanMackey2012} to the \scentral{} and \baryonsd{} data yields
\begin{equation}
\scentral{} = (7.6 \pm 0.9)
\left( \frac{ \Sigma_\mathrm{baryon} } {10^7 \, \msun{} \, \mathrm{kpc}^{-2}} \right)^{0.13 \pm 0.01},
\end{equation}
with a slope three times shallower than expected.
This discrepancy may be due to the influence of the dark matter component within the disk, or to variable scale heights in our sample.
Using the equation for scale height given by \citet{Ott2001}:
\begin{equation}
h_z = 579 \; \mathrm{pc} \;
\left( \frac{\sigma_\mathrm{gas}}{10 \; \mathrm{km \; s}^{-1} } \right)^2
\left( \frac{N_\mathrm{HI}}{10^{21} \; \mathrm{cm}^{-2} } \right)^{-1}
\left( \frac{\rho_\mathrm{HI}}{\rho_t} \right),
\end{equation}
where $\sigma_\mathrm{gas}$ is the velocity dispersion, $N_\mathrm{HI}$ is the \hi{} column density, and $\rho_\mathrm{HI}$ and $\rho_t$ are the \hi{} and total mass densities of the disk,
we derive scale heights ranging between $\sim 50 - 900$~pc for the subregions.
The idea of a ``scale height'' may not be appropriate for the smallest dwarfs in our sample, which show very irregular \hi{} distributions and high $\sigma / w_{20}$ ratios for disks.
We note that this slope remains the same if we use the \sfrsd{} superprofiles instead of the radially-resolved ones.
Both the fact that gravitational drivers are inefficient in these systems \citepalias{StilpGlobal} as well as the mismatch between the expected and measured scaling of \scentral{} with $\Sigma_\mathrm{baryon}$ imply that gravity is not a strong factor in driving \hi{} turbulence.

\subsubsection{Other Possible Turbulent Drivers}
\label{resolved::sec:discussion--correlations--other-drivers}

Many of the other proposed non-stellar drivers of turbulence, such as the magnetorotational instability (MRI), require shear from differential rotation to extract energy from the gravitational potential and convert it to turbulence.
Unfortunately, the rotation curves of dwarf galaxies are often solid-body and therefore lack the necessary shear.
We note that \hi{} in dwarf galaxies shows similar velocity dispersions to those measured in the outer regions of sprials, where shear is present.
Because the MRI is expected to be less efficient in this regime, but \hi{} velocity dispersions are similar, one can question the idea that MRI is a driver of \hi{} velocity dispersions in spirals as well.
On the other hand, non-circular motions in galaxies could potentially provide a source of shear to extract energy for turbulence, but the average amplitude of non-circular motions \citep[$\lesssim 5$~\kms{};][]{Trachternach2008, Oh2011} is smaller than the typical \hi{} velocity dispersions, thus requiring efficiencies in excess of 100\%.

Recently, mass accretion has been proposed as a driver of turbulence in galaxies \citep{Klessen2010}.
In this scenario, infalling gas converts its kinetic energy to turbulent energy with some efficiency:
\begin{equation}
\epsilon_\mathrm{acc} \equiv \left| \frac{\dot{E}_\mathrm{decay}}{\dot{E}_\mathrm{in}} \right| =  \frac{M_\mathrm{gas} \sigma^3}{\lambda \dot{M}_\mathrm{in} v_\mathrm{in}^2},
\end{equation}
where $\dot{E}_\mathrm{decay}$ and $\dot{E}_\mathrm{in}$ are the rates of turbulent energy decay and energy input, respectively.
Here, $M_\mathrm{gas}$ is the mass of turbulent gas, $\sigma$ is the turbulent velocity dispersion, and $\lambda$ the turbulent decay length.
$\dot{M}_\mathrm{in}$ and $v_\mathrm{in}$ represent the mass accretion rate and the velocity of infalling gas.
The authors calculated the required coupling efficiencies for accretion-driven turbulence in 11 THINGS galaxies, under the assumption that $\dot{M}_\mathrm{in} = \dot{M}_\mathrm{SF}$ (i.e., that all infalling gas is being converted to stars) and $v_\mathrm{in} = v_\mathrm{rot}$.
They found that the efficiencies were $\epsilon \lesssim 10$\% for the 8 spirals in their sample, including the outer disks, but the 3 of the 4 dwarfs that overlap with our sample required $\epsilon > 1$ to drive observed levels of turbulence (IC~2574, NGC~4214, and Ho~I).
While this method may be a viable source of energy for turbulence in the outer disks of spirals, it does not appear to function effectively in dwarfs.

We note that SN bubbles or \hii{} regions could induce expanding gas motions, resulting in infalling material that could drive turbulence through a similar mechanism \citep[``galactic fountain''; e.g.,][]{Shapiro1976}.
In spiral galaxies, this idea is supported by the existence of extraplanar gas at anomalous velocities in some systems \citeeg{Fraternali2001, Fraternali2002, Barbieri2005, Boomsma2008}.
Expanding ISM material has also been detected in both typical and starbursting dwarf galaxies \citep[e.g., NGC~2366, NGC~4861, NGC~1569, NGC~4214;][]{Schwartz2004, vanEymeren2009, vanEymeren2009b, vanEymeren2010}.
The timescale for this expanding gas to fall back to the disk is unclear, as ionized gas structures can be found in regions far from any current star formation \citeeg{Hunter1993}, thus hampering any energy calculations.
In addition, the fact that less than half the dwarfs in the \citet{Hunter1993} sample showed evidence of expanding shells indicates that galactic fountains are unlikely to be responsible for accretion-driven turbulence in all dwarf galaxies.

Dynamical interactions with dark matter subhalos have recently been proposed as a cause for \hi{} holes or both gaseous and stellar substructure in extended galaxy disks \citeeg{Bekki2006, Kazantzidis2008}.
These interactions could potentially be a source of energy for turbulence in low \sfrsd{} regions if energy could be extracted and transferred to the gas.
Because the \hi{} gas will dissipate its turbulent energy over $\sim 10$~Myr, the interactions must be at least this frequent to sustain observed levels of turbulence.
Followup simulations have concluded that impacts from dark matter subhalos are an unlikely cause of \hi{} holes \citeeg{Kannan2012}, but interactions that increase velocity dispersion are not necessarily restricted only to the energetic events necessary for hole creation .
Halo impacts on the disk can increase gas velocity perpendicular the disk of order $\sim 5 - 10$ \kms{}, but the effect is local \citeeg{Kannan2012}.

As a first test of this idea, we compare the gravitational potential energy of the galaxy, $E_\mathrm{GPE}$, to the energy required to maintain the observed velocity dispersion, $E_\mathrm{turb} = \dot E_\mathrm{turb} \tau_D$, since a redshift of $z = 1$.
To first order, we can estimate the gravitational potential energy of a halo with an isothermal density profile as:
\begin{equation}
E_\mathrm{GPE} = \frac{G M_{200}^2}{R_{200}} = (10 G H )^{2/3} M_{200}^{5/3},
\label{resolved::eqn:gpe}
\end{equation}
where $G$ is the gravitational constant, $H$ is the Hubble constant \citep[70 \kms{} \per{Mpc};][]{Komatsu2011}, and $M_\mathrm{200}$ and $R_\mathrm{200}$ are the mass and radius of the halo when the average density reaches $200$ times the critical density of the universe.
The dissipation rate of turbulent energy associated with a mass $M_\mathrm{HI}$ for a given velocity dispersion $\sigma$ is
\begin{equation}
\dot E_\mathrm{turb} = \eta_v k M_\mathrm{HI} \sigma^3,
\label{resolved::eqn:dot-e-turb}
\end{equation}
where $\eta_v = 0.21 / \pi$ and $k$ is related to the driving length scale of turbulence by $k = 2 \pi / \lambda$ \citep{MacLow1999}.

We can compare the energy required to maintain a velocity dispersion of $10$ \kms{} since a redshift of $z = 1$ to the gravitational potential energy of the halo by taking the ratio of $E_\mathrm{turb}$ to $E_\mathrm{GPE}$.
If we assume that $M_\mathrm{HI} = f_\mathrm{gas} M_\mathrm{baryon}$ and $M_\mathrm{200} = M_\mathrm{baryon} / f_\mathrm{baryon}$, we can re-write the ratio $\dot E_\mathrm{turb} \times t / E_\mathrm{GPE}$ as:
\begin{equation}
\epsilon = 0.078
\left( \frac{f_\mathrm{gas}}{0.5} \right)
\left( \frac{f_\mathrm{baryon}}{0.0066} \right)^{5/3}
\left( \frac{M_\mathrm{baryon}}{10^8 \; \msun{}} \right)^{-2/3}
\left( \frac{\sigma}{10 \; \mathrm{km \, s}^{-1}} \right)^3
\left( \frac{\lambda}{100 \; \mathrm{pc}} \right).
\label{resolved::eqn:subhalo-ratio}
\end{equation}
In this case, we have assumed $f_\mathrm{gas} = 0.5$, a value compatible with the \m{HI} and \m{baryon} measurements of our sample.
Estimates for $f_\mathrm{baryon}$ for the range of \m{star} in our sample are $0.01 - 0.1 f_\mathrm{cosmic}$ \citep{Trujillo2011}, where $f_\mathrm{cosmic}$ is the cosmic baryon fraction of the universe (i.e., $\Omega_\mathrm{baryon} / \Omega_\mathrm{matter}$); we choose the median $f_\mathrm{baryon} = 0.05 f_\mathrm{cosmic}$.
We use $H = 70$ \kms{} \per{Mpc}, as consistent with the seventh year WMAP results \citep{Komatsu2011}.

The ratio of $E_\mathrm{turb}$ to $E_\mathrm{GPE}$ is $0.078$ for halos with $\m{baryon} = 10^8$ \msun{}, the mean stellar mass of our sample, and increases for the lower-mass dwarfs in our sample with smaller \m{baryon}.
If energy can be extracted from the gravitational potential of the galaxy (i.e., due to interactions between the disk and dark matter subhalos), the ratio given in Equation~\ref{resolved::eqn:subhalo-ratio} must at the very least be less than the canonical fraction of mass in subhalos smaller than a given halo mass \citep[10\%; e.g.,][]{Klypin1999, Ghigna2000}.
Since dark matter is collisionless and can only affect the disk gravitationally, one would expect the efficiency of extracting energy to be very low.
It would also depend on the interaction rate between subhalos and the gaseous disk, as not all subhalos have orbits that pass through the \hi{} disk \citeeg{Kannan2012}.
This method may be potentially feasible for massive galaxies, with larger gravitational potentials, but is unlikely to provide enough energy in the lower-mass galaxies where other potential drivers of turbulence (e.g., the MRI or gravitational instabilities) are also unable to function.


\subsubsection{Implications}

\hi{} velocity dispersions in dwarf galaxies are typically comparable to those in spirals.
In this paper, we have examined a number of potential sources of energy for turbulent velocity dispersions (star formation, known gravitational instabilities, the MRI, and accretion-driven turbulence) and found that no single one is able to consistently sustain the observed velocity dispersions in all regions of dwarf galaxies.
If some instability could extract energy from the rotational energy of the galaxies, it would require efficiencies on the order of $\epsilon \sim (\sigma_\mathrm{central} / v_\mathrm{rot})^2$, which ranges from $\epsilon \sim 0.4 - 10$\% for a median $\scentral{} = 7.8$ \kms{} and the range of $v_\mathrm{rot} \sim w_\mathrm{20} / 2$ values of our sample.
These efficiencies are not unreasonable, but the mechanism to extract this energy is unknown.

In addition, it is not necessary to limit the energy for turbulence to a single source. Instead, it may come from a number of sources, e.g.:
\begin{equation}
E_\mathrm{total} = \epsilon_\mathrm{SF} E_\mathrm{SF} + \epsilon_\mathrm{grav} E_\mathrm{grav} + \epsilon_\mathrm{accr} E_\mathrm{accr} + ...
\end{equation}
for contributions from star formation, gravitational, accretion energies, and other sources, respectively, but disentangling the required efficiencies for each energetic component is beyond the scope of this paper.

It is also possible that the observed velocity dispersions are thermal in nature, instead of turbulent.
A UV background could potentially drive thermal velocity dispersions to $\sim 6$ \kms{} \citep{Schaye2004, Tamburro2009}, but is unlikely to explain the higher velocity dispersions of $\sim 10$ \kms{}.
If some other mechanism is able to heat \hi{} to higher temperatures (i.e., the decay of previous turbulent energy into thermal energy), the combination of low metallicity from inefficient SF and the low pressure due to low surface density means that \hi{} is likely to exist primarily in the warm phase.
We can test this by assuming that existing SF couples to the surrounding \hi{} gas at a 10\% efficiency, producing \hi{} with a turbulent velocity dispersion $\sigma_\mathrm{turb}$, with thermal broadening at a level $\sigma_\mathrm{thermal}$ making up the remainder of the observed velocity dispersion.
By adding the two velocity dispersions in quadrature to obtain observed \scentral{} values (i.e., $\sigma_\mathrm{central}^2 = \sigma_\mathrm{thermal}^2 + \sigma_\mathrm{turb}^2$), we estimate the necessary $\sigma_\mathrm{thermal}$ as:
\begin{equation}
\sigma_\mathrm{thermal} = 
\left(
\sigma_\mathrm{central}^2 - 0.1 \frac{2}{3} \frac{\Sigma_\mathrm{E,SF}}{\Sigma_\mathrm{HI}}
\right)^{1/2}.
\label{resolved::eqn:sigma-thermal}
\end{equation}
Here, $0.1$ represents a fiducial coupling efficiency between SF energy and \hi{} turbulent energy.
This approximation is valid only in low \sfrsd{} regimes, as observed efficiencies can be $\lesssim 0.1$ in higher \sfrsd{} regions.

\begin{figure}
\centering
\includegraphics[height=5in]{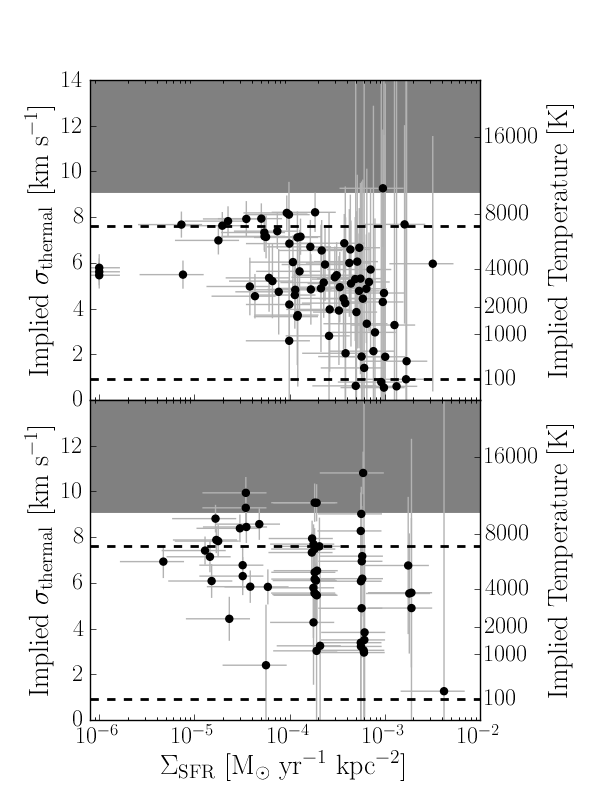}
\caption[Implied thermal velocity dispersions and temperatures]{
Implied thermal velocity dispersions and corresponding temperatures for superprofiles in regions of constant radius (top panel) and constant \sfrsd{} (bottom panel), assuming a coupling efficiency of 0.1 between star formation and \hi{} energy.
The horizontal dashed lines are the two stable temperatures expected for \hi{} \citep{Wolfire1995}.
We have shaded $T > 10^4$~K, where \hi{} is not expected to exist.
\label{resolved::fig:sigma-thermal}}
\end{figure}

In Figure~\ref{resolved::fig:sigma-thermal} we plot the implied $\sigma_\mathrm{thermal}$ and corresponding implied temperature $T$ as a function of \sfrsd{} for superprofiles in regions of constant radius (top) and constant \sfrsd{} (bottom).
The implied $T$ values span a wide range, from $100 - 16000$ K, with many $\sigma_\mathrm{thermal}$ values requiring temperatures in the unstable regime of the phase diagram \citeeg{Wolfire1995}.
In the Milky Way, roughly $50$\% of \hi{} has been observed to have temperatures in this unstable regime \citep{Heiles2001, Heiles2003}, so it may not be unrealistic to expect similar results in other galaxies.
A small number of superprofiles have implied temperatures greater than $10^4$~K, approximately the maximum temperature expected for \hi{} \citeeg{Wolfire1995}.
The Mach numbers implied by $\sigma_\mathrm{thermal}$ and $\sigma_\mathrm{turb}$ are typically subsonic for estimates of $\sigma_\mathrm{thermal}$ from Equation~\ref{resolved::eqn:sigma-thermal} or from assuming that all \hi{} has $T = 7000$~K.
If the \hi{} were cold ($T = 100$~K), the Mach numbers indicate supersonic turbulent motions for the superprofiles.

For the low column densities and larger scale heights of dwarf galaxies, we might expect the majority of \hi{} to be in the warm phase, with $T \sim 10^4$~K.
The low metallicity characteristic of these galaxies plus lower gas densities means that $n^2$ cooling processes are less efficient, and thus gas will remain in the warm phase for longer times.
Therefore, it may not be unreasonable to expect that \hi{} velocity dispersions are primarily thermal in low \sfrsd{} regions.
However, implied temperatures greater than $10^4$~K suggest that both turbulence due to star formation and thermal broadening are unable to explain the velocity dispersion of \emph{all} \hi{} line profiles.

In summary, we have determined that \emph{none} of the proposed drivers of turbulence alone can function effectively in dwarf galaxies where the typical \sfrsd{} is low.
In all cases, more energy is required than can be provided by any driver alone.
If the velocity dispersions are indeed due to turbulence, the only apparent solution is that the turbulent energy is not being dissipated at the expected rate.
It is therefore unclear what mechanism is driving the turbulence in the low-\sfrsd{} regime.
Thermal broadening can potentially provide broadening at the observed levels of velocity dispersion, but the connection between thermal velocity dispersions and star formation is not necessarily clear when both high and low \sfrsd{} regions are taken into account.

\subsection{The Superprofile Wings}
\label{resolved::sec:discussion--shape}

We find that only a small fraction ($\lesssim 15$\%) of the \hi{} gas exists at high velocities compared to \scentral{}.
These small fractions may either indicate small systematic non-Gaussianity inherent in \hi{} line profiles \citeeg{Petric2007} or be due to small amounts of \hi{} gas accelerated to high velocities by SN feedback.
We see no significant trend between \fw{} and the local star formation intensity.
The only possible trend is an indication that \fw{} has lower scatter at higher star formation intensities, which may be due only to the small number of measurements at these extreme values.
While the lack of strong correlations suggests only a tenuous connection between high velocity gas and star formation or a poor mismatch of the spatial and temporal scales to which are measurements are sensitive, the lack of a clear lifetime for high velocity gas makes any firm conclusions difficult, given that the presence of high velocity gas may be restricted to timescales either much shorter or longer than the timescale associated with FUV+24\um{} SFR estimates.
We investigate this possibility further in 
\ifthesis
Chapter~\ref{angst} of this thesis.
\else
\citetalias{StilpTimescales} of this series.
\fi

Like the width of the central peak of the superprofiles, \swing{} also tends to increase with increasing measures of mass surface density and/or star formation.
If this trend is interpreted physically, this behavior is consistent with the idea that star formation drives bulk \hi{} motions to faster velocities, for some fraction of the \hi{} disk.
However, \swing{} is correlated with \scentral{} by definition (Equation~\ref{resolved::eqn:swing}), so superprofiles with higher \scentral{} values must necessarily have higher \swing{} values as well.
The correlation between \swing{} and measures of mass or star formation surface density may be due to an underlying correlation with \scentral{}.

In \citetalias{StilpGlobal}, we proposed that the asymmetry of the wings had more scatter in lower mass galaxies due to star formation.
In that scenario, individual star formation regions drove asymmetric \hi{} motions, so galaxies with fewer star formation regions had a higher chance of retaining asymmetry in the superprofile.
If that were the case, we might expect to see increased asymmetry in the spatially resolved superprofiles with increased measures of star formation, as traced by either \sfrsd{} or $\sfrsd{} / \hisd{}$.
In general, the \afull{} parameter does not show the expected trends, though some individual galaxies do show increasing \afull{} with increasing \sfrsd{} (i.e., IC~2574, NGC~4214, DDO~53).

The asymmetry of the superprofiles may therefore be due to other factors, such as infalling or outflowing \hi{}.
The trend of decreasing scatter in \aw{} with increasing SFR found in \citetalias{StilpGlobal} could also be attributed to the fact that galaxies with lower total SFRs tend to be lower mass, with shallower potentials that are more conducive to sloshing motions that could induce the observed asymmetry.
For example, the velocity field of of UGC~4483 has been shown to be consistent with rotation plus $\sim 5$ \kms{} radial motions \citep{Lelli2012}.
As stated before, average non-circular motions are $\lesssim 5$ \kms{} in THINGS dwarf galaxies in our sample \citep{Trachternach2008, Oh2011}.
If these motions are not uniform across the galaxy disk, they could produce asymmetries that propagate to the superprofiles.

\subsection{Universality in \hi{} profile shapes}
\label{sec:discussion--universal-shape}

\begin{figure}
\centering
\includegraphics[width=2.5in]{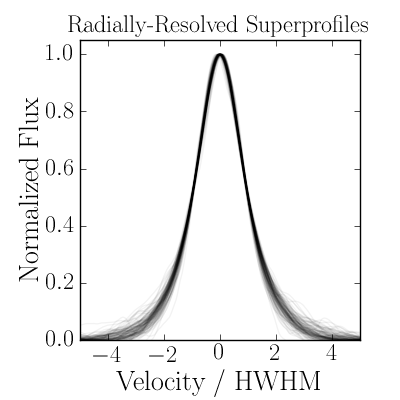}
\includegraphics[width=2.5in]{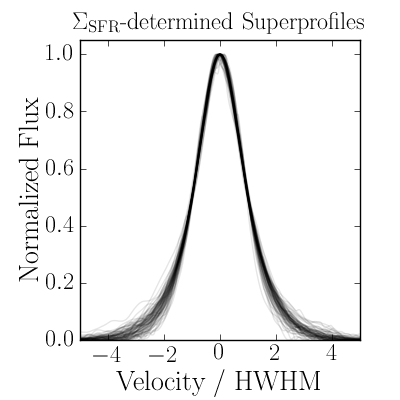}
\caption[Scaled \hi{} superprofiles]{
Superprofiles scaled to the same amplitude and HWHM for our sample.
The left panel shows all superprofiles from radial subregions, while the right shows the all superprofiles from subregions of constant \sfrsd{}.
Each line represents a single superprofile, and has been plotted with a transparency value.
Darker regions indicate more overlapping superprofiles.
In both cases, the superprofiles have very similar shapes when scaled to the same HWHM, with the majority of fluctuations in the wing regions.
The similarity of average \hi{} line shapes suggests that the intrinsic \hi{} profiles may not be Gaussian.
\label{resolved::fig:sps-scaled}}
\end{figure}

As proposed in \citet{Petric2007} and discussed in \citetalias{StilpGlobal}, average \hi{} line profiles tend to show very similar shapes when normalized to the same width.
We now determine if the spatially-resolved superprofiles also have similar shapes.
In Figure~\ref{resolved::fig:sps-scaled}, we show the resolved superprofiles after scaling to the same amplitude and HWHM.
Both radial superprofiles (left) and \sfrsd{} superprofiles (right) are shown.
We find similar behavior for the spatially-resolved superprofiles as in \citetalias{StilpGlobal} and \citet{Petric2007}.
The primary differences manifest in the wing regions, with some galaxies showing more flux in the wings compared to others.
This finding supports the idea that the \hi{} emission tends to follow a non-Gaussian profile shape.

\section{Conclusions}
\label{resolved::sec:conclusions}

We have presented average \hi{} spectra on spatially-resolved scales in a number of nearby dwarf galaxies from VLA-ANGST and THINGS.
To produce these superprofiles, we have co-added \hi{} line-of-sight spectra after removal of the rotational velocity for regions determined both by radius and by the local \sfrsd{}.
Like the global superprofiles presented in \citetalias{StilpGlobal}, the spatially-resolved superprofiles typically exhibit a narrow central peak with higher velocity wings to either side.

As with the global superprofiles, the shape of the spatially-resolved \hi{} superprofiles, when scaled to the same HWHM, is very similar from galaxy to galaxy.
The majority of the differences in shape are in the wings of the superprofiles.
The similarity in shape supports the idea discussed in \citetalias{StilpGlobal} that average \hi{} profiles are non-Gaussian.

We follow the interpretation in \citetalias{StilpGlobal} and characterize the central peak as turbulent with a width \scentral{}.
We interpret the higher-velocity wings as representing anomalous \hi{} motions, and characterize their \emph{rms} velocity (\swing{}) and the fraction of \hi{} in the wings (\fw{}).
We also measure the asymmetry of the full superprofile, \afull{}.
By comparing these parameters to the spatially-resolved physical properties, we have found a number of correlations.

\begin{itemize}

\item \scentral{}: The width of the central peak shows correlations with all physical properties we have measured (\hisd{}, \baryonsd{}, \sfrsd{}, and $\sfrsd{} / \hisd{}$).
The correlations with star formation are strongly driven by the highest \sfrsd{} regions in our sample.
At $\sfrsd{} < 10^{-3}$ \msun{} \per{yr} \per[2]{kpc}, \hi{} velocity dispersions do not appear to be connected to \sfrsd{} and can even be as high as 10 \kms{}.
However, star formation does appear to set a lower threshold below which velocity dispersions cannot fall.

\item \swing{}: The \emph{rms} velocity of the wings is also correlated with all physical properties we have measured.
This behavior could indicate that star formation can drive surrounding \hi{} to faster velocities, but it may also be caused by correlations between our parameters; higher \scentral{} values produce higher \swing{} values by definition.

\item \fw{}: We do not find strong correlations between \fw{} and physical properties, which may indicate that the wings of the \hi{} profiles are not due to star formation but instead reflect an intrinsic \hi{} profile shape.

\item \afull{}: The asymmetry of the full superprofiles does not appear to be connected with local ISM properties, and therefore may be due to other effects, such as warps or inflowing gas.

\end{itemize}

We have also compared the energy from star formation to the \hi{} kinetic energy in the central turbulent peak of the superprofiles over one turbulent timescale.
As previously observed, the coupling efficiency must increase with radius in large dwarfs if star formation is the sole driver of \hi{} kinematics.
Similarly,  the coupling efficiency between star formation energy and \hi{} kinetic energy decreases as a function of \sfrsd{}.
Otherwise, the efficiencies are realistic ($\epsilon < 0.1$) only in regions where $\sfrsd{} > 5 \times 10^{-4}$ \msun{} \per{yr} \per[2]{kpc}.

Star formation therefore does not appear to be the sole driver of \hi{} kinematics in dwarf galaxies, though our data suggest that it does influence them.
Star formation may also provide a lower threshold to the surrounding \hi{} velocity dispersions.
It is therefore likely that some other physical mechanism is inducing turbulent motions in regions of low star formation, or that the velocity dispersions are thermal in nature.
Many of the proposed drivers of turbulence cannot function efficiently in the systems studied in this paper, but simulations of dwarf galaxies are necessary to estimate the effectiveness of other instabilities in a non-shearing, low \sfrsd{} regime.

\begin{acknowledgements}
We thank the referee for helpful comments that improved this paper.
We thank the THINGS team for providing additional data sets used in this paper, as well as Cliff Johnson and Daniel Dale for generously allowing us to use their 3.6\um{} point subtracted maps.
The National Radio Astronomy Observatory is a facility of the National Science foundation operated under cooperative agreement by Associated Universities, Inc.  
Support for this work was provided by the National Science Foundation collaborative research grant ``Star Formation, Feedback, and the ISM: Time Resolved Constraints from a Large VLA Survey of Nearby Galaxies,'' grant number AST-0807710.  
This material is based on work supported by the National Science Foundation under grant No. DGE-0718124 as awarded to A.M.S.  
\end{acknowledgements}

\FloatBarrier
\clearpage

\begin{appendix}
\ifthesis
\subsection{Robustness and Verification}
\else
\section{Robustness and Verification}
\fi
\label{resolved::sec:robustness}

We performed a number of tests in \citetalias{StilpGlobal} to ensure that the superprofiles do not exhibit artificial signals.
In particular, we examined the effects of finite spatial resolution; finite velocity resolution; uncertainties in \vp{}; and noise.
Of these effects, the only one whose uncertainties should not change when we move to spatially-resolved superprofiles is that for finite velocity resolution; we therefore adopt uncertainties from \citetalias{StilpGlobal} for this effect.
For the other three effects, we perform the same tests described in \citetalias{StilpGlobal} on the spatially-resolved superprofiles.
We discuss those tests and their results in this section.

\ifthesis
\subsubsection{Finite Spatial Resolution}
\else
\subsection{Finite Spatial Resolution}
\fi
\label{resolved::sec:robustness--models}

Line widths in the central regions of galaxies can be affected by finite spatial resolution, as the velocity gradient across a single resolution element can be appreciable, leading to ``beam smearing.''
This affect is most problematic in massive galaxies with steeply rising rotation curves or high inclinations.
In contrast, our sample is composed primarily of dwarfs with relatively small rotational velocities and shallow velocity gradients in the central regions.
We therefore expect that beam smearing should not strongly influence our results, but the central regions of the more massive galaxies may be susceptible to increased effects from finite spatial resolution.
As a check, we use models of NGC~2366 that includes the effects of beam smearing from \citetalias{StilpGlobal}.
This galaxy has both a well-measured rotation curve \citep{Oh2008} as well as a high inclination ($63$\degrees{}).
It should therefore exhibit some of the strongest effects of beam smearing in our sample.

For this test, we follow the same procedure outlined in \citetalias{StilpGlobal} to produce model data cubes both with and without the presence of beam smearing (``true'' and ``convolved,'' respectively).
We use two models; both have a rotational curve that mimics the observed rotation curve, with a linear rise to $r_\mathrm{flat} = 1.9$~kpc and $v_\mathrm{flat} = 60$ \kms{}, as measured by \citet{Oh2008}.
For the first model, we assume that all line-of-sight spectra are Gaussian with an intrinsic velocity dispersion of 6 \kms{}.
In the second model, we again assume that all line-of-sight spectra are Gaussian. 
To determine the Gaussian width of a line-of-sight spectrum at a given radius, we fit an exponential to the average second moment value in radial annuli and adopt that scaling as the velocity dispersion.

\begin{figure}
\centering
\includegraphics[height=5in]{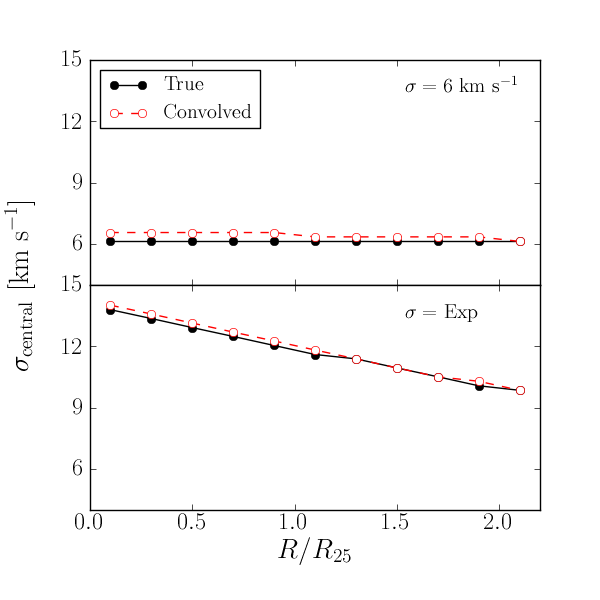}
\caption[Effects of beam smearing on radial superprofiles]{ The effects of beam smearing on radial superprofiles for a model galaxy based on NGC~2366.
The upper panel shows the effects on a model data cube with 6 \kms{} intrinsic velocity dispersion, and the lower is for one with an exponentially-declining intrinsic velocity dispersion.
The black solid line with filled circles shows the ``true'' cube, unaffected by beam smearing.
The red dashed line with unfilled circles shows the ``convolved'' cube, which has the effects of beam smearing included.
We have scaled the $y$-axis to show the approximate range of measured \scentral{} values in our sample.
At all radii, beam smearing increases \scentral{} by $< 10$\%.
It therefore cannot account for the larger variation of observed \scentral{} values.
\label{resolved::fig:models} }
\end{figure}

Using these model cubes, we can generate superprofiles in radial bins from both the ``true'' and ``convolved'' cubes to assess the effects that beam smearing has at each radius.
We show the difference between \scentral{} for the ``convolved'' and the ``true'' cubes in Figure~\ref{resolved::fig:models}. 
The upper panel shows \scentral{} values for the model with fixed velocity dispersion, and the lower panel shows \scentral{} values for the model with exponentially-declining velocity dispersion.
Within each panel, the black line with filled markers represents \scentral{} from the ``true'' model in each radial bin.
The dashed red line shows \scentral{} for the ``convolved'' data cube in the same radial bins.
We have scaled the $y$-axis to show the approximate range of \scentral{} measured in our sample.
We find that beam smearing has a $< 10$\% effect on the width of line profiles for the first model, and $< 5$\% for the second model.
Finite spatial resolution also does not contribute a significant fraction of flux to the wings, so \swing{} and \fw{} are not affected by beam smearing.

Because a good understanding of the rotation curve is necessary to assess the detailed effects of beam smearing in this fashion, we adopt the 0.5~\kms{} uncertainty on \scentral{} from \citetalias{StilpGlobal}.

\ifthesis
\subsubsection{Uncertainties in \vp{}}
\else
\subsection{Uncertainties in \vp{}}
\fi
\label{resolved::sec:robustness--vp}

The regions with the lowest \sfrsd{} are often in the outskirts of galaxies where \hi{} surface densities are also lower.
At low \hi{} surface densities, individual line-of-sight spectra tend to have lower $S/N$, therefore hindering the determination of \vp{}.
We have repeated the test detailed in \citetalias{StilpGlobal} to assess the effects of \vp{} uncertainties on the superprofiles as a function of radius.
In this test, we use the Monte Carlo assessment of \vp{} uncertainties for a representative sample of galaxies (NGC 2366, Sextans A, DDO 190, GR 8, and UGC 4483).
For each selected pixel, we generate a fake superprofile ($S_\mathrm{fake}(v)$) by co-adding Gaussian profiles with amplitude $A_i$ determined by the $S/N$ ratio of each pixel, offset $\mu_i$ drawn randomly from a Gaussian distribution with that pixel's \vp{} uncertainty as its standard deviation, and width of $\sigma = 6$ \kms{}:
\begin{equation}
S_\mathrm{fake}(v) = \sum_i A_i \; \mathrm{exp} 
\left[ \frac{1}{2} 
        \left( \frac{ v - \mu_i  }{ 6 \; \mathrm{km \; s^{-1}}} 
        \right)^2 
\right]
\end{equation}
We then compare this superprofile to what would have been obtained in the absence of any \vp{} uncertainties.
We find that the differences between the ``observed'' fake superprofile and the ``true'' fake superprofile are $< 2$\% in all cases.
Therefore, \vp{} uncertainties do not strongly affect the observed superprofiles, even in low $S/N$ regions; we do not include uncertainties from this effect.

\ifthesis
\subsubsection{Noise}
\else
\subsection{Noise}
\fi
\label{resolved::sec:robustness--noise}

The noise of the final superprofile can also influence the determined parameters.
We use our estimate of noise (Equation~\ref{resolved::eqn:noise-jvm}) to assess the approximate uncertainties on each superprofile parameter.
For this test, we assume that the measured superprofile is the true superprofile.
We then add an estimate of noise to each pixel, drawn from a Gaussian distribution whose width is determined by the estimated noise of that point.
For this ``noisy'' superprofile, we again measure the \scentral{}, \swing{}, \fw{}, and \afull{} parameters.
We repeat this process 10,000 times and fit a Gaussian function to the range of allowed ``noisy'' parameters. We adopt the width of this Gaussian as the uncertainty due to noise.

\ifthesis
\subsubsection{Final Parameter Uncertainties}
\else
\subsection{Final Parameter Uncertainties}
\fi
\label{resolved::sec:robustness--final-uncertainties}

We include effects from finite velocity resolution using the tests presented in \citetalias{StilpGlobal}.
These uncertainties should not change when we move to spatially-resolved scales.

For each superprofile, the final uncertainty for \scentral{} is given by:
\begin{equation}
\Delta \scentral{} = 
\sqrt{
(0.5 \; \mathrm{km \; s^{-1}})^2 +
(0.17 \; \mathrm{km \; s^{-1}})^2 +
(\Delta \sigma_\mathrm{central, noise})^2
}.
\end{equation}

Here, the first term represents the uncertainty due to spatial resolution; we have adopted the more conservative uncertainty from \citetalias{StilpGlobal}. The second term is taken from the finite velocity resolution tests in \citetalias{StilpGlobal}.

The final uncertainty for \swing{} is:
\begin{equation}
\Delta \swing{} = \sqrt{
(0.13 \; \mathrm{km \; s^{-1}})^2 +
(\Delta \sigma_\mathrm{wings, noise})^2
}.
\end{equation}
The $0.13$ \kms{} uncertainty is due to the effects of finite velocity resolution from \citetalias{StilpGlobal}.

The final uncertainty for \fw{} is:
\begin{equation}
\Delta \fw{} = \Delta f_\mathrm{wings,noise}.
\end{equation}
Neither finite spatial resolution or finite velocity resolution \citetalias[from ][]{StilpGlobal} affect \fw{} by $>0.01$.

The final uncertainty for \afull{} is:
\begin{equation}
\Delta \afull{} = \sqrt{ 
(0.003)^2 + 
(\Delta a_\mathrm{full, noise})^2 },
\end{equation}
where we have recalculated the standard deviation due to velocity resolution on $\Delta \afull{}$ instead of \aw{} based on the finite velocity resolution tests from \citetalias{StilpGlobal}.


\section{Spatially-Resolved Superprofiles for the Entire Sample}
\label{resolved::sec:sp-figures}

In this section, we present the spatially-resolved superprofiles for the entire sample.
In
\ifthesis
\S~
\else
Appendix~
\fi
\ref{resolved::sec:sp-figures--radial}, we show the superprofiles derived in subregions of constant radius, and in 
\ifthesis
\S~
\else
Appendix~
\fi
\ref{resolved::sec:sp-figures--sfrsd} we show the superprofiles derived in subregions of constant \sfrsd{}.
In both sections, the figures are ordered by decreasing galaxy $M_\mathrm{baryon,tot}$.

\subsection{Radial Superprofiles}
\label{resolved::sec:sp-figures--radial}

In the following pages, we present the superprofiles generated in subregions of constant radius.
For a single galaxy, the figures are the same as Figures~\ref{resolved::fig:superprofiles-radial-n7793-a} - \ref{resolved::fig:superprofiles-radial-n7793-c}.
Galaxies are shown in order of decreasing $M_\mathrm{baryon,tot}$, with the exception of NGC~7993, which was previously shown in the text.

\ifthesis
\renewcommand{\thefigure}{\arabic{chapter}.\arabic{figure}\alph{subfig}}
\else
\renewcommand{\thefigure}{\arabic{figure}\alph{subfig}}
\fi

\setcounter{subfig}{1}
\begin{figure}[p]
\begin{leftfullpage}
\centering
\includegraphics[width=4in]{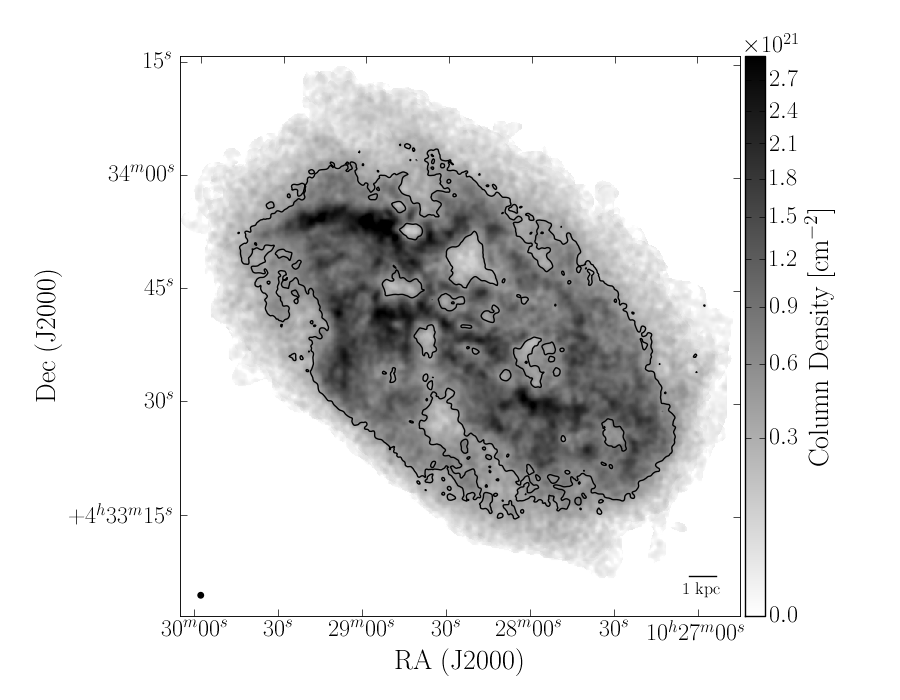}
\includegraphics[width=4in]{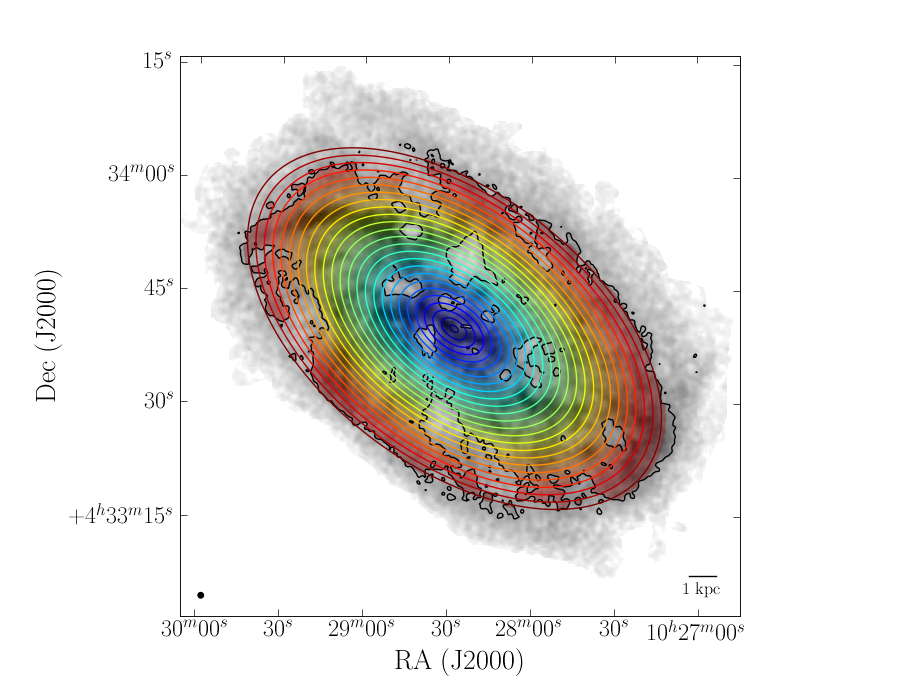}
\caption[Superprofile radial annuli for IC~2574]{Radial annuli in which radial superprofiles are generated for IC2574. 
In both panels, the background greyscale shows \hisd{}, and the solid black line represents the $S/N > 5$ threshold where we can accurately measure \vp{}. 
In the lower panel, the colored solid lines represent the average radius of each annulus, and the corresponding shaded regions of the same color indicate which pixels have contributed to each radial superprofile.
\label{resolved::fig:superprofiles-radial-ic2574-a} }
\end{leftfullpage}
\end{figure}
\addtocounter{figure}{-1}
\addtocounter{subfig}{1}
\begin{figure}
\centering
\includegraphics[height=2.7in]{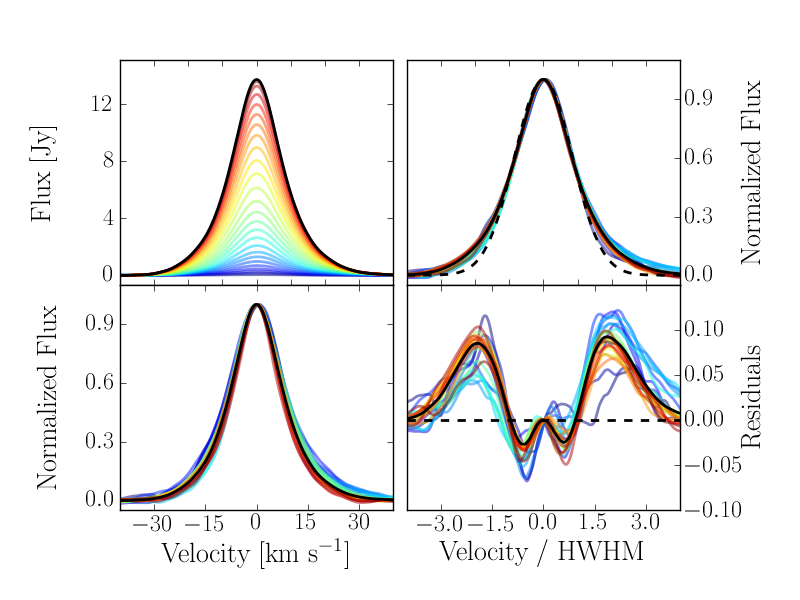}
\caption[Radial superprofiles in IC~2574]{The radial superprofiles in IC~2574, where colors indicate the corresponding radial annuli in the previous figure.
The left hand panels show the raw superprofiles (upper left) and the superprofiles normalized to the same peak flux (lower left).
The right hand panels show the flux-normalized superprofiles scaled by the HWHM (upper right) and the flux-normalized superprofiles minus the model of the Gaussian core (lower right). In all panels, the solid black line represents the global superprofile. In the left panels, we have shown the HWHM-scaled Gaussian model as the dashed black line.
\label{resolved::fig:superprofiles-radial-ic2574-b}
}
\end{figure}
\addtocounter{figure}{-1}
\addtocounter{subfig}{1}
\begin{figure}
\centering
\includegraphics[height=2.7in]{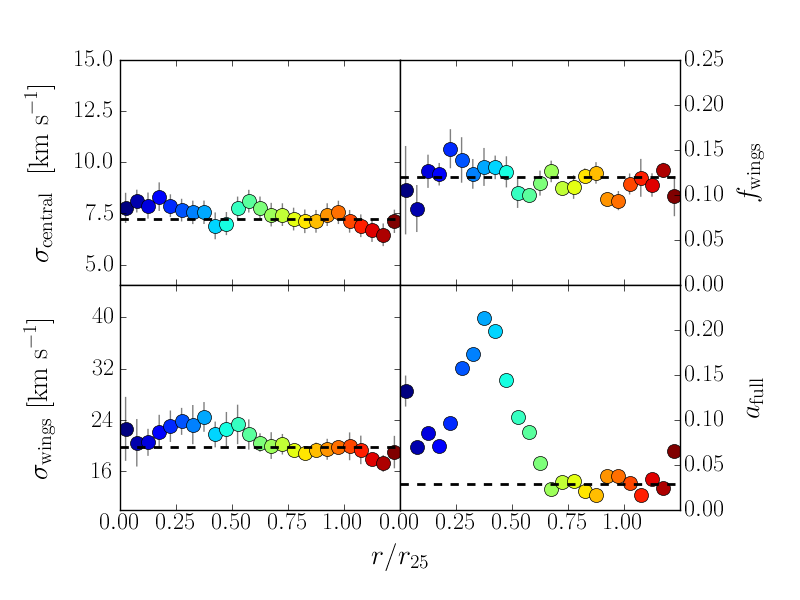}
\caption[Variation of radial superprofile parameters for IC~2574]{Variation of the superprofile parameters as a function of normalized radius for IC~2574.
The solid dashed line shows the parameter value for the global superprofile \citepalias{StilpGlobal}.
The left panels show \scentral{} (upper) and \swing{} (lower), and the right panels show \fw{} (upper) and \afull{} (lower).
\label{resolved::fig:superprofiles-radial-ic2574-c}
}
\end{figure}
\clearpage

\setcounter{subfig}{1}
\begin{figure}[p]
\begin{leftfullpage}
\centering
\includegraphics[width=4in]{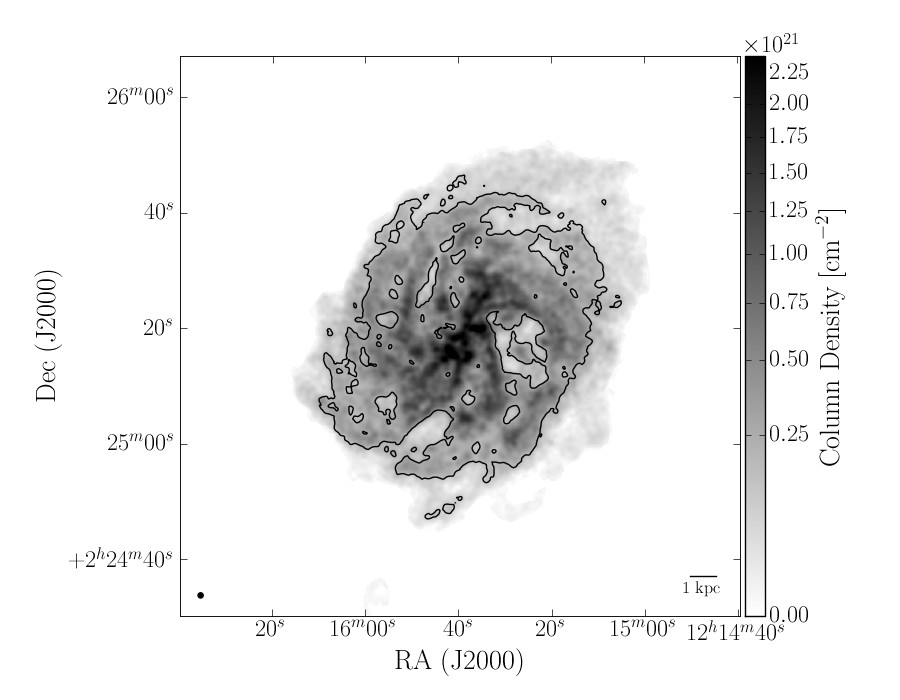}
\includegraphics[width=4in]{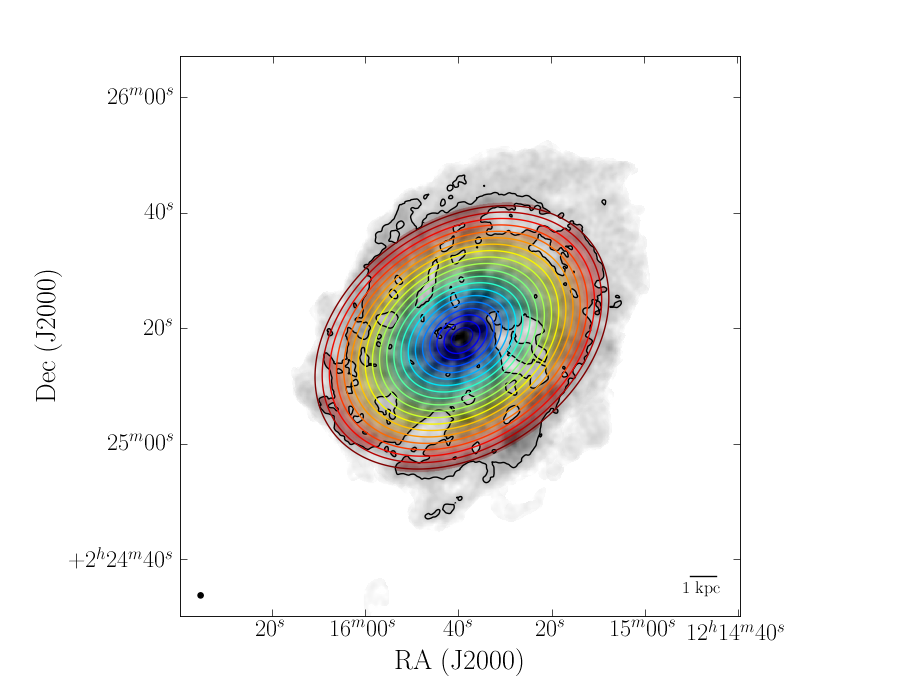}
\caption[Superprofile radial annuli for NGC~4214]{Radial annuli in which radial superprofiles are generated for NGC~4214. 
In both panels, the background greyscale shows \hisd{}, and the solid black line represents the $S/N > 5$ threshold where we can accurately measure \vp{}. 
In the lower panel, the colored solid lines represent the average radius of each annulus, and the corresponding shaded regions of the same color indicate which pixels have contributed to each radial superprofile.
\label{resolved::fig:superprofiles-radial-n4214-a} }
\end{leftfullpage}
\end{figure}
\addtocounter{figure}{-1}
\addtocounter{subfig}{1}
\begin{figure}
\centering
\includegraphics[height=2.7in]{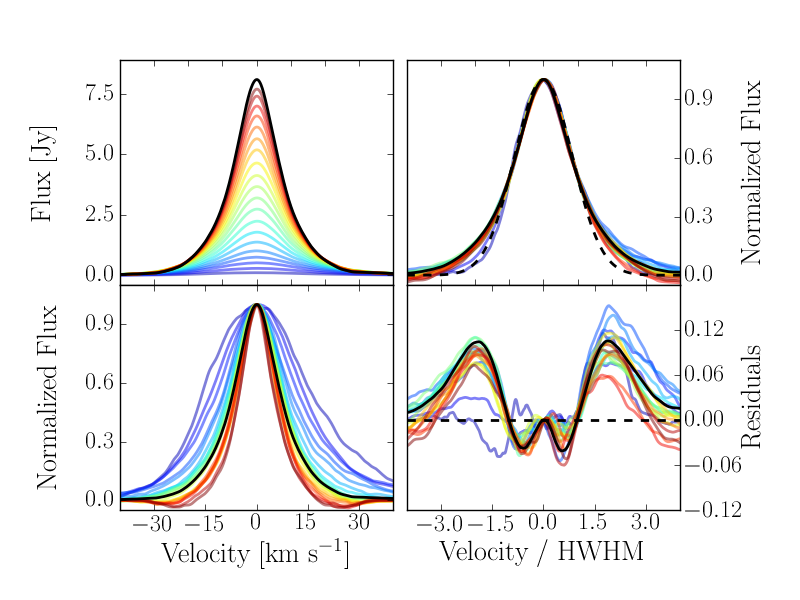}
\caption[Radial superprofiles in NGC~4214]{The radial superprofiles in NGC~4214, where colors indicate the corresponding radial annuli in the previous figure.
The left hand panels show the raw superprofiles (upper left) and the superprofiles normalized to the same peak flux (lower left).
The right hand panels show the flux-normalized superprofiles scaled by the HWHM (upper right) and the flux-normalized superprofiles minus the model of the Gaussian core (lower right). In all panels, the solid black line represents the global superprofile. In the left panels, we have shown the HWHM-scaled Gaussian model as the dashed black line.
\label{resolved::fig:superprofiles-radial-n4214-b}
}
\end{figure}
\addtocounter{figure}{-1}
\addtocounter{subfig}{1}
\begin{figure}
\centering
\includegraphics[height=2.7in]{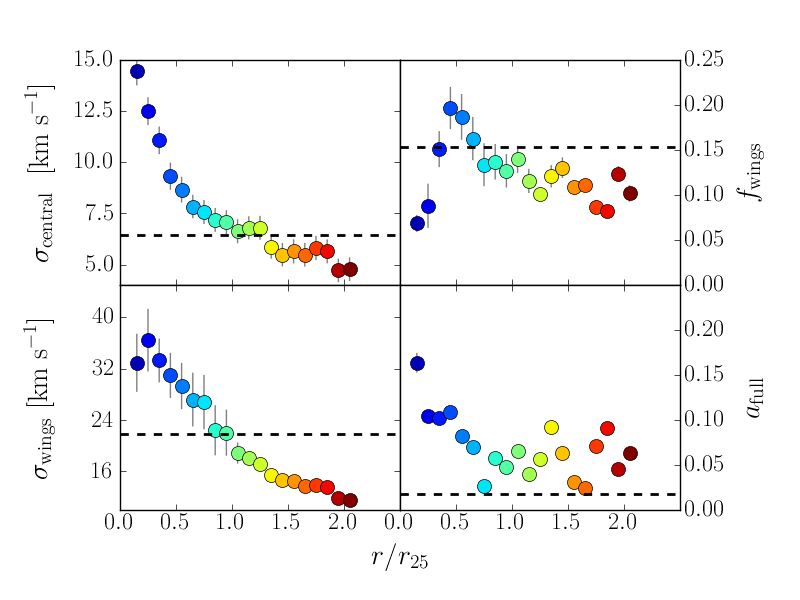}
\caption[Variation of radial superprofile parameters for NGC~4214]{Variation of the superprofile parameters as a function of normalized radius for NGC~4214.
The solid dashed line shows the parameter value for the global superprofile \citepalias{StilpGlobal}.
The left panels show \scentral{} (upper) and \swing{} (lower), and the right panels show \fw{} (upper) and \afull{} (lower).
\label{resolved::fig:superprofiles-radial-n4214-c}
}
\end{figure}
\clearpage

\setcounter{subfig}{1}
\begin{figure}[p]
\begin{leftfullpage}
\centering
\includegraphics[width=4in]{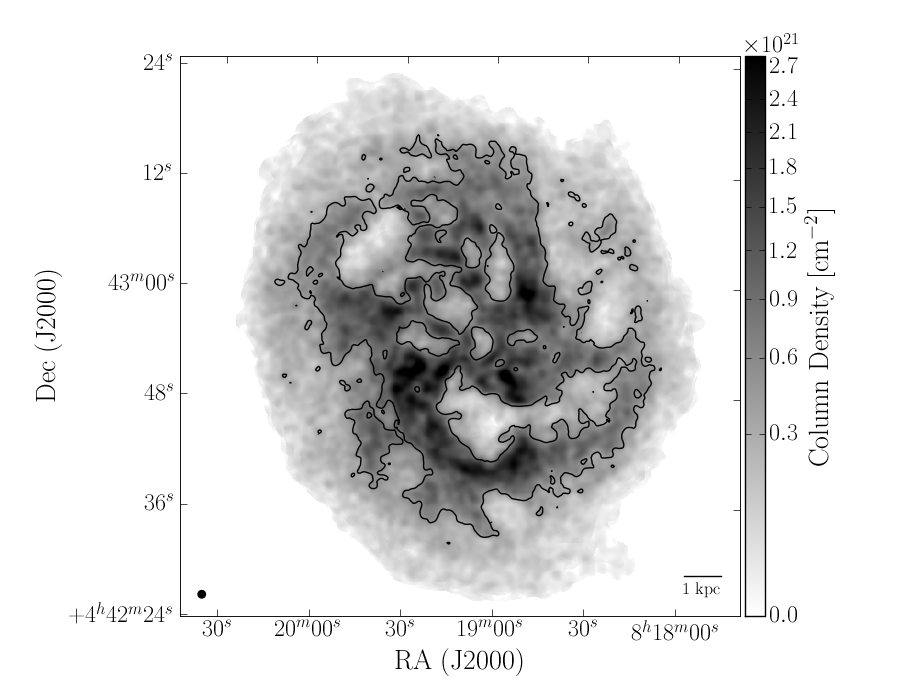}
\includegraphics[width=4in]{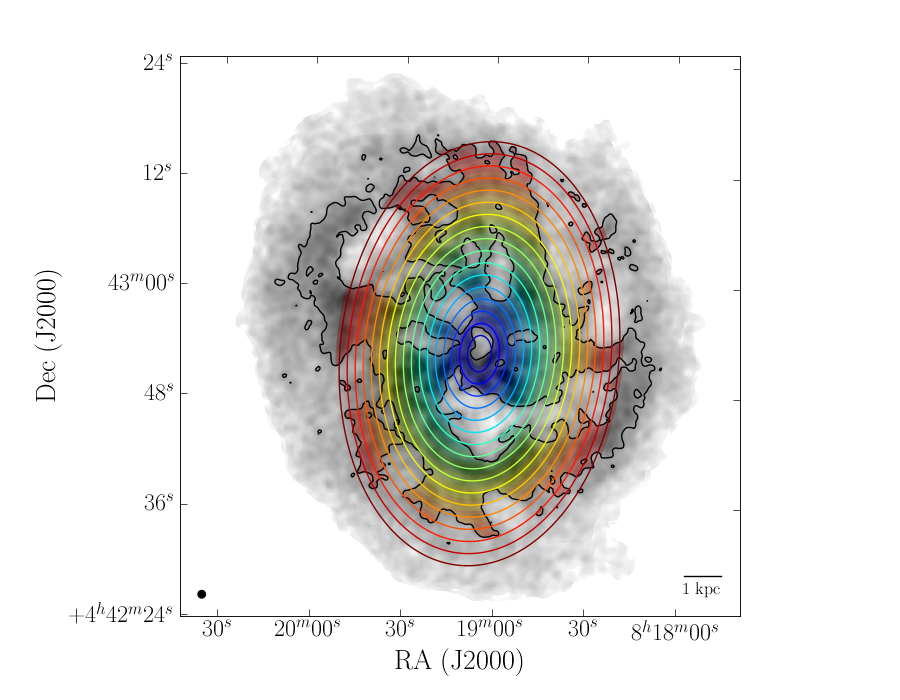}
\caption[Superprofile radial annuli for Ho~II]{Radial annuli in which radial superprofiles are generated for Ho~II. 
In both panels, the background greyscale shows \hisd{}, and the solid black line represents the $S/N > 5$ threshold where we can accurately measure \vp{}. 
In the lower panel, the colored solid lines represent the average radius of each annulus, and the corresponding shaded regions of the same color indicate which pixels have contributed to each radial superprofile.
\label{resolved::fig:superprofiles-radial-hoii-a} }
\end{leftfullpage}
\end{figure}
\addtocounter{figure}{-1}
\addtocounter{subfig}{1}
\begin{figure}
\centering
\includegraphics[height=2.7in]{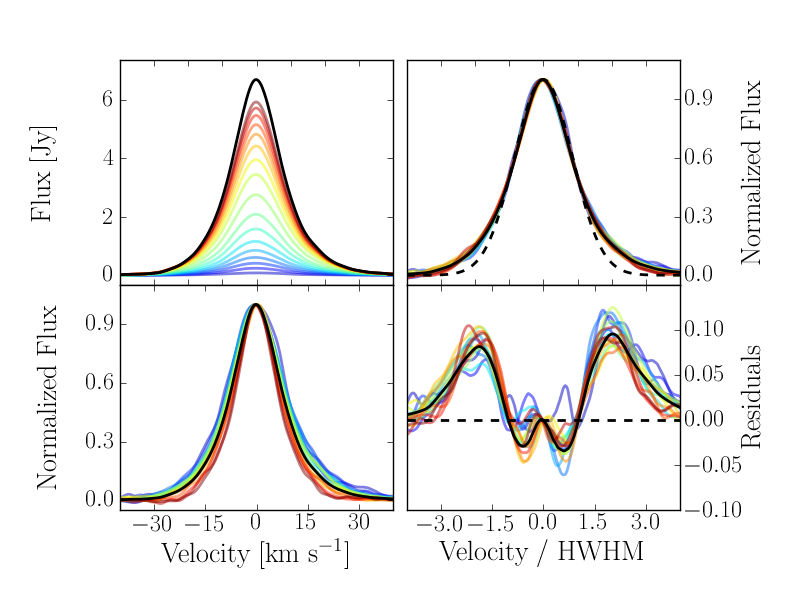}
\caption[Radial superprofiles in Ho~II]{The radial superprofiles in Ho~II, where colors indicate the corresponding radial annuli in the previous figure.
The left hand panels show the raw superprofiles (upper left) and the superprofiles normalized to the same peak flux (lower left).
The right hand panels show the flux-normalized superprofiles scaled by the HWHM (upper right) and the flux-normalized superprofiles minus the model of the Gaussian core (lower right). In all panels, the solid black line represents the global superprofile. In the left panels, we have shown the HWHM-scaled Gaussian model as the dashed black line.
\label{resolved::fig:superprofiles-radial-hoii-b}
}
\end{figure}
\addtocounter{figure}{-1}
\addtocounter{subfig}{1}
\begin{figure}
\centering
\includegraphics[height=2.7in]{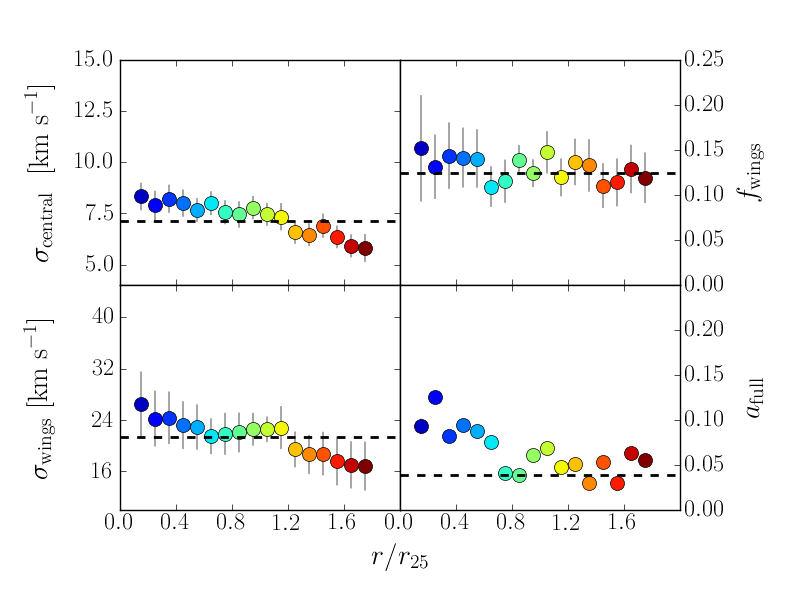}
\caption[Variation of radial superprofile parameters for Ho~II]{Variation of the superprofile parameters as a function of normalized radius for Ho~II.
The solid dashed line shows the parameter value for the global superprofile \citepalias{StilpGlobal}.
The left panels show \scentral{} (upper) and \swing{} (lower), and the right panels show \fw{} (upper) and \afull{} (lower).
\label{resolved::fig:superprofiles-radial-hoii-c}
}
\end{figure}
\clearpage

\setcounter{subfig}{1}
\begin{figure}[p]
\begin{leftfullpage}
\centering
\includegraphics[width=4in]{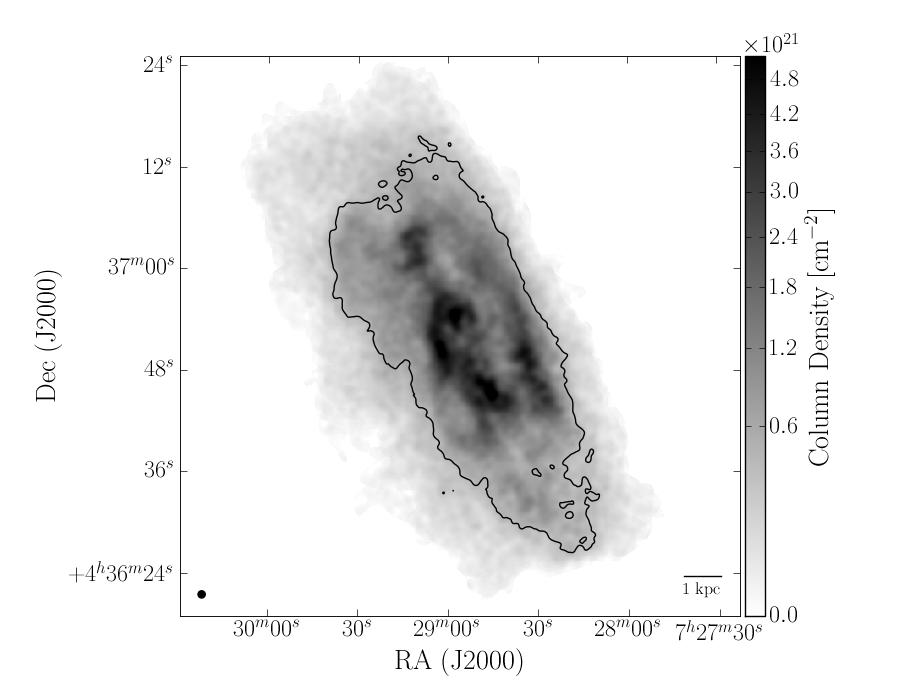}
\includegraphics[width=4in]{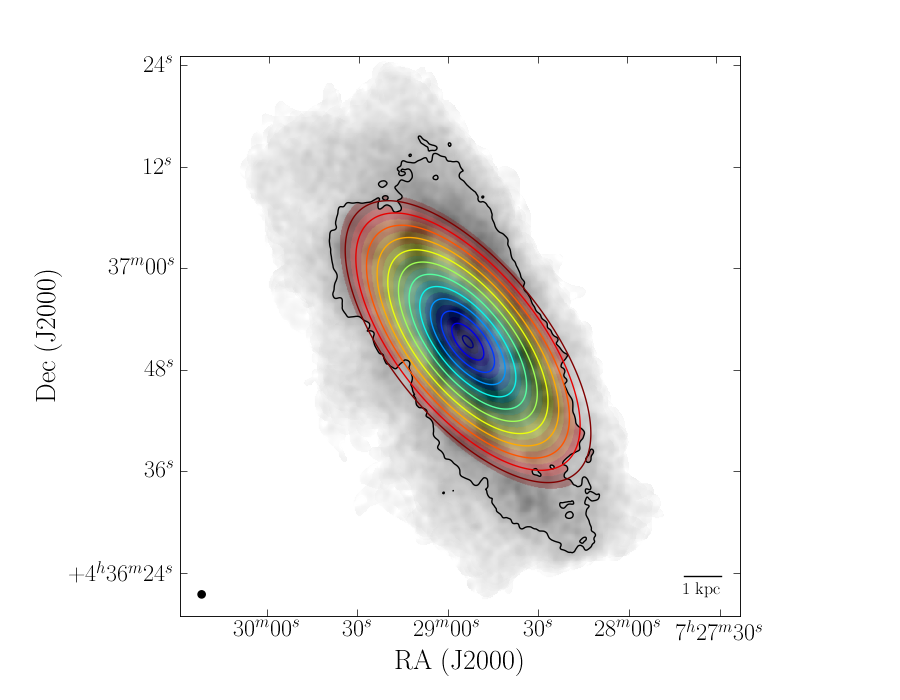}
\caption[Superprofile radial annuli for NGC~2366]{Radial annuli in which radial superprofiles are generated for NGC~2366. 
In both panels, the background greyscale shows \hisd{}, and the solid black line represents the $S/N > 5$ threshold where we can accurately measure \vp{}. 
In the lower panel, the colored solid lines represent the average radius of each annulus, and the corresponding shaded regions of the same color indicate which pixels have contributed to each radial superprofile.
\label{resolved::fig:superprofiles-radial-n2366-a} }
\end{leftfullpage}
\end{figure}
\addtocounter{figure}{-1}
\addtocounter{subfig}{1}
\begin{figure}
\centering
\includegraphics[height=2.7in]{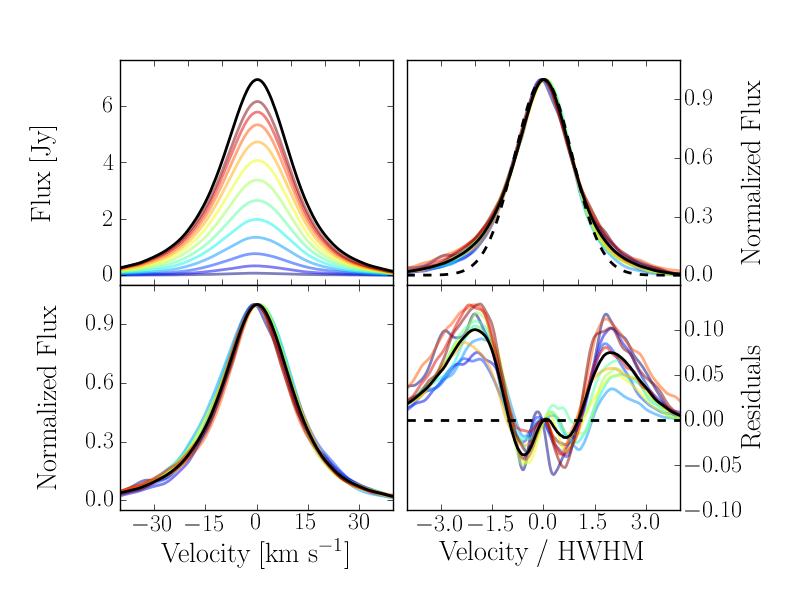}
\caption[Radial superprofiles in NGC~2366]{The radial superprofiles in NGC~2366, where colors indicate the corresponding radial annuli in the previous figure.
The left hand panels show the raw superprofiles (upper left) and the superprofiles normalized to the same peak flux (lower left).
The right hand panels show the flux-normalized superprofiles scaled by the HWHM (upper right) and the flux-normalized superprofiles minus the model of the Gaussian core (lower right). In all panels, the solid black line represents the global superprofile. In the left panels, we have shown the HWHM-scaled Gaussian model as the dashed black line.
\label{resolved::fig:superprofiles-radial-n2366-b}
}
\end{figure}
\addtocounter{figure}{-1}
\addtocounter{subfig}{1}
\begin{figure}
\centering
\includegraphics[height=2.7in]{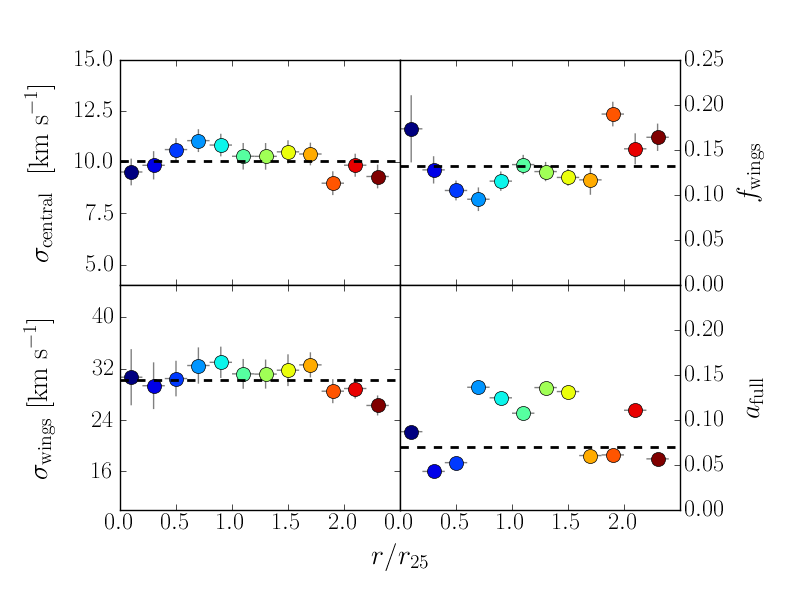}
\caption[Variation of radial superprofile parameters for NGC~2366]{Variation of the superprofile parameters as a function of normalized radius for NGC~2366.
The solid dashed line shows the parameter value for the global superprofile \citepalias{StilpGlobal}.
The left panels show \scentral{} (upper) and \swing{} (lower), and the right panels show \fw{} (upper) and \afull{} (lower).
\label{resolved::fig:superprofiles-radial-n2366-c}
}
\end{figure}
\clearpage

\setcounter{subfig}{1}
\begin{figure}[p]
\begin{leftfullpage}
\centering
\includegraphics[width=4in]{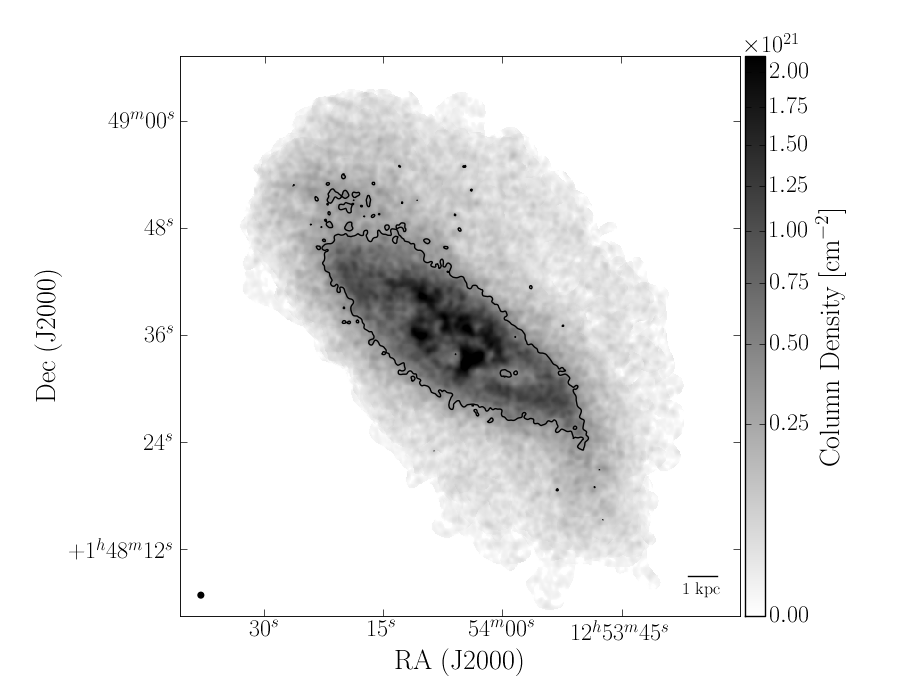}
\includegraphics[width=4in]{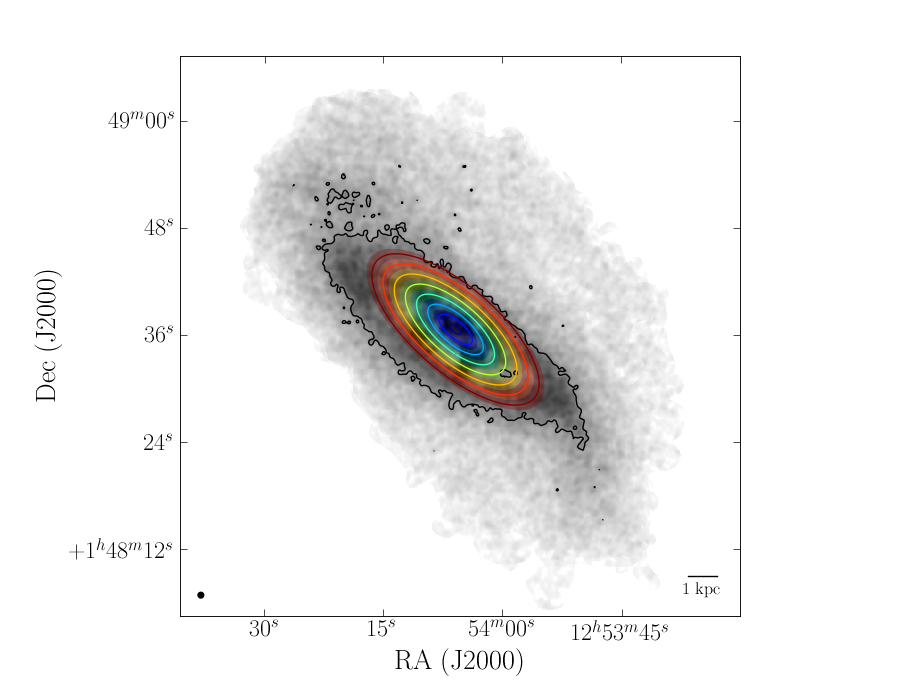}
\caption[Superprofile radial annuli for DDO~154]{Radial annuli in which radial superprofiles are generated for DDO~154. 
In both panels, the background greyscale shows \hisd{}, and the solid black line represents the $S/N > 5$ threshold where we can accurately measure \vp{}. 
In the lower panel, the colored solid lines represent the average radius of each annulus, and the corresponding shaded regions of the same color indicate which pixels have contributed to each radial superprofile.
\label{resolved::fig:superprofiles-radial-ddo154-a} }
\end{leftfullpage}
\end{figure}
\addtocounter{figure}{-1}
\addtocounter{subfig}{1}
\begin{figure}
\centering
\includegraphics[height=2.7in]{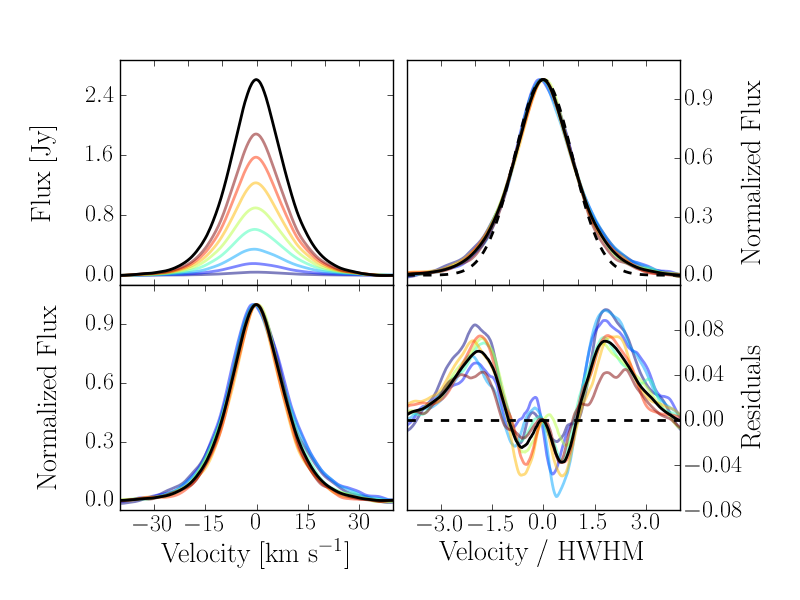}
\caption[Radial superprofiles in DDO~154]{The radial superprofiles in DDO~154, where colors indicate the corresponding radial annuli in the previous figure.
The left hand panels show the raw superprofiles (upper left) and the superprofiles normalized to the same peak flux (lower left).
The right hand panels show the flux-normalized superprofiles scaled by the HWHM (upper right) and the flux-normalized superprofiles minus the model of the Gaussian core (lower right). In all panels, the solid black line represents the global superprofile. In the left panels, we have shown the HWHM-scaled Gaussian model as the dashed black line.
\label{resolved::fig:superprofiles-radial-ddo154-b}
}
\end{figure}
\addtocounter{figure}{-1}
\addtocounter{subfig}{1}
\begin{figure}
\centering
\includegraphics[height=2.7in]{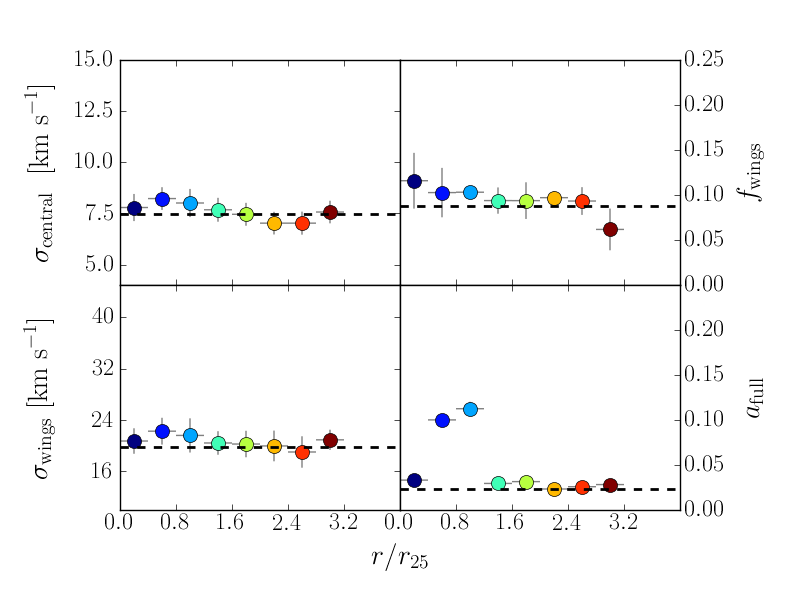}
\caption[Variation of radial superprofile parameters for DDO~154]{Variation of the superprofile parameters as a function of normalized radius for DDO~154.
The solid dashed line shows the parameter value for the global superprofile \citepalias{StilpGlobal}.
The left panels show \scentral{} (upper) and \swing{} (lower), and the right panels show \fw{} (upper) and \afull{} (lower).
\label{resolved::fig:superprofiles-radial-ddo154-c}
}
\end{figure}
\clearpage

\setcounter{subfig}{1}
\begin{figure}
\begin{leftfullpage}
\centering
\includegraphics[width=4in]{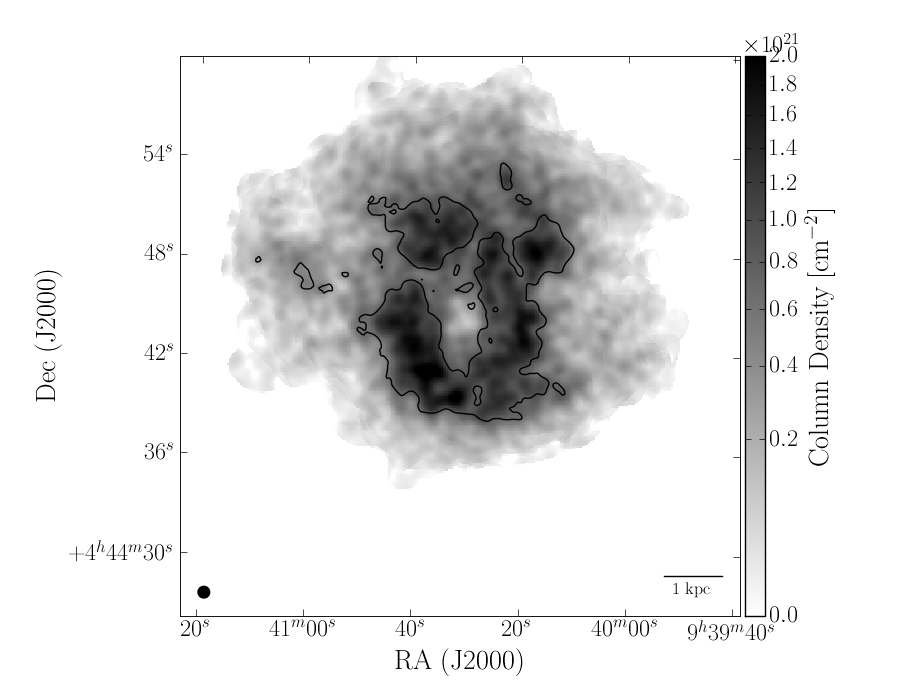}
\includegraphics[width=4in]{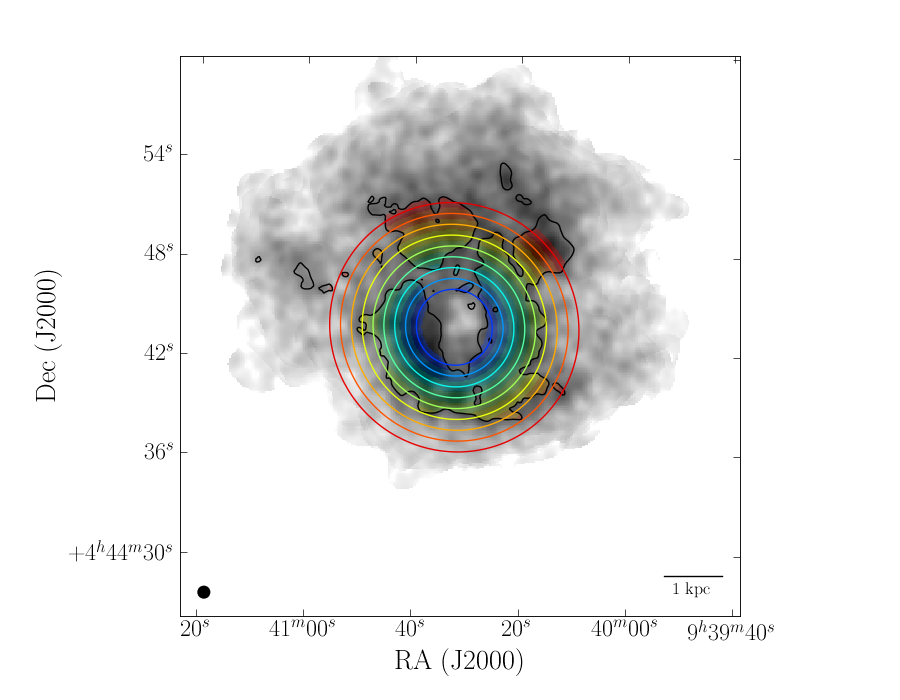}
\caption[Superprofile radial annuli for Ho~I]{Radial annuli in which radial superprofiles are generated for Ho~I. 
In both panels, the background greyscale shows \hisd{}, and the solid black line represents the $S/N > 5$ threshold where we can accurately measure \vp{}. 
In the lower panel, the colored solid lines represent the average radius of each annulus, and the corresponding shaded regions of the same color indicate which pixels have contributed to each radial superprofile.
\label{resolved::fig:superprofiles-radial-hoi-a} }
\end{leftfullpage}
\end{figure}
\addtocounter{figure}{-1}
\addtocounter{subfig}{1}
\begin{figure}
\centering
\includegraphics[height=2.7in]{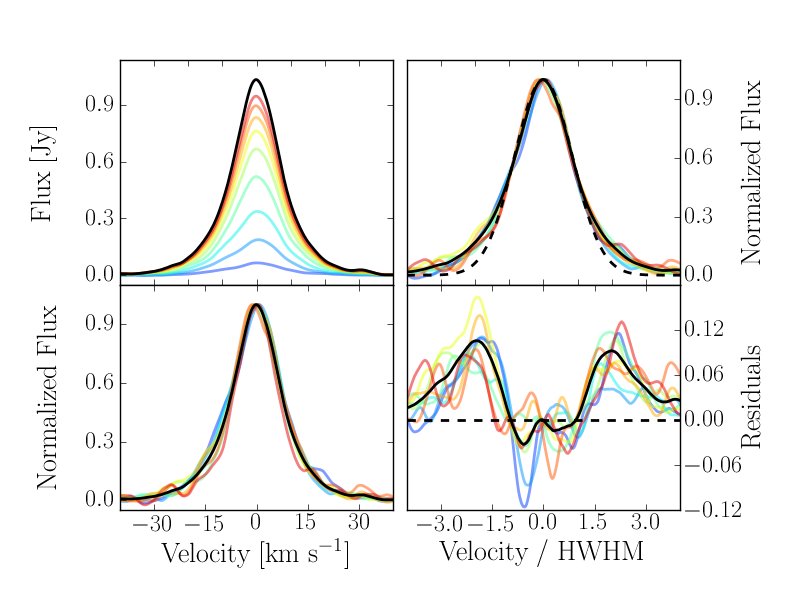}
\caption[Radial superprofiles in Ho~I]{The radial superprofiles in Ho~I, where colors indicate the corresponding radial annuli in the previous figure.
The left hand panels show the raw superprofiles (upper left) and the superprofiles normalized to the same peak flux (lower left).
The right hand panels show the flux-normalized superprofiles scaled by the HWHM (upper right) and the flux-normalized superprofiles minus the model of the Gaussian core (lower right). In all panels, the solid black line represents the global superprofile. In the left panels, we have shown the HWHM-scaled Gaussian model as the dashed black line.
\label{resolved::fig:superprofiles-radial-hoi-b}
}
\end{figure}
\addtocounter{figure}{-1}
\addtocounter{subfig}{1}
\begin{figure}
\centering
\includegraphics[height=2.7in]{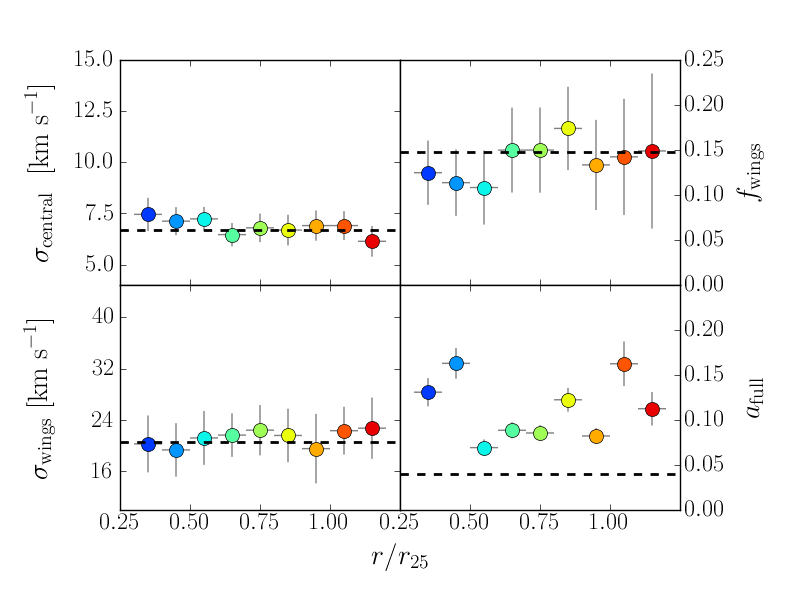}
\caption[Variation of radial superprofile parameters for Ho~I]{Variation of the superprofile parameters as a function of normalized radius for Ho~I.
The solid dashed line shows the parameter value for the global superprofile \citepalias{StilpGlobal}.
The left panels show \scentral{} (upper) and \swing{} (lower), and the right panels show \fw{} (upper) and \afull{} (lower).
\label{resolved::fig:superprofiles-radial-hoi-c}
}
\end{figure}
\clearpage

\setcounter{subfig}{1}
\begin{figure}
\begin{leftfullpage}
\centering
\includegraphics[width=4in]{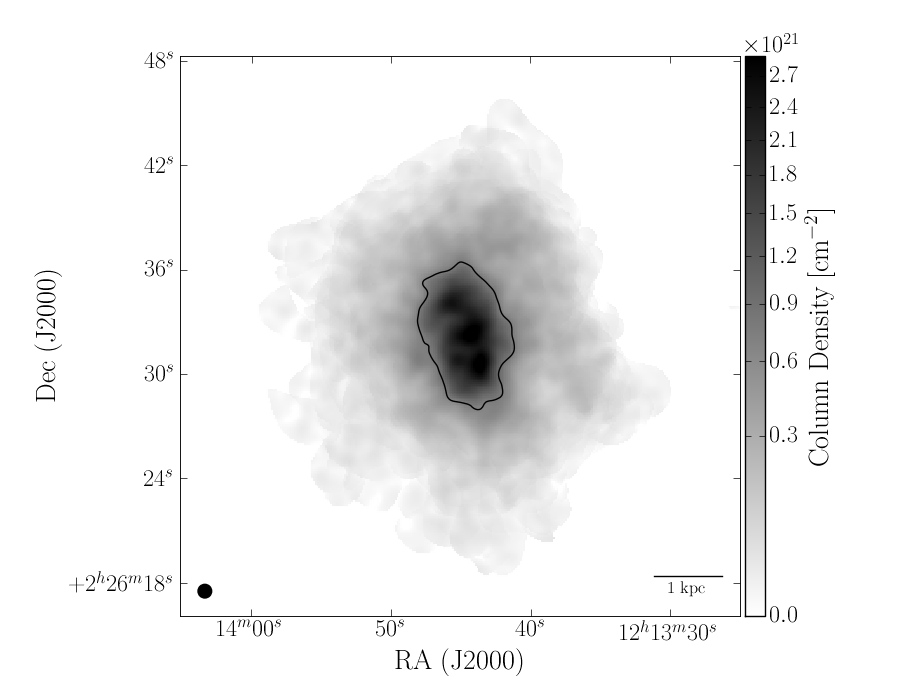}
\includegraphics[width=4in]{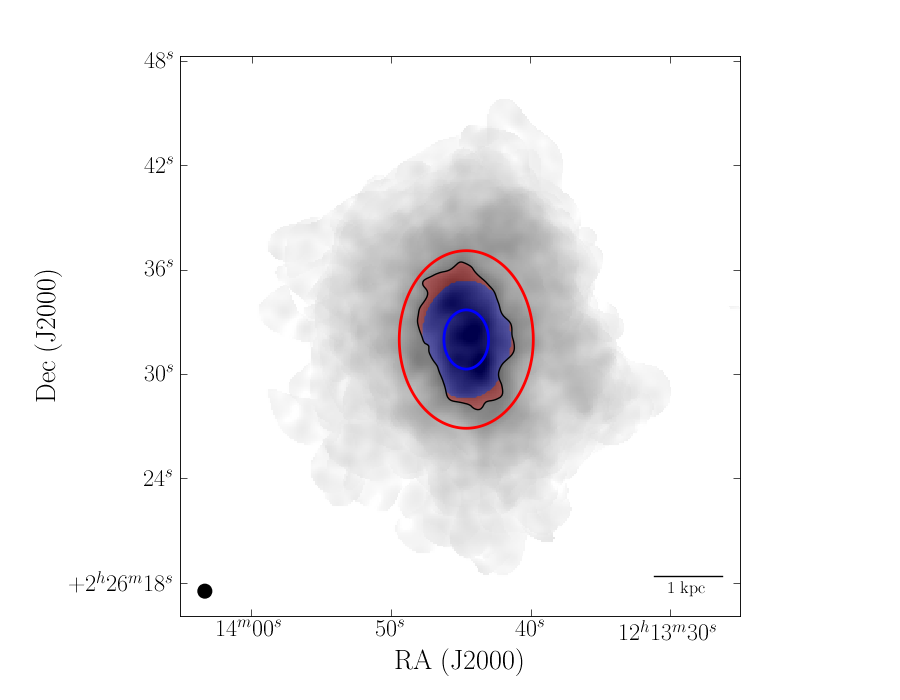}
\caption[Superprofile radial annuli for NGC~4190]{Radial annuli in which radial superprofiles are generated for NGC~4190. 
In both panels, the background greyscale shows \hisd{}, and the solid black line represents the $S/N > 5$ threshold where we can accurately measure \vp{}. 
In the lower panel, the colored solid lines represent the average radius of each annulus, and the corresponding shaded regions of the same color indicate which pixels have contributed to each radial superprofile.
\label{resolved::fig:superprofiles-radial-n4190-a} }
\end{leftfullpage}
\end{figure}
\addtocounter{figure}{-1}
\addtocounter{subfig}{1}
\begin{figure}
\centering
\includegraphics[height=2.7in]{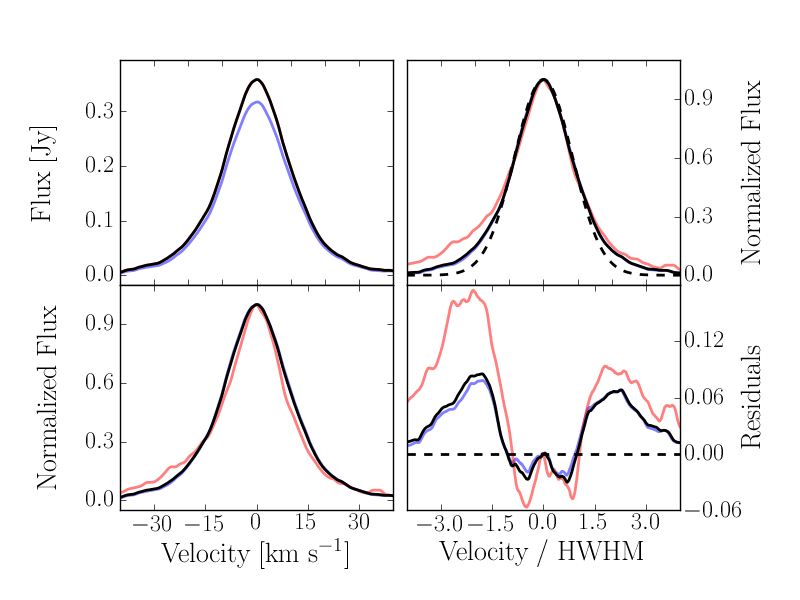}
\caption[Radial superprofiles in NGC~4190]{The radial superprofiles in NGC~4190, where colors indicate the corresponding radial annuli in the previous figure.
The left hand panels show the raw superprofiles (upper left) and the superprofiles normalized to the same peak flux (lower left).
The right hand panels show the flux-normalized superprofiles scaled by the HWHM (upper right) and the flux-normalized superprofiles minus the model of the Gaussian core (lower right). In all panels, the solid black line represents the global superprofile. In the left panels, we have shown the HWHM-scaled Gaussian model as the dashed black line.
\label{resolved::fig:superprofiles-radial-n4190-b}
}
\end{figure}
\addtocounter{figure}{-1}
\addtocounter{subfig}{1}
\begin{figure}
\centering
\includegraphics[height=2.7in]{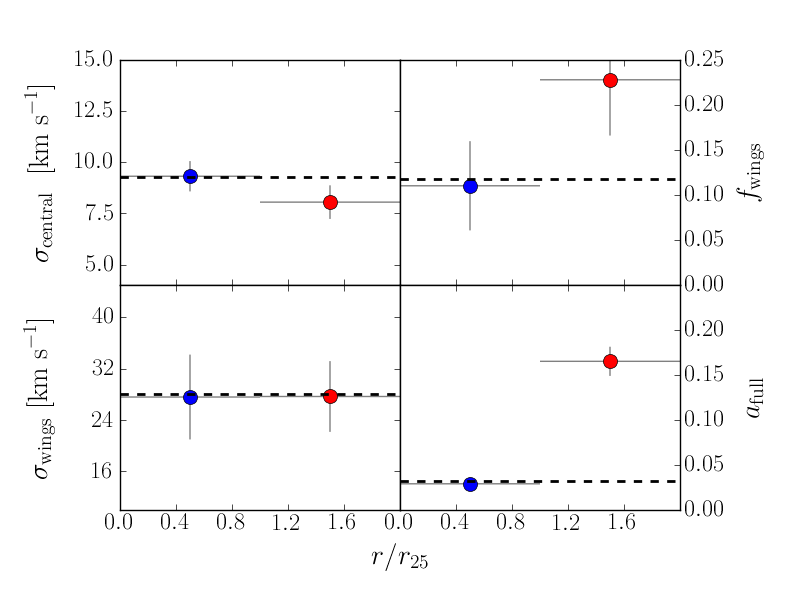}
\caption[Variation of radial superprofile parameters for NGC~4190]{Variation of the superprofile parameters as a function of normalized radius for NGC~4190.
The solid dashed line shows the parameter value for the global superprofile \citepalias{StilpGlobal}.
The left panels show \scentral{} (upper) and \swing{} (lower), and the right panels show \fw{} (upper) and \afull{} (lower).
\label{resolved::fig:superprofiles-radial-n4190-c}
}
\end{figure}
\clearpage

\setcounter{subfig}{1}
\begin{figure}[p]
\begin{leftfullpage}
\centering
\includegraphics[width=4in]{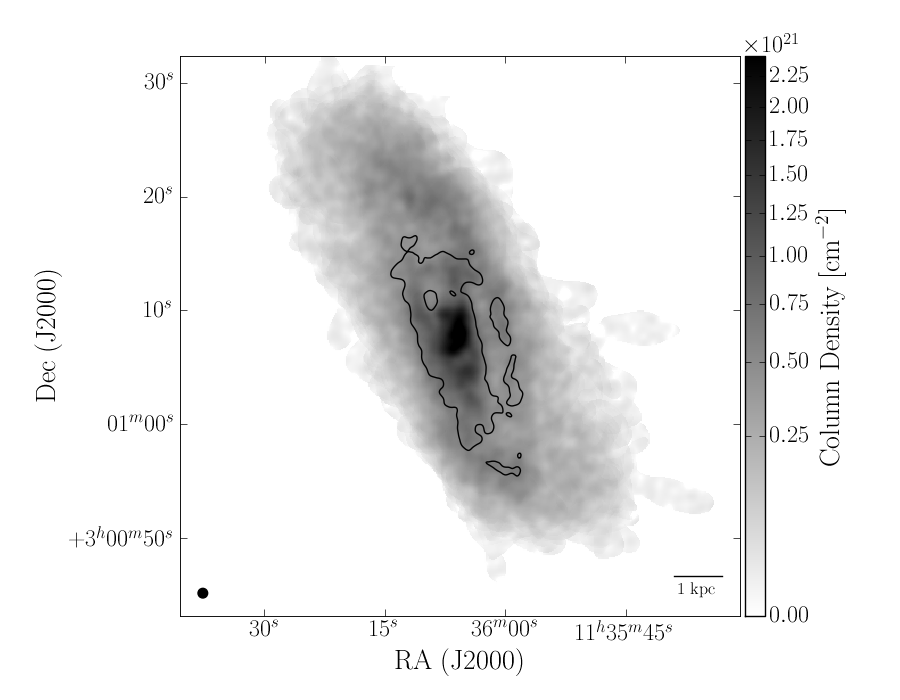}
\includegraphics[width=4in]{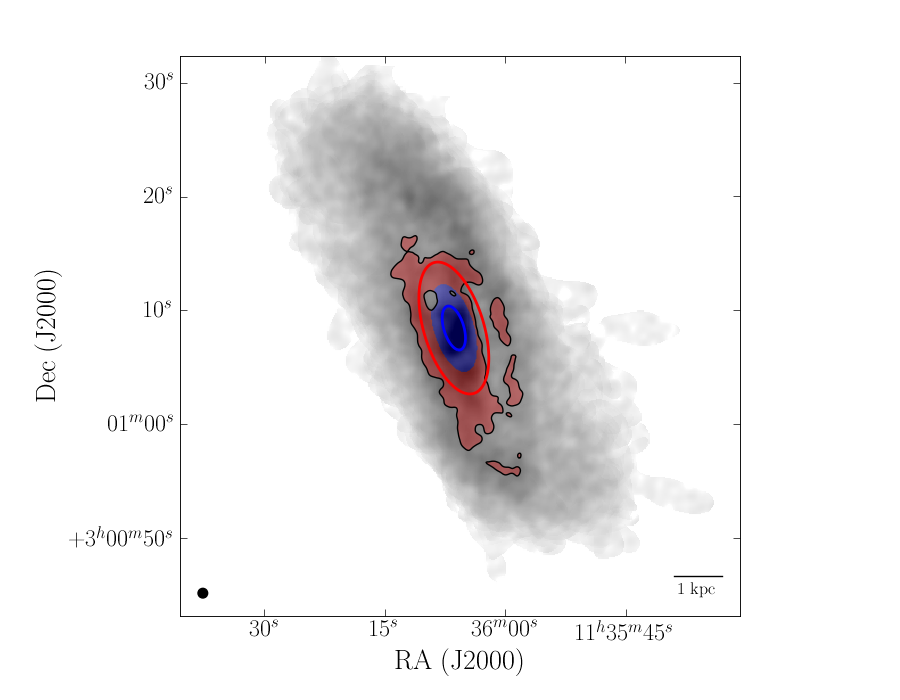}
\caption[Superprofile radial annuli for NGC~3741]{Radial annuli in which radial superprofiles are generated for NGC~3741. 
In both panels, the background greyscale shows \hisd{}, and the solid black line represents the $S/N > 5$ threshold where we can accurately measure \vp{}. 
In the lower panel, the colored solid lines represent the average radius of each annulus, and the corresponding shaded regions of the same color indicate which pixels have contributed to each radial superprofile.
\label{resolved::fig:superprofiles-radial-n3741-a} }
\end{leftfullpage}
\end{figure}
\addtocounter{figure}{-1}
\addtocounter{subfig}{1}
\begin{figure}
\centering
\includegraphics[height=2.7in]{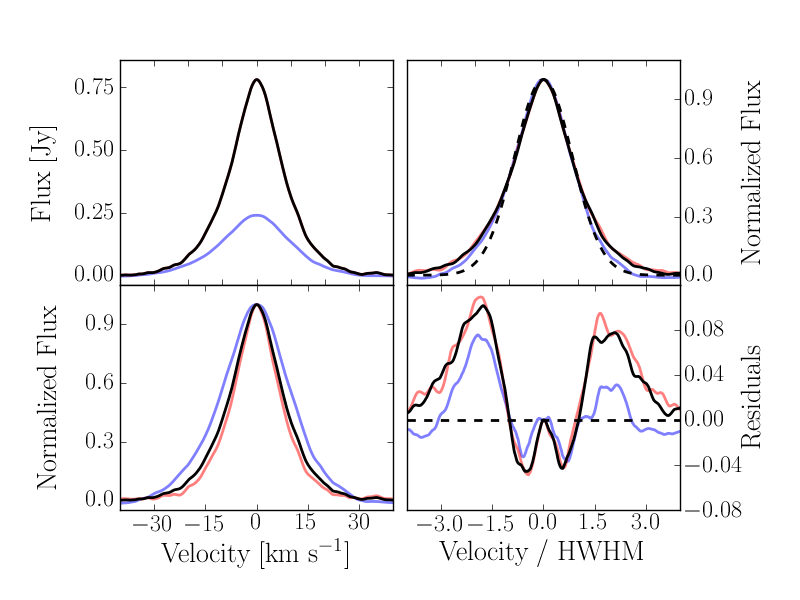}
\caption[Radial superprofiles in NGC~3741]{The radial superprofiles in NGC~3741, where colors indicate the corresponding radial annuli in the previous figure.
The left hand panels show the raw superprofiles (upper left) and the superprofiles normalized to the same peak flux (lower left).
The right hand panels show the flux-normalized superprofiles scaled by the HWHM (upper right) and the flux-normalized superprofiles minus the model of the Gaussian core (lower right). In all panels, the solid black line represents the global superprofile. In the left panels, we have shown the HWHM-scaled Gaussian model as the dashed black line.
\label{resolved::fig:superprofiles-radial-n3741-b}
}
\end{figure}
\addtocounter{figure}{-1}
\addtocounter{subfig}{1}
\begin{figure}
\centering
\includegraphics[height=2.7in]{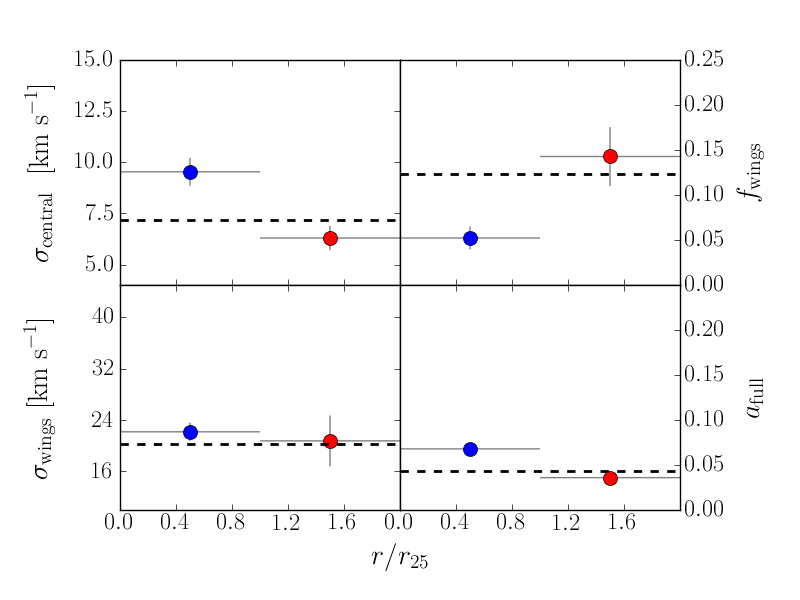}
\caption[Variation of radial superprofile parameters for NGC~3741]{Variation of the superprofile parameters as a function of normalized radius for NGC~3741.
The solid dashed line shows the parameter value for the global superprofile \citepalias{StilpGlobal}.
The left panels show \scentral{} (upper) and \swing{} (lower), and the right panels show \fw{} (upper) and \afull{} (lower).
\label{resolved::fig:superprofiles-radial-n3741-c}
}
\end{figure}
\clearpage

\setcounter{subfig}{1}
\begin{figure}[p]
\begin{leftfullpage}
\centering
\includegraphics[width=4in]{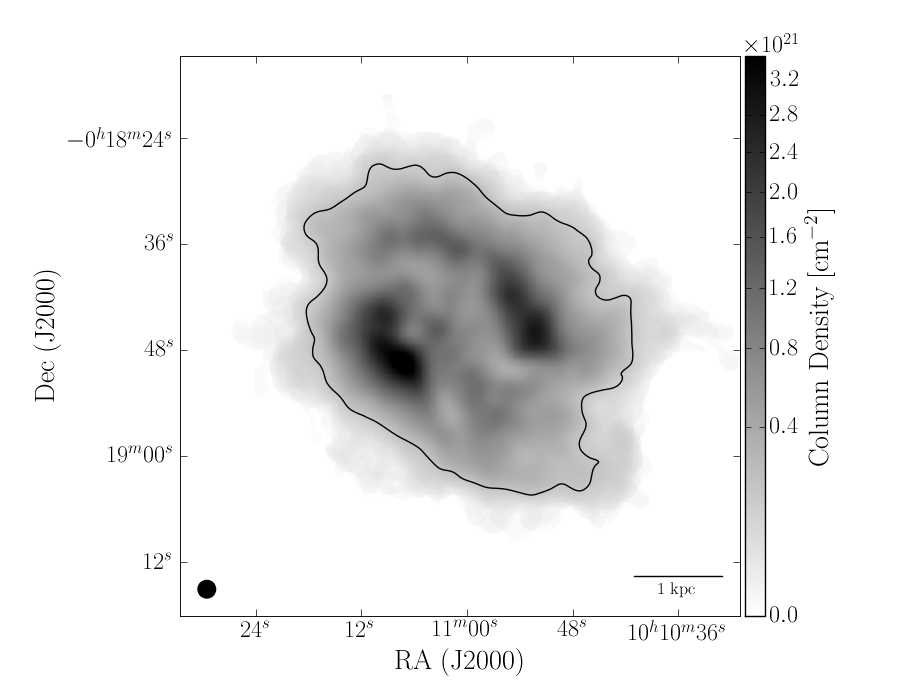}
\includegraphics[width=4in]{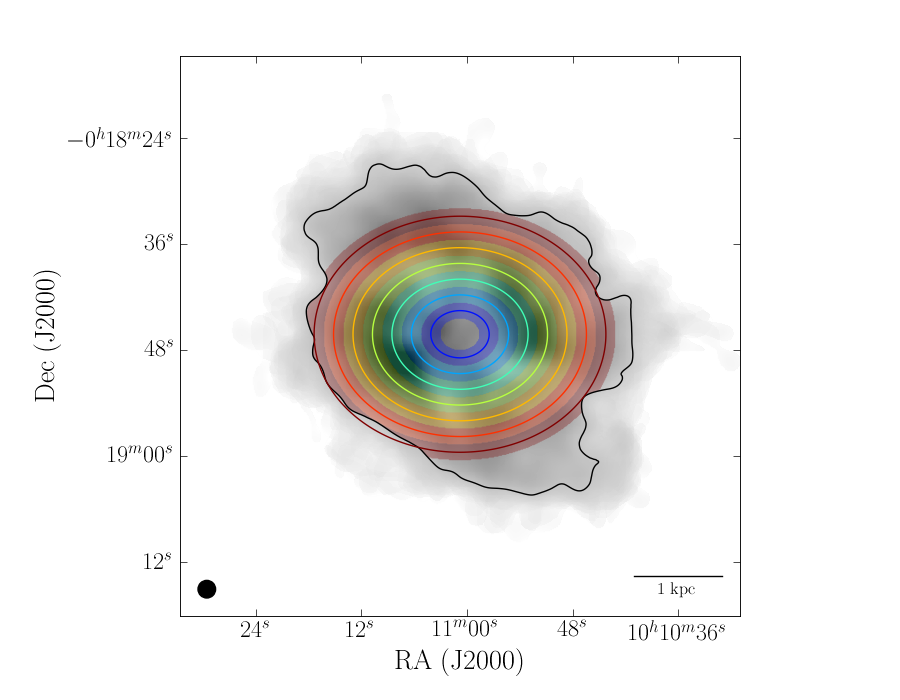}
\caption[Superprofile radial annuli for Sextans A]{Radial annuli in which radial superprofiles are generated for Sextans A. 
In both panels, the background greyscale shows \hisd{}, and the solid black line represents the $S/N > 5$ threshold where we can accurately measure \vp{}. 
In the lower panel, the colored solid lines represent the average radius of each annulus, and the corresponding shaded regions of the same color indicate which pixels have contributed to each radial superprofile.
\label{resolved::fig:superprofiles-radial-sexa-a} }
\end{leftfullpage}
\end{figure}
\addtocounter{figure}{-1}
\addtocounter{subfig}{1}
\begin{figure}
\centering
\includegraphics[height=2.7in]{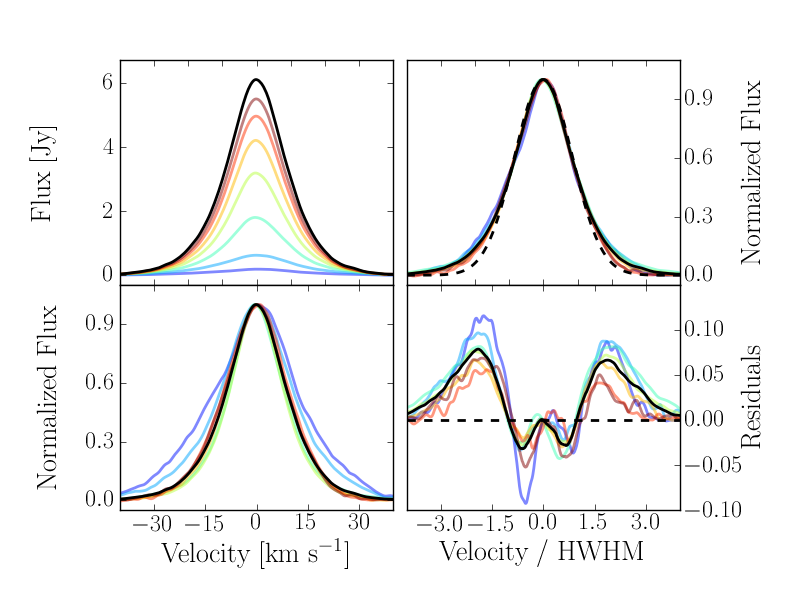}
\caption[Radial superprofiles in Sextans A]{The radial superprofiles in Sextans A, where colors indicate the corresponding radial annuli in the previous figure.
The left hand panels show the raw superprofiles (upper left) and the superprofiles normalized to the same peak flux (lower left).
The right hand panels show the flux-normalized superprofiles scaled by the HWHM (upper right) and the flux-normalized superprofiles minus the model of the Gaussian core (lower right). In all panels, the solid black line represents the global superprofile. In the left panels, we have shown the HWHM-scaled Gaussian model as the dashed black line.
\label{resolved::fig:superprofiles-radial-sexa-b}
}
\end{figure}
\addtocounter{figure}{-1}
\addtocounter{subfig}{1}
\begin{figure}
\centering
\includegraphics[height=2.7in]{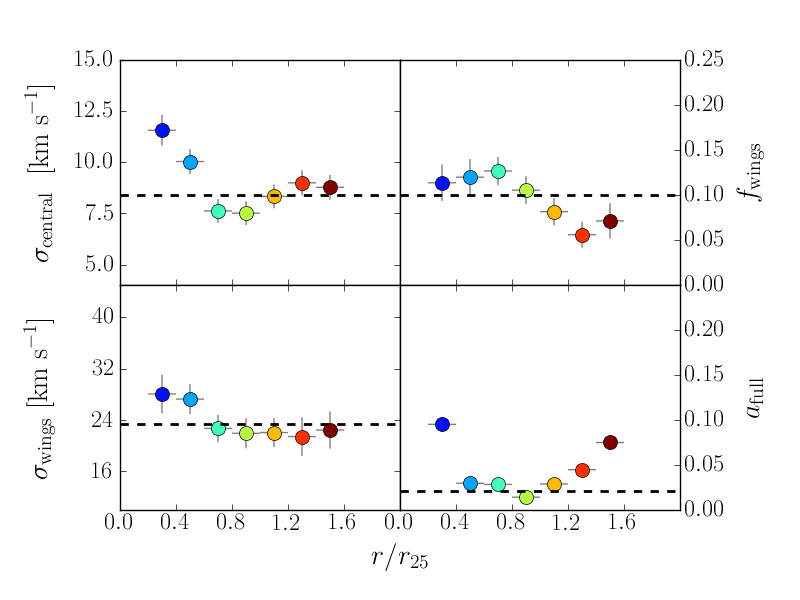}
\caption[Variation of radial superprofile parameters for Sextans A]{Variation of the superprofile parameters as a function of normalized radius for Sextans A.
The solid dashed line shows the parameter value for the global superprofile \citepalias{StilpGlobal}.
The left panels show \scentral{} (upper) and \swing{} (lower), and the right panels show \fw{} (upper) and \afull{} (lower).
\label{resolved::fig:superprofiles-radial-sexa-c}
}
\end{figure}
\clearpage

\setcounter{subfig}{1}
\begin{figure}[p]
\begin{leftfullpage}
\centering
\includegraphics[width=4in]{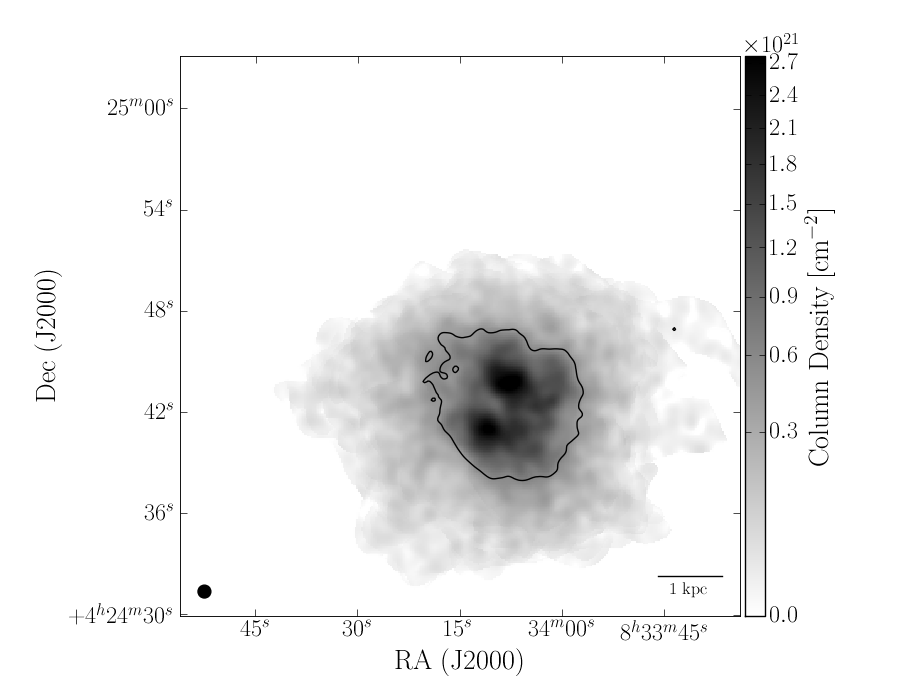}
\includegraphics[width=4in]{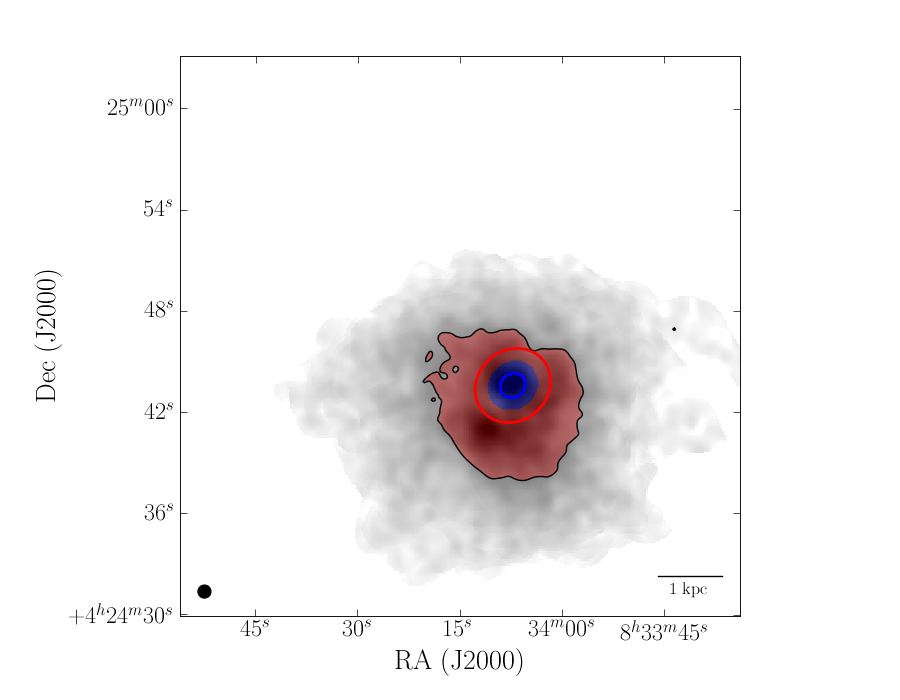}
\caption[Superprofile radial annuli for DDO~53]{Radial annuli in which radial superprofiles are generated for DDO~53. 
In both panels, the background greyscale shows \hisd{}, and the solid black line represents the $S/N > 5$ threshold where we can accurately measure \vp{}. 
In the lower panel, the colored solid lines represent the average radius of each annulus, and the corresponding shaded regions of the same color indicate which pixels have contributed to each radial superprofile.
\label{resolved::fig:superprofiles-radial-ddo53-a} }
\end{leftfullpage}
\end{figure}
\addtocounter{figure}{-1}
\addtocounter{subfig}{1}
\begin{figure}
\centering
\includegraphics[height=2.7in]{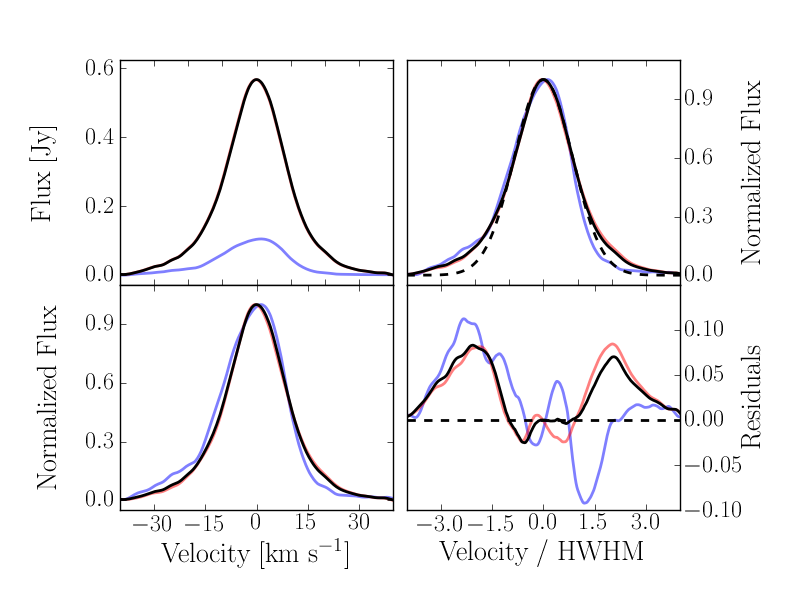}
\caption[Radial superprofiles in DDO~53]{The radial superprofiles in DDO~53, where colors indicate the corresponding radial annuli in the previous figure.
The left hand panels show the raw superprofiles (upper left) and the superprofiles normalized to the same peak flux (lower left).
The right hand panels show the flux-normalized superprofiles scaled by the HWHM (upper right) and the flux-normalized superprofiles minus the model of the Gaussian core (lower right). In all panels, the solid black line represents the global superprofile. In the left panels, we have shown the HWHM-scaled Gaussian model as the dashed black line.
\label{resolved::fig:superprofiles-radial-ddo53-b}
}
\end{figure}
\addtocounter{figure}{-1}
\addtocounter{subfig}{1}
\begin{figure}
\centering
\includegraphics[height=2.7in]{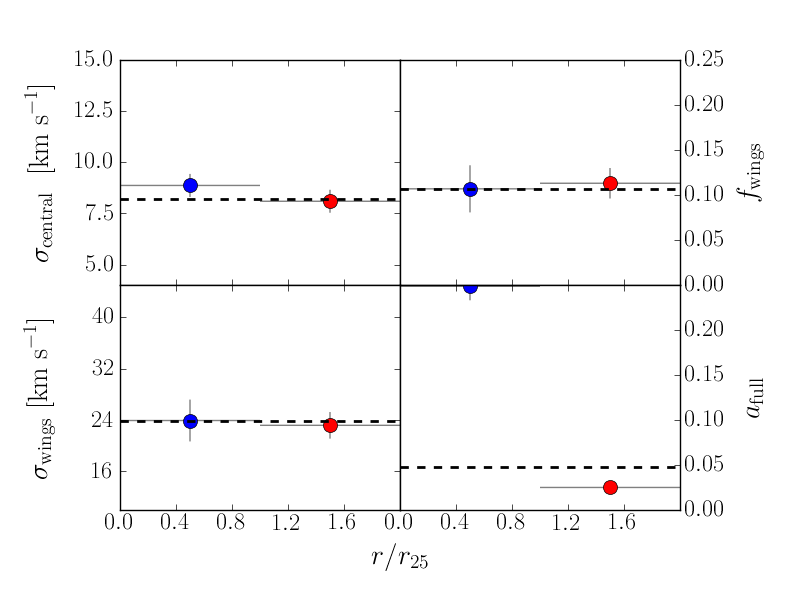}
\caption[Variation of radial superprofile parameters for DDO~53]{Variation of the superprofile parameters as a function of normalized radius for DDO~53.
The solid dashed line shows the parameter value for the global superprofile \citepalias{StilpGlobal}.
The left panels show \scentral{} (upper) and \swing{} (lower), and the right panels show \fw{} (upper) and \afull{} (lower).
\label{resolved::fig:superprofiles-radial-ddo53-c}
}
\end{figure}
\clearpage

\setcounter{subfig}{1}
\begin{figure}[p]
\begin{leftfullpage}
\centering
\includegraphics[width=4in]{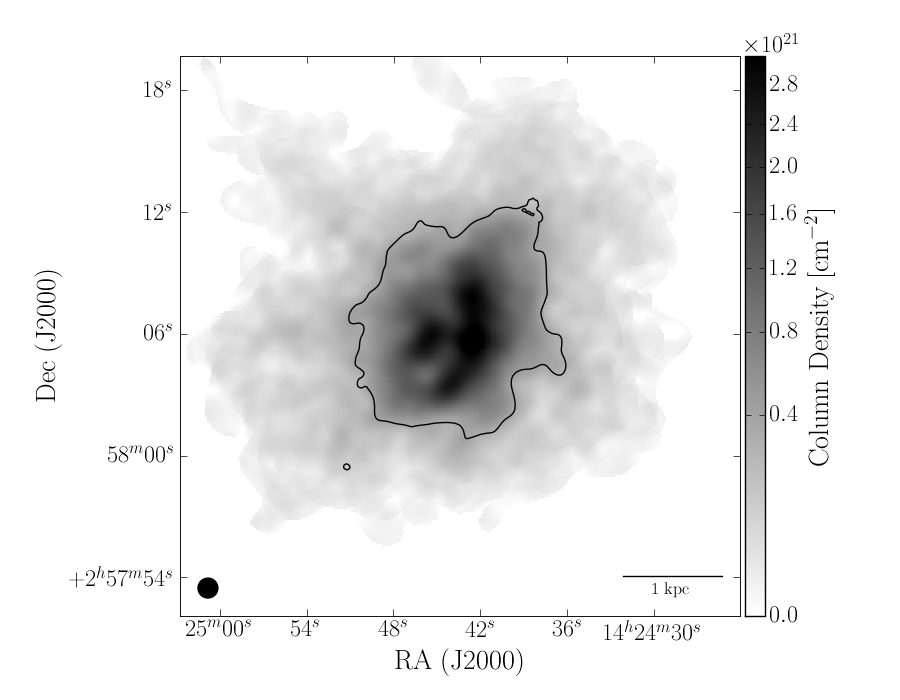}
\includegraphics[width=4in]{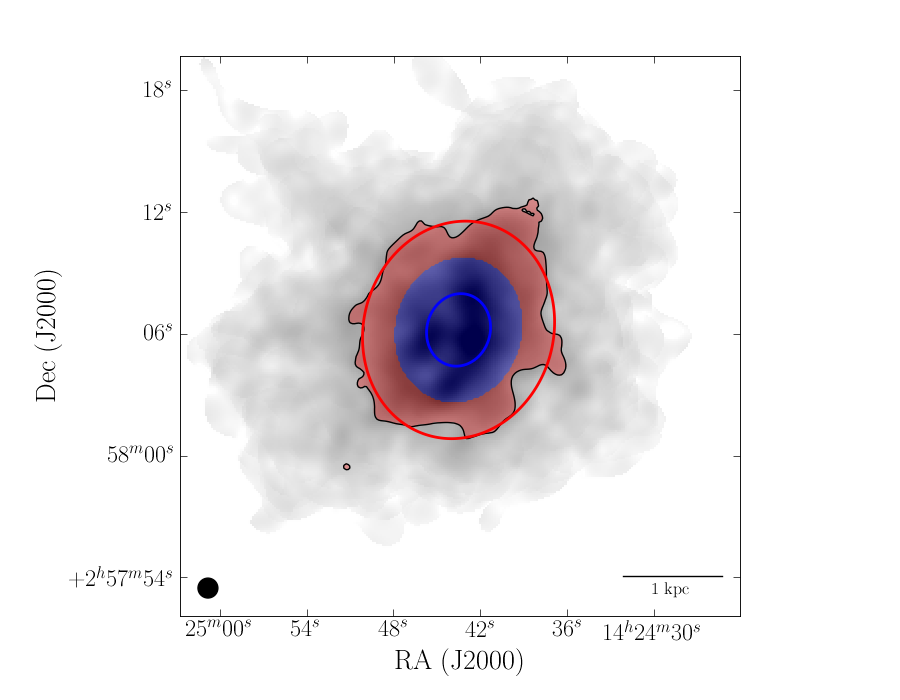}
\caption[Superprofile radial annuli for DDO~190]{Radial annuli in which radial superprofiles are generated for DDO~190. 
In both panels, the background greyscale shows \hisd{}, and the solid black line represents the $S/N > 5$ threshold where we can accurately measure \vp{}. 
In the lower panel, the colored solid lines represent the average radius of each annulus, and the corresponding shaded regions of the same color indicate which pixels have contributed to each radial superprofile.
\label{resolved::fig:superprofiles-radial-ddo190-a} }
\end{leftfullpage}
\end{figure}
\addtocounter{figure}{-1}
\addtocounter{subfig}{1}
\begin{figure}
\centering
\includegraphics[height=2.7in]{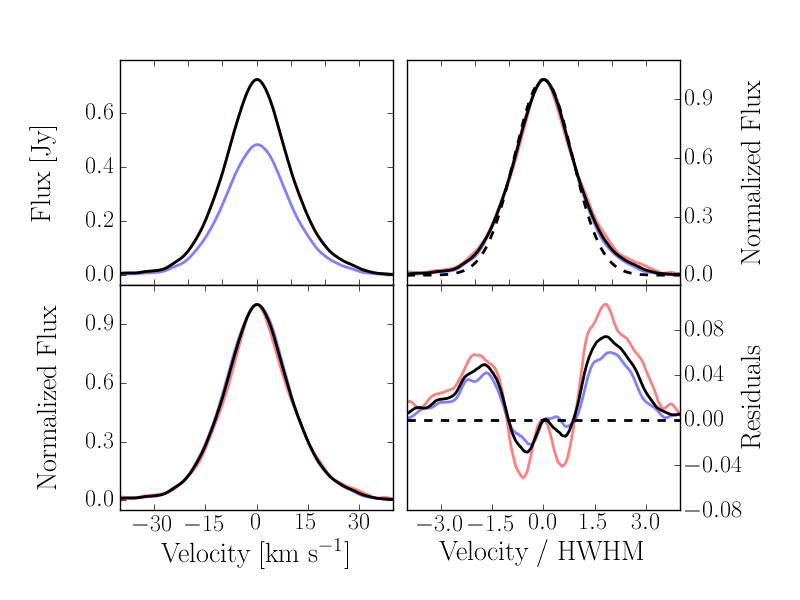}
\caption[Radial superprofiles in DDO~190]{The radial superprofiles in DDO~190, where colors indicate the corresponding radial annuli in the previous figure.
The left hand panels show the raw superprofiles (upper left) and the superprofiles normalized to the same peak flux (lower left).
The right hand panels show the flux-normalized superprofiles scaled by the HWHM (upper right) and the flux-normalized superprofiles minus the model of the Gaussian core (lower right). In all panels, the solid black line represents the global superprofile. In the left panels, we have shown the HWHM-scaled Gaussian model as the dashed black line.
\label{resolved::fig:superprofiles-radial-ddo190-b}
}
\end{figure}
\addtocounter{figure}{-1}
\addtocounter{subfig}{1}
\begin{figure}
\centering
\includegraphics[height=2.7in]{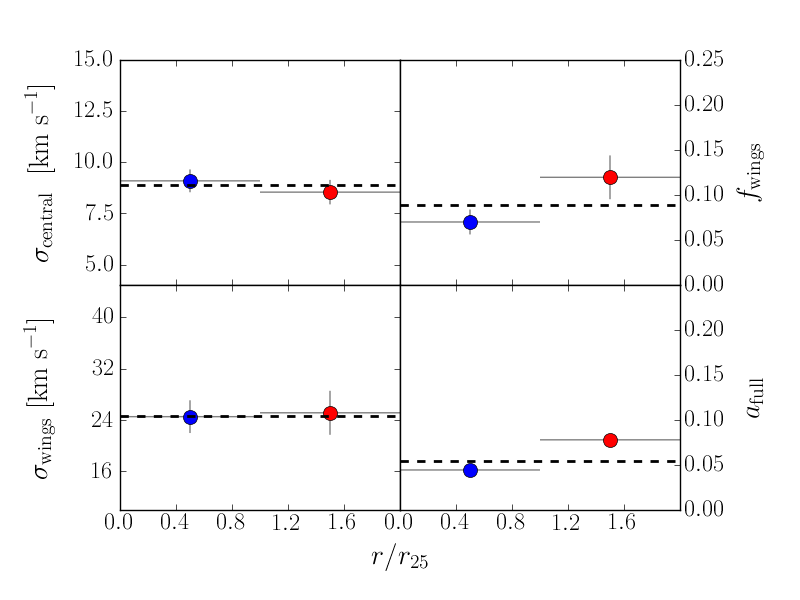}
\caption[Variation of radial superprofile parameters for DDO~190]{Variation of the superprofile parameters as a function of normalized radius for DDO~190.
The solid dashed line shows the parameter value for the global superprofile \citepalias{StilpGlobal}.
The left panels show \scentral{} (upper) and \swing{} (lower), and the right panels show \fw{} (upper) and \afull{} (lower).
\label{resolved::fig:superprofiles-radial-ddo190-c}
}
\end{figure}
\clearpage

\setcounter{subfig}{1}
\begin{figure}[p]
\begin{leftfullpage}
\centering
\includegraphics[width=4in]{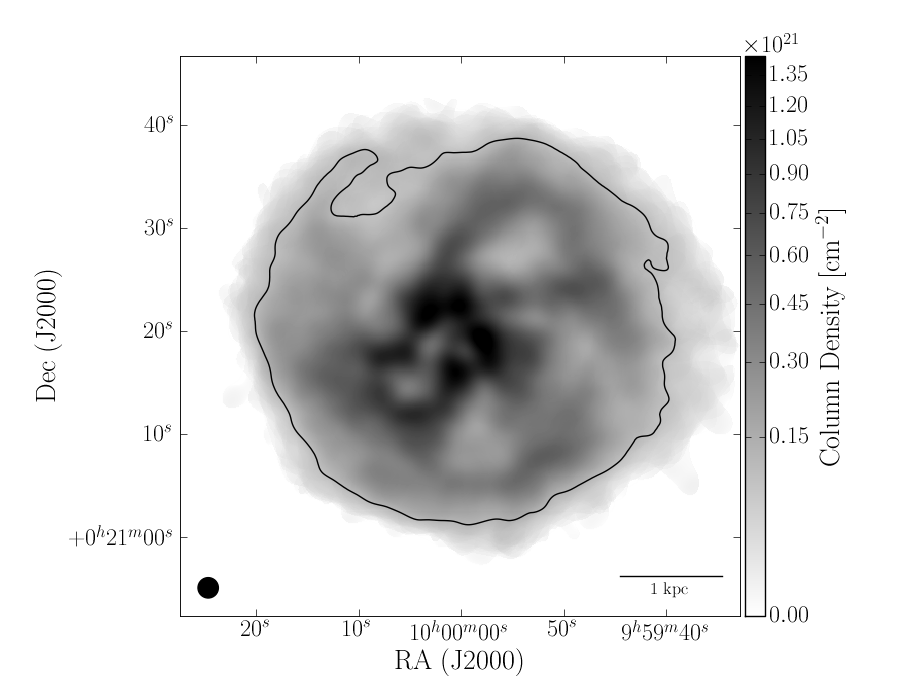}
\includegraphics[width=4in]{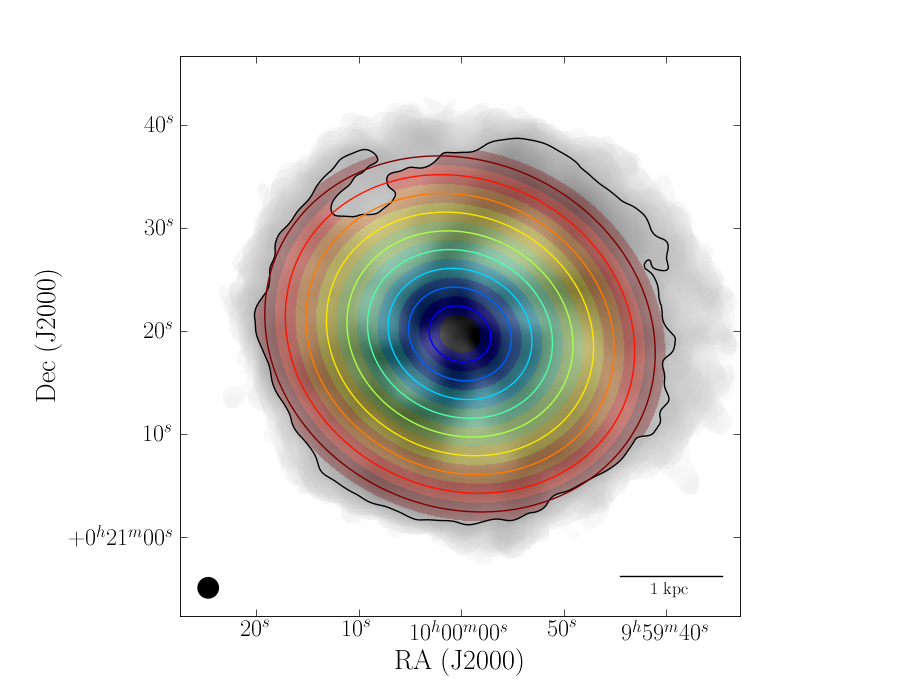}
\caption[Superprofile radial annuli for Sextans B]{Radial annuli in which radial superprofiles are generated for Sextans B. 
In both panels, the background greyscale shows \hisd{}, and the solid black line represents the $S/N > 5$ threshold where we can accurately measure \vp{}. 
In the lower panel, the colored solid lines represent the average radius of each annulus, and the corresponding shaded regions of the same color indicate which pixels have contributed to each radial superprofile.
\label{resolved::fig:superprofiles-radial-sexb-a} }
\end{leftfullpage}
\end{figure}
\addtocounter{figure}{-1}
\addtocounter{subfig}{1}
\begin{figure}
\centering
\includegraphics[height=2.7in]{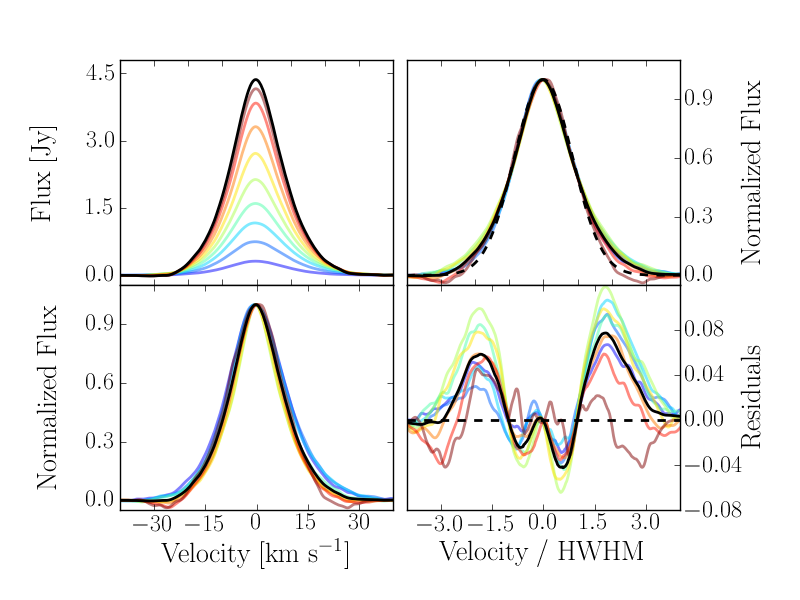}
\caption[Radial superprofiles in Sextans B]{The radial superprofiles in Sextans B, where colors indicate the corresponding radial annuli in the previous figure.
The left hand panels show the raw superprofiles (upper left) and the superprofiles normalized to the same peak flux (lower left).
The right hand panels show the flux-normalized superprofiles scaled by the HWHM (upper right) and the flux-normalized superprofiles minus the model of the Gaussian core (lower right). In all panels, the solid black line represents the global superprofile. In the left panels, we have shown the HWHM-scaled Gaussian model as the dashed black line.
\label{resolved::fig:superprofiles-radial-sexb-b}
}
\end{figure}
\addtocounter{figure}{-1}
\addtocounter{subfig}{1}
\begin{figure}
\centering
\includegraphics[height=2.7in]{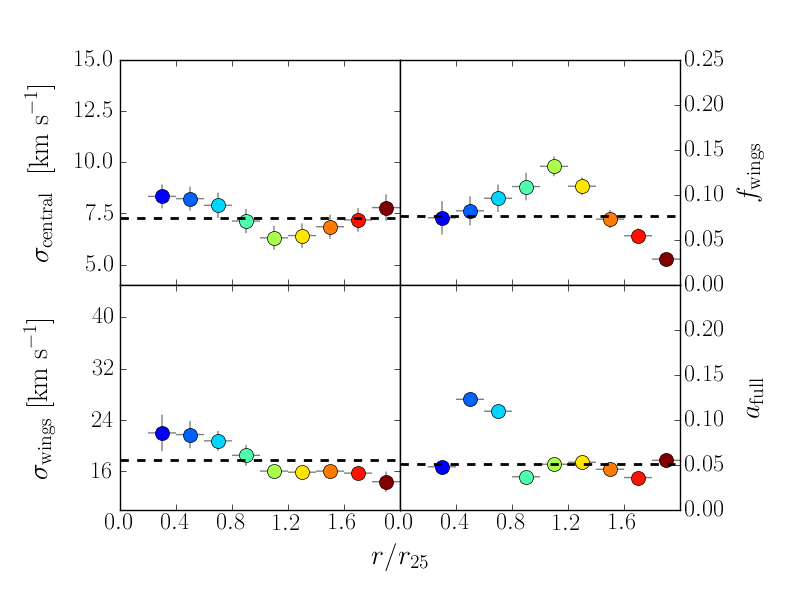}
\caption[Variation of radial superprofile parameters for Sextans B]{Variation of the superprofile parameters as a function of normalized radius for Sextans B.
The solid dashed line shows the parameter value for the global superprofile \citepalias{StilpGlobal}.
The left panels show \scentral{} (upper) and \swing{} (lower), and the right panels show \fw{} (upper) and \afull{} (lower).
\label{resolved::fig:superprofiles-radial-sexb-c}
}
\end{figure}
\clearpage

\setcounter{subfig}{1}
\begin{figure}[p]
\begin{leftfullpage}
\centering
\includegraphics[width=4in]{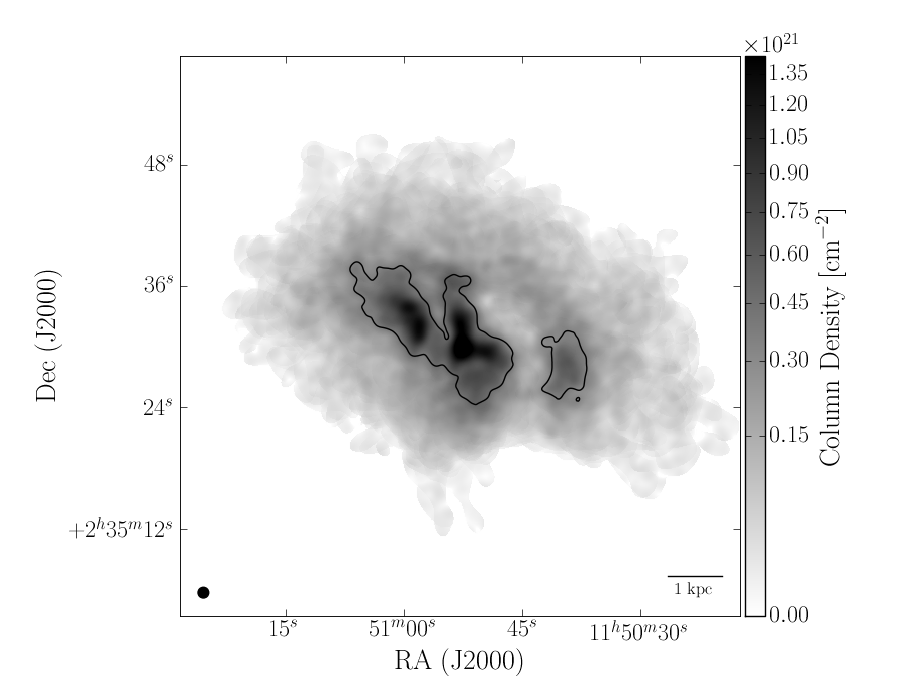}
\includegraphics[width=4in]{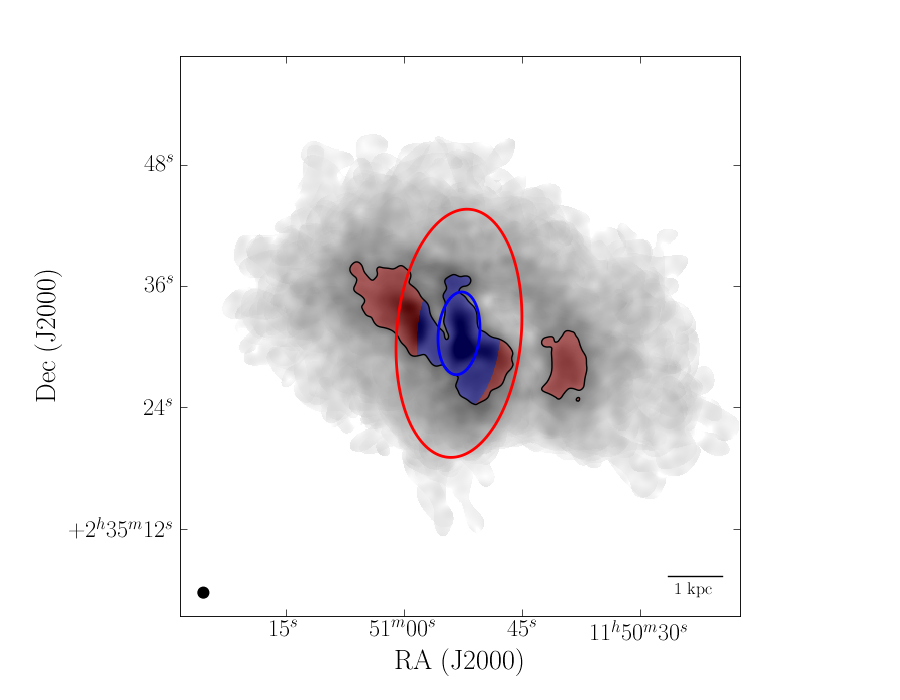}
\caption[Superprofile radial annuli for DDO~99]{Radial annuli in which radial superprofiles are generated for DDO~99. 
In both panels, the background greyscale shows \hisd{}, and the solid black line represents the $S/N > 5$ threshold where we can accurately measure \vp{}. 
In the lower panel, the colored solid lines represent the average radius of each annulus, and the corresponding shaded regions of the same color indicate which pixels have contributed to each radial superprofile.
\label{resolved::fig:superprofiles-radial-ddo99-a} }
\end{leftfullpage}
\end{figure}
\addtocounter{figure}{-1}
\addtocounter{subfig}{1}
\begin{figure}
\centering
\includegraphics[height=2.7in]{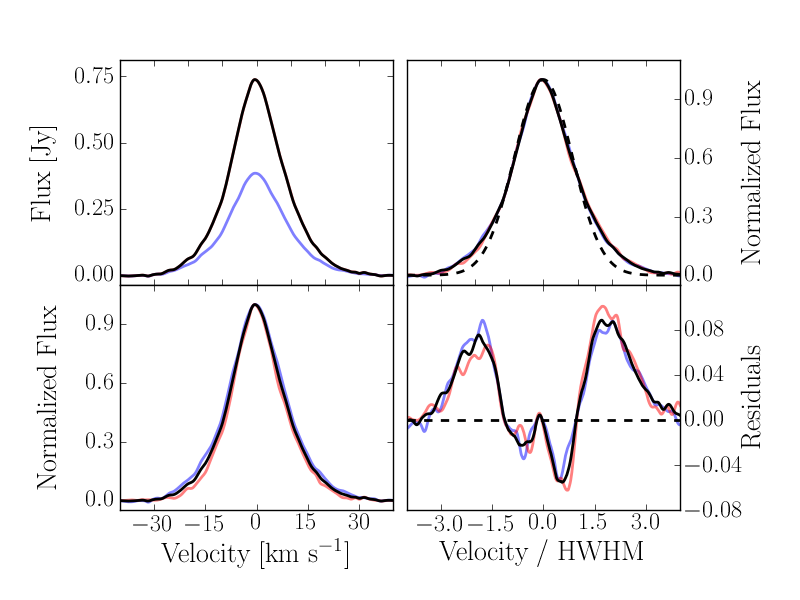}
\caption[Radial superprofiles in DDO~99]{The radial superprofiles in DDO~99, where colors indicate the corresponding radial annuli in the previous figure.
The left hand panels show the raw superprofiles (upper left) and the superprofiles normalized to the same peak flux (lower left).
The right hand panels show the flux-normalized superprofiles scaled by the HWHM (upper right) and the flux-normalized superprofiles minus the model of the Gaussian core (lower right). In all panels, the solid black line represents the global superprofile. In the left panels, we have shown the HWHM-scaled Gaussian model as the dashed black line.
\label{resolved::fig:superprofiles-radial-ddo99-b}
}
\end{figure}
\addtocounter{figure}{-1}
\addtocounter{subfig}{1}
\begin{figure}
\centering
\includegraphics[height=2.7in]{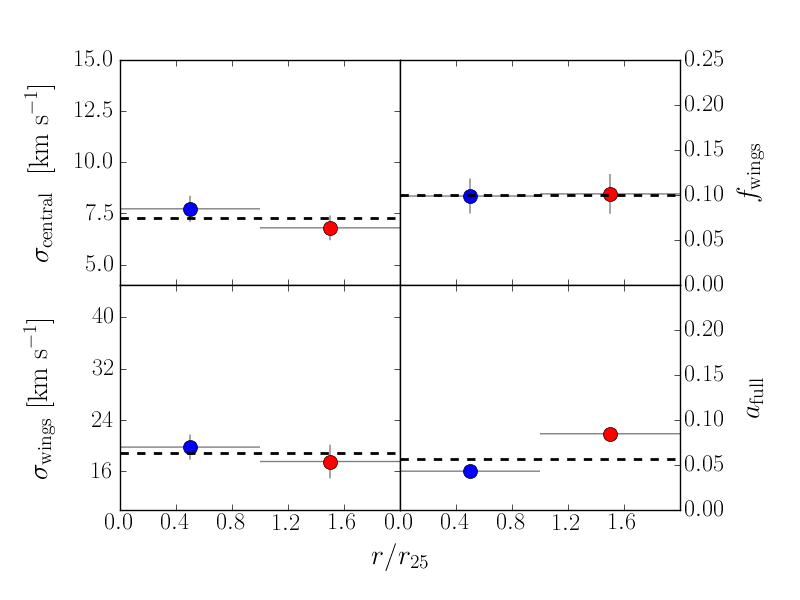}
\caption[Variation of radial superprofile parameters for DDO~99]{Variation of the superprofile parameters as a function of normalized radius for DDO~99.
The solid dashed line shows the parameter value for the global superprofile \citepalias{StilpGlobal}.
The left panels show \scentral{} (upper) and \swing{} (lower), and the right panels show \fw{} (upper) and \afull{} (lower).
\label{resolved::fig:superprofiles-radial-ddo99-c}
}
\end{figure}
\clearpage

\setcounter{subfig}{1}
\begin{figure}[p]
\begin{leftfullpage}
\centering
\includegraphics[width=4in]{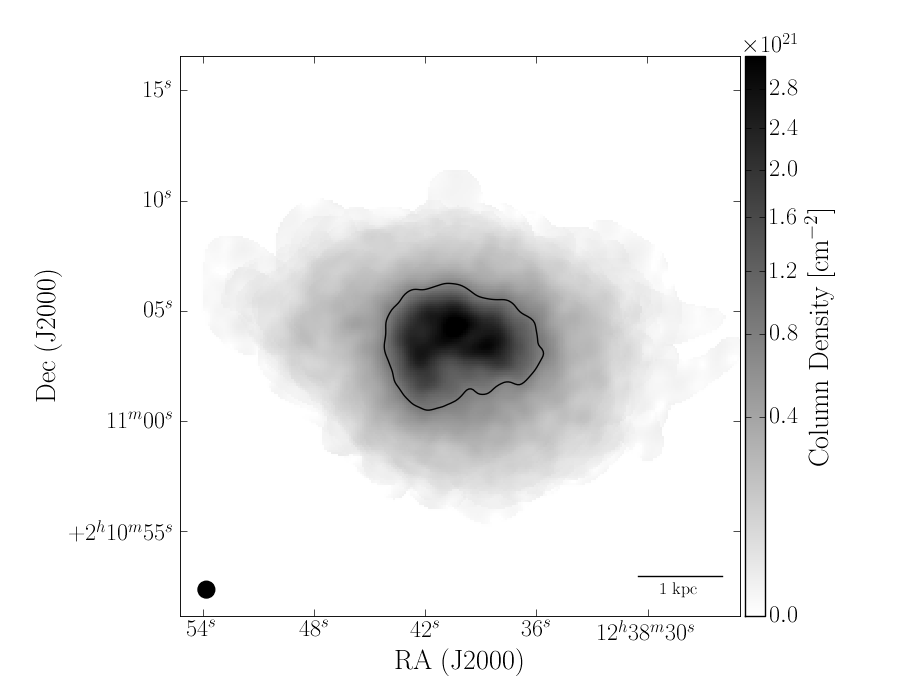}
\includegraphics[width=4in]{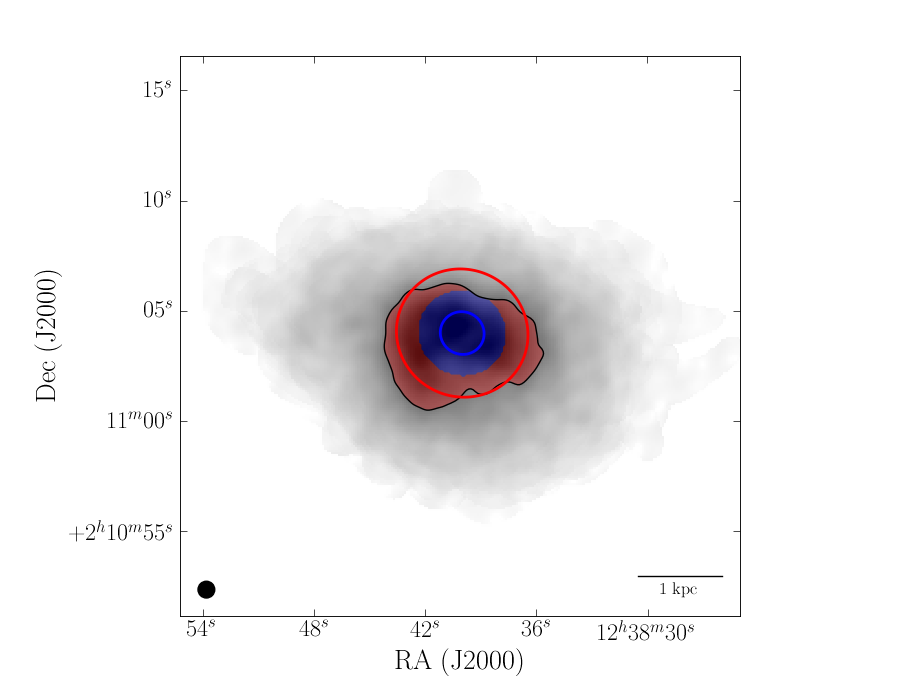}
\caption[Superprofile radial annuli for UGCA~292]{Radial annuli in which radial superprofiles are generated for UGCA~292. 
In both panels, the background greyscale shows \hisd{}, and the solid black line represents the $S/N > 5$ threshold where we can accurately measure \vp{}. 
In the lower panel, the colored solid lines represent the average radius of each annulus, and the corresponding shaded regions of the same color indicate which pixels have contributed to each radial superprofile.
\label{resolved::fig:superprofiles-radial-ua292-a} }
\end{leftfullpage}
\end{figure}
\addtocounter{figure}{-1}
\addtocounter{subfig}{1}
\begin{figure}
\centering
\includegraphics[height=2.7in]{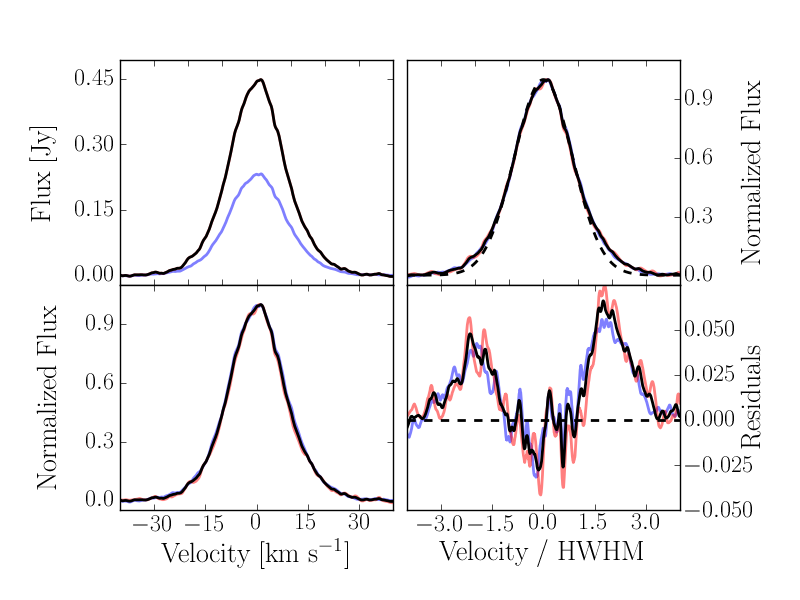}
\caption[Radial superprofiles in UGCA~292]{The radial superprofiles in UGCA~292, where colors indicate the corresponding radial annuli in the previous figure.
The left hand panels show the raw superprofiles (upper left) and the superprofiles normalized to the same peak flux (lower left).
The right hand panels show the flux-normalized superprofiles scaled by the HWHM (upper right) and the flux-normalized superprofiles minus the model of the Gaussian core (lower right). In all panels, the solid black line represents the global superprofile. In the left panels, we have shown the HWHM-scaled Gaussian model as the dashed black line.
\label{resolved::fig:superprofiles-radial-ua292-b}
}
\end{figure}
\addtocounter{figure}{-1}
\addtocounter{subfig}{1}
\begin{figure}
\centering
\includegraphics[height=2.7in]{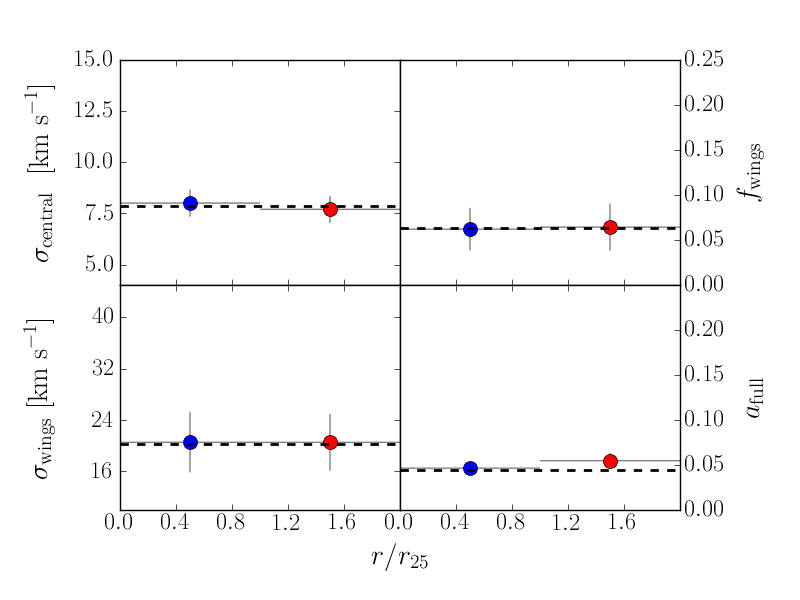}
\caption[Variation of radial superprofile parameters for UGCA~292]{Variation of the superprofile parameters as a function of normalized radius for UGCA~292.
The solid dashed line shows the parameter value for the global superprofile \citepalias{StilpGlobal}.
The left panels show \scentral{} (upper) and \swing{} (lower), and the right panels show \fw{} (upper) and \afull{} (lower).
\label{resolved::fig:superprofiles-radial-ua292-c}
}
\end{figure}
\clearpage

\setcounter{subfig}{1}
\begin{figure}[p]
\begin{leftfullpage}
\centering
\includegraphics[width=4in]{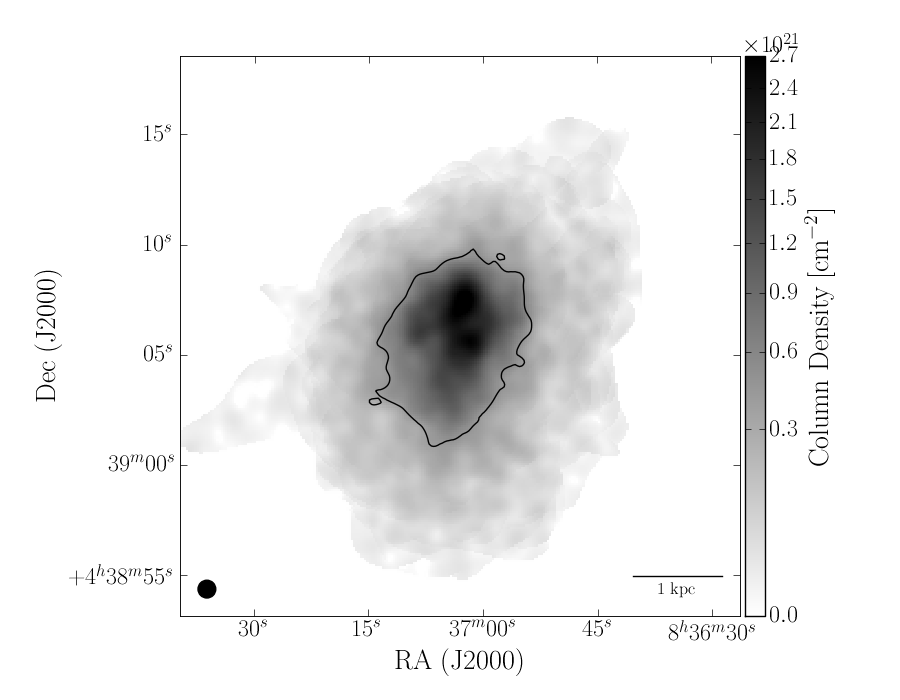}
\includegraphics[width=4in]{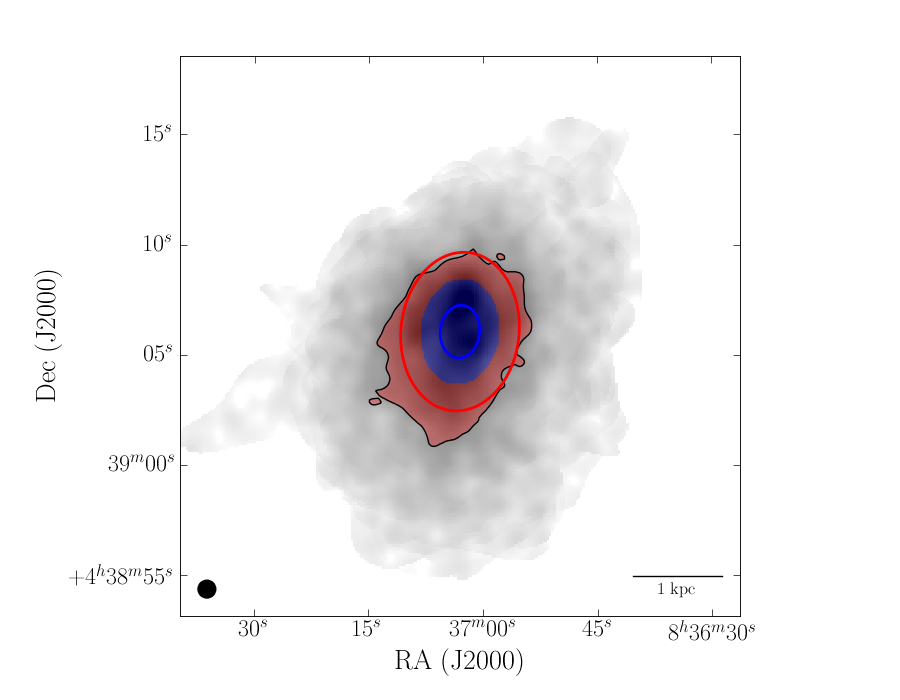}
\caption[Superprofile radial annuli for UGC~4483]{Radial annuli in which radial superprofiles are generated for UGC~4483. 
In both panels, the background greyscale shows \hisd{}, and the solid black line represents the $S/N > 5$ threshold where we can accurately measure \vp{}. 
In the lower panel, the colored solid lines represent the average radius of each annulus, and the corresponding shaded regions of the same color indicate which pixels have contributed to each radial superprofile.
\label{resolved::fig:superprofiles-radial-u4483-a} }
\end{leftfullpage}
\end{figure}
\addtocounter{figure}{-1}
\addtocounter{subfig}{1}
\begin{figure}
\centering
\includegraphics[height=2.7in]{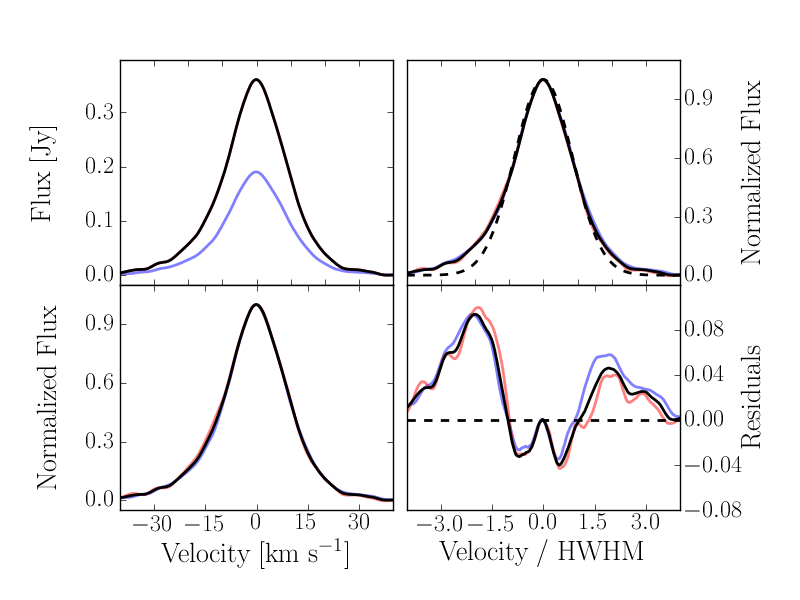}
\caption[Radial superprofiles in UGC~4483]{The radial superprofiles in UGC~4483, where colors indicate the corresponding radial annuli in the previous figure.
The left hand panels show the raw superprofiles (upper left) and the superprofiles normalized to the same peak flux (lower left).
The right hand panels show the flux-normalized superprofiles scaled by the HWHM (upper right) and the flux-normalized superprofiles minus the model of the Gaussian core (lower right). In all panels, the solid black line represents the global superprofile. In the left panels, we have shown the HWHM-scaled Gaussian model as the dashed black line.
\label{resolved::fig:superprofiles-radial-u4483-b}
}
\end{figure}
\addtocounter{figure}{-1}
\addtocounter{subfig}{1}
\begin{figure}
\centering
\includegraphics[height=2.7in]{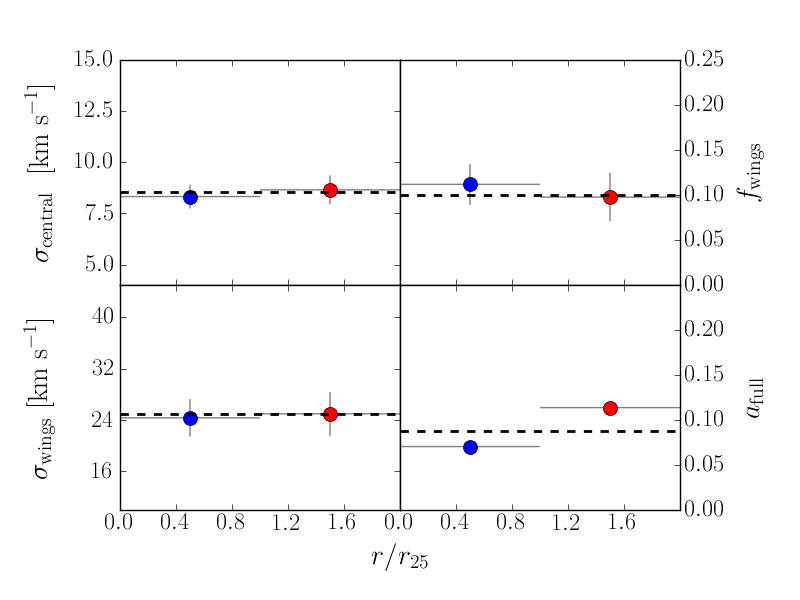}
\caption[Variation of radial superprofile parameters for UGC~4483]{Variation of the superprofile parameters as a function of normalized radius for UGC~4483.
The solid dashed line shows the parameter value for the global superprofile \citepalias{StilpGlobal}.
The left panels show \scentral{} (upper) and \swing{} (lower), and the right panels show \fw{} (upper) and \afull{} (lower).
\label{resolved::fig:superprofiles-radial-u4483-c}
}
\end{figure}
\clearpage

\setcounter{subfig}{1}
\begin{figure}[p]
\begin{leftfullpage}
\centering
\includegraphics[width=4in]{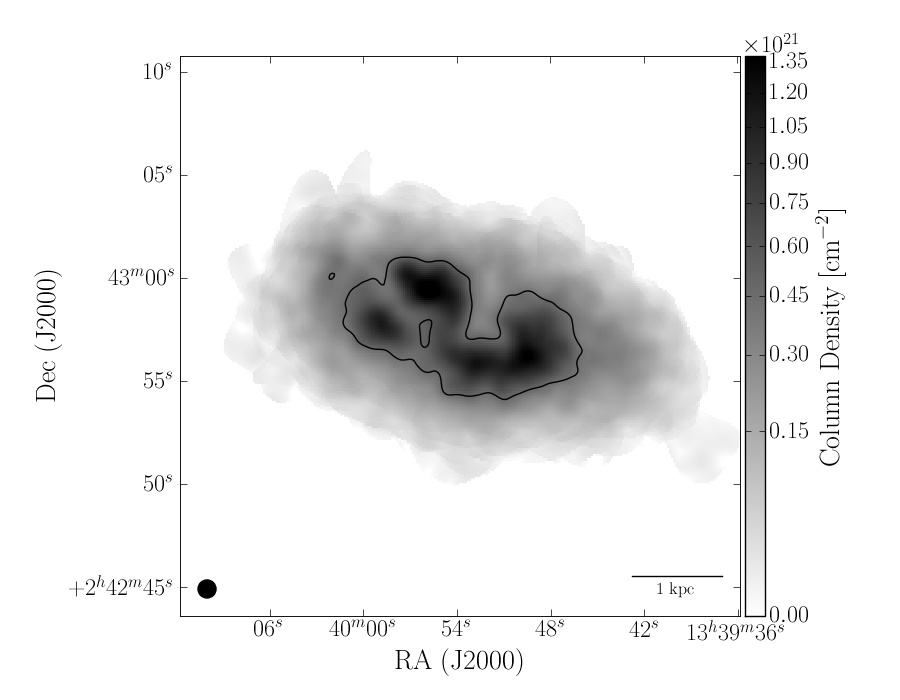}
\includegraphics[width=4in]{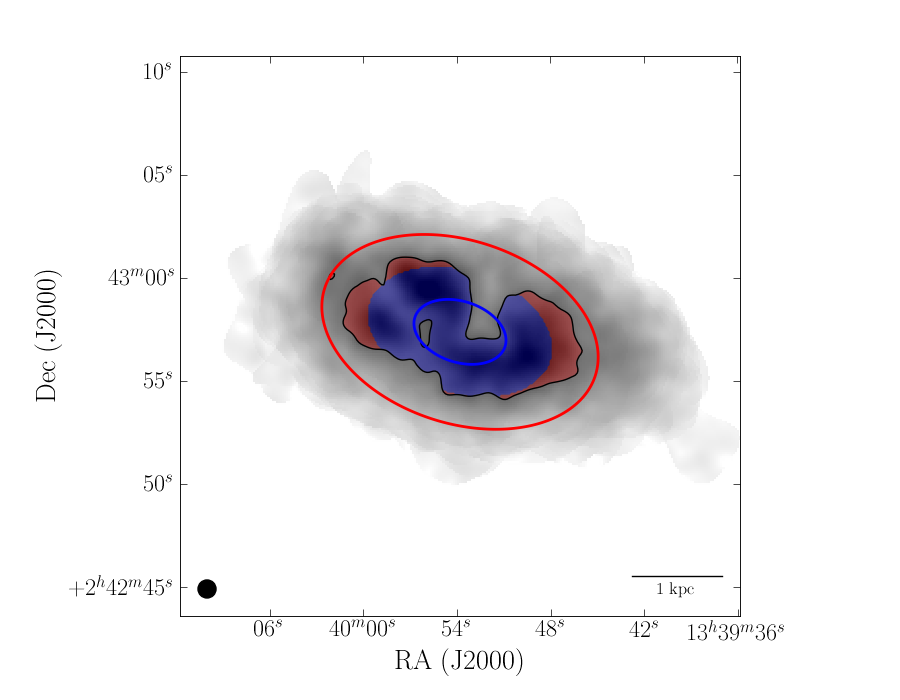}
\caption[Superprofile radial annuli for DDO~181]{Radial annuli in which radial superprofiles are generated for DDO~181. 
In both panels, the background greyscale shows \hisd{}, and the solid black line represents the $S/N > 5$ threshold where we can accurately measure \vp{}. 
In the lower panel, the colored solid lines represent the average radius of each annulus, and the corresponding shaded regions of the same color indicate which pixels have contributed to each radial superprofile.
\label{resolved::fig:superprofiles-radial-ddo181-a} }
\end{leftfullpage}
\end{figure}
\addtocounter{figure}{-1}
\addtocounter{subfig}{1}
\begin{figure}
\centering
\includegraphics[height=2.7in]{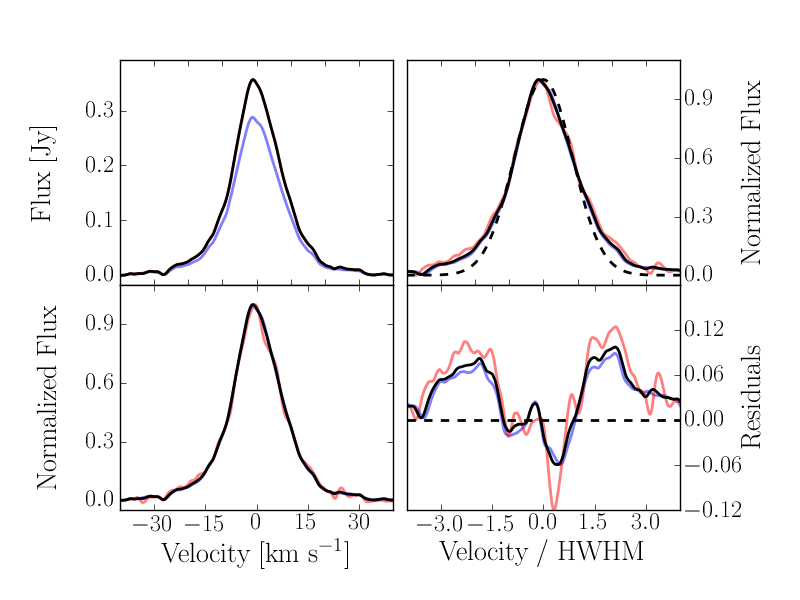}
\caption[Radial superprofiles in DDO~181]{The radial superprofiles in DDO~181, where colors indicate the corresponding radial annuli in the previous figure.
The left hand panels show the raw superprofiles (upper left) and the superprofiles normalized to the same peak flux (lower left).
The right hand panels show the flux-normalized superprofiles scaled by the HWHM (upper right) and the flux-normalized superprofiles minus the model of the Gaussian core (lower right). In all panels, the solid black line represents the global superprofile. In the left panels, we have shown the HWHM-scaled Gaussian model as the dashed black line.
\label{resolved::fig:superprofiles-radial-ddo181-b}
}
\end{figure}
\addtocounter{figure}{-1}
\addtocounter{subfig}{1}
\begin{figure}
\centering
\includegraphics[height=2.7in]{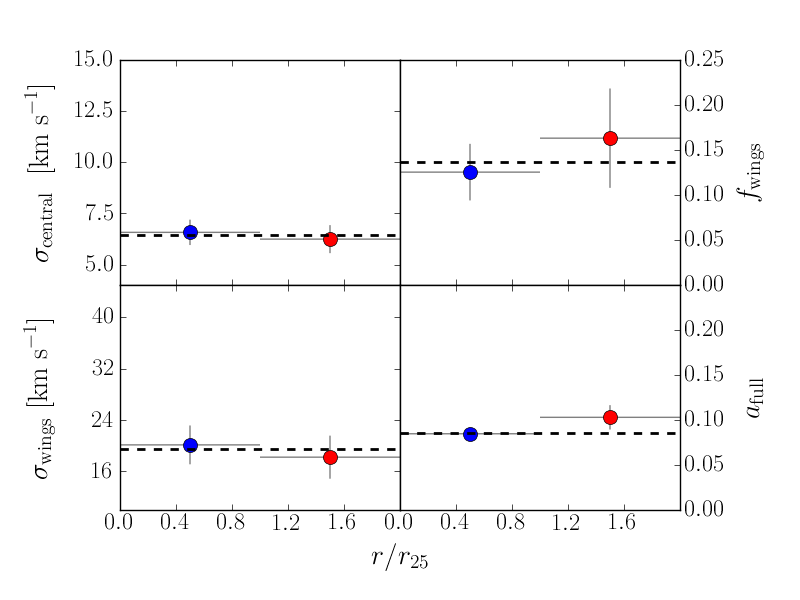}
\caption[Variation of radial superprofile parameters for DDO~181]{Variation of the superprofile parameters as a function of normalized radius for DDO~181.
The solid dashed line shows the parameter value for the global superprofile \citepalias{StilpGlobal}.
The left panels show \scentral{} (upper) and \swing{} (lower), and the right panels show \fw{} (upper) and \afull{} (lower).
\label{resolved::fig:superprofiles-radial-ddo181-c}
}
\end{figure}
\clearpage

\setcounter{subfig}{1}
\begin{figure}[p]
\begin{leftfullpage}
\centering
\includegraphics[width=4in]{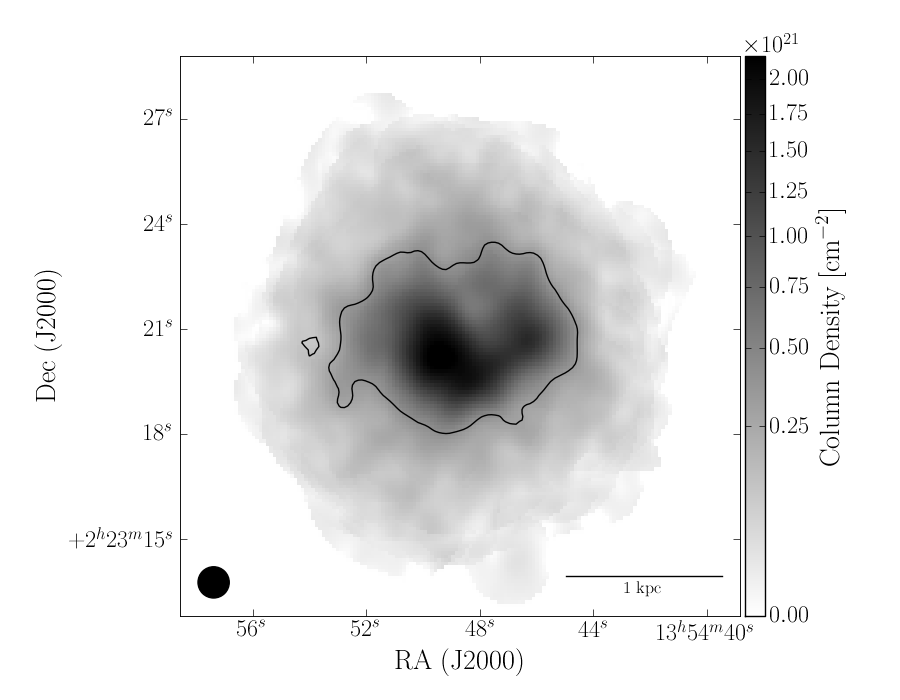}
\includegraphics[width=4in]{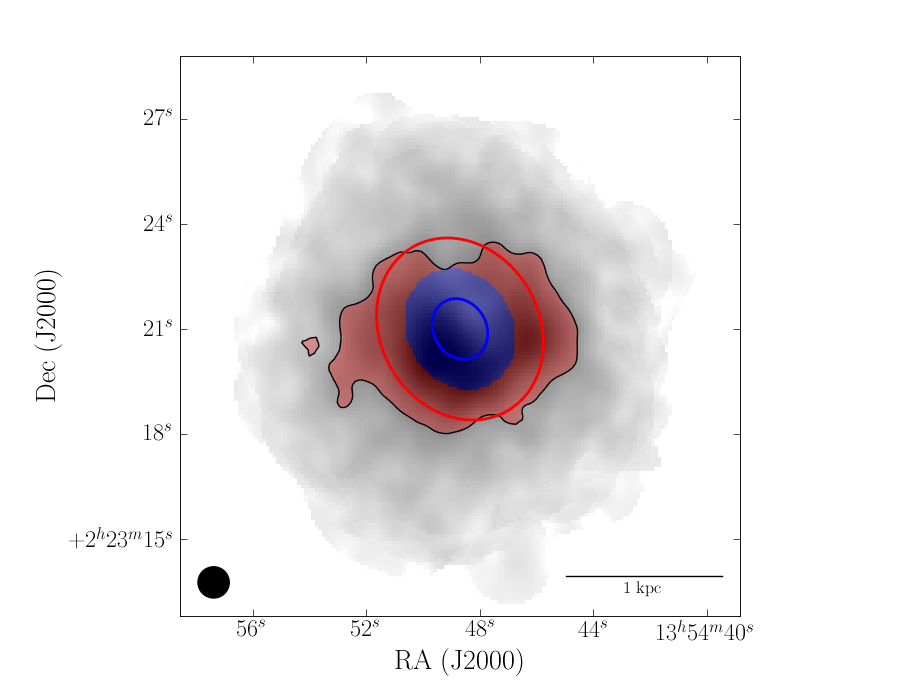}
\caption[Superprofile radial annuli for UGC~8833]{Radial annuli in which radial superprofiles are generated for UGC~8833. 
In both panels, the background greyscale shows \hisd{}, and the solid black line represents the $S/N > 5$ threshold where we can accurately measure \vp{}. 
In the lower panel, the colored solid lines represent the average radius of each annulus, and the corresponding shaded regions of the same color indicate which pixels have contributed to each radial superprofile.
\label{resolved::fig:superprofiles-radial-u8833-a} }
\end{leftfullpage}
\end{figure}
\addtocounter{figure}{-1}
\addtocounter{subfig}{1}
\begin{figure}
\centering
\includegraphics[height=2.7in]{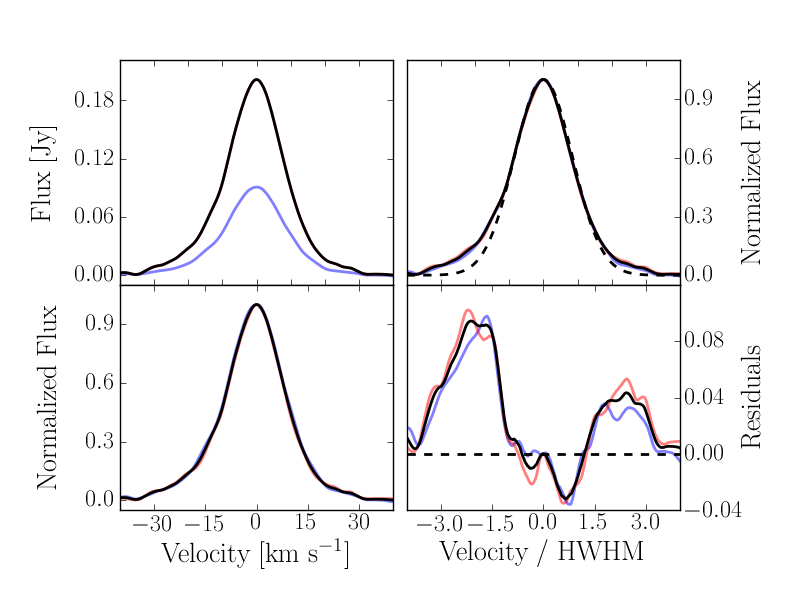}
\caption[Radial superprofiles in UGC~8833]{The radial superprofiles in UGC~8833, where colors indicate the corresponding radial annuli in the previous figure.
The left hand panels show the raw superprofiles (upper left) and the superprofiles normalized to the same peak flux (lower left).
The right hand panels show the flux-normalized superprofiles scaled by the HWHM (upper right) and the flux-normalized superprofiles minus the model of the Gaussian core (lower right). In all panels, the solid black line represents the global superprofile. In the left panels, we have shown the HWHM-scaled Gaussian model as the dashed black line.
\label{resolved::fig:superprofiles-radial-u8833-b}
}
\end{figure}
\addtocounter{figure}{-1}
\addtocounter{subfig}{1}
\begin{figure}
\centering
\includegraphics[height=2.7in]{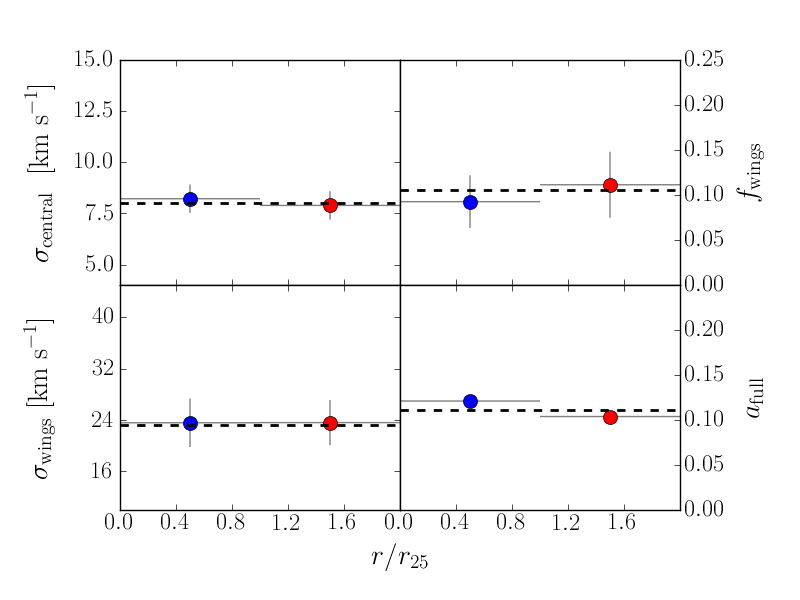}
\caption[Variation of radial superprofile parameters for UGC~8833]{Variation of the superprofile parameters as a function of normalized radius for UGC~8833.
The solid dashed line shows the parameter value for the global superprofile \citepalias{StilpGlobal}.
The left panels show \scentral{} (upper) and \swing{} (lower), and the right panels show \fw{} (upper) and \afull{} (lower).
\label{resolved::fig:superprofiles-radial-u8833-c}
}
\end{figure}
\clearpage

\setcounter{subfig}{1}
\begin{figure}[p]
\begin{leftfullpage}
\centering
\includegraphics[width=4in]{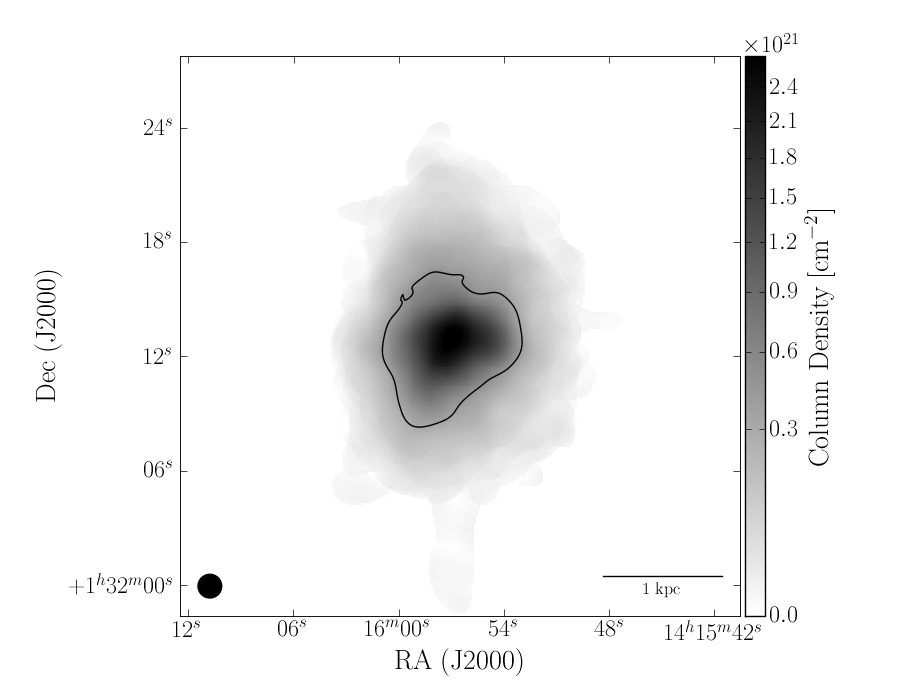}
\includegraphics[width=4in]{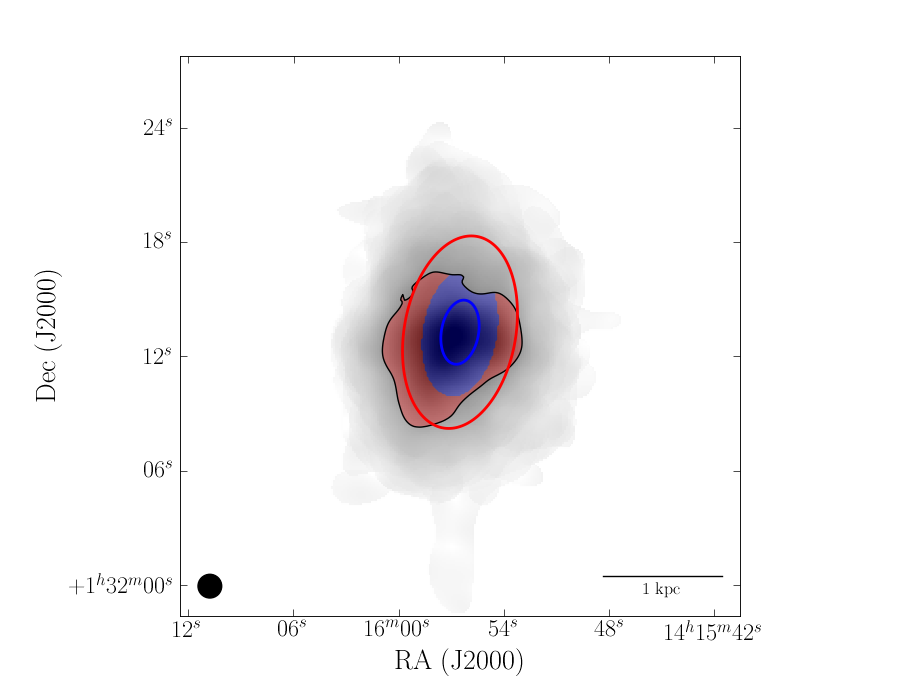}
\caption[Superprofile radial annuli for DDO~187]{Radial annuli in which radial superprofiles are generated for DDO~187. 
In both panels, the background greyscale shows \hisd{}, and the solid black line represents the $S/N > 5$ threshold where we can accurately measure \vp{}. 
In the lower panel, the colored solid lines represent the average radius of each annulus, and the corresponding shaded regions of the same color indicate which pixels have contributed to each radial superprofile.
\label{resolved::fig:superprofiles-radial-ddo187-a} }
\end{leftfullpage}
\end{figure}
\addtocounter{figure}{-1}
\addtocounter{subfig}{1}
\begin{figure}
\centering
\includegraphics[height=2.7in]{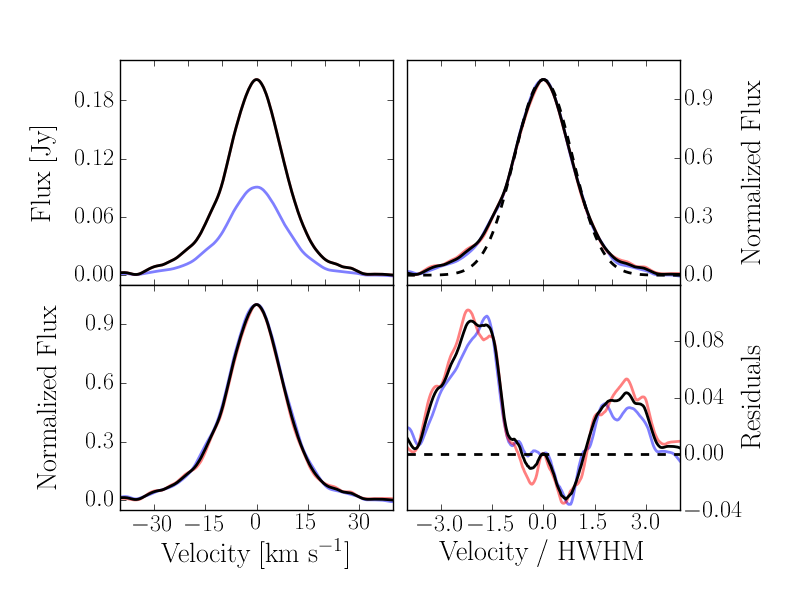}
\caption[Radial superprofiles in DDO~187]{The radial superprofiles in DDO~187, where colors indicate the corresponding radial annuli in the previous figure.
The left hand panels show the raw superprofiles (upper left) and the superprofiles normalized to the same peak flux (lower left).
The right hand panels show the flux-normalized superprofiles scaled by the HWHM (upper right) and the flux-normalized superprofiles minus the model of the Gaussian core (lower right). In all panels, the solid black line represents the global superprofile. In the left panels, we have shown the HWHM-scaled Gaussian model as the dashed black line.
\label{resolved::fig:superprofiles-radial-ddo187-b}
}
\end{figure}
\addtocounter{figure}{-1}
\addtocounter{subfig}{1}
\begin{figure}
\centering
\includegraphics[height=2.7in]{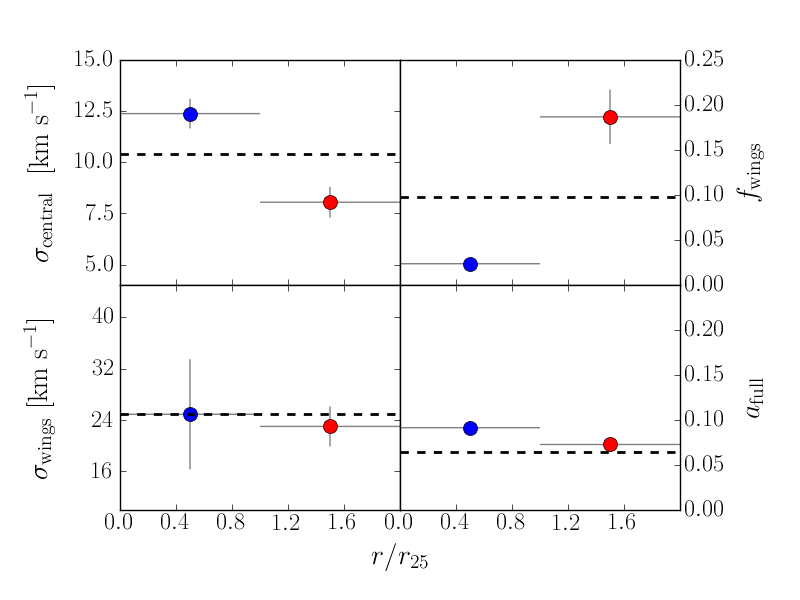}
\caption[Variation of radial superprofile parameters for DDO~187]{Variation of the superprofile parameters as a function of normalized radius for DDO~187.
The solid dashed line shows the parameter value for the global superprofile \citepalias{StilpGlobal}.
The left panels show \scentral{} (upper) and \swing{} (lower), and the right panels show \fw{} (upper) and \afull{} (lower).
\label{resolved::fig:superprofiles-radial-ddo187-c}
}
\end{figure}

\FloatBarrier

\subsection{\sfrsd{} Superprofiles}
\label{resolved::sec:sp-figures--sfrsd}

In the following pages, we present the superprofiles generated in subregions of constant \sfrsd{}.
For a single galaxy, the figures are the same as Figures~\ref{resolved::fig:superprofiles-sfr-n7793-a} and \ref{resolved::fig:superprofiles-sfr-n7793-c}.
Galaxies are shown in order of decreasing $M_\mathrm{baryon,tot}$, with the exception of NGC~7993, which was previously shown in the text.

\ifthesis
\renewcommand{\thefigure}{\arabic{chapter}.\arabic{figure}\alph{subfig}}
\else
\renewcommand{\thefigure}{\arabic{figure}\alph{subfig}}
\fi

\setcounter{subfig}{1}
\begin{figure}[p]
\begin{leftfullpage}
\centering
\includegraphics[width=4in]{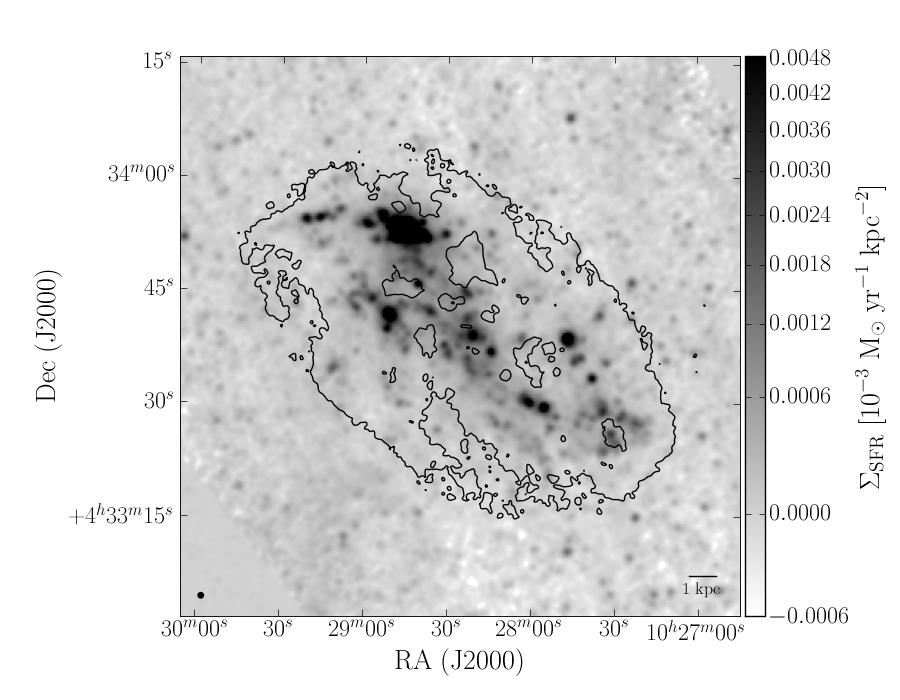}
\includegraphics[width=4in]{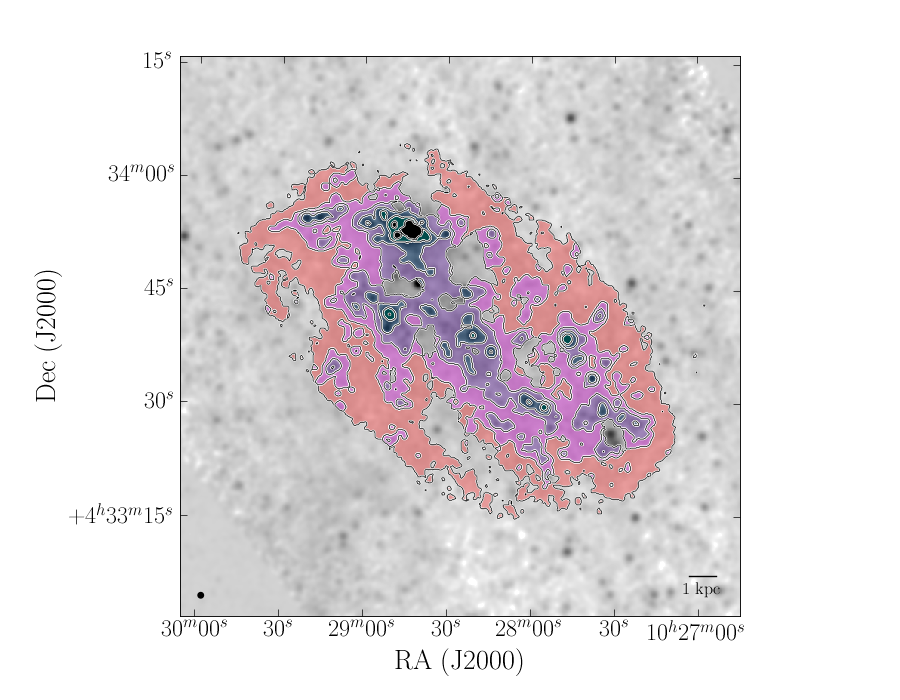}
\caption[Superprofile \sfrsd{} regions for IC~2574]{Subregions in which \sfrsd{} superprofiles are generated for IC2574. 
In both panels, the background greyscale shows \sfrsd{}, and the solid black line represents the $S/N > 5$ threshold where we can accurately measure \vp{}. 
In the lower panel, the colored regions show which pixels have contributed to each \sfrsd{} superprofile.
\label{resolved::fig:superprofiles-sfr-ic2574-a} }
\end{leftfullpage}
\end{figure}
\addtocounter{figure}{-1}
\addtocounter{subfig}{1}
\begin{figure}
\centering
\includegraphics[height=2.7in]{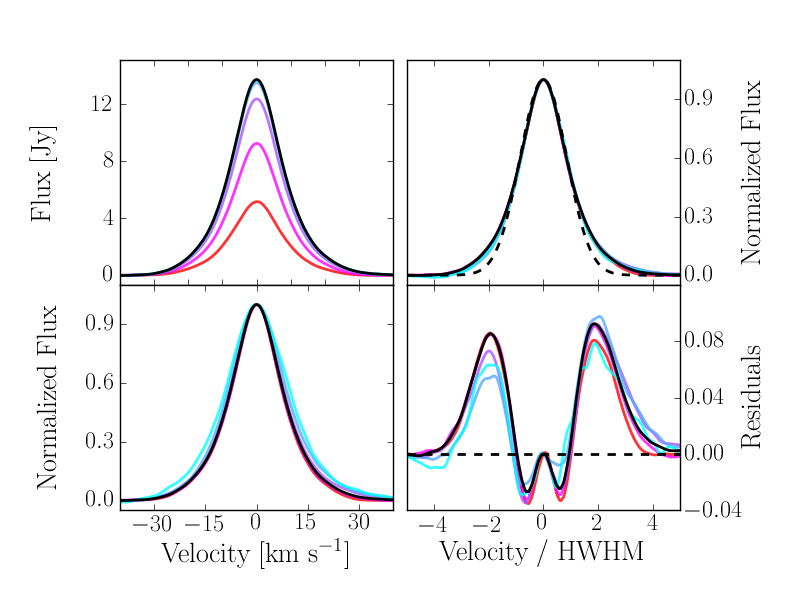}
\caption[The \sfrsd{} superprofiles in IC~2574]{The \sfrsd{} superprofiles in IC~2574, where colors indicate the corresponding \sfrsd{} regions in the previous figure.
The left hand panels show the raw superprofiles (upper left) and the superprofiles normalized to the same peak flux (lower left).
The right hand panels show the flux-normalized superprofiles scaled by the HWHM (upper right) and the flux-normalized superprofiles minus the model of the Gaussian core (lower right). In all panels, the solid black line represents the global superprofile. In the left panels, we have shown the HWHM-scaled Gaussian model as the dashed black line.
\label{resolved::fig:superprofiles-sfr-ic2574-b}
}
\end{figure}
\addtocounter{figure}{-1}
\addtocounter{subfig}{1}
\begin{figure}
\centering
\includegraphics[height=2.7in]{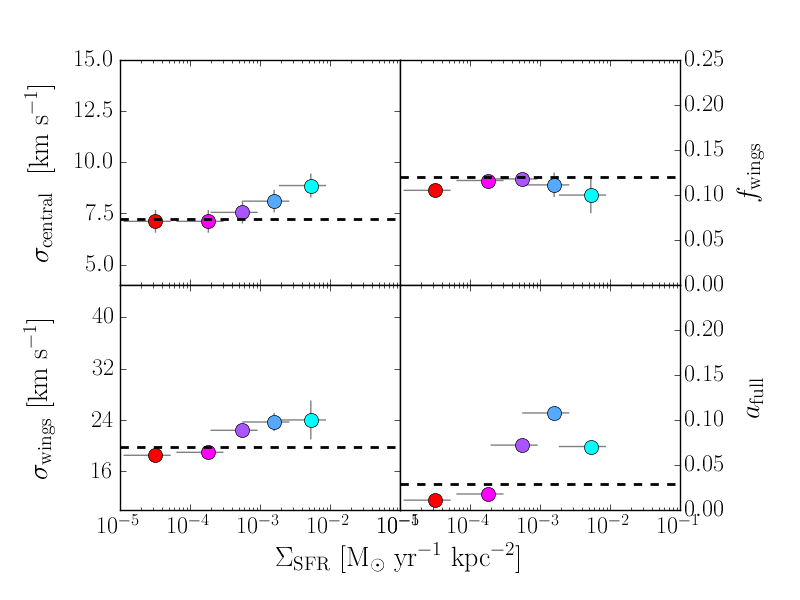}
\caption[Variation of \sfrsd{} superprofile parameters for IC~2574]{Variation of the superprofile parameters as a function of \sfrsd{} for IC~2574.
The solid dashed line shows the parameter value for the global superprofile \citepalias{StilpGlobal}.
The left panels show \scentral{} (upper) and \swing{} (lower), and the right panels show \fw{} (upper) and \afull{} (lower).
\label{resolved::fig:superprofiles-sfr-ic2574-c}
}
\end{figure}
\clearpage

\setcounter{subfig}{1}
\begin{figure}[p]
\begin{leftfullpage}
\centering
\includegraphics[width=4in]{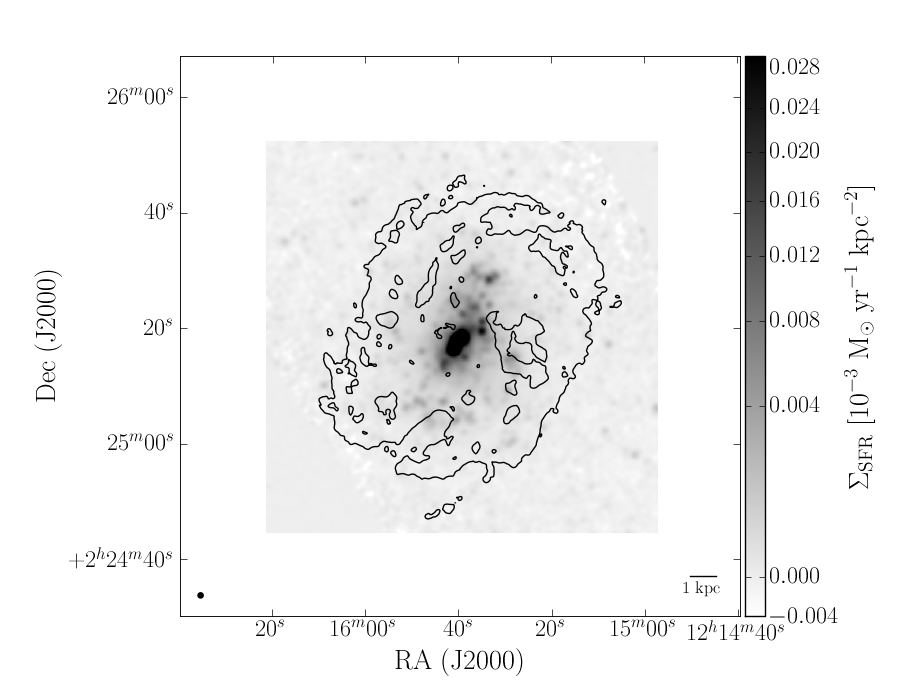}
\includegraphics[width=4in]{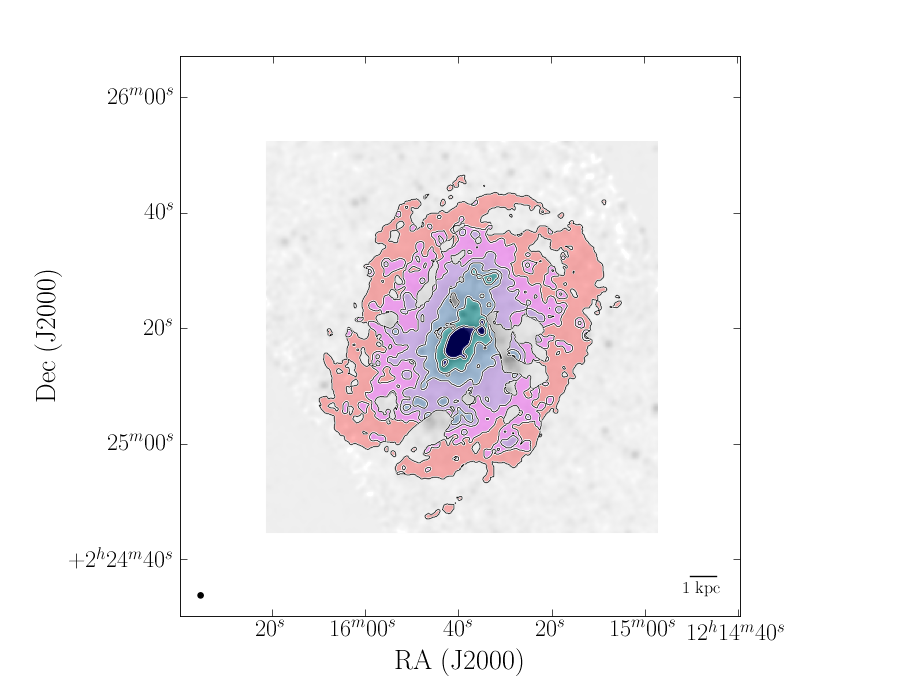}
\caption[Superprofile \sfrsd{} regions for NGC~4214]{\sfrsd{} regions in which radial superprofiles are generated for NGC~4214. 
In both panels, the background greyscale shows \sfrsd{}, and the solid black line represents the $S/N > 5$ threshold where we can accurately measure \vp{}. 
In the lower panel, the colored regions show which pixels have contributed to each \sfrsd{} superprofile.
\label{resolved::fig:superprofiles-sfr-n4214-a} }
\end{leftfullpage}
\end{figure}
\addtocounter{figure}{-1}
\addtocounter{subfig}{1}
\begin{figure}
\centering
\includegraphics[height=2.7in]{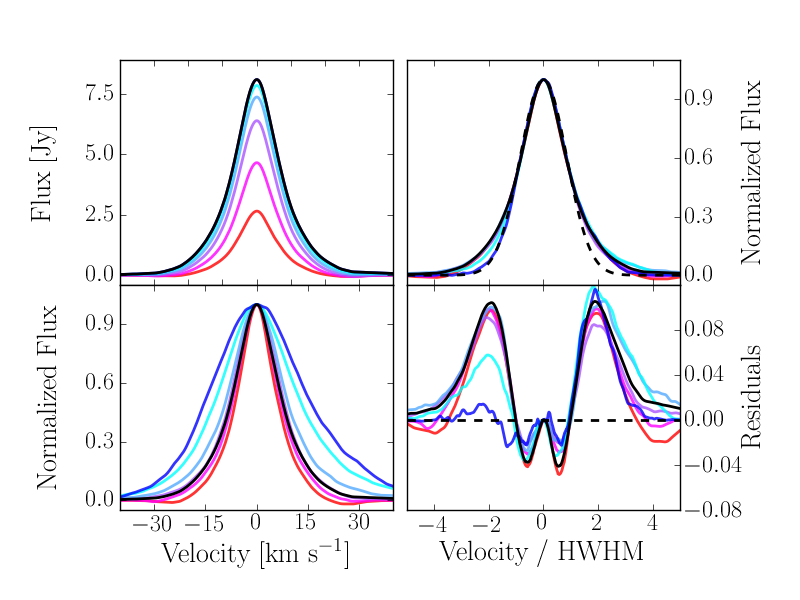}
\caption[The \sfrsd{} superprofiles in NGC~4214]{The \sfrsd{} superprofiles in NGC~4214, where colors indicate the corresponding \sfrsd{} regions in the previous figure.
The left hand panels show the raw superprofiles (upper left) and the superprofiles normalized to the same peak flux (lower left).
The right hand panels show the flux-normalized superprofiles scaled by the HWHM (upper right) and the flux-normalized superprofiles minus the model of the Gaussian core (lower right). In all panels, the solid black line represents the global superprofile. In the left panels, we have shown the HWHM-scaled Gaussian model as the dashed black line.
\label{resolved::fig:superprofiles-sfr-n4214-b}
}
\end{figure}
\addtocounter{figure}{-1}
\addtocounter{subfig}{1}
\begin{figure}
\centering
\includegraphics[height=2.7in]{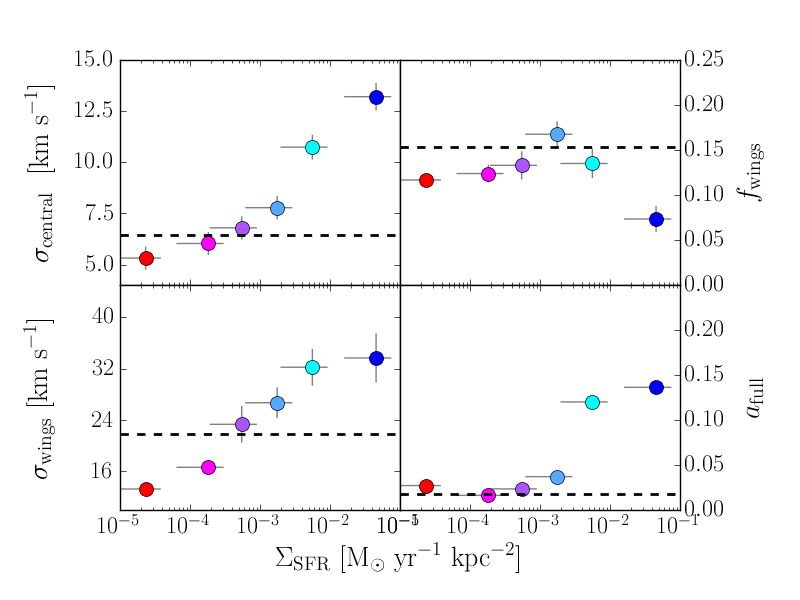}
\caption[Variation of \sfrsd{} superprofile parameters for NGC~4214]{Variation of the superprofile parameters as a function of \sfrsd{} for NGC~4214.
The solid dashed line shows the parameter value for the global superprofile \citepalias{StilpGlobal}.
The left panels show \scentral{} (upper) and \swing{} (lower), and the right panels show \fw{} (upper) and \afull{} (lower).
\label{resolved::fig:superprofiles-sfr-n4214-c}
}
\end{figure}
\clearpage

\setcounter{subfig}{1}
\begin{figure}[p]
\begin{leftfullpage}
\centering
\includegraphics[width=4in]{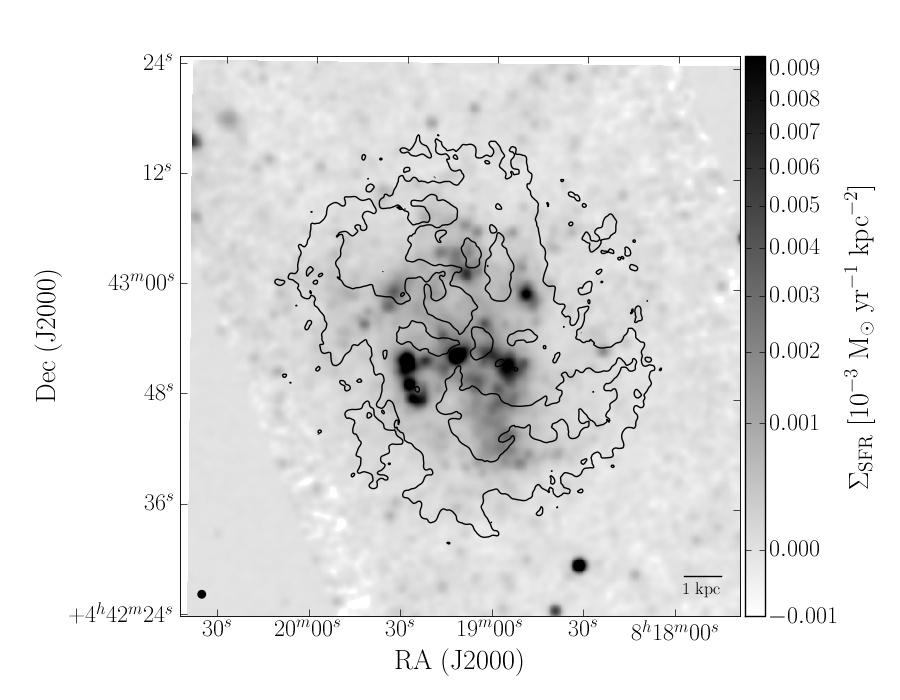}
\includegraphics[width=4in]{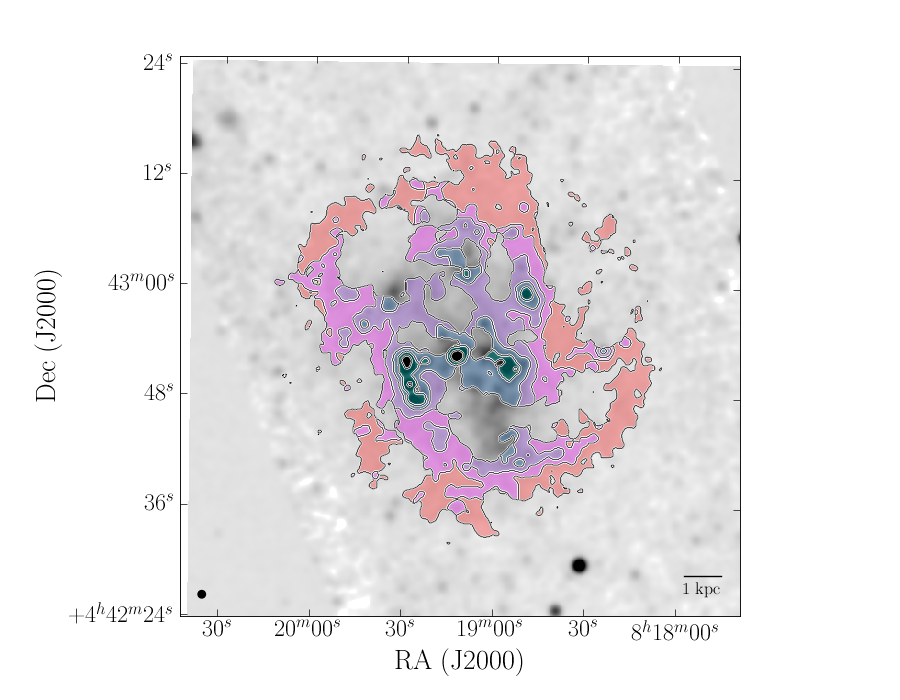}
\caption[Superprofile \sfrsd{} regions for Ho~II]{Subregions in which \sfrsd{} superprofiles are generated for Ho~II. 
In both panels, the background greyscale shows \sfrsd{}, and the solid black line represents the $S/N > 5$ threshold where we can accurately measure \vp{}. 
In the lower panel, the colored regions show which pixels have contributed to each \sfrsd{} superprofile.
\label{resolved::fig:superprofiles-sfr-hoii-a} }
\end{leftfullpage}
\end{figure}
\addtocounter{figure}{-1}
\addtocounter{subfig}{1}
\begin{figure}
\centering
\includegraphics[height=2.7in]{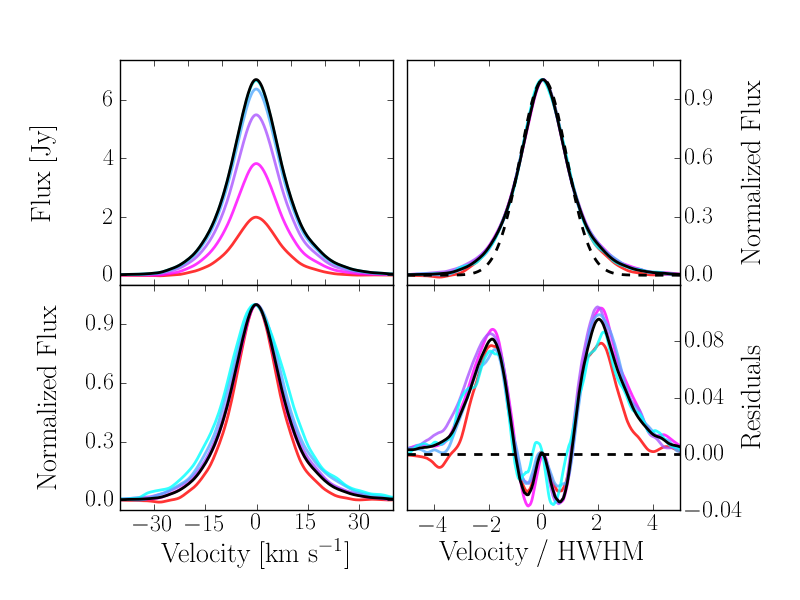}
\caption[The \sfrsd{} superprofiles in Ho~II]{The \sfrsd{} superprofiles in Ho~II, where colors indicate the corresponding \sfrsd{} regions in the previous figure.
The left hand panels show the raw superprofiles (upper left) and the superprofiles normalized to the same peak flux (lower left).
The right hand panels show the flux-normalized superprofiles scaled by the HWHM (upper right) and the flux-normalized superprofiles minus the model of the Gaussian core (lower right). In all panels, the solid black line represents the global superprofile. In the left panels, we have shown the HWHM-scaled Gaussian model as the dashed black line.
\label{resolved::fig:superprofiles-sfr-hoii-b}
}
\end{figure}
\addtocounter{figure}{-1}
\addtocounter{subfig}{1}
\begin{figure}
\centering
\includegraphics[height=2.7in]{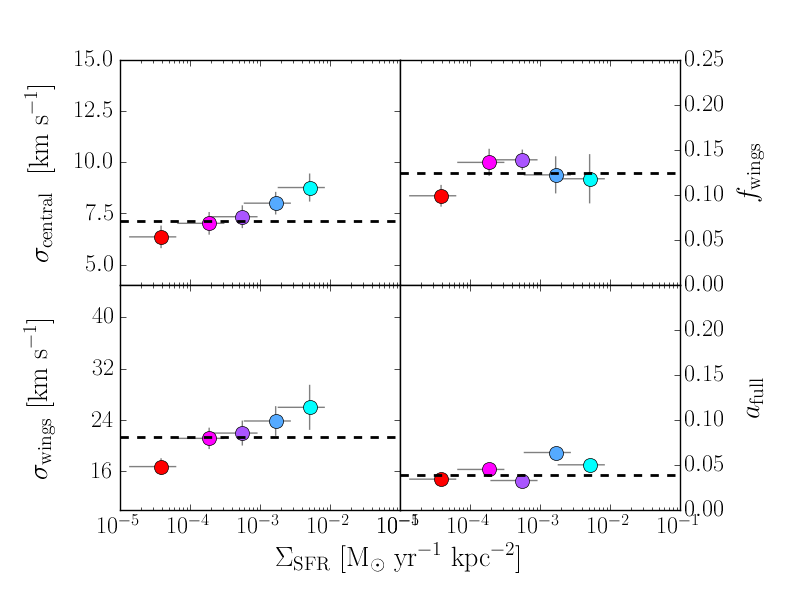}
\caption[Variation of \sfrsd{} superprofile parameters for Ho~II]{Variation of the superprofile parameters as a function of \sfrsd{} for Ho~II.
The solid dashed line shows the parameter value for the global superprofile \citepalias{StilpGlobal}.
The left panels show \scentral{} (upper) and \swing{} (lower), and the right panels show \fw{} (upper) and \afull{} (lower).
\label{resolved::fig:superprofiles-sfr-hoii-c}
}
\end{figure}
\clearpage

\setcounter{subfig}{1}
\begin{figure}[p]
\begin{leftfullpage}
\centering
\includegraphics[width=4in]{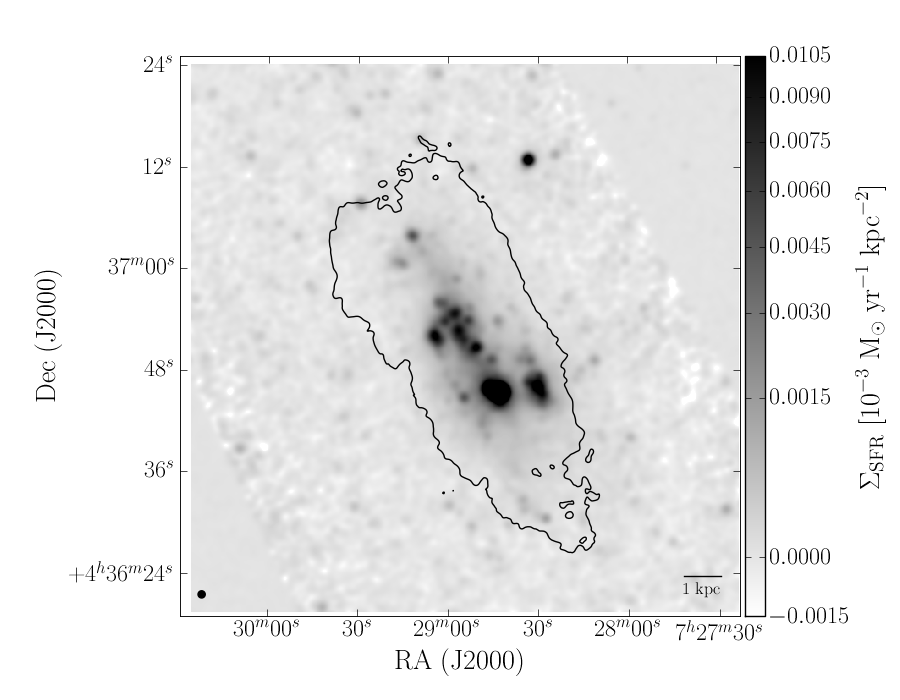}
\includegraphics[width=4in]{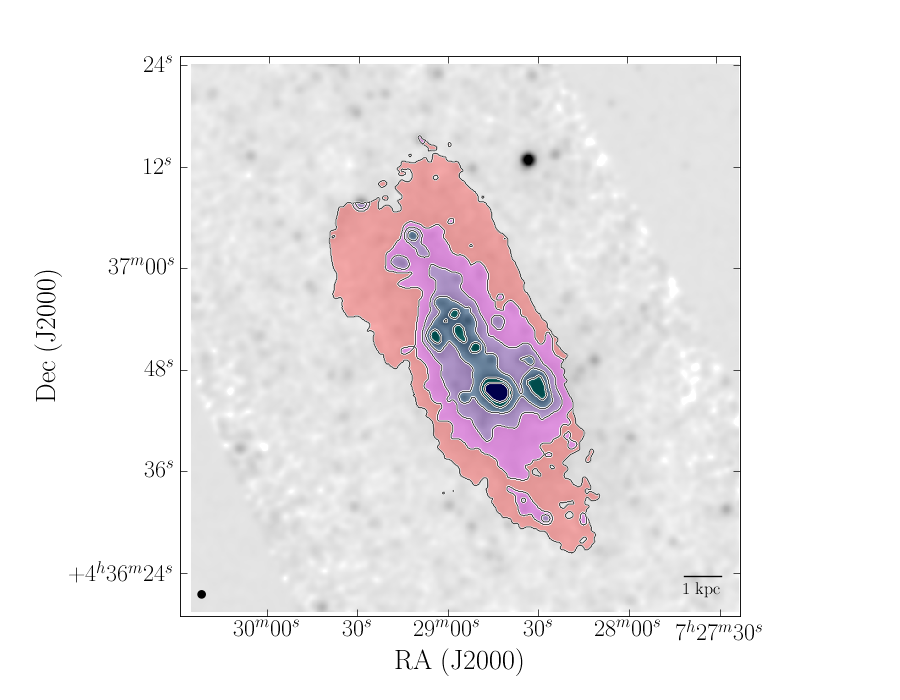}
\caption[Superprofile \sfrsd{} regions for NGC~2366]{Subregions in which \sfrsd{} superprofiles are generated for NGC~2366. 
In both panels, the background greyscale shows \sfrsd{}, and the solid black line represents the $S/N > 5$ threshold where we can accurately measure \vp{}. 
In the lower panel, the colored regions show which pixels have contributed to each \sfrsd{} superprofile.
\label{resolved::fig:superprofiles-sfr-n2366-a} }
\end{leftfullpage}
\end{figure}
\addtocounter{figure}{-1}
\addtocounter{subfig}{1}
\begin{figure}
\centering
\includegraphics[height=2.7in]{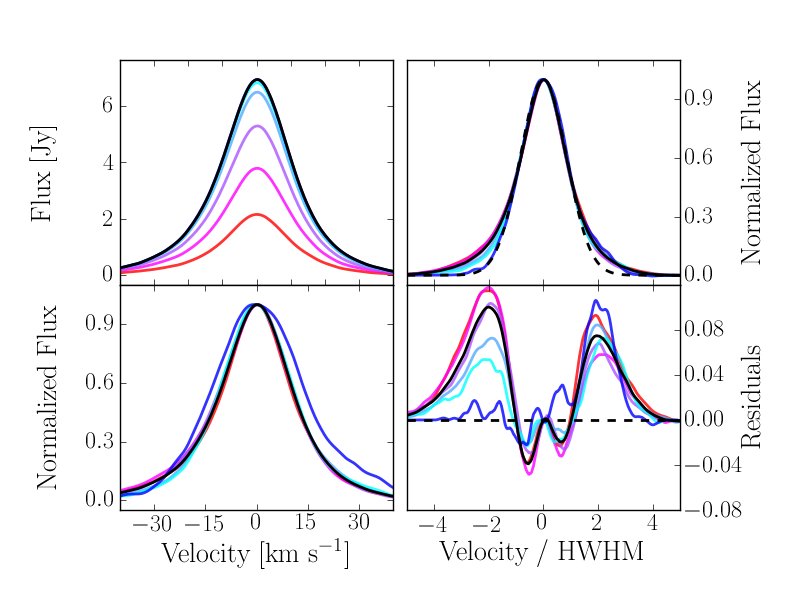}
\caption[The \sfrsd{} superprofiles in NGC~2366]{The \sfrsd{} superprofiles in NGC~2366, where colors indicate the corresponding \sfrsd{} regions in the previous figure.
The left hand panels show the raw superprofiles (upper left) and the superprofiles normalized to the same peak flux (lower left).
The right hand panels show the flux-normalized superprofiles scaled by the HWHM (upper right) and the flux-normalized superprofiles minus the model of the Gaussian core (lower right). In all panels, the solid black line represents the global superprofile. In the left panels, we have shown the HWHM-scaled Gaussian model as the dashed black line.
\label{resolved::fig:superprofiles-sfr-n2366-b}
}
\end{figure}
\addtocounter{figure}{-1}
\addtocounter{subfig}{1}
\begin{figure}
\centering
\includegraphics[height=2.7in]{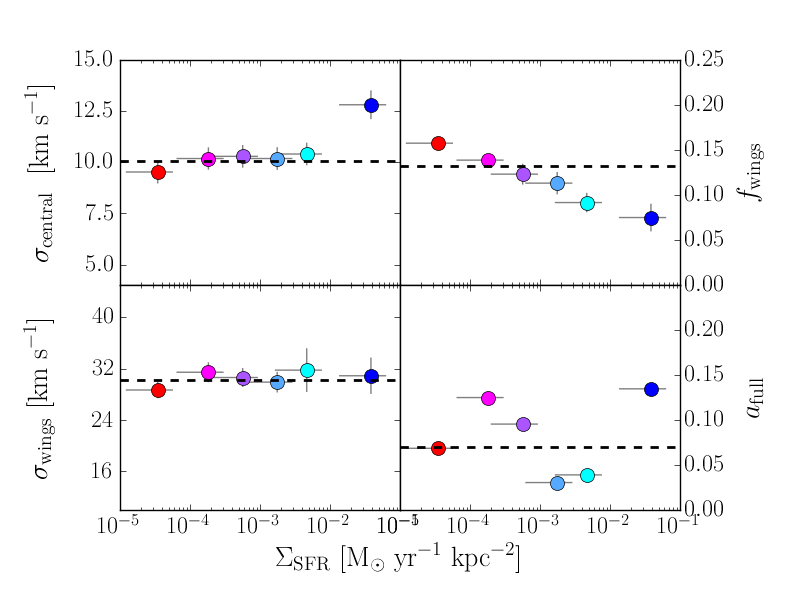}
\caption[Variation of \sfrsd{} superprofile parameters for NGC~2366]{Variation of the superprofile parameters as a function of \sfrsd{} for NGC~2366.
The solid dashed line shows the parameter value for the global superprofile \citepalias{StilpGlobal}.
The left panels show \scentral{} (upper) and \swing{} (lower), and the right panels show \fw{} (upper) and \afull{} (lower).
\label{resolved::fig:superprofiles-sfr-n2366-c}
}
\end{figure}
\clearpage

\setcounter{subfig}{1}
\begin{figure}[p]
\begin{leftfullpage}
\centering
\includegraphics[width=4in]{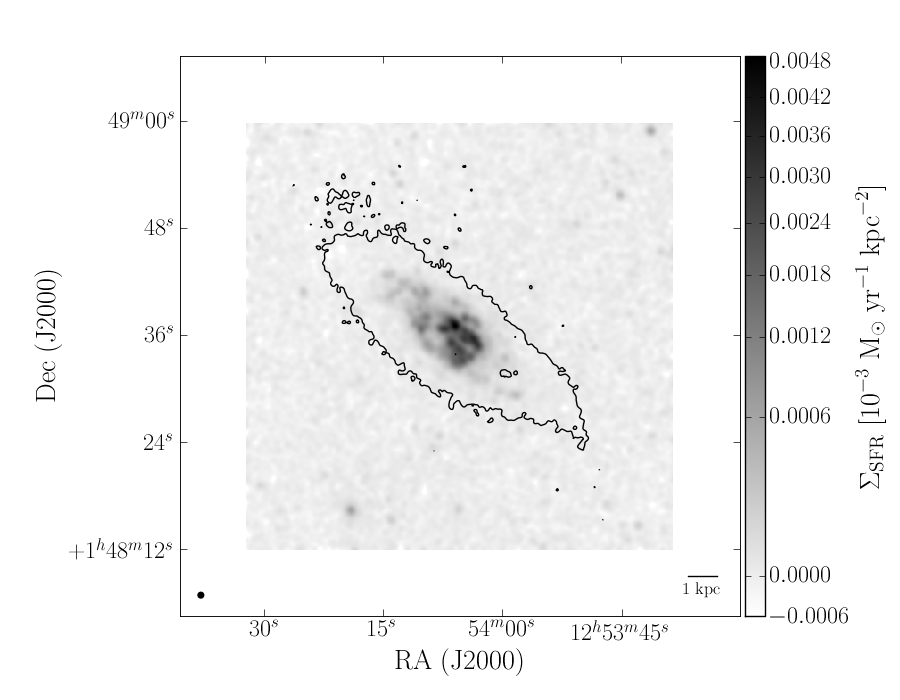}
\includegraphics[width=4in]{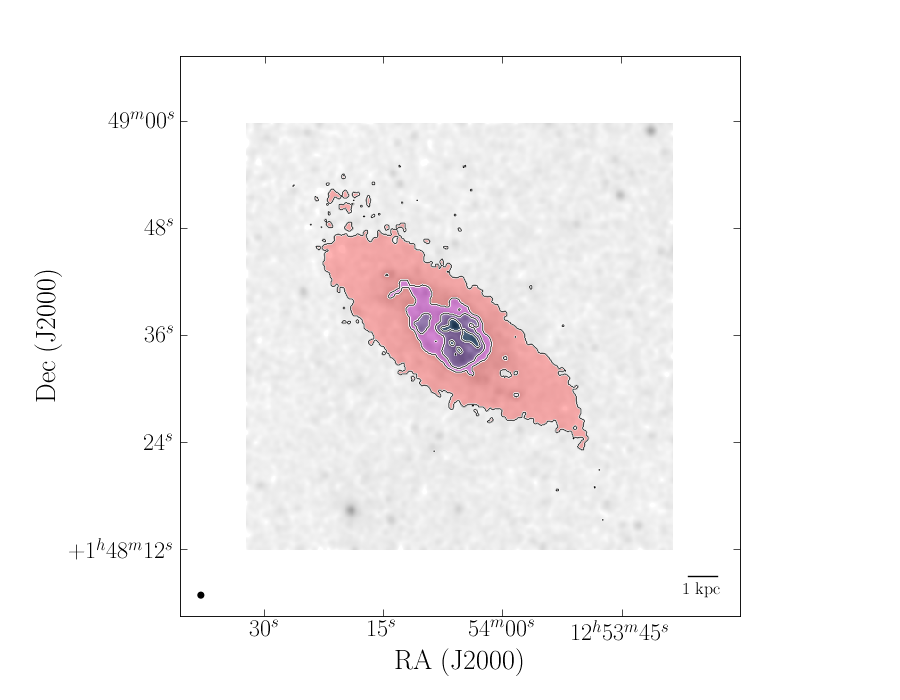}
\caption[Superprofile \sfrsd{} regions for DDO~154]{Subregions in which \sfrsd{} superprofiles are generated for DDO~154. 
In both panels, the background greyscale shows \sfrsd{}, and the solid black line represents the $S/N > 5$ threshold where we can accurately measure \vp{}. 
In the lower panel, the colored regions show which pixels have contributed to each \sfrsd{} superprofile.
\label{resolved::fig:superprofiles-sfr-ddo154-a} }
\end{leftfullpage}
\end{figure}
\addtocounter{figure}{-1}
\addtocounter{subfig}{1}
\begin{figure}
\centering
\includegraphics[height=2.7in]{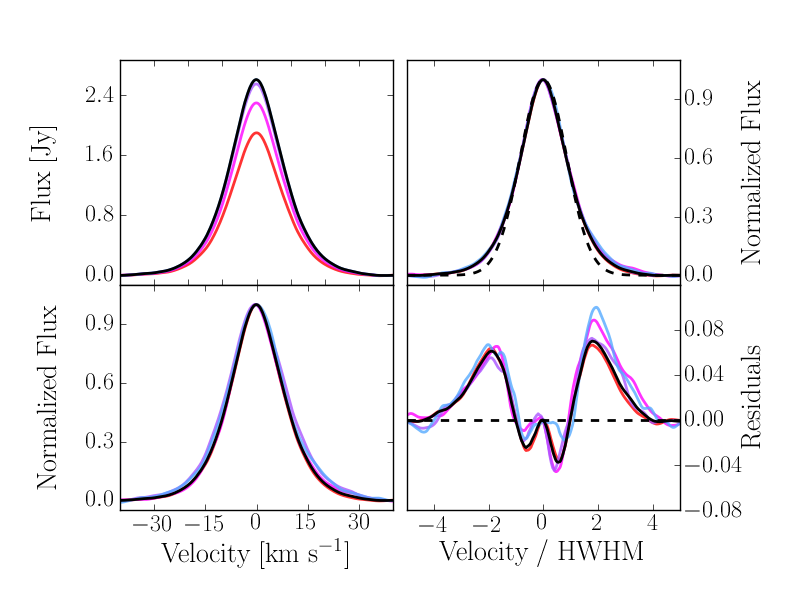}
\caption[The \sfrsd{} superprofiles in DDO~154]{The \sfrsd{} superprofiles in DDO~154, where colors indicate the corresponding \sfrsd{} regions in the previous figure.
The left hand panels show the raw superprofiles (upper left) and the superprofiles normalized to the same peak flux (lower left).
The right hand panels show the flux-normalized superprofiles scaled by the HWHM (upper right) and the flux-normalized superprofiles minus the model of the Gaussian core (lower right). In all panels, the solid black line represents the global superprofile. In the left panels, we have shown the HWHM-scaled Gaussian model as the dashed black line.
\label{resolved::fig:superprofiles-sfr-ddo154-b}
}
\end{figure}
\addtocounter{figure}{-1}
\addtocounter{subfig}{1}
\begin{figure}
\centering
\includegraphics[height=2.7in]{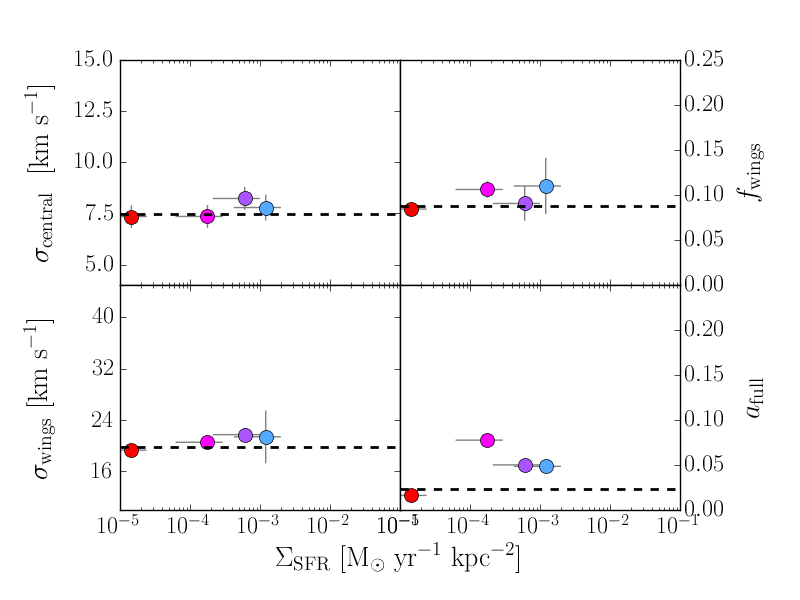}
\caption[Variation of \sfrsd{} superprofile parameters for DDO~154]{Variation of the superprofile parameters as a function of \sfrsd{} for DDO~154.
The solid dashed line shows the parameter value for the global superprofile \citepalias{StilpGlobal}.
The left panels show \scentral{} (upper) and \swing{} (lower), and the right panels show \fw{} (upper) and \afull{} (lower).
\label{resolved::fig:superprofiles-sfr-ddo154-c}
}
\end{figure}
\clearpage

\setcounter{subfig}{1}
\begin{figure}
\begin{leftfullpage}
\centering
\includegraphics[width=4in]{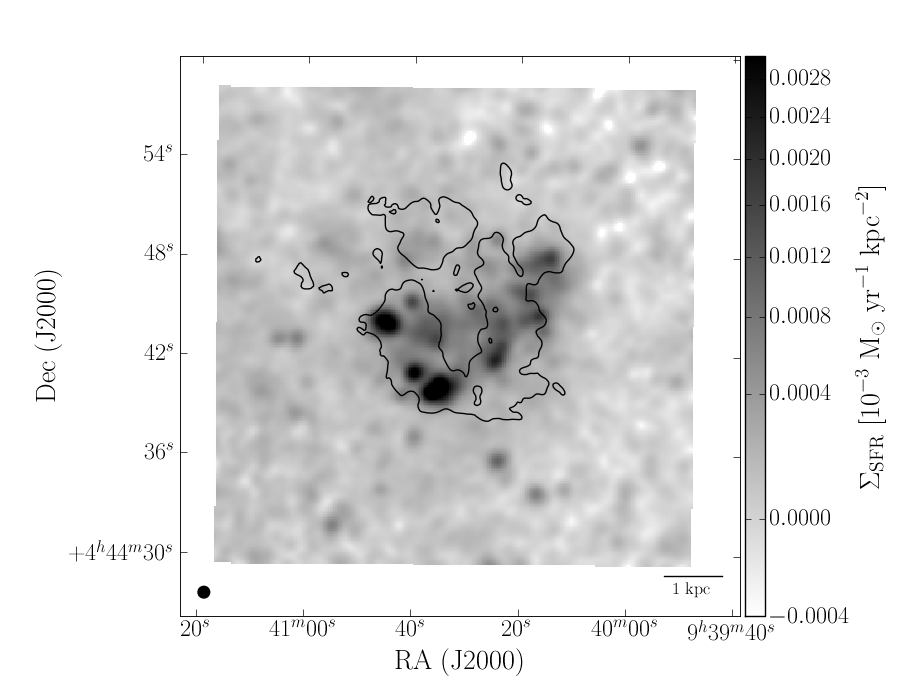}
\includegraphics[width=4in]{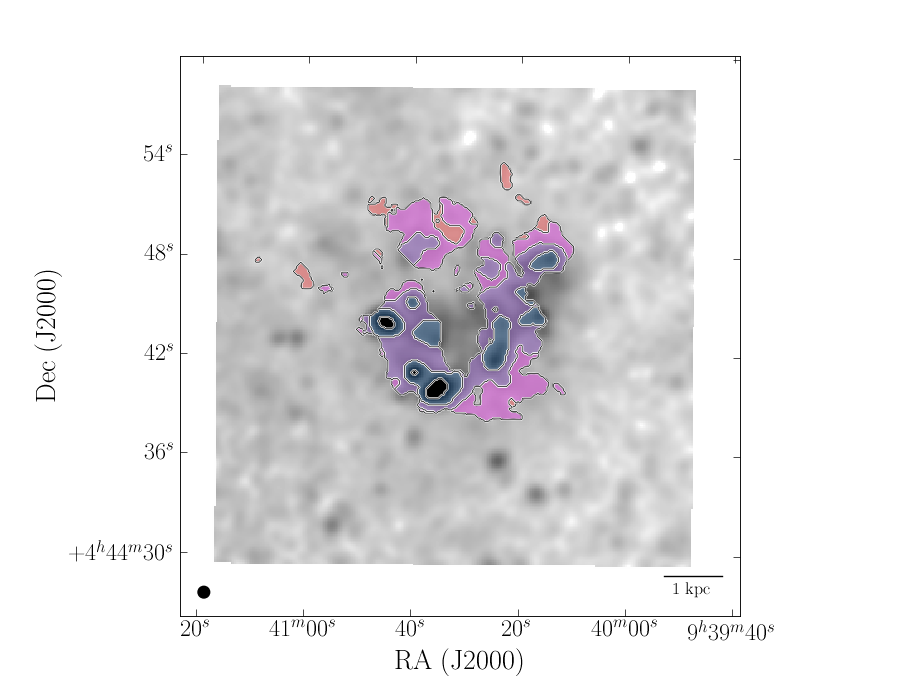}
\caption[Superprofile \sfrsd{} regions for Ho~I]{Subregions in which \sfrsd{} superprofiles are generated for Ho~I. 
In both panels, the background greyscale shows \sfrsd{}, and the solid black line represents the $S/N > 5$ threshold where we can accurately measure \vp{}. 
In the lower panel, the colored regions show which pixels have contributed to each \sfrsd{} superprofile.
\label{resolved::fig:superprofiles-sfr-hoi-a} }
\end{leftfullpage}
\end{figure}
\addtocounter{figure}{-1}
\addtocounter{subfig}{1}
\begin{figure}
\centering
\includegraphics[height=2.7in]{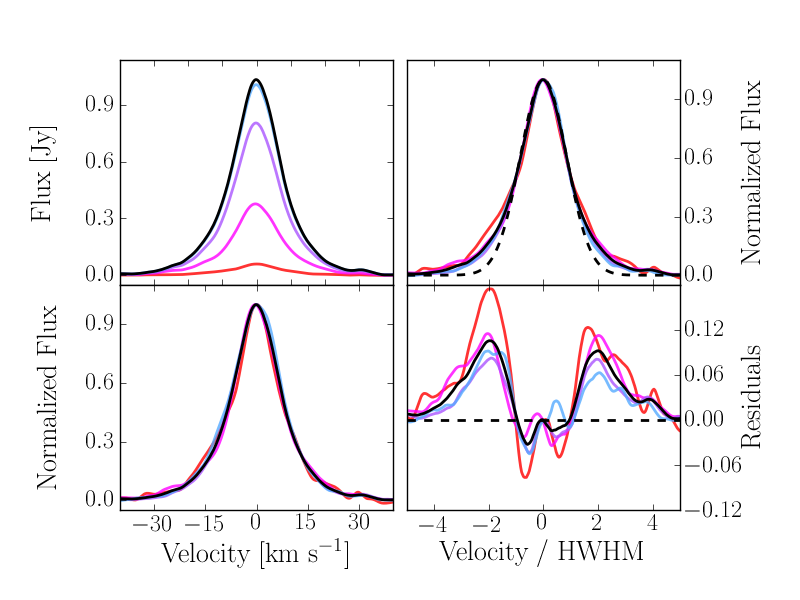}
\caption[The \sfrsd{} superprofiles in Ho~I]{The \sfrsd{} superprofiles in Ho~I, where colors indicate the corresponding \sfrsd{} regions in the previous figure.
The left hand panels show the raw superprofiles (upper left) and the superprofiles normalized to the same peak flux (lower left).
The right hand panels show the flux-normalized superprofiles scaled by the HWHM (upper right) and the flux-normalized superprofiles minus the model of the Gaussian core (lower right). In all panels, the solid black line represents the global superprofile. In the left panels, we have shown the HWHM-scaled Gaussian model as the dashed black line.
\label{resolved::fig:superprofiles-sfr-hoi-b}
}
\end{figure}
\addtocounter{figure}{-1}
\addtocounter{subfig}{1}
\begin{figure}
\centering
\includegraphics[height=2.7in]{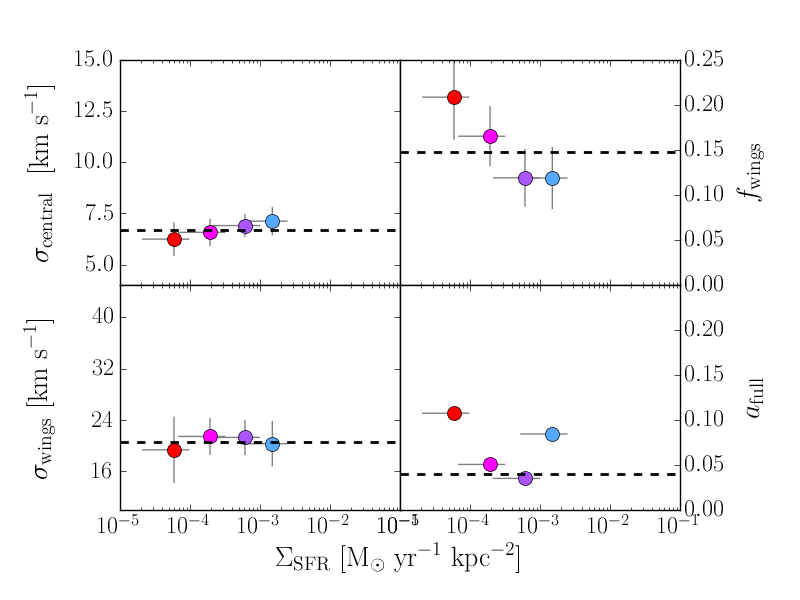}
\caption[Variation of \sfrsd{} superprofile parameters for Ho~I]{Variation of the superprofile parameters as a function of \sfrsd{} for Ho~I.
The solid dashed line shows the parameter value for the global superprofile \citepalias{StilpGlobal}.
The left panels show \scentral{} (upper) and \swing{} (lower), and the right panels show \fw{} (upper) and \afull{} (lower).
\label{resolved::fig:superprofiles-sfr-hoi-c}
}
\end{figure}
\clearpage

\setcounter{subfig}{1}
\begin{figure}
\begin{leftfullpage}
\centering
\includegraphics[width=4in]{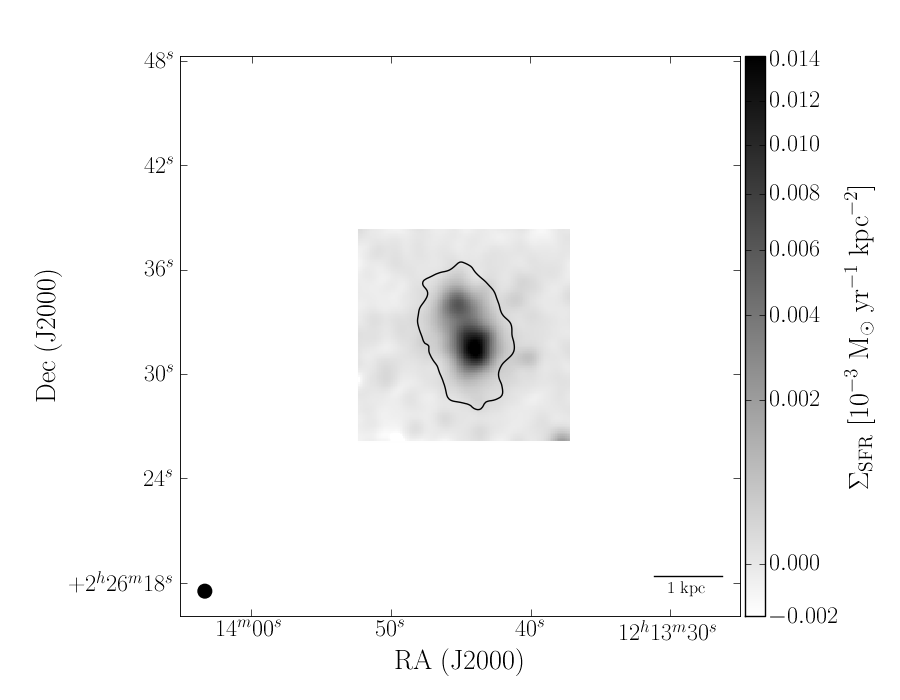}
\includegraphics[width=4in]{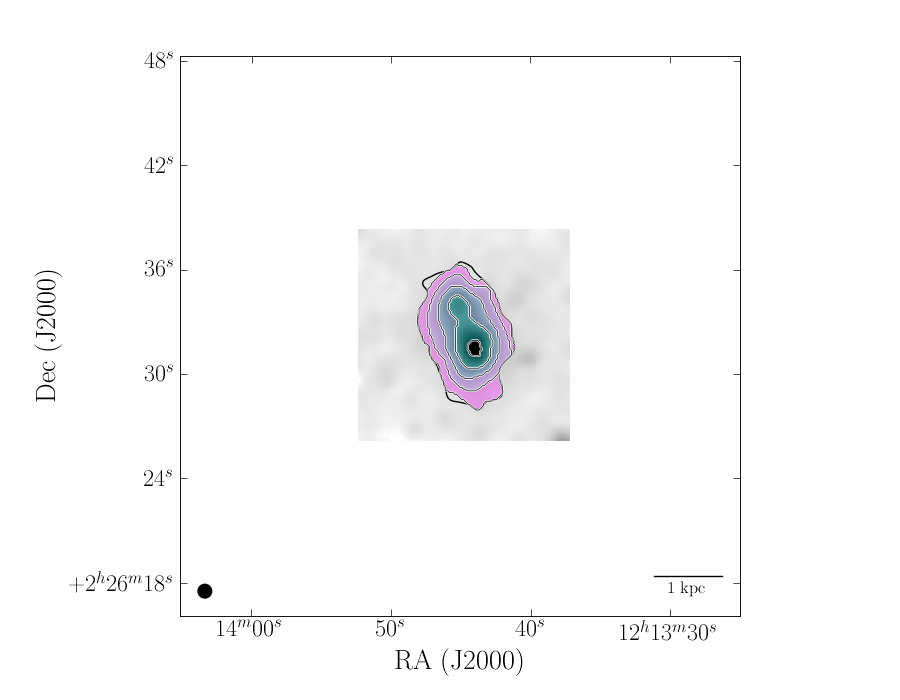}
\caption[Superprofile \sfrsd{} regions for NGC~4190]{Subregions in which \sfrsd{} superprofiles are generated for NGC~4190. 
In both panels, the background greyscale shows \sfrsd{}, and the solid black line represents the $S/N > 5$ threshold where we can accurately measure \vp{}. 
In the lower panel, the colored regions show which pixels have contributed to each \sfrsd{} superprofile.
\label{resolved::fig:superprofiles-sfr-n4190-a} }
\end{leftfullpage}
\end{figure}
\addtocounter{figure}{-1}
\addtocounter{subfig}{1}
\begin{figure}
\centering
\includegraphics[height=2.7in]{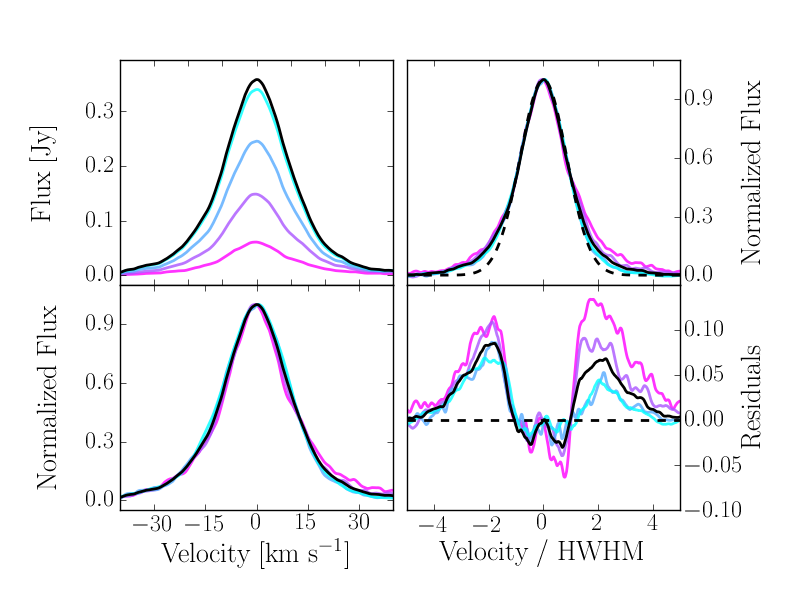}
\caption[The \sfrsd{} superprofiles in NGC~4190]{The \sfrsd{} superprofiles in NGC~4190, where colors indicate the corresponding \sfrsd{} regions in the previous figure.
The left hand panels show the raw superprofiles (upper left) and the superprofiles normalized to the same peak flux (lower left).
The right hand panels show the flux-normalized superprofiles scaled by the HWHM (upper right) and the flux-normalized superprofiles minus the model of the Gaussian core (lower right). In all panels, the solid black line represents the global superprofile. In the left panels, we have shown the HWHM-scaled Gaussian model as the dashed black line.
\label{resolved::fig:superprofiles-sfr-n4190-b}
}
\end{figure}
\addtocounter{figure}{-1}
\addtocounter{subfig}{1}
\begin{figure}
\centering
\includegraphics[height=2.7in]{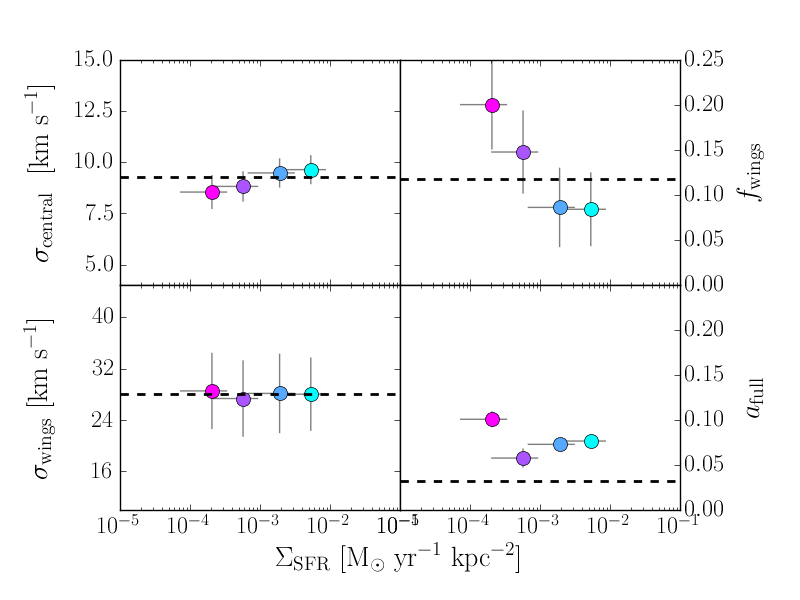}
\caption[Variation of \sfrsd{} superprofile parameters for NGC~4190]{Variation of the superprofile parameters as a function of \sfrsd{} for NGC~4190.
The solid dashed line shows the parameter value for the global superprofile \citepalias{StilpGlobal}.
The left panels show \scentral{} (upper) and \swing{} (lower), and the right panels show \fw{} (upper) and \afull{} (lower).
\label{resolved::fig:superprofiles-sfr-n4190-c}
}
\end{figure}
\clearpage

\setcounter{subfig}{1}
\begin{figure}[p]
\begin{leftfullpage}
\centering
\includegraphics[width=4in]{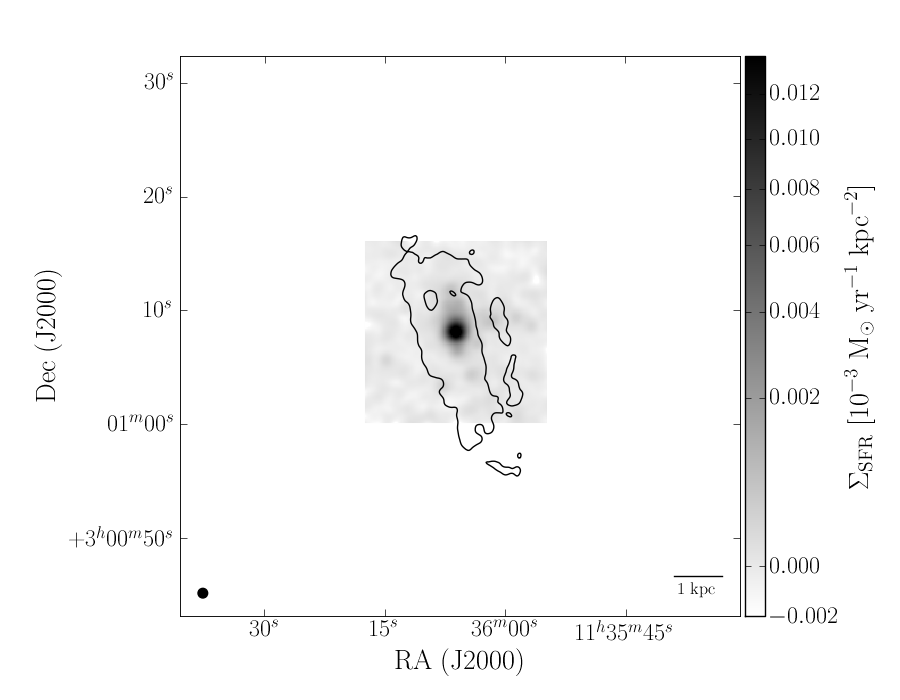}
\includegraphics[width=4in]{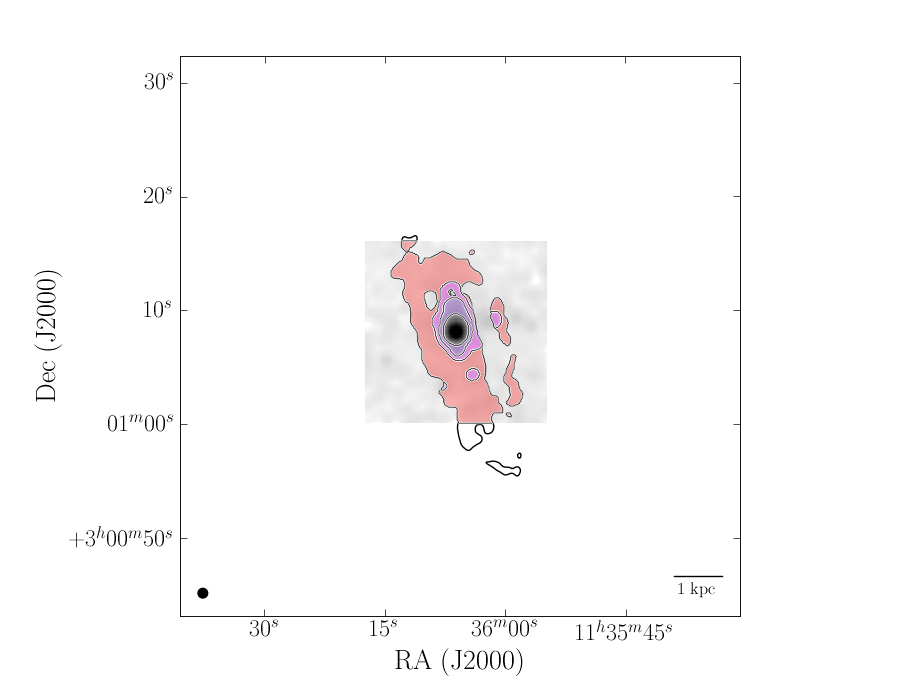}
\caption[Superprofile \sfrsd{} regions for NGC~3741]{Subregions in which \sfrsd{} superprofiles are generated for NGC~3741. 
In both panels, the background greyscale shows \sfrsd{}, and the solid black line represents the $S/N > 5$ threshold where we can accurately measure \vp{}. 
In the lower panel, the colored regions show which pixels have contributed to each \sfrsd{} superprofile.
\label{resolved::fig:superprofiles-sfr-n3741-a} }
\end{leftfullpage}
\end{figure}
\addtocounter{figure}{-1}
\addtocounter{subfig}{1}
\begin{figure}
\centering
\includegraphics[height=2.7in]{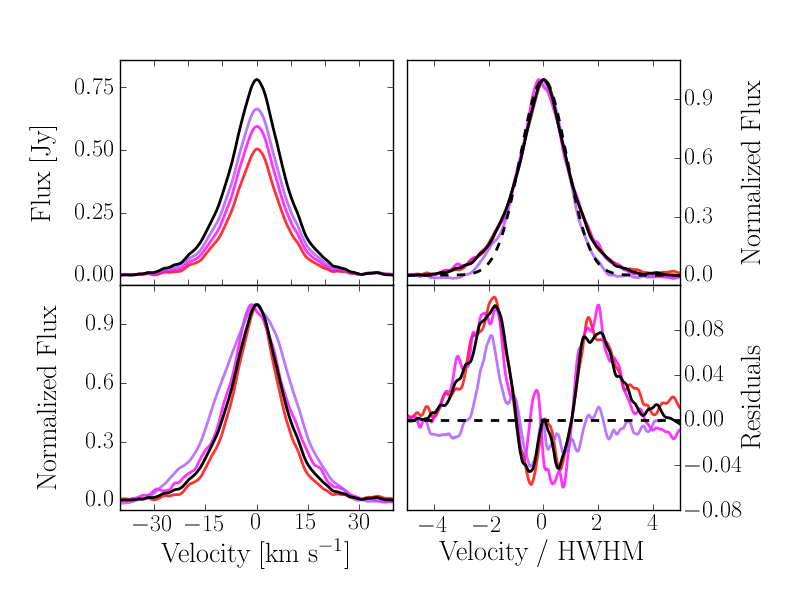}
\caption[The \sfrsd{} superprofiles in NGC~3741]{The \sfrsd{} superprofiles in NGC~3741, where colors indicate the corresponding \sfrsd{} regions in the previous figure.
The left hand panels show the raw superprofiles (upper left) and the superprofiles normalized to the same peak flux (lower left).
The right hand panels show the flux-normalized superprofiles scaled by the HWHM (upper right) and the flux-normalized superprofiles minus the model of the Gaussian core (lower right). In all panels, the solid black line represents the global superprofile. In the left panels, we have shown the HWHM-scaled Gaussian model as the dashed black line.
\label{resolved::fig:superprofiles-sfr-n3741-b}
}
\end{figure}
\addtocounter{figure}{-1}
\addtocounter{subfig}{1}
\begin{figure}
\centering
\includegraphics[height=2.7in]{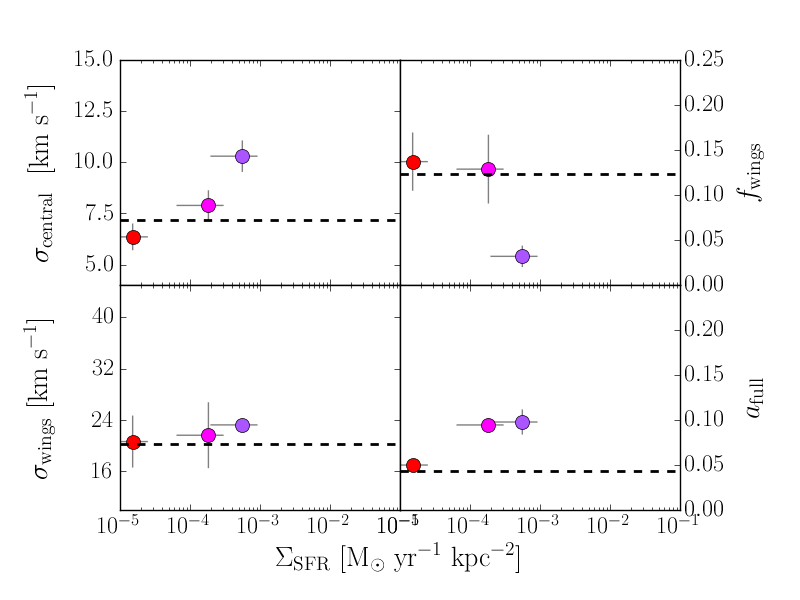}
\caption[Variation of \sfrsd{} superprofile parameters for NGC~3741]{Variation of the superprofile parameters as a function of \sfrsd{} for NGC~3741.
The solid dashed line shows the parameter value for the global superprofile \citepalias{StilpGlobal}.
The left panels show \scentral{} (upper) and \swing{} (lower), and the right panels show \fw{} (upper) and \afull{} (lower).
\label{resolved::fig:superprofiles-sfr-n3741-c}
}
\end{figure}
\clearpage

\setcounter{subfig}{1}
\begin{figure}[p]
\begin{leftfullpage}
\centering
\includegraphics[width=4in]{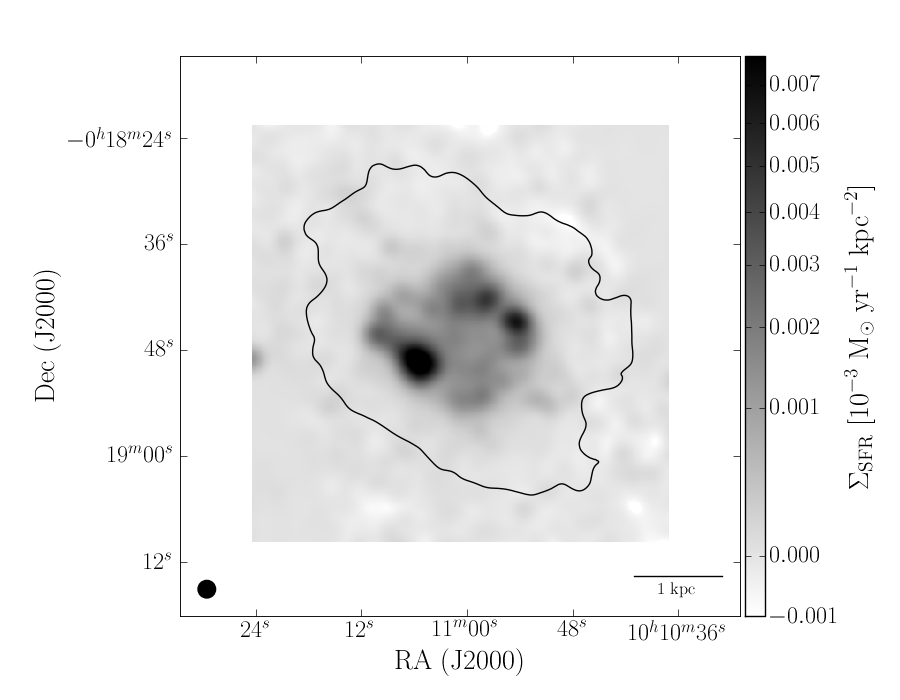}
\includegraphics[width=4in]{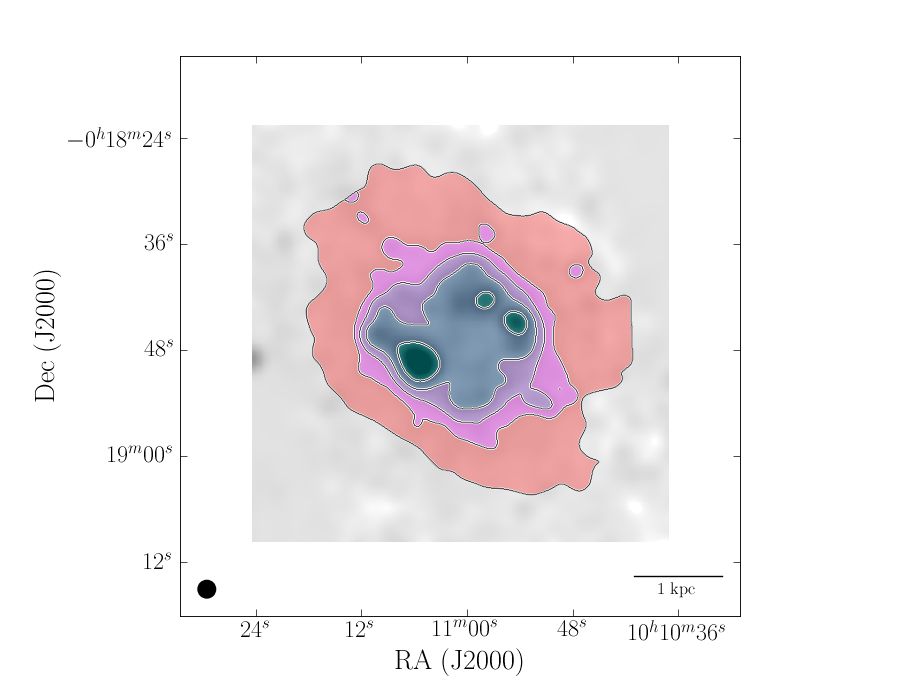}
\caption[Superprofile \sfrsd{} regions for Sextans A]{Subregions in which \sfrsd{} superprofiles are generated for Sextans A. 
In both panels, the background greyscale shows \sfrsd{}, and the solid black line represents the $S/N > 5$ threshold where we can accurately measure \vp{}. 
In the lower panel, the colored regions show which pixels have contributed to each \sfrsd{} superprofile.
\label{resolved::fig:superprofiles-sfr-sexa-a} }
\end{leftfullpage}
\end{figure}
\addtocounter{figure}{-1}
\addtocounter{subfig}{1}
\begin{figure}
\centering
\includegraphics[height=2.7in]{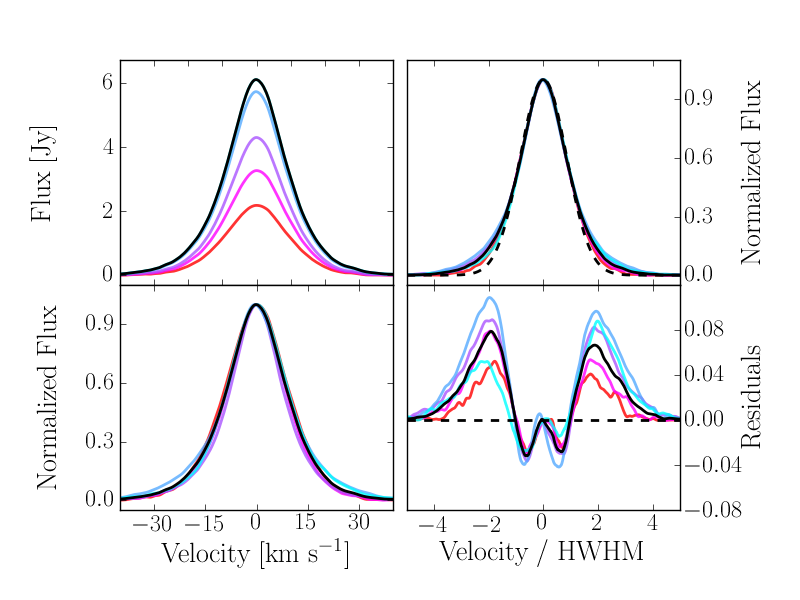}
\caption[The \sfrsd{} superprofiles in Sextans A]{The \sfrsd{} superprofiles in Sextans A, where colors indicate the corresponding \sfrsd{} regions in the previous figure.
The left hand panels show the raw superprofiles (upper left) and the superprofiles normalized to the same peak flux (lower left).
The right hand panels show the flux-normalized superprofiles scaled by the HWHM (upper right) and the flux-normalized superprofiles minus the model of the Gaussian core (lower right). In all panels, the solid black line represents the global superprofile. In the left panels, we have shown the HWHM-scaled Gaussian model as the dashed black line.
\label{resolved::fig:superprofiles-sfr-sexa-b}
}
\end{figure}
\addtocounter{figure}{-1}
\addtocounter{subfig}{1}
\begin{figure}
\centering
\includegraphics[height=2.7in]{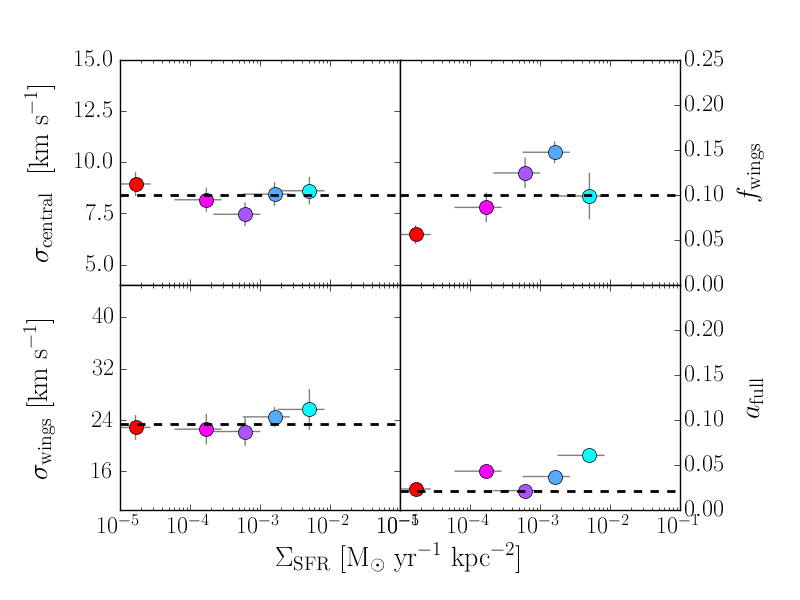}
\caption[Variation of \sfrsd{} superprofile parameters for Sextans A]{Variation of the superprofile parameters as a function of \sfrsd{} for Sextans A.
The solid dashed line shows the parameter value for the global superprofile \citepalias{StilpGlobal}.
The left panels show \scentral{} (upper) and \swing{} (lower), and the right panels show \fw{} (upper) and \afull{} (lower).
\label{resolved::fig:superprofiles-sfr-sexa-c}
}
\end{figure}
\clearpage

\setcounter{subfig}{1}
\begin{figure}[p]
\begin{leftfullpage}
\centering
\includegraphics[width=4in]{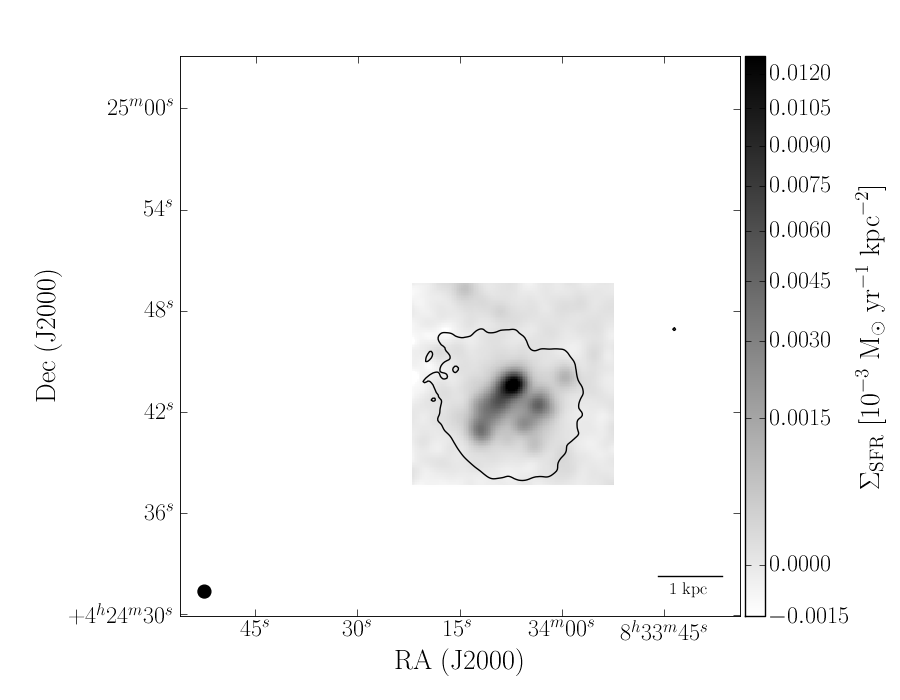}
\includegraphics[width=4in]{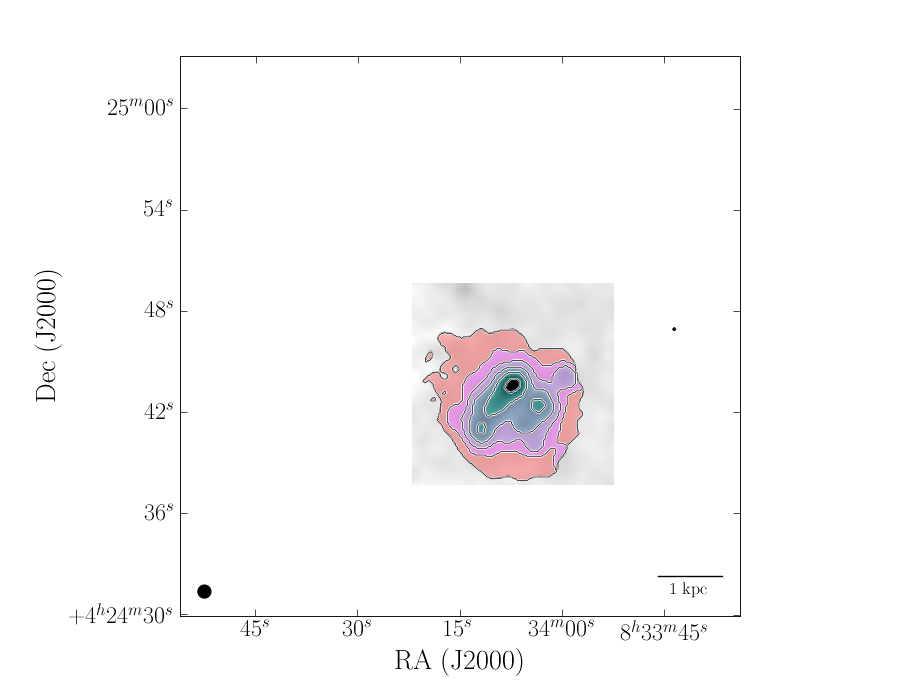}
\caption[Superprofile \sfrsd{} regions for DDO~53]{Subregions in which \sfrsd{} superprofiles are generated for DDO~53. 
In both panels, the background greyscale shows \sfrsd{}, and the solid black line represents the $S/N > 5$ threshold where we can accurately measure \vp{}. 
In the lower panel, the colored regions show which pixels have contributed to each \sfrsd{} superprofile.
\label{resolved::fig:superprofiles-sfr-ddo53-a} }
\end{leftfullpage}
\end{figure}
\addtocounter{figure}{-1}
\addtocounter{subfig}{1}
\begin{figure}
\centering
\includegraphics[height=2.7in]{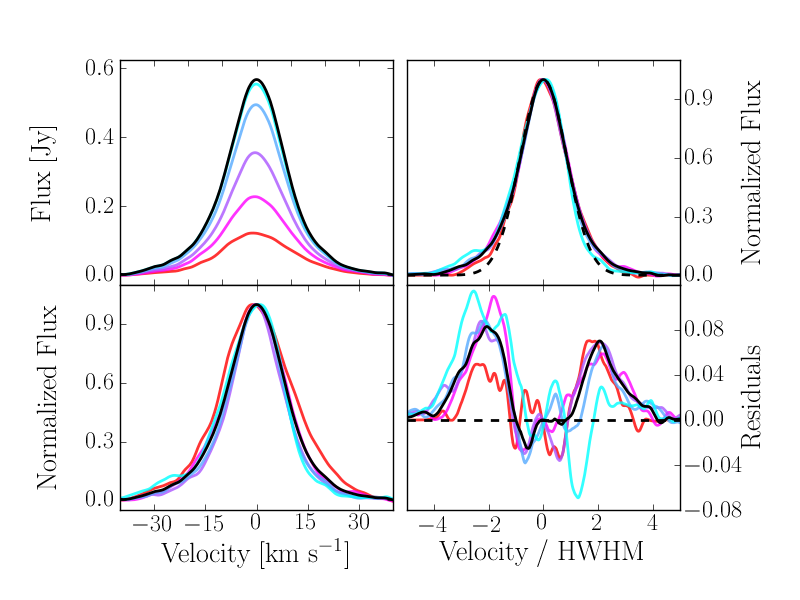}
\caption[The \sfrsd{} superprofiles in DDO~53]{The \sfrsd{} superprofiles in DDO~53, where colors indicate the corresponding \sfrsd{} regions in the previous figure.
The left hand panels show the raw superprofiles (upper left) and the superprofiles normalized to the same peak flux (lower left).
The right hand panels show the flux-normalized superprofiles scaled by the HWHM (upper right) and the flux-normalized superprofiles minus the model of the Gaussian core (lower right). In all panels, the solid black line represents the global superprofile. In the left panels, we have shown the HWHM-scaled Gaussian model as the dashed black line.
\label{resolved::fig:superprofiles-sfr-ddo53-b}
}
\end{figure}
\addtocounter{figure}{-1}
\addtocounter{subfig}{1}
\begin{figure}
\centering
\includegraphics[height=2.7in]{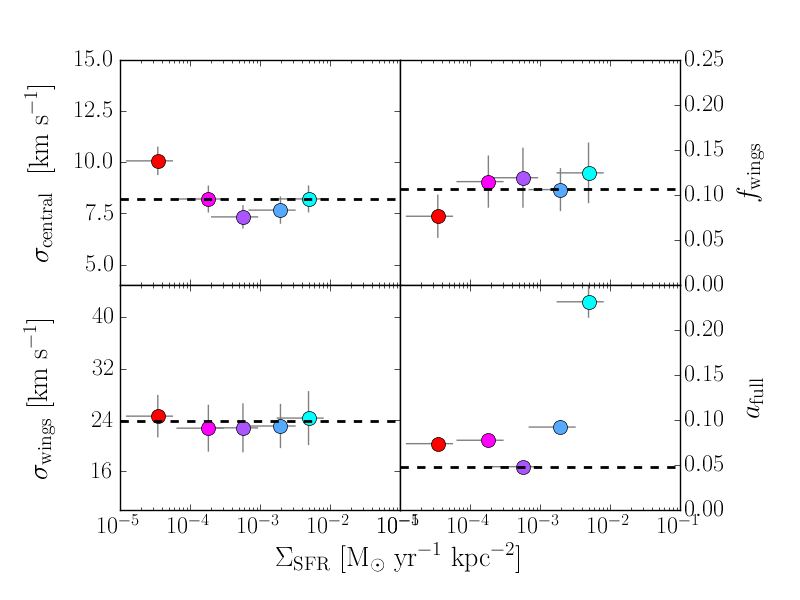}
\caption[Variation of \sfrsd{} superprofile parameters for DDO~53]{Variation of the superprofile parameters as a function of \sfrsd{} for DDO~53.
The solid dashed line shows the parameter value for the global superprofile \citepalias{StilpGlobal}.
The left panels show \scentral{} (upper) and \swing{} (lower), and the right panels show \fw{} (upper) and \afull{} (lower).
\label{resolved::fig:superprofiles-sfr-ddo53-c}
}
\end{figure}
\clearpage

\setcounter{subfig}{1}
\begin{figure}[p]
\begin{leftfullpage}
\centering
\includegraphics[width=4in]{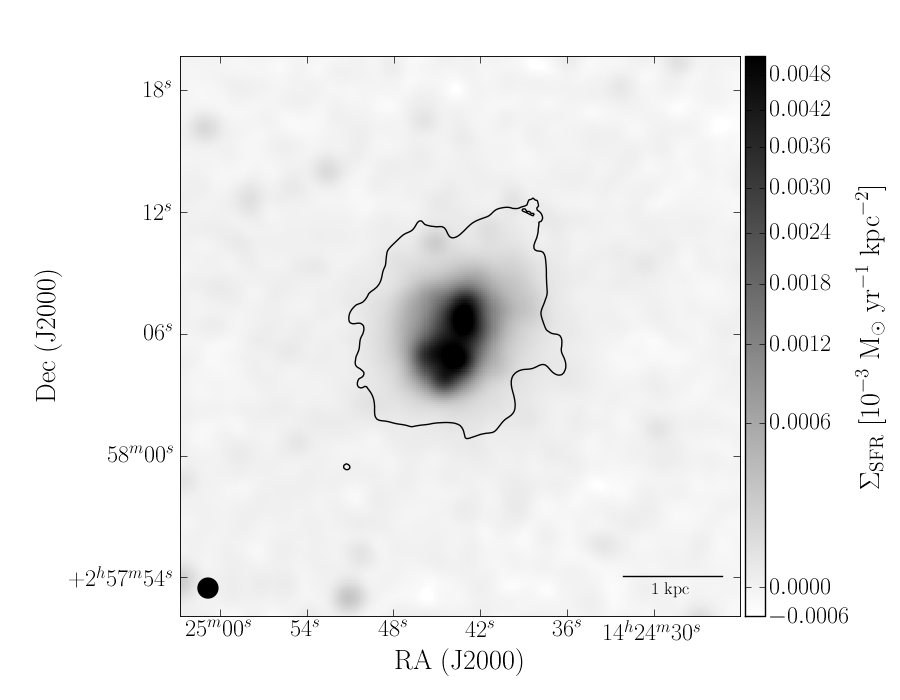}
\includegraphics[width=4in]{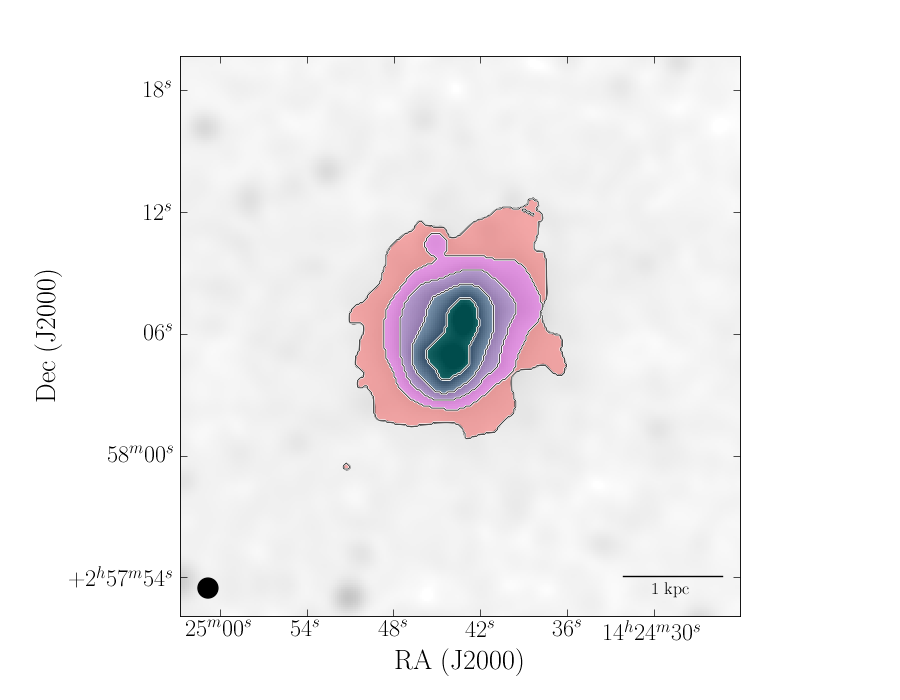}
\caption[Superprofile \sfrsd{} regions for DDO~190]{Subregions in which \sfrsd{} superprofiles are generated for DDO~190. 
In both panels, the background greyscale shows \sfrsd{}, and the solid black line represents the $S/N > 5$ threshold where we can accurately measure \vp{}. 
In the lower panel, the colored regions show which pixels have contributed to each \sfrsd{} superprofile.
\label{resolved::fig:superprofiles-sfr-ddo190-a} }
\end{leftfullpage}
\end{figure}
\addtocounter{figure}{-1}
\addtocounter{subfig}{1}
\begin{figure}
\centering
\includegraphics[height=2.7in]{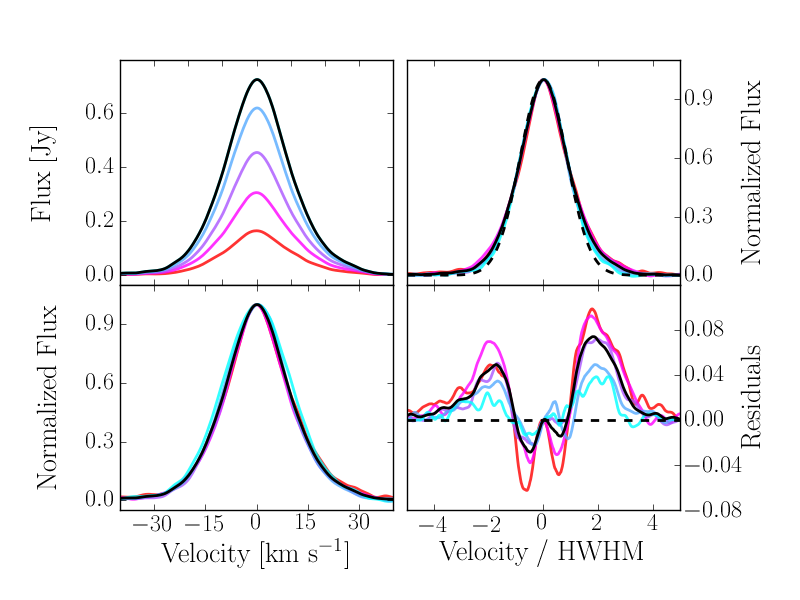}
\caption[The \sfrsd{} superprofiles in DDO~190]{The \sfrsd{} superprofiles in DDO~190, where colors indicate the corresponding \sfrsd{} regions in the previous figure.
The left hand panels show the raw superprofiles (upper left) and the superprofiles normalized to the same peak flux (lower left).
The right hand panels show the flux-normalized superprofiles scaled by the HWHM (upper right) and the flux-normalized superprofiles minus the model of the Gaussian core (lower right). In all panels, the solid black line represents the global superprofile. In the left panels, we have shown the HWHM-scaled Gaussian model as the dashed black line.
\label{resolved::fig:superprofiles-sfr-ddo190-b}
}
\end{figure}
\addtocounter{figure}{-1}
\addtocounter{subfig}{1}
\begin{figure}
\centering
\includegraphics[height=2.7in]{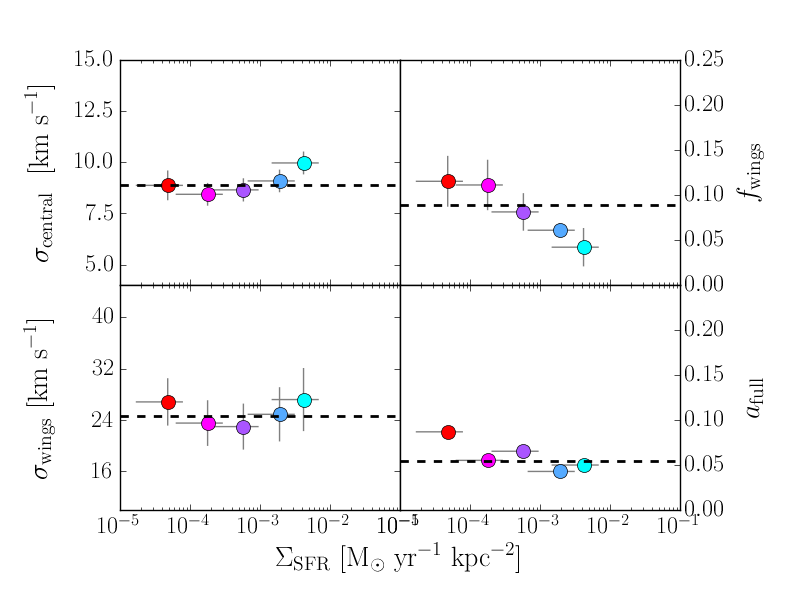}
\caption[Variation of \sfrsd{} superprofile parameters for DDO~190]{Variation of the superprofile parameters as a function of \sfrsd{} for DDO~190.
The solid dashed line shows the parameter value for the global superprofile \citepalias{StilpGlobal}.
The left panels show \scentral{} (upper) and \swing{} (lower), and the right panels show \fw{} (upper) and \afull{} (lower).
\label{resolved::fig:superprofiles-sfr-ddo190-c}
}
\end{figure}
\clearpage

\setcounter{subfig}{1}
\begin{figure}[p]
\begin{leftfullpage}
\centering
\includegraphics[width=4in]{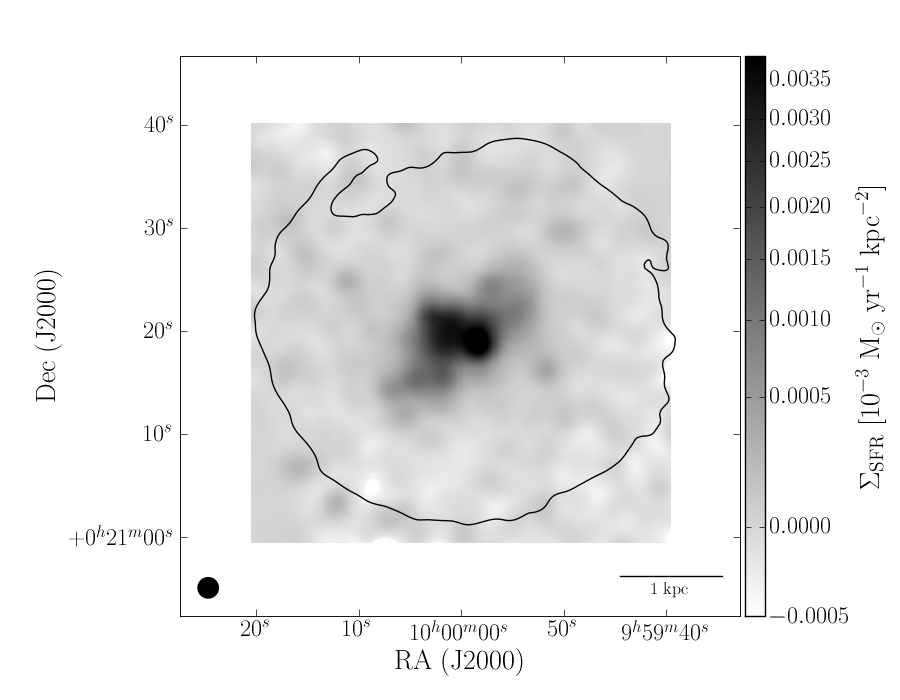}
\includegraphics[width=4in]{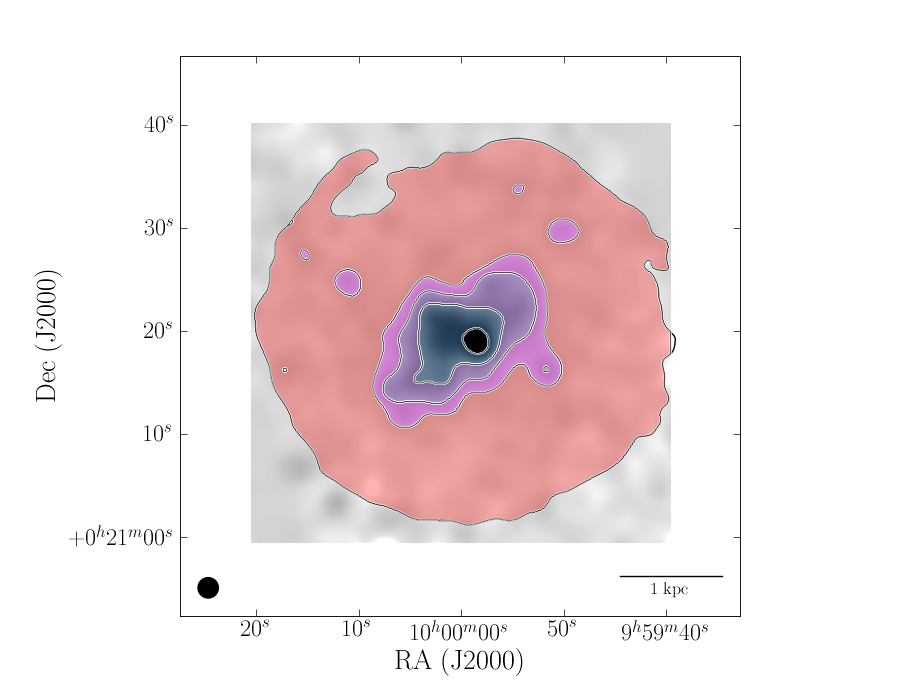}
\caption[Superprofile \sfrsd{} regions for Sextans B]{Subregions in which \sfrsd{} superprofiles are generated for Sextans B. 
In both panels, the background greyscale shows \sfrsd{}, and the solid black line represents the $S/N > 5$ threshold where we can accurately measure \vp{}. 
In the lower panel, the colored regions show which pixels have contributed to each \sfrsd{} superprofile.
\label{resolved::fig:superprofiles-sfr-sexb-a} }
\end{leftfullpage}
\end{figure}
\addtocounter{figure}{-1}
\addtocounter{subfig}{1}
\begin{figure}
\centering
\includegraphics[height=2.7in]{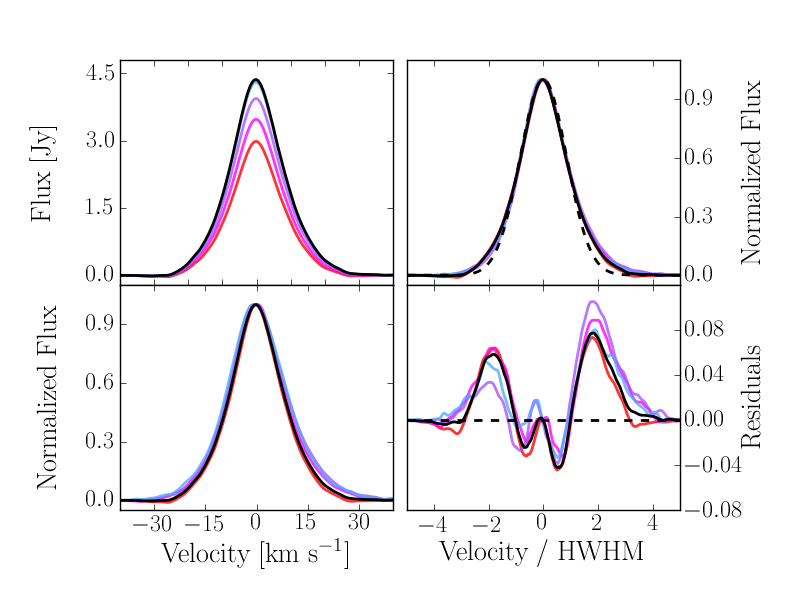}
\caption[The \sfrsd{} superprofiles in Sextans B]{The \sfrsd{} superprofiles in Sextans B, where colors indicate the corresponding \sfrsd{} regions in the previous figure.
The left hand panels show the raw superprofiles (upper left) and the superprofiles normalized to the same peak flux (lower left).
The right hand panels show the flux-normalized superprofiles scaled by the HWHM (upper right) and the flux-normalized superprofiles minus the model of the Gaussian core (lower right). In all panels, the solid black line represents the global superprofile. In the left panels, we have shown the HWHM-scaled Gaussian model as the dashed black line.
\label{resolved::fig:superprofiles-sfr-sexb-b}
}
\end{figure}
\addtocounter{figure}{-1}
\addtocounter{subfig}{1}
\begin{figure}
\centering
\includegraphics[height=2.7in]{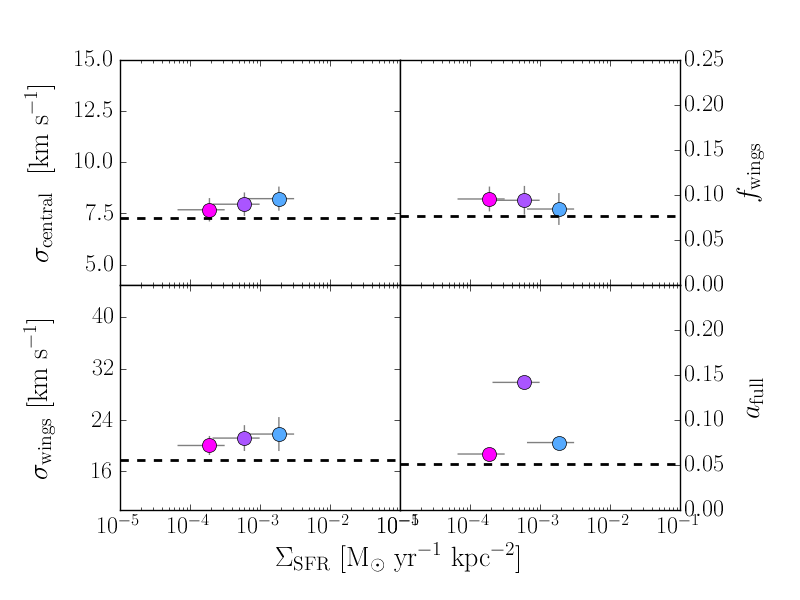}
\caption[Variation of \sfrsd{} superprofile parameters for Sextans B]{Variation of the superprofile parameters as a function of \sfrsd{} for Sextans B.
The solid dashed line shows the parameter value for the global superprofile \citepalias{StilpGlobal}.
The left panels show \scentral{} (upper) and \swing{} (lower), and the right panels show \fw{} (upper) and \afull{} (lower).
\label{resolved::fig:superprofiles-sfr-sexb-c}
}
\end{figure}
\clearpage

\setcounter{subfig}{1}
\begin{figure}[p]
\begin{leftfullpage}
\centering
\includegraphics[width=4in]{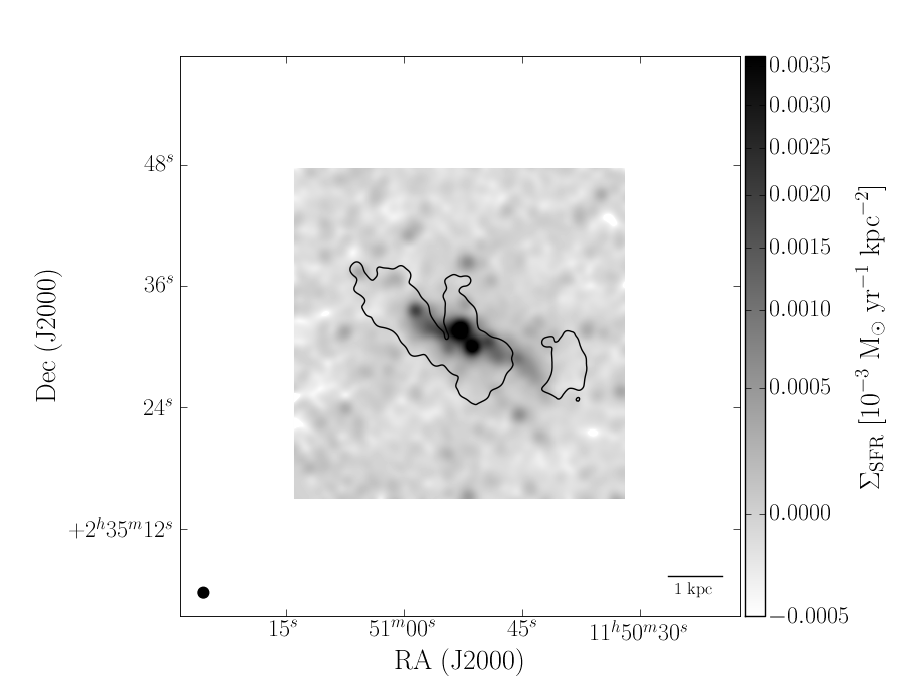}
\includegraphics[width=4in]{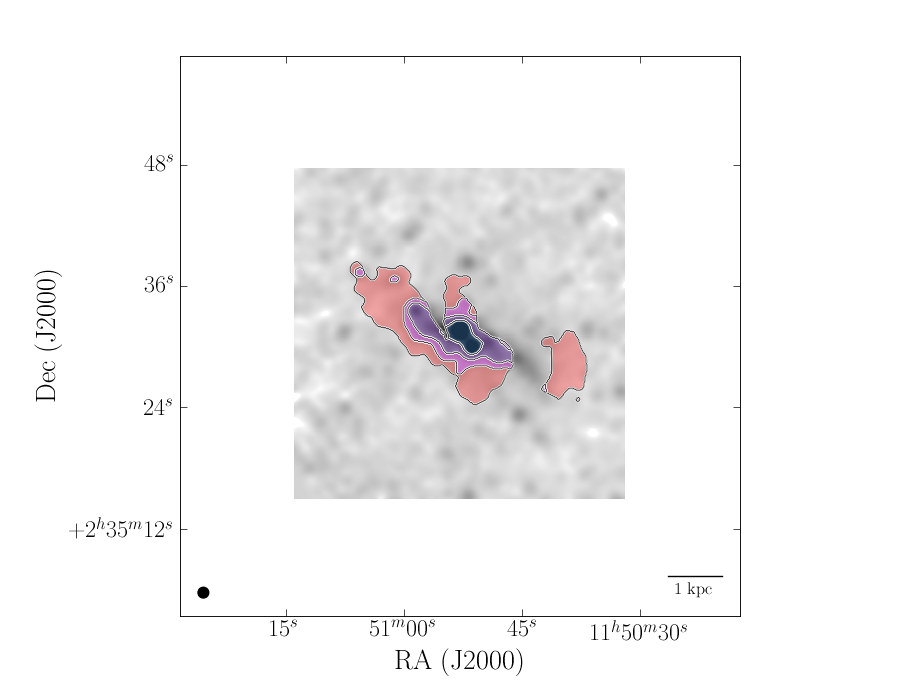}
\caption[Superprofile \sfrsd{} regions for DDO~99]{Subregions in which \sfrsd{} superprofiles are generated for DDO~99. 
In both panels, the background greyscale shows \sfrsd{}, and the solid black line represents the $S/N > 5$ threshold where we can accurately measure \vp{}. 
In the lower panel, the colored regions show which pixels have contributed to each \sfrsd{} superprofile.
\label{resolved::fig:superprofiles-sfr-ddo99-a} }
\end{leftfullpage}
\end{figure}
\addtocounter{figure}{-1}
\addtocounter{subfig}{1}
\begin{figure}
\centering
\includegraphics[height=2.7in]{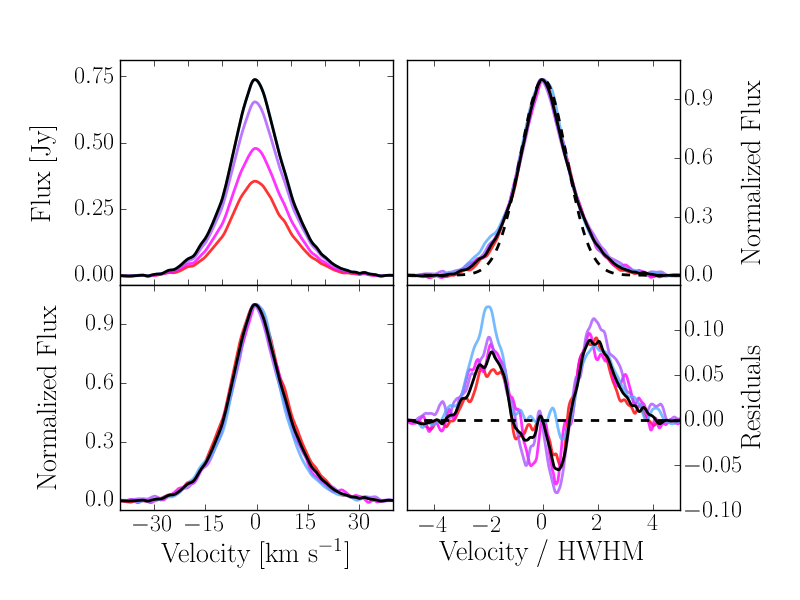}
\caption[The \sfrsd{} superprofiles in DDO~99]{The \sfrsd{} superprofiles in DDO~99, where colors indicate the corresponding \sfrsd{} regions in the previous figure.
The left hand panels show the raw superprofiles (upper left) and the superprofiles normalized to the same peak flux (lower left).
The right hand panels show the flux-normalized superprofiles scaled by the HWHM (upper right) and the flux-normalized superprofiles minus the model of the Gaussian core (lower right). In all panels, the solid black line represents the global superprofile. In the left panels, we have shown the HWHM-scaled Gaussian model as the dashed black line.
\label{resolved::fig:superprofiles-sfr-ddo99-b}
}
\end{figure}
\addtocounter{figure}{-1}
\addtocounter{subfig}{1}
\begin{figure}
\centering
\includegraphics[height=2.7in]{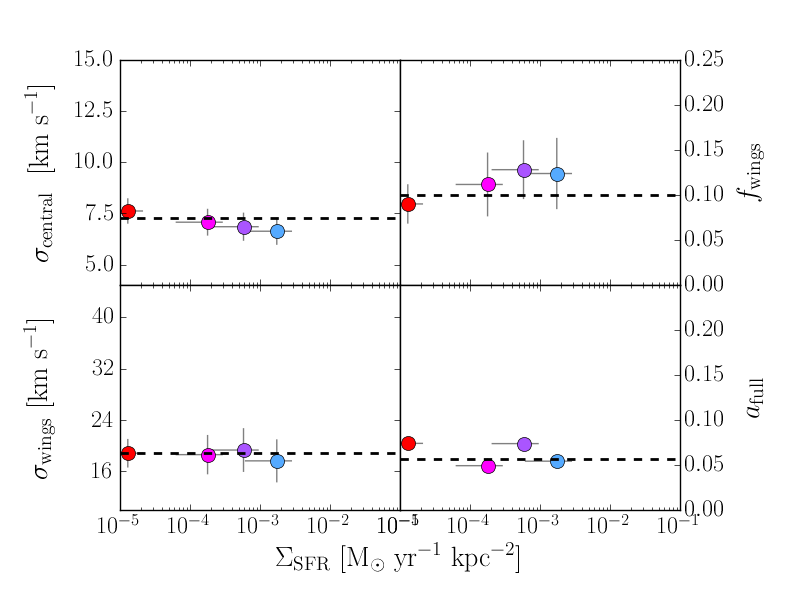}
\caption[Variation of \sfrsd{} superprofile parameters for DDO~99]{Variation of the superprofile parameters as a function of \sfrsd{} for DDO~99.
The solid dashed line shows the parameter value for the global superprofile \citepalias{StilpGlobal}.
The left panels show \scentral{} (upper) and \swing{} (lower), and the right panels show \fw{} (upper) and \afull{} (lower).
\label{resolved::fig:superprofiles-sfr-ddo99-c}
}
\end{figure}
\clearpage

\setcounter{subfig}{1}
\begin{figure}[p]
\begin{leftfullpage}
\centering
\includegraphics[width=4in]{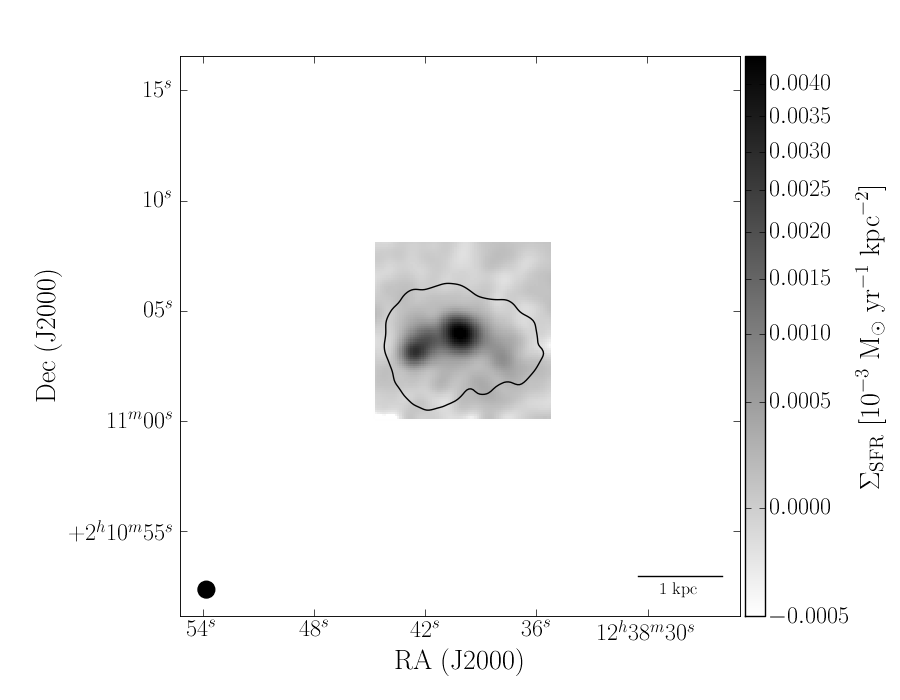}
\includegraphics[width=4in]{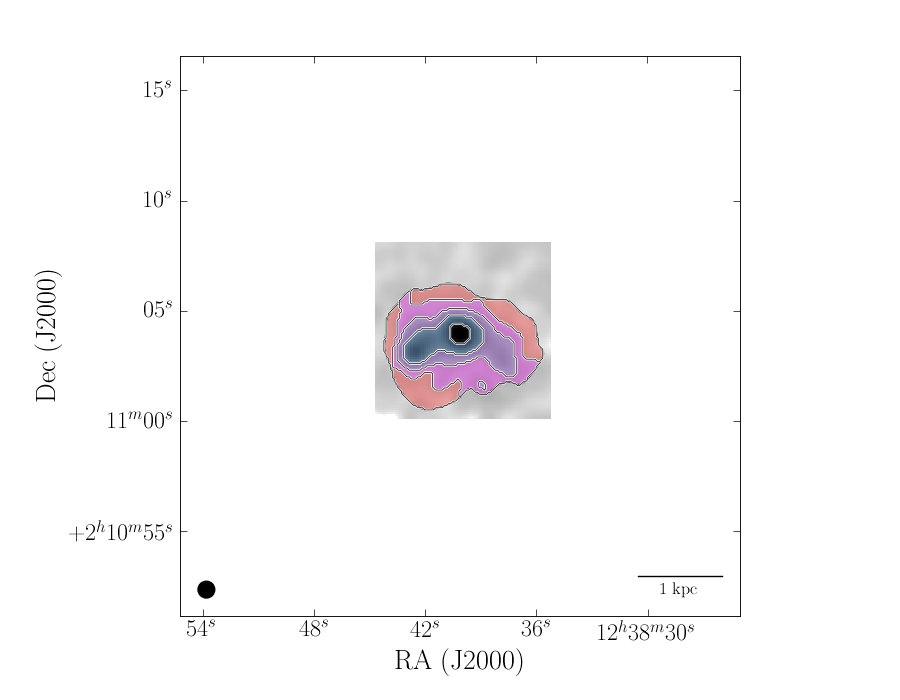}
\caption[Superprofile \sfrsd{} regions for UGCA~292]{Subregions in which \sfrsd{} superprofiles are generated for UGCA~292. 
In both panels, the background greyscale shows \sfrsd{}, and the solid black line represents the $S/N > 5$ threshold where we can accurately measure \vp{}. 
In the lower panel, the colored regions show which pixels have contributed to each \sfrsd{} superprofile.
\label{resolved::fig:superprofiles-sfr-ua292-a} }
\end{leftfullpage}
\end{figure}
\addtocounter{figure}{-1}
\addtocounter{subfig}{1}
\begin{figure}
\centering
\includegraphics[height=2.7in]{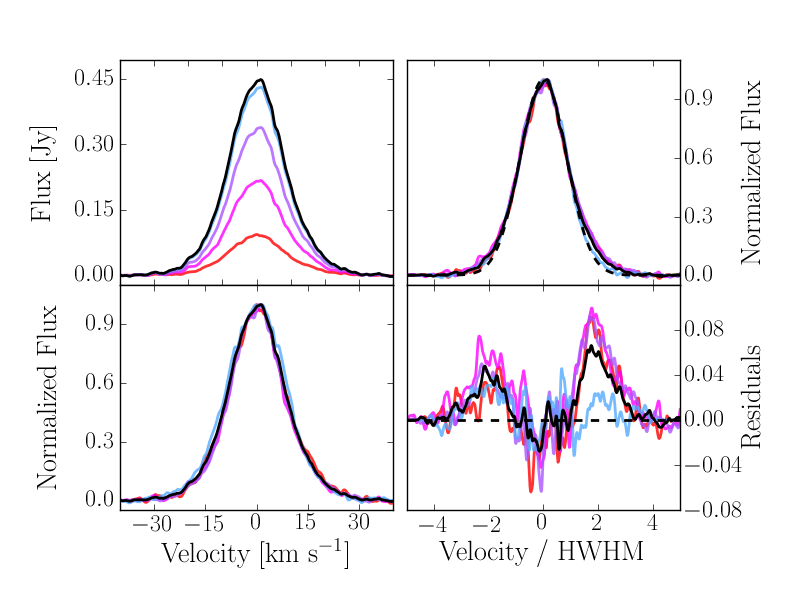}
\caption[The \sfrsd{} superprofiles in UGCA~292]{The \sfrsd{} superprofiles in UGCA~292, where colors indicate the corresponding \sfrsd{} regions in the previous figure.
The left hand panels show the raw superprofiles (upper left) and the superprofiles normalized to the same peak flux (lower left).
The right hand panels show the flux-normalized superprofiles scaled by the HWHM (upper right) and the flux-normalized superprofiles minus the model of the Gaussian core (lower right). In all panels, the solid black line represents the global superprofile. In the left panels, we have shown the HWHM-scaled Gaussian model as the dashed black line.
\label{resolved::fig:superprofiles-sfr-ua292-b}
}
\end{figure}
\addtocounter{figure}{-1}
\addtocounter{subfig}{1}
\begin{figure}
\centering
\includegraphics[height=2.7in]{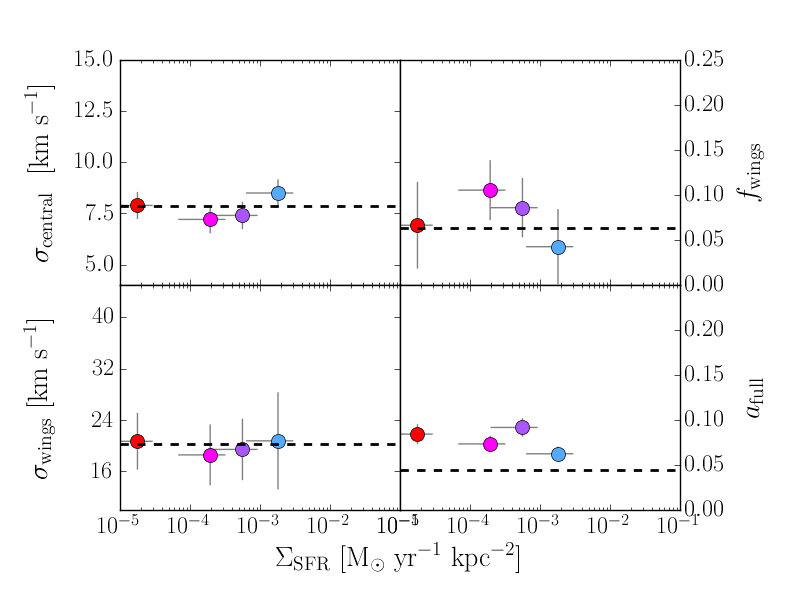}
\caption[Variation of \sfrsd{} superprofile parameters for UGCA~292]{Variation of the superprofile parameters as a function of \sfrsd{} for UGCA~292.
The solid dashed line shows the parameter value for the global superprofile \citepalias{StilpGlobal}.
The left panels show \scentral{} (upper) and \swing{} (lower), and the right panels show \fw{} (upper) and \afull{} (lower).
\label{resolved::fig:superprofiles-sfr-ua292-c}
}
\end{figure}
\clearpage

\setcounter{subfig}{1}
\begin{figure}[p]
\begin{leftfullpage}
\centering
\includegraphics[width=4in]{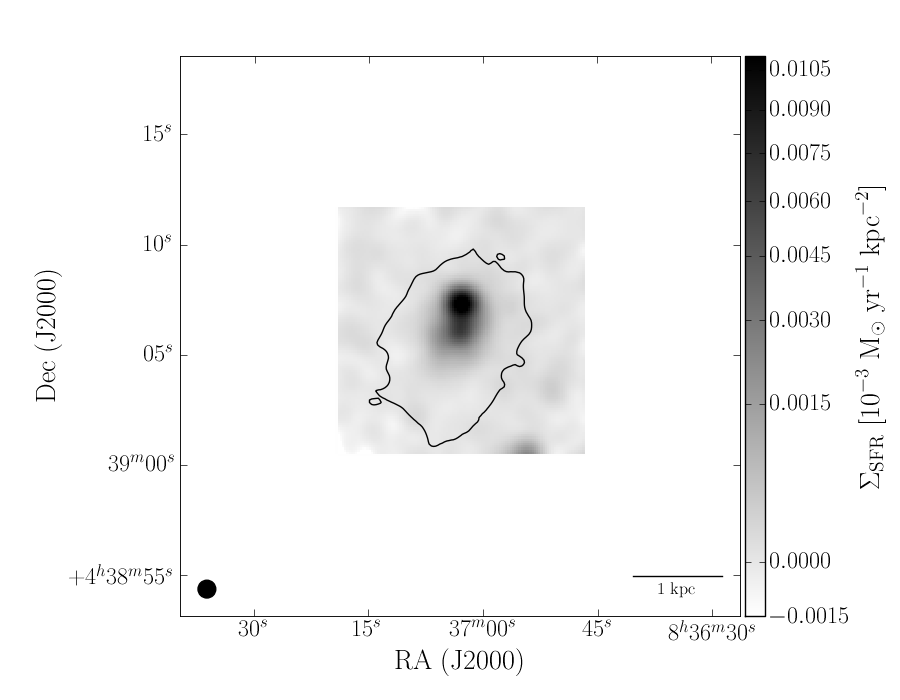}
\includegraphics[width=4in]{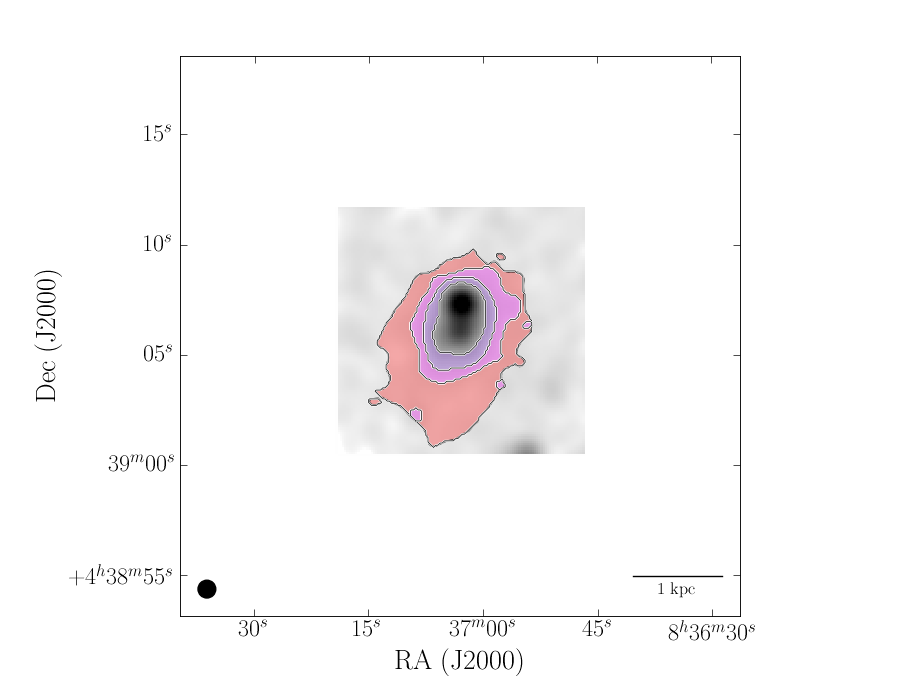}
\caption[Superprofile \sfrsd{} regions for UGC~4483]{Subregions in which \sfrsd{} superprofiles are generated for UGC~4483. 
In both panels, the background greyscale shows \sfrsd{}, and the solid black line represents the $S/N > 5$ threshold where we can accurately measure \vp{}. 
In the lower panel, the colored regions show which pixels have contributed to each \sfrsd{} superprofile.
\label{resolved::fig:superprofiles-sfr-u4483-a} }
\end{leftfullpage}
\end{figure}
\addtocounter{figure}{-1}
\addtocounter{subfig}{1}
\begin{figure}
\centering
\includegraphics[height=2.7in]{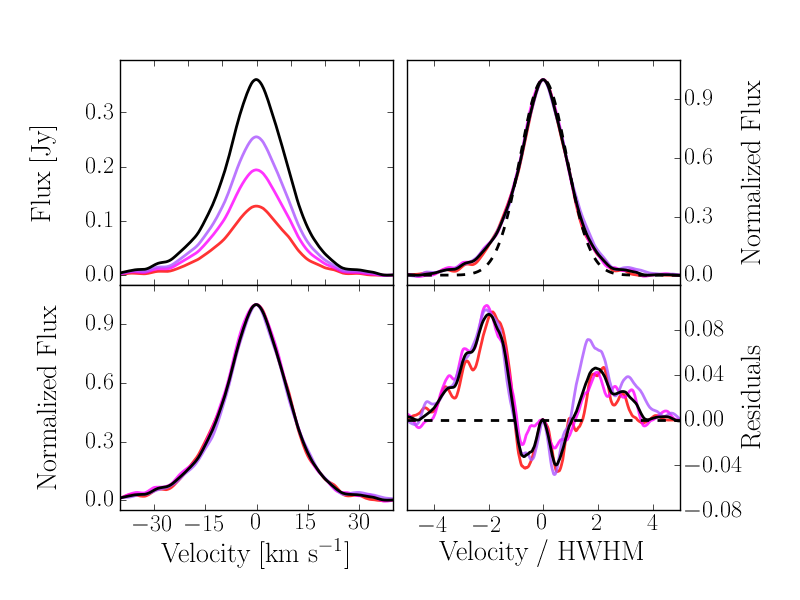}
\caption[The \sfrsd{} superprofiles in UGC~4483]{The \sfrsd{} superprofiles in UGC~4483, where colors indicate the corresponding \sfrsd{} regions in the previous figure.
The left hand panels show the raw superprofiles (upper left) and the superprofiles normalized to the same peak flux (lower left).
The right hand panels show the flux-normalized superprofiles scaled by the HWHM (upper right) and the flux-normalized superprofiles minus the model of the Gaussian core (lower right). In all panels, the solid black line represents the global superprofile. In the left panels, we have shown the HWHM-scaled Gaussian model as the dashed black line.
\label{resolved::fig:superprofiles-sfr-u4483-b}
}
\end{figure}
\addtocounter{figure}{-1}
\addtocounter{subfig}{1}
\begin{figure}
\centering
\includegraphics[height=2.7in]{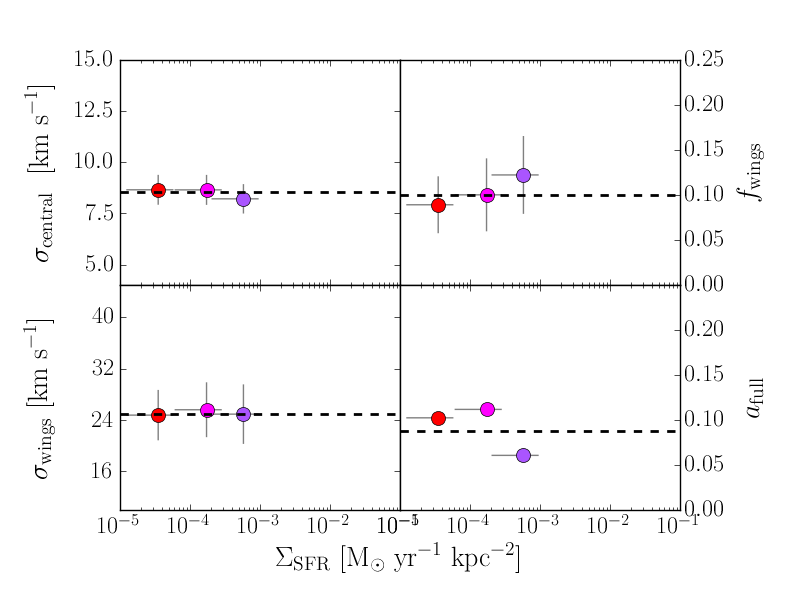}
\caption[Variation of \sfrsd{} superprofile parameters for UGC~4483]{Variation of the superprofile parameters as a function of \sfrsd{} for UGC~4483.
The solid dashed line shows the parameter value for the global superprofile \citepalias{StilpGlobal}.
The left panels show \scentral{} (upper) and \swing{} (lower), and the right panels show \fw{} (upper) and \afull{} (lower).
\label{resolved::fig:superprofiles-sfr-u4483-c}
}
\end{figure}
\clearpage

\setcounter{subfig}{1}
\begin{figure}[p]
\begin{leftfullpage}
\centering
\includegraphics[width=4in]{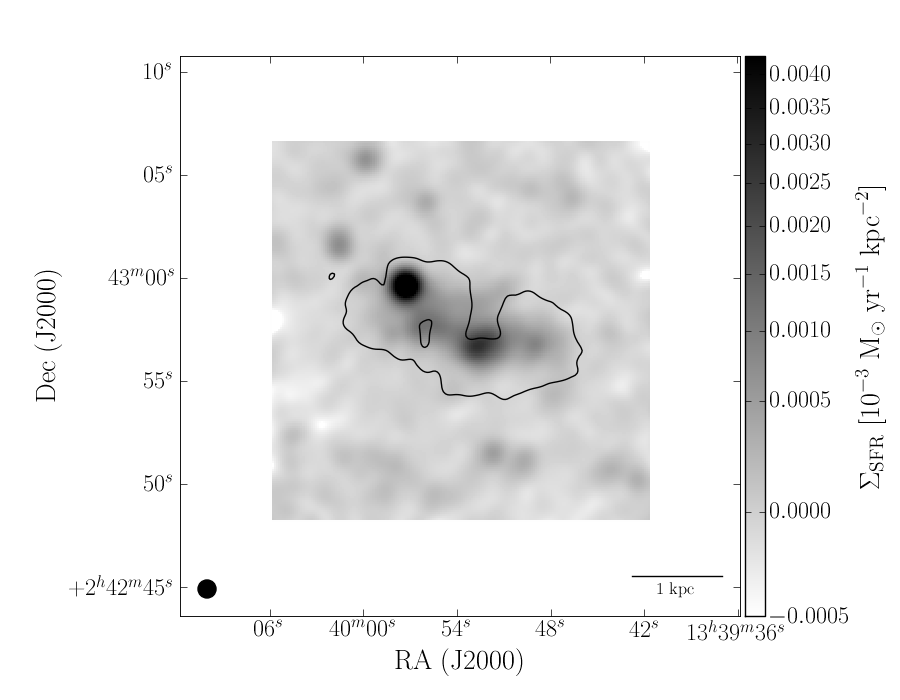}
\includegraphics[width=4in]{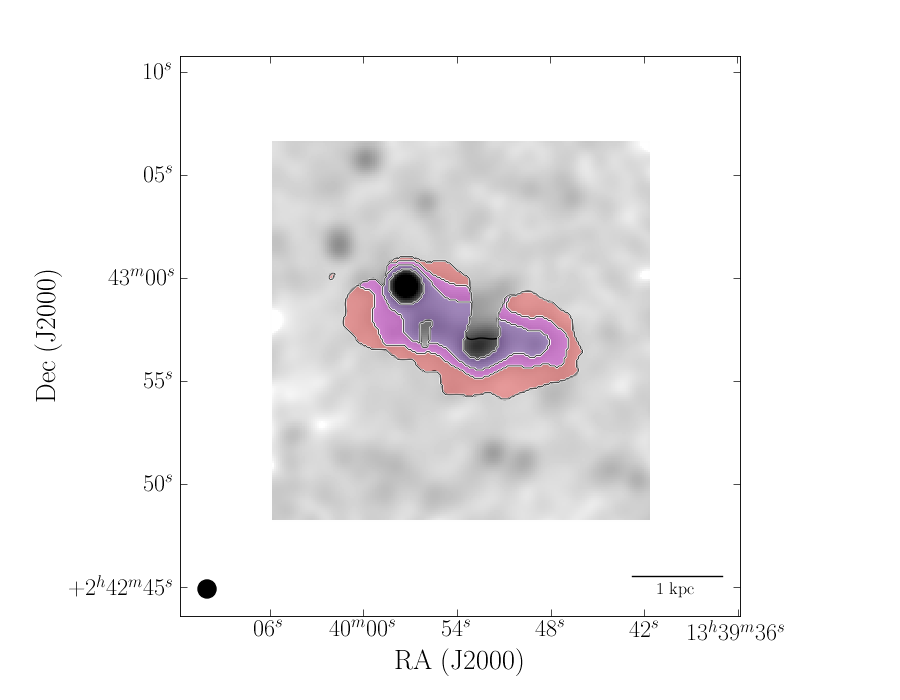}
\caption[Superprofile \sfrsd{} regions for DDO~181]{Subregions in which \sfrsd{} superprofiles are generated for DDO~181. 
In both panels, the background greyscale shows \sfrsd{}, and the solid black line represents the $S/N > 5$ threshold where we can accurately measure \vp{}. 
In the lower panel, the colored regions show which pixels have contributed to each \sfrsd{} superprofile.
\label{resolved::fig:superprofiles-sfr-ddo181-a} }
\end{leftfullpage}
\end{figure}
\addtocounter{figure}{-1}
\addtocounter{subfig}{1}
\begin{figure}
\centering
\includegraphics[height=2.7in]{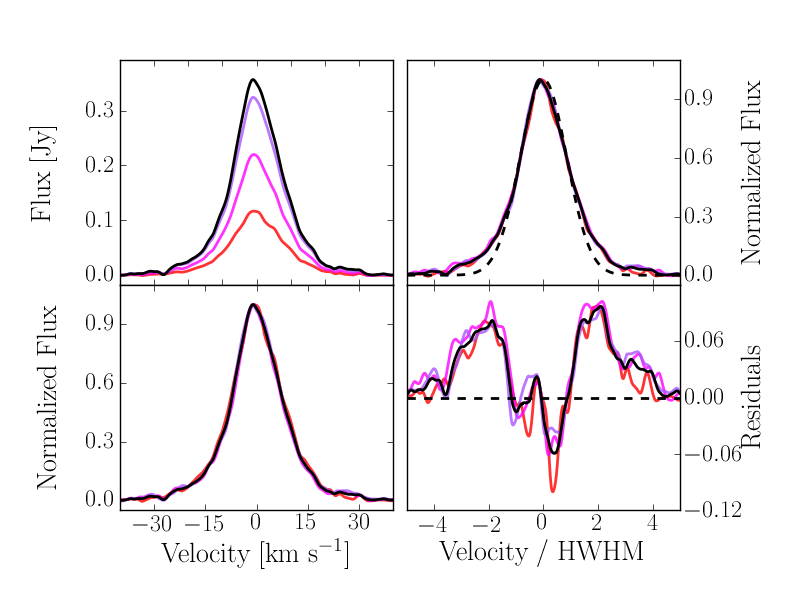}
\caption[The \sfrsd{} superprofiles in DDO~181]{The \sfrsd{} superprofiles in DDO~181, where colors indicate the corresponding \sfrsd{} regions in the previous figure.
The left hand panels show the raw superprofiles (upper left) and the superprofiles normalized to the same peak flux (lower left).
The right hand panels show the flux-normalized superprofiles scaled by the HWHM (upper right) and the flux-normalized superprofiles minus the model of the Gaussian core (lower right). In all panels, the solid black line represents the global superprofile. In the left panels, we have shown the HWHM-scaled Gaussian model as the dashed black line.
\label{resolved::fig:superprofiles-sfr-ddo181-b}
}
\end{figure}
\addtocounter{figure}{-1}
\addtocounter{subfig}{1}
\begin{figure}
\centering
\includegraphics[height=2.7in]{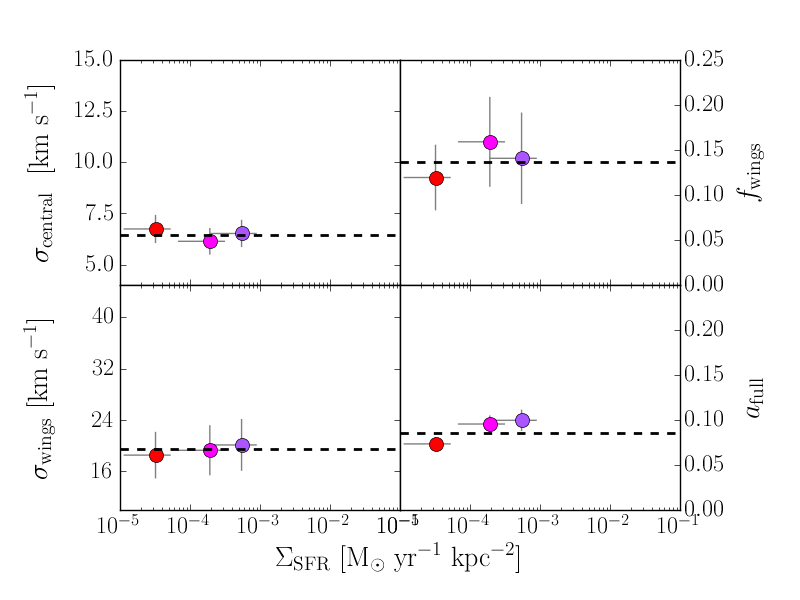}
\caption[Variation of \sfrsd{} superprofile parameters for DDO~181]{Variation of the superprofile parameters as a function of \sfrsd{} for DDO~181.
The solid dashed line shows the parameter value for the global superprofile \citepalias{StilpGlobal}.
The left panels show \scentral{} (upper) and \swing{} (lower), and the right panels show \fw{} (upper) and \afull{} (lower).
\label{resolved::fig:superprofiles-sfr-ddo181-c}
}
\end{figure}
\clearpage

\setcounter{subfig}{1}
\begin{figure}[p]
\begin{leftfullpage}
\centering
\includegraphics[width=4in]{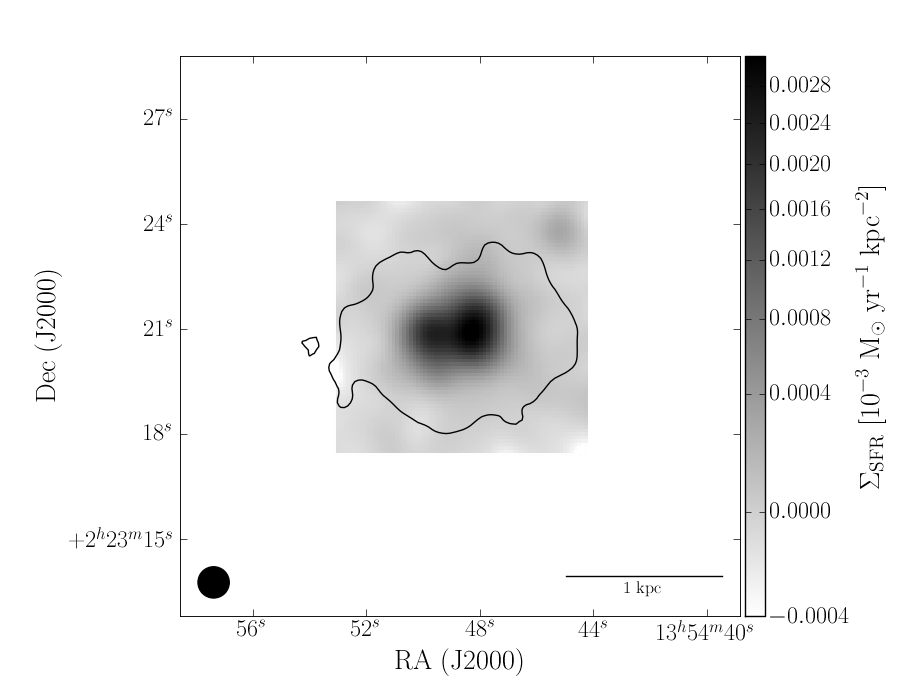}
\includegraphics[width=4in]{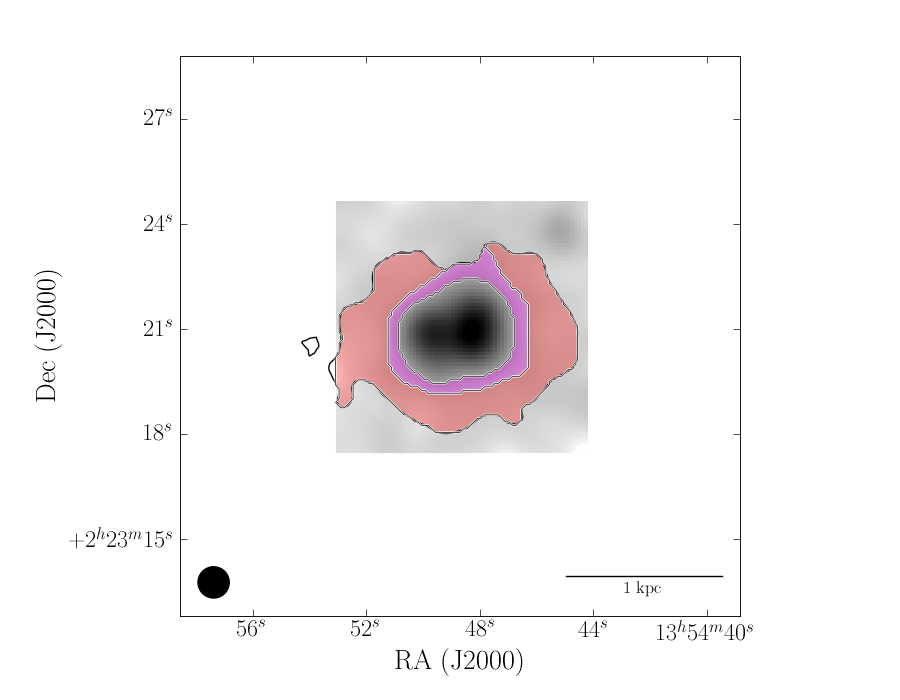}
\caption[Superprofile \sfrsd{} regions for UGC~8833]{Subregions in which \sfrsd{} superprofiles are generated for UGC~8833. 
In both panels, the background greyscale shows \sfrsd{}, and the solid black line represents the $S/N > 5$ threshold where we can accurately measure \vp{}. 
In the lower panel, the colored regions show which pixels have contributed to each \sfrsd{} superprofile.
\label{resolved::fig:superprofiles-sfr-u8833-a} }
\end{leftfullpage}
\end{figure}
\addtocounter{figure}{-1}
\addtocounter{subfig}{1}
\begin{figure}
\centering
\includegraphics[height=2.7in]{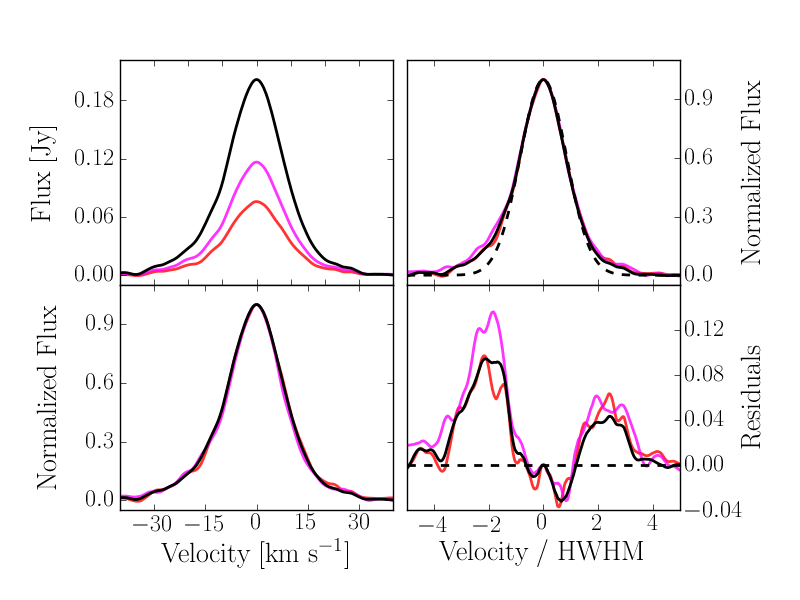}
\caption[The \sfrsd{} superprofiles in UGC~8833]{The \sfrsd{} superprofiles in UGC~8833, where colors indicate the corresponding \sfrsd{} regions in the previous figure.
The left hand panels show the raw superprofiles (upper left) and the superprofiles normalized to the same peak flux (lower left).
The right hand panels show the flux-normalized superprofiles scaled by the HWHM (upper right) and the flux-normalized superprofiles minus the model of the Gaussian core (lower right). In all panels, the solid black line represents the global superprofile. In the left panels, we have shown the HWHM-scaled Gaussian model as the dashed black line.
\label{resolved::fig:superprofiles-sfr-u8833-b}
}
\end{figure}
\addtocounter{figure}{-1}
\addtocounter{subfig}{1}
\begin{figure}
\centering
\includegraphics[height=2.7in]{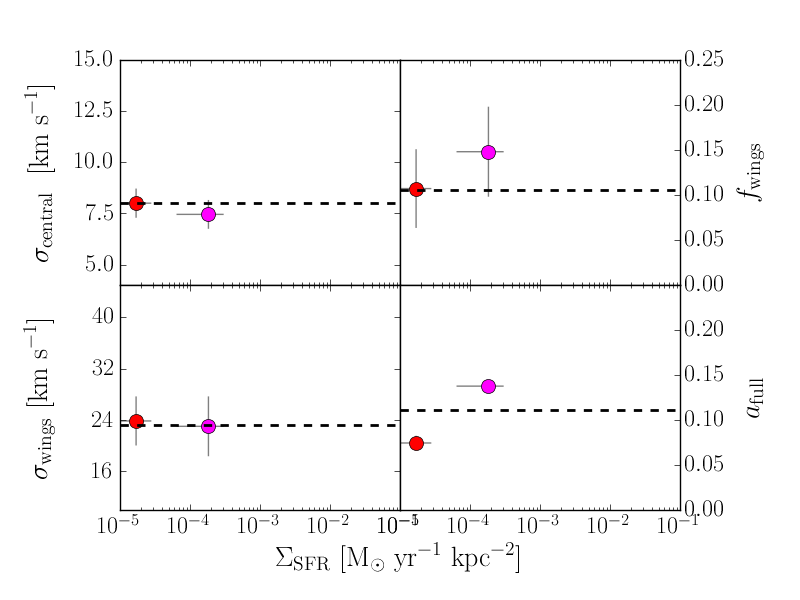}
\caption[Variation of \sfrsd{} superprofile parameters for UGC~8833]{Variation of the superprofile parameters as a function of \sfrsd{} for UGC~8833.
The solid dashed line shows the parameter value for the global superprofile \citepalias{StilpGlobal}.
The left panels show \scentral{} (upper) and \swing{} (lower), and the right panels show \fw{} (upper) and \afull{} (lower).
\label{resolved::fig:superprofiles-sfr-u8833-c}
}
\end{figure}
\clearpage

\setcounter{subfig}{1}
\begin{figure}[p]
\begin{leftfullpage}
\centering
\includegraphics[width=4in]{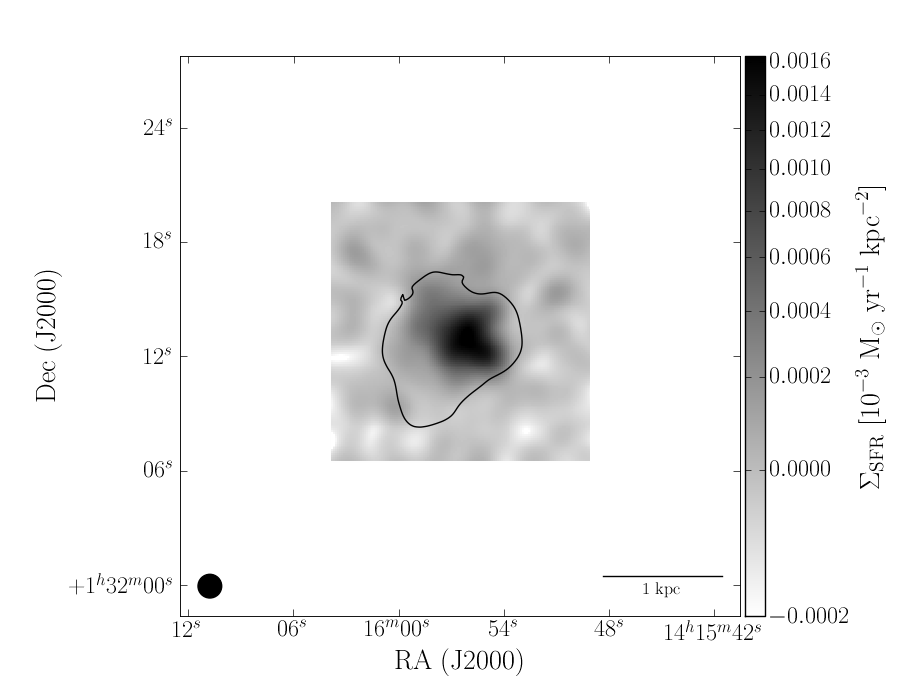}
\includegraphics[width=4in]{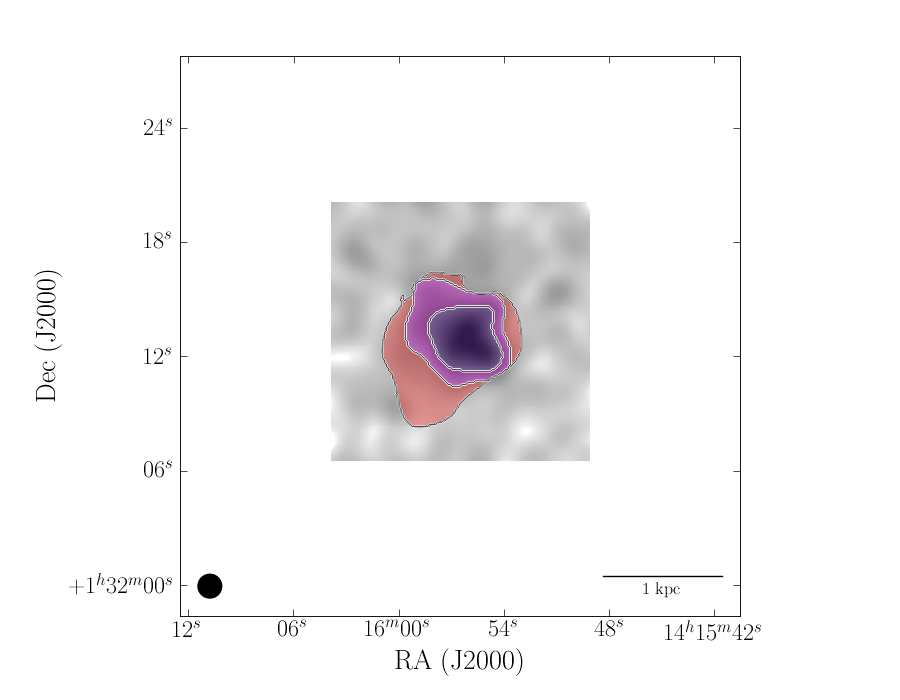}
\caption[Superprofile \sfrsd{} regions for DDO~187]{Subregions in which \sfrsd{} superprofiles are generated for DDO~187. 
In both panels, the background greyscale shows \sfrsd{}, and the solid black line represents the $S/N > 5$ threshold where we can accurately measure \vp{}. 

In the lower panel, the colored regions show which pixels have contributed to each \sfrsd{} superprofile.
\label{resolved::fig:superprofiles-sfr-ddo187-a} }
\end{leftfullpage}
\end{figure}
\addtocounter{figure}{-1}
\addtocounter{subfig}{1}
\begin{figure}
\centering
\includegraphics[height=2.7in]{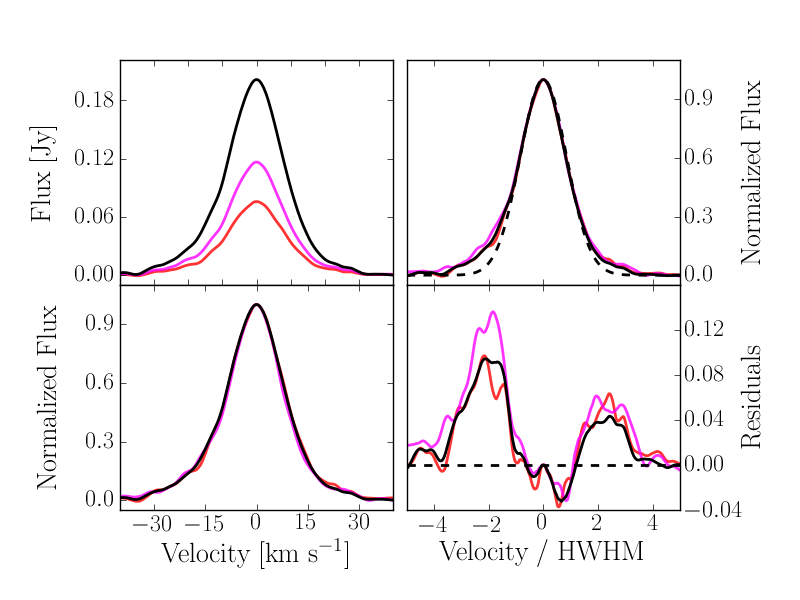}
\caption[The \sfrsd{} superprofiles in DDO~187]{The \sfrsd{} superprofiles in DDO~187, where colors indicate the corresponding \sfrsd{} regions in the previous figure.
The left hand panels show the raw superprofiles (upper left) and the superprofiles normalized to the same peak flux (lower left).
The right hand panels show the flux-normalized superprofiles scaled by the HWHM (upper right) and the flux-normalized superprofiles minus the model of the Gaussian core (lower right). In all panels, the solid black line represents the global superprofile. In the left panels, we have shown the HWHM-scaled Gaussian model as the dashed black line.
\label{resolved::fig:superprofiles-sfr-ddo187-b}
}
\end{figure}
\addtocounter{figure}{-1}
\addtocounter{subfig}{1}
\begin{figure}
\centering
\includegraphics[height=2.7in]{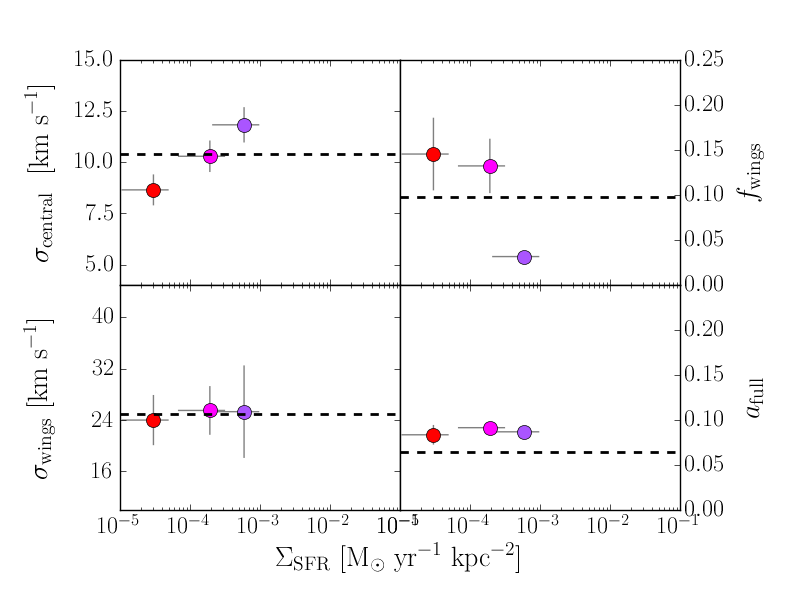}
\caption[Variation of \sfrsd{} superprofile parameters for DDO~187]{Variation of the superprofile parameters as a function of \sfrsd{} for DDO~187.
The solid dashed line shows the parameter value for the global superprofile \citepalias{StilpGlobal}.
The left panels show \scentral{} (upper) and \swing{} (lower), and the right panels show \fw{} (upper) and \afull{} (lower).
\label{resolved::fig:superprofiles-sfr-ddo187-c}
}
\end{figure}
\clearpage

\ifthesis
\renewcommand{\thefigure}{\arabic{chapter}.\arabic{figure}}
\else
\renewcommand{\thefigure}{\arabic{figure}}
\fi

\end{appendix}

\bibliographystyle{apj}
\bibliography{resolved_paper}

\newcommand{\noop}[1]{}
\begin{thebibliography}{74}
\expandafter\ifx\csname natexlab\endcsname\relax\def\natexlab#1{#1}\fi

\bibitem[{{Agertz} {et~al.}(2009){Agertz}, {Lake}, {Teyssier}, {Moore},
  {Mayer}, \& {Romeo}}]{Agertz2009}
{Agertz}, O., {Lake}, G., {Teyssier}, R., {Moore}, B., {Mayer}, L., \& {Romeo},
  A.~B. 2009, \mnras, 392, 294

\bibitem[{{Banerjee} {et~al.}(2011){Banerjee}, {Jog}, {Brinks}, \&
  {Bagetakos}}]{Banerjee2011}
{Banerjee}, A., {Jog}, C.~J., {Brinks}, E., \& {Bagetakos}, I. 2011, \mnras,
  415, 687

\bibitem[{{Barbieri} {et~al.}(2005){Barbieri}, {Fraternali}, {Oosterloo},
  {Bertin}, {Boomsma}, \& {Sancisi}}]{Barbieri2005}
{Barbieri}, C.~V., {Fraternali}, F., {Oosterloo}, T., {Bertin}, G., {Boomsma},
  R., \& {Sancisi}, R. 2005, \aap, 439, 947

\bibitem[{{Bekki} \& {Chiba}(2006)}]{Bekki2006}
{Bekki}, K., \& {Chiba}, M. 2006, \apjl, 637, L97

\bibitem[{{Bolatto} {et~al.}(2011){Bolatto}, {Leroy}, {Jameson}, {Ostriker},
  {Gordon}, {Lawton}, {Stanimirovi{\'c}}, {Israel}, {Madden}, {Hony},
  {Sandstrom}, {Bot}, {Rubio}, {Winkler}, {Roman-Duval}, {van Loon},
  {Oliveira}, \& {Indebetouw}}]{Bolatto2011}
{Bolatto}, A.~D. {et~al.} 2011, \apj, 741, 12

\bibitem[{{Boomsma} {et~al.}(2008){Boomsma}, {Oosterloo}, {Fraternali}, {van
  der Hulst}, \& {Sancisi}}]{Boomsma2008}
{Boomsma}, R., {Oosterloo}, T.~A., {Fraternali}, F., {van der Hulst}, J.~M., \&
  {Sancisi}, R. 2008, \aap, 490, 555

\bibitem[{{Boulanger} \& {Viallefond}(1992)}]{Boulanger1992}
{Boulanger}, F., \& {Viallefond}, F. 1992, \aap, 266, 37

\bibitem[{{Dalcanton} {et~al.}(2009){Dalcanton}, {Williams}, {Seth}, {Dolphin},
  {Holtzman}, {Rosema}, {Skillman}, {Cole}, {Girardi}, {Gogarten},
  {Karachentsev}, {Olsen}, {Weisz}, {Christensen}, {Freeman}, {Gilbert},
  {Gallart}, {Harris}, {Hodge}, {de Jong}, {Karachentseva}, {Mateo}, {Stetson},
  {Tavarez}, {Zaritsky}, {Governato}, \& {Quinn}}]{Dalcanton2009}
{Dalcanton}, J.~J. {et~al.} 2009, \apjs, 183, 67

\bibitem[{{Dale} {et~al.}(2009){Dale}, {Cohen}, {Johnson}, {Schuster},
  {Calzetti}, {Engelbracht}, {Gil de Paz}, {Kennicutt}, {Lee}, {Begum},
  {Block}, {Dalcanton}, {Funes}, {Gordon}, {Johnson}, {Marble}, {Sakai},
  {Skillman}, {van Zee}, {Walter}, {Weisz}, {Williams}, {Wu}, \&
  {Wu}}]{Dale2009}
{Dale}, D.~A. {et~al.} 2009, \apj, 703, 517

\bibitem[{{Dekel} \& {Silk}(1986)}]{Dekel1986}
{Dekel}, A., \& {Silk}, J. 1986, \apj, 303, 39

\bibitem[{{Dickey} {et~al.}(1990){Dickey}, {Hanson}, \& {Helou}}]{Dickey1990}
{Dickey}, J.~M., {Hanson}, M.~M., \& {Helou}, G. 1990, \apj, 352, 522

\bibitem[{{Dickey} {et~al.}(2000){Dickey}, {Mebold}, {Stanimirovic}, \&
  {Staveley-Smith}}]{Dickey2000}
{Dickey}, J.~M., {Mebold}, U., {Stanimirovic}, S., \& {Staveley-Smith}, L.
  2000, \apj, 536, 756

\bibitem[{{Foreman-Mackey} {et~al.}(2012){Foreman-Mackey}, {Hogg}, {Lang}, \&
  {Goodman}}]{ForemanMackey2012}
{Foreman-Mackey}, D., {Hogg}, D.~W., {Lang}, D., \& {Goodman}, J. 2012, ArXiv
  e-prints

\bibitem[{{Fraternali} {et~al.}(2001){Fraternali}, {Oosterloo}, {Sancisi}, \&
  {van Moorsel}}]{Fraternali2001}
{Fraternali}, F., {Oosterloo}, T., {Sancisi}, R., \& {van Moorsel}, G. 2001,
  \apjl, 562, L47

\bibitem[{{Fraternali} {et~al.}(2002){Fraternali}, {van Moorsel}, {Sancisi}, \&
  {Oosterloo}}]{Fraternali2002}
{Fraternali}, F., {van Moorsel}, G., {Sancisi}, R., \& {Oosterloo}, T. 2002,
  \aj, 123, 3124

\bibitem[{{Ghigna} {et~al.}(2000){Ghigna}, {Moore}, {Governato}, {Lake},
  {Quinn}, \& {Stadel}}]{Ghigna2000}
{Ghigna}, S., {Moore}, B., {Governato}, F., {Lake}, G., {Quinn}, T., \&
  {Stadel}, J. 2000, \apj, 544, 616

\bibitem[{{Heiles}(2001)}]{Heiles2001}
{Heiles}, C. 2001, \apjl, 551, L105

\bibitem[{{Heiles} \& {Troland}(2003)}]{Heiles2003}
{Heiles}, C., \& {Troland}, T.~H. 2003, \apj, 586, 1067

\bibitem[{{Hunter} {et~al.}(2001){Hunter}, {Elmegreen}, \& {van
  Woerden}}]{Hunter2001}
{Hunter}, D.~A., {Elmegreen}, B.~G., \& {van Woerden}, H. 2001, \apj, 556, 773

\bibitem[{{Hunter} {et~al.}(1993){Hunter}, {Hawley}, \&
  {Gallagher}}]{Hunter1993}
{Hunter}, D.~A., {Hawley}, W.~N., \& {Gallagher}, III, J.~S. 1993, \aj, 106,
  1797

\bibitem[{{Hunter} {et~al.}(1999){Hunter}, {van Woerden}, \&
  {Gallagher}}]{Hunter1999}
{Hunter}, D.~A., {van Woerden}, H., \& {Gallagher}, J.~S. 1999, \aj, 118, 2184

\bibitem[{{Ianjamasimanana} {et~al.}(2012){Ianjamasimanana}, {de Blok},
  {Walter}, \& {Heald}}]{Ianj2012}
{Ianjamasimanana}, R., {de Blok}, W.~J.~G., {Walter}, F., \& {Heald}, G.~H.
  2012, \aj, 144, 96

\bibitem[{{Joung} {et~al.}(2009){Joung}, {Mac Low}, \& {Bryan}}]{Joung2009}
{Joung}, M.~R., {Mac Low}, M.-M., \& {Bryan}, G.~L. 2009, \apj, 704, 137

\bibitem[{{Kannan} {et~al.}(2012){Kannan}, {Macci{\`o}}, {Pasquali}, {Moster},
  \& {Walter}}]{Kannan2012}
{Kannan}, R., {Macci{\`o}}, A.~V., {Pasquali}, A., {Moster}, B.~P., \&
  {Walter}, F. 2012, \apj, 746, 10

\bibitem[{{Kazantzidis} {et~al.}(2008){Kazantzidis}, {Bullock}, {Zentner},
  {Kravtsov}, \& {Moustakas}}]{Kazantzidis2008}
{Kazantzidis}, S., {Bullock}, J.~S., {Zentner}, A.~R., {Kravtsov}, A.~V., \&
  {Moustakas}, L.~A. 2008, \apj, 688, 254

\bibitem[{{Kennicutt} \& {Evans}(2012)}]{Kennicutt2012}
{Kennicutt}, R.~C., \& {Evans}, N.~J. 2012, \araa, 50, 531

\bibitem[{{Kim} \& {Ostriker}(2007)}]{Kim2007}
{Kim}, W.-T., \& {Ostriker}, E.~C. 2007, \apj, 660, 1232

\bibitem[{{Klessen} \& {Hennebelle}(2010)}]{Klessen2010}
{Klessen}, R.~S., \& {Hennebelle}, P. 2010, \aap, 520, A17

\bibitem[{{Klypin} {et~al.}(1999){Klypin}, {Kravtsov}, {Valenzuela}, \&
  {Prada}}]{Klypin1999}
{Klypin}, A., {Kravtsov}, A.~V., {Valenzuela}, O., \& {Prada}, F. 1999, \apj,
  522, 82

\bibitem[{{Komatsu} {et~al.}(2011){Komatsu}, {Smith}, {Dunkley}, {Bennett},
  {Gold}, {Hinshaw}, {Jarosik}, {Larson}, {Nolta}, {Page}, {Spergel},
  {Halpern}, {Hill}, {Kogut}, {Limon}, {Meyer}, {Odegard}, {Tucker}, {Weiland},
  {Wollack}, \& {Wright}}]{Komatsu2011}
{Komatsu}, E. {et~al.} 2011, \apjs, 192, 18

\bibitem[{{Kroupa}(2001)}]{Kroupa2001}
{Kroupa}, P. 2001, \mnras, 322, 231

\bibitem[{{Lee} {et~al.}(2011){Lee}, {Gil de Paz}, {Kennicutt}, {Bothwell},
  {Dalcanton}, {Jos{\'e} G.~Funes S.}, {Johnson}, {Sakai}, {Skillman},
  {Tremonti}, \& {van Zee}}]{Lee2011}
{Lee}, J.~C. {et~al.} 2011, \apjs, 192, 6

\bibitem[{{Lelli} {et~al.}(2012){Lelli}, {Verheijen}, {Fraternali}, \&
  {Sancisi}}]{Lelli2012}
{Lelli}, F., {Verheijen}, M., {Fraternali}, F., \& {Sancisi}, R. 2012, \aap,
  544, A145

\bibitem[{{Leroy} {et~al.}(2008){Leroy}, {Walter}, {Brinks}, {Bigiel}, {de
  Blok}, {Madore}, \& {Thornley}}]{Leroy2008}
{Leroy}, A.~K., {Walter}, F., {Brinks}, E., {Bigiel}, F., {de Blok}, W.~J.~G.,
  {Madore}, B., \& {Thornley}, M.~D. 2008, \aj, 136, 2782

\bibitem[{{Lin} \& {Shu}(1964)}]{Lin1964}
{Lin}, C.~C., \& {Shu}, F.~H. 1964, \apj, 140, 646

\bibitem[{{Lopez} {et~al.}(2011){Lopez}, {Krumholz}, {Bolatto}, {Prochaska}, \&
  {Ramirez-Ruiz}}]{Lopez2011}
{Lopez}, L.~A., {Krumholz}, M.~R., {Bolatto}, A.~D., {Prochaska}, J.~X., \&
  {Ramirez-Ruiz}, E. 2011, \apj, 731, 91

\bibitem[{{Mac Low}(1999)}]{MacLow1999}
{Mac Low}, M.-M. 1999, \apj, 524, 169

\bibitem[{{Mac Low} \& {Klessen}(2004)}]{MacLow2004}
{Mac Low}, M.-M., \& {Klessen}, R.~S. 2004, Reviews of Modern Physics, 76, 125

\bibitem[{{Martin}(1998)}]{Martin1998}
{Martin}, C.~L. 1998, \apj, 506, 222

\bibitem[{{McQuinn} {et~al.}(2010){McQuinn}, {Skillman}, {Cannon}, {Dalcanton},
  {Dolphin}, {Hidalgo-Rodr{\'{\i}}guez}, {Holtzman}, {Stark}, {Weisz}, \&
  {Williams}}]{McQuinn2010}
{McQuinn}, K.~B.~W. {et~al.} 2010, \apj, 721, 297

\bibitem[{{Moiseev} \& {Lozinskaya}(2012)}]{Moiseev2012}
{Moiseev}, A.~V., \& {Lozinskaya}, T.~A. 2012, \mnras, 423, 1831

\bibitem[{{Monaco}(2004)}]{Monaco2004}
{Monaco}, P. 2004, \mnras, 352, 181

\bibitem[{{Oh} {et~al.}(2011){Oh}, {de Blok}, {Brinks}, {Walter}, \&
  {Kennicutt}}]{Oh2011}
{Oh}, S.-H., {de Blok}, W.~J.~G., {Brinks}, E., {Walter}, F., \& {Kennicutt},
  Jr., R.~C. 2011, \aj, 141, 193

\bibitem[{{Oh} {et~al.}(2008){Oh}, {de Blok}, {Walter}, {Brinks}, \&
  {Kennicutt}}]{Oh2008}
{Oh}, S.-H., {de Blok}, W.~J.~G., {Walter}, F., {Brinks}, E., \& {Kennicutt},
  Jr., R.~C. 2008, \aj, 136, 2761

\bibitem[{{Ott} {et~al.}(2012){Ott}, {Stilp}, {Warren}, {Skillman},
  {Dalcanton}, {Walter}, {de Blok}, {Koribalski}, \& {West}}]{Ott2012}
{Ott}, J. {et~al.} 2012, \aj, 144, 123

\bibitem[{{Ott} {et~al.}(2001){Ott}, {Walter}, {Brinks}, {Van Dyk}, {Dirsch},
  \& {Klein}}]{Ott2001}
{Ott}, J., {Walter}, F., {Brinks}, E., {Van Dyk}, S.~D., {Dirsch}, B., \&
  {Klein}, U. 2001, \aj, 122, 3070

\bibitem[{{Petric} \& {Rupen}(2007)}]{Petric2007}
{Petric}, A.~O., \& {Rupen}, M.~P. 2007, \aj, 134, 1952

\bibitem[{{Roberts}(1969)}]{Roberts1969}
{Roberts}, W.~W. 1969, \apj, 158, 123

\bibitem[{{Rossa} \& {Dettmar}(2003)}]{Rossa2003}
{Rossa}, J., \& {Dettmar}, R.-J. 2003, \aap, 406, 505

\bibitem[{{Schaye}(2004)}]{Schaye2004}
{Schaye}, J. 2004, \apj, 609, 667

\bibitem[{{Schwartz} \& {Martin}(2004)}]{Schwartz2004}
{Schwartz}, C.~M., \& {Martin}, C.~L. 2004, \apj, 610, 201

\bibitem[{{Sellwood} \& {Balbus}(1999)}]{Sellwood1999}
{Sellwood}, J.~A., \& {Balbus}, S.~A. 1999, \apj, 511, 660

\bibitem[{{Shapiro} \& {Field}(1976)}]{Shapiro1976}
{Shapiro}, P.~R., \& {Field}, G.~B. 1976, \apj, 205, 762

\bibitem[{{Stilp} {et~al.}(2013a){Stilp}, {Dalcanton}, {Warren}, {Skillman},
  {Ott}, \& {Koribalski}}]{StilpGlobal}
{Stilp}, A.~M., {Dalcanton}, J.~J., {Warren}, S.~R., {Skillman}, E., {Ott}, J.,
  \& {Koribalski}, B. 2013a, \apj, 765, 136

\bibitem[{{Stilp} {et~al.}(2013c){Stilp}, {Dalcanton}, {Weisz}, {Warren},
  {Skillman}, {Ott}, {Williams}, \& {Dolphin}}]{StilpTimescales}
{Stilp}, A.~M., {Dalcanton}, J.~J., {Weisz}, D., {Warren}, S.~R., {Skillman},
  E., {Ott}, J., {Williams}, B.~F., \& {Dolphin}, A.~E. 2013c, submitted to
  \apj

\bibitem[{{Tamburro} {et~al.}(2009){Tamburro}, {Rix}, {Leroy}, {Mac Low},
  {Walter}, {Kennicutt}, {Brinks}, \& {de Blok}}]{Tamburro2009}
{Tamburro}, D., {Rix}, H.-W., {Leroy}, A.~K., {Mac Low}, M.-M., {Walter}, F.,
  {Kennicutt}, R.~C., {Brinks}, E., \& {de Blok}, W.~J.~G. 2009, \aj, 137, 4424

\bibitem[{{Tenorio-Tagle} {et~al.}(1991){Tenorio-Tagle}, {Rozyczka}, {Franco},
  \& {Bodenheimer}}]{TenorioTagle1991}
{Tenorio-Tagle}, G., {Rozyczka}, M., {Franco}, J., \& {Bodenheimer}, P. 1991,
  \mnras, 251, 318

\bibitem[{{Thornton} {et~al.}(1998){Thornton}, {Gaudlitz}, {Janka}, \&
  {Steinmetz}}]{Thornton1998}
{Thornton}, K., {Gaudlitz}, M., {Janka}, H.-T., \& {Steinmetz}, M. 1998, \apj,
  500, 95

\bibitem[{{Trachternach} {et~al.}(2008){Trachternach}, {de Blok}, {Walter},
  {Brinks}, \& {Kennicutt}}]{Trachternach2008}
{Trachternach}, C., {de Blok}, W.~J.~G., {Walter}, F., {Brinks}, E., \&
  {Kennicutt}, Jr., R.~C. 2008, \aj, 136, 2720

\bibitem[{{Trujillo-Gomez} {et~al.}(2011){Trujillo-Gomez}, {Klypin}, {Primack},
  \& {Romanowsky}}]{Trujillo2011}
{Trujillo-Gomez}, S., {Klypin}, A., {Primack}, J., \& {Romanowsky}, A.~J. 2011,
  \apj, 742, 16

\bibitem[{{van der Kruit}(1981)}]{vanDerKruit1981}
{van der Kruit}, P.~C. 1981, \aap, 99, 298

\bibitem[{{van Eymeren} {et~al.}(2010){van Eymeren}, {Koribalski},
  {L{\'o}pez-S{\'a}nchez}, {Dettmar}, \& {Bomans}}]{vanEymeren2010}
{van Eymeren}, J., {Koribalski}, B.~S., {L{\'o}pez-S{\'a}nchez}, {\'A}.~R.,
  {Dettmar}, R.-J., \& {Bomans}, D.~J. 2010, \mnras, 407, 113

\bibitem[{{van Eymeren} {et~al.}(2009{\natexlab{a}}){van Eymeren}, {Marcelin},
  {Koribalski}, {Dettmar}, {Bomans}, {Gach}, \& {Balard}}]{vanEymeren2009}
{van Eymeren}, J., {Marcelin}, M., {Koribalski}, B., {Dettmar}, R.-J.,
  {Bomans}, D.~J., {Gach}, J.-L., \& {Balard}, P. 2009{\natexlab{a}}, \aap,
  493, 511

\bibitem[{{van Eymeren} {et~al.}(2009{\natexlab{b}}){van Eymeren}, {Marcelin},
  {Koribalski}, {Dettmar}, {Bomans}, {Gach}, \& {Balard}}]{vanEymeren2009b}
{van Eymeren}, J., {Marcelin}, M., {Koribalski}, B.~S., {Dettmar}, R.-J.,
  {Bomans}, D.~J., {Gach}, J.-L., \& {Balard}, P. 2009{\natexlab{b}}, \aap,
  505, 105

\bibitem[{{van Zee} \& {Bryant}(1999)}]{vanZee1999}
{van Zee}, L., \& {Bryant}, J. 1999, \aj, 118, 2172

\bibitem[{{Wada} {et~al.}(2002){Wada}, {Meurer}, \& {Norman}}]{Wada2002}
{Wada}, K., {Meurer}, G., \& {Norman}, C.~A. 2002, \apj, 577, 197

\bibitem[{{Walter} {et~al.}(2008){Walter}, {Brinks}, {de Blok}, {Bigiel},
  {Kennicutt}, {Thornley}, \& {Leroy}}]{Walter2008}
{Walter}, F., {Brinks}, E., {de Blok}, W.~J.~G., {Bigiel}, F., {Kennicutt},
  Jr., R.~C., {Thornley}, M.~D., \& {Leroy}, A. 2008, \aj, 136, 2563

\bibitem[{{Warren} {et~al.}(2012){Warren}, {Skillman}, {Stilp}, {Dalcanton},
  {Ott}, {Walter}, {Petersen}, {Koribalski}, \& {West}}]{Warren2012}
{Warren}, S.~R. {et~al.} 2012, \apj, 757, 84

\bibitem[{{Weisz} {et~al.}(2011){Weisz}, {Dalcanton}, {Williams}, {Gilbert},
  {Skillman}, {Seth}, {Dolphin}, {McQuinn}, {Gogarten}, {Holtzman}, {Rosema},
  {Cole}, {Karachentsev}, \& {Zaritsky}}]{Weisz2011}
{Weisz}, D.~R. {et~al.} 2011, \apj, 739, 5

\bibitem[{{Weisz} {et~al.}(2012){Weisz}, {Johnson}, {Johnson}, {Skillman},
  {Lee}, {Kennicutt}, {Calzetti}, {van Zee}, {Bothwell}, {Dalcanton}, {Dale},
  \& {Williams}}]{Weisz2012}
---. 2012, \apj, 744, 44

\bibitem[{{Weisz} {et~al.}(2009){Weisz}, {Skillman}, {Cannon}, {Dolphin},
  {Kennicutt}, {Lee}, \& {Walter}}]{Weisz2009}
{Weisz}, D.~R., {Skillman}, E.~D., {Cannon}, J.~M., {Dolphin}, A.~E.,
  {Kennicutt}, Jr., R.~C., {Lee}, J., \& {Walter}, F. 2009, \apj, 704, 1538

\bibitem[{{Wolfire} {et~al.}(1995){Wolfire}, {Hollenbach}, {McKee}, {Tielens},
  \& {Bakes}}]{Wolfire1995}
{Wolfire}, M.~G., {Hollenbach}, D., {McKee}, C.~F., {Tielens}, A.~G.~G.~M., \&
  {Bakes}, E.~L.~O. 1995, \apj, 443, 152

\bibitem[{{Young} {et~al.}(2003){Young}, {van Zee}, {Lo}, {Dohm-Palmer}, \&
  {Beierle}}]{Young2003}
{Young}, L.~M., {van Zee}, L., {Lo}, K.~Y., {Dohm-Palmer}, R.~C., \& {Beierle},
  M.~E. 2003, \apj, 592, 111

\bibitem[{{Zhang} {et~al.}(2012){Zhang}, {Hunter}, \& {Elmegreen}}]{Zhang2012}
{Zhang}, H.-X., {Hunter}, D.~A., \& {Elmegreen}, B.~G. 2012, \apj, 754, 29

\end{thebibliography}

\end{document}